UNIVERSITÉ DE NICE-SOPHIA ANTIPOLIS - UFR Sciences
SAPIENZA UNIVERSITÀ DI ROMA
UNIVERSITÉ FRANCO-ITALIENNE

Ecole Doctorale de Sciences Fondamentales et Appliquées
Laboratoire Hippolite Fizeau
International Relativistic Astrophysics PhD

# T H E S E

pour obtenir le titre de

## Docteur en Sciences

de l'UNIVERSITÉ de Nice-Sophia Antipolis

Spécialité: Astrophysique Rélativiste

Présentée et soutenue par

***Costantino SIGISMONDI***

# High precision ground-based measurements of solar diameter in support of PICARD mission

Thèse dirigée par **Marianne** ***FAUROBERT***
et par **Alberto** ***EGIDI***

soutenue le *12 décembre 2011*

Devant le jury composé de:

M. Paolo DE BERNARDIS, Président du jury, PROFESSEUR

M. Michele BIANDA, Rapporteur du jury, DOCTEUR

M. Serge KOUTCHMY, Rapporteur du jury, DIRECTEUR DE RECHERCHE CNRS

M. Roberto CAPUZZO DOLCETTA, Membre du jury, PROFESSEUR

M. Alessandro MELCHIORRI, Membre du jury, PROFESSEUR

M. Alberto EGIDI, Co-Directeur, PROFESSEUR

Mme Marianne FAUROBERT, Directeur de thèse, PROFESSEUR

A 14h 00 au Département de Physique de l'Université de Rome «La Sapienza»

**Abstract:** The theme of the measurement of the solar diameter is introduced in the wider framework of solar variability, and, consequently of the influences of the Sun upon the Earth's climate.
It is possible to measure the solar diameter with an accuracy enough good for studies on climate changes and irradiation variations using ancient data on total eclipses. This would permit to extend the knowledge of the solar luminosity back to three centuries, through the knowledge of the parameter W=dLogR/dLog L.
The method of eclipses and Baily beads is widely discussed, and a significant improvement with respect to the last 40 years, has been obtained by reconstructing the Limb Darkening Function from the Baily's bead light curve, and the search of its inflexion point.
The case of the last annular eclipse (Jan 15, 2010) has been studied in more detail, while the atlas of Baily's beads has been published with all the observations made worldwide by IOTA members, along with an analysis of the solar diameter during the eclipse of 2006.
The work of this thesis has been developped during the transition between the photographic atlas of the lunar limb published by Watts in 1963 and corrected in the following decades by thousands observations, and the laser-altimeter map made by the Kaguya lunar probe and published in November 2009.
The other method for the accurate measurement of the solar diameter alternative to the Picard / Picard-sol mission is the drift-scan method used either by the solar astrolabes either by larger telescopes. The observatories of Locarno and Paris have started an observational program of the Sun with this method with encouraging results. For the first time an image motion of the whole Sun has been detected over frequencies of 1/100 Hz. This may start explain the puzzling results of the observational campaigns made in Greenwich and Rome from 1850 to 1955. This sub-Hertz activity of the atmospheric seeing is another fundamental achievement in this research field.
A giant pinhole telescope as the meridian line of Santa Maria degli Angeli in Rome, permits to introduce almost all the arguments of classical astrometry presented in this thesis. In this consists the didactic outreach as a complement of this thesis.

**Résumé**: l'argument de la mesure du diamètre solaire est introduit dans le cadre plus ample de la variabilité solaire, et, en conséquence, de l'influence du Soleil sur le climat de la Terre.
Il est possible de mesurer le diamètre solaire avec une précision suffisamment bonne pour des études de changements climatiques et de variations de irradiance en utilisant des données anciennes d'éclipses totales. Cela permettrait l'extension de la connaissance de la luminosité solaire en arrière de trois siècle, grâce à la connaissance du paramètre W=dLogR/dLog L.
La méthode des éclipses et des grains de Baily est largement traité, et un avancement significatif par rapport aux derniers 40 ans, a été obtenu avec la reconstruction de la fonction d'assombrissement du limb solaire à partir de la lumière des grain de Baily, et la recherche du point d'inflexion.
Le cas de la dernière éclipse annulaire (15 de Janvier 2010) a été étudié en détail, tandis que l'atlas des grains de Baily a été publié à partir de toutes les observations faites dans le Monde par les membres du IOTA, aussi avec l'analyse du diamètre solaire pendant l'éclipse du 2006.
Le travail de thèse a été développé pendant la transition entre l'atlas photographique du limb lunaire publié par Watts en 1963, et corrigé dans les années suivantes par milliers d'observations, et la carte realisée avec le laser-altimètre de la mission lunaire japonaise Kaguya, publiée en Novembre 2009.
L'autre méthode pour une mesure précise du diamètre solaire en alternative de la mission Picard / Picard-sol c'est le «drift-scan» utilisé soit par les astrolabes solaires, soit par de télescopes plus grands. Les observatoires de Locarno et Paris ont commencé des observations du Soleil avec cette méthode avec des résultés encourageants. Pour la première fois le mouvement de l'image du Soleil entier a été détectée sur des fréquences de 1/100 Hz. Cela peut commencer à expliquer les résultats des campagnes observatives de Greenwich et Rome du 1850 au 1955.
Cette activité «sous-Hertz» de la turbulence atmosphérique est un autre avancement significatif dans ce domaine de recherche. Le «télescope géant à trou sténopeïque» de la ligne méridienne de Santa Maria degli Angeli en Rome, permet l'introduction didactique de presque tous les arguments d'astrométrie classique traités dans cette thèse. Cet essay didactique complète la thèse.

# HIGH PRECISION GROUND-BASED MEASUREMENTS OF SOLAR DIAMETER

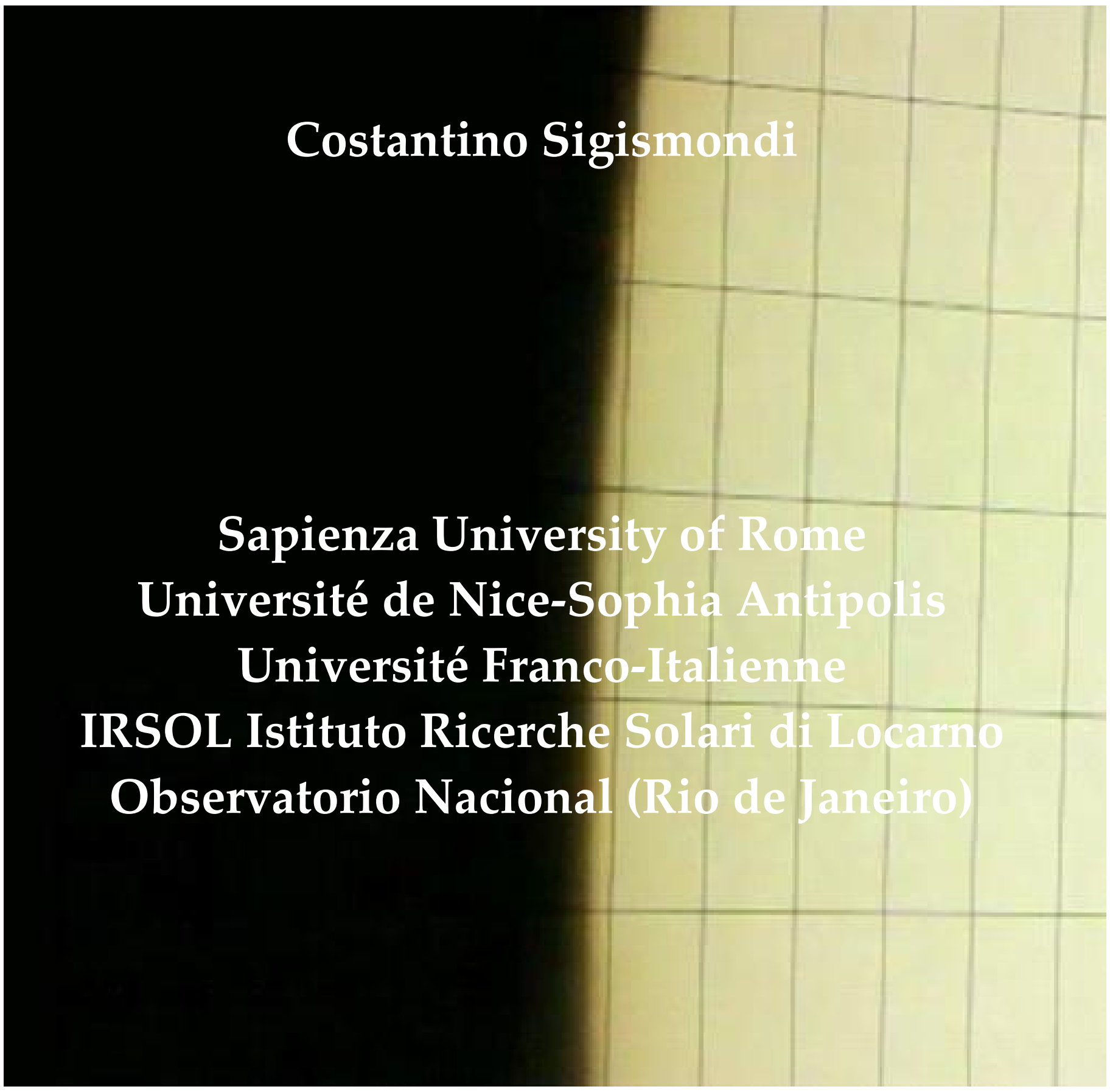

Image: Solar limb observed at the Gregory-Coudé telescope of Istituto Ricerche Solari di Locarno (Switzerland) on 9 August 2008.

*First revision: 23 November 2012*

in memory of Jean Arnaud

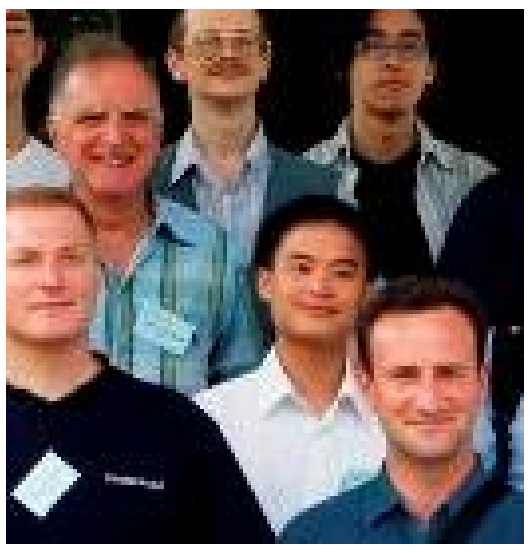

Jean Arnaud (top left) and myself (bottom right)
during the meeting "Light of the Dark Universe"
held in Taipei in end May 2008.

and Alessandro Cacciani

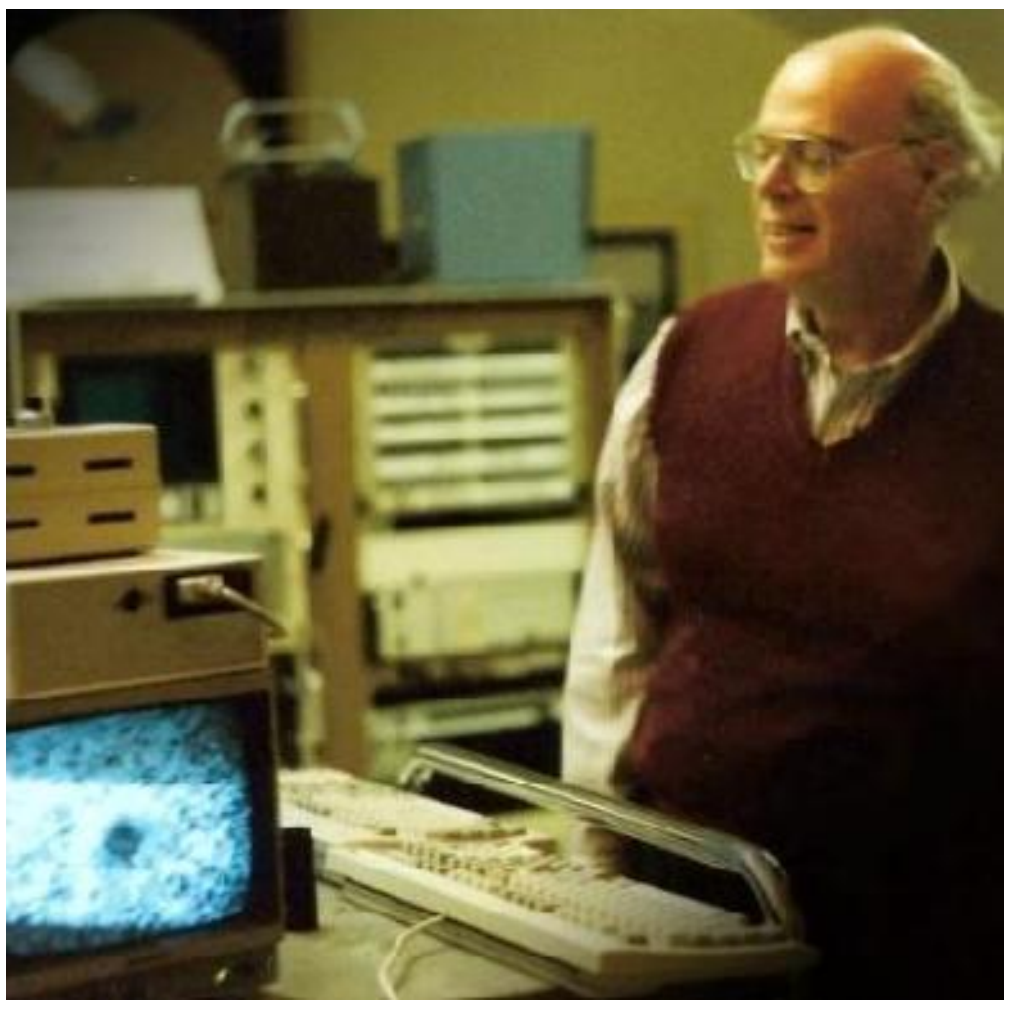

and with the deepest gratitude to my father Camillo

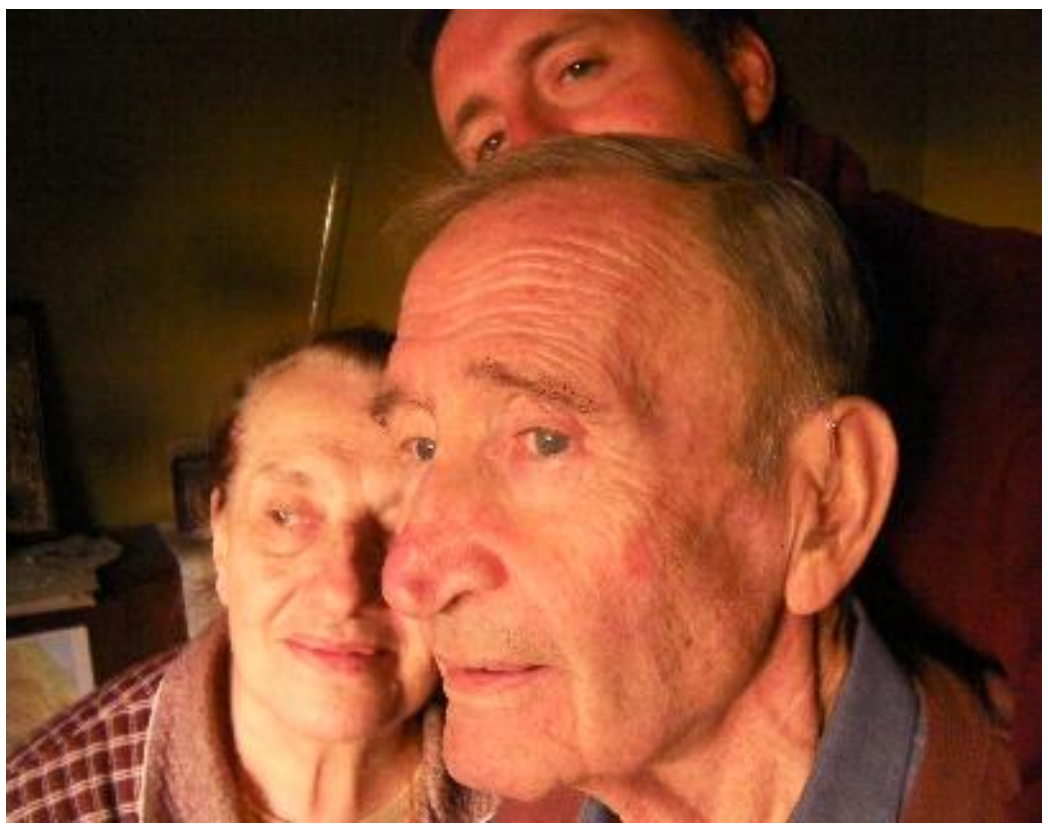

with whom I have seen the better sunsets from the Eternal City
May the Lord welcome them in the eternal life

# Chapter 0. Introduction

## 0.1 Structure of the book

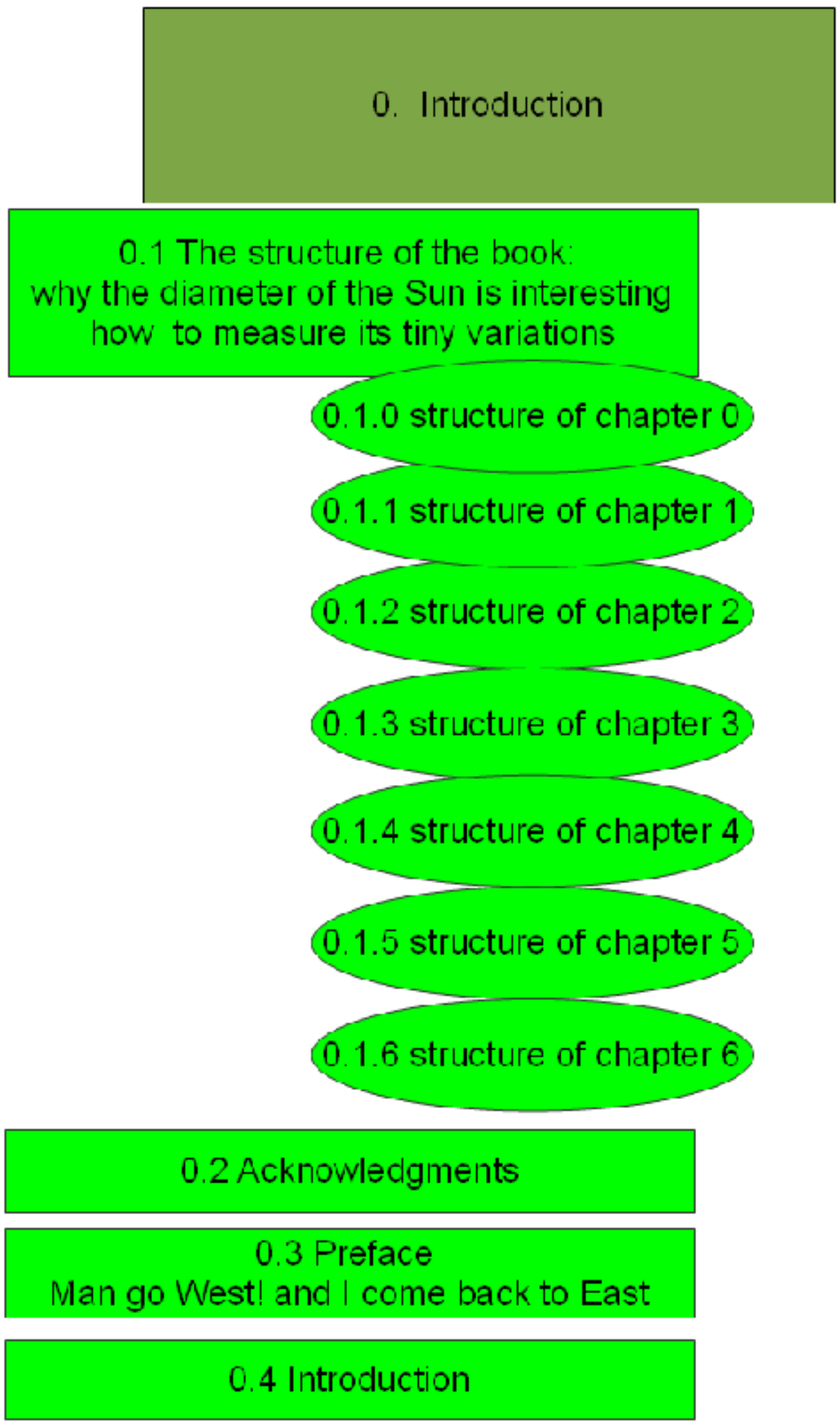

All chapters are presented with their topics, in order to have a conceptual map of the argumentations.

### 0.1.0 Chapter 0: Introduction

#### 0.1 The structure of the book

The first chapter is devoted to present the conceptual map of the book, divided into seven chapters numbered from 0 to 6, in the paragraphs 0.1.0 to 0.1.6.

#### 0.2 Acknowledgments

The success of a work is due to a network of positive human relationships.
I don't like autistic personalities even if capable of great achievements in science.

#### 0.3 Preface

My roadmap towards solar astrometry lasted more than a decade, and it is drafted in this preface.

#### 0.4 Introduction "from Picard the abbot, to the satellite"

The theme of the measurement of solar diameter is presented form an historical point of view, driving the focus from the early measurements made by the abbot Jean Picard (1620-1682) to the satellite bearing his name launched on 15 July 2010.

During the XIX century (see the Nautical Almanac, quoted by Angelo Secchi and Simon Newcomb) the standard value for the angular solar diameter at one astronomical unit has been changed a few times by about 1". These changes may reflect the use of different methods, instruments as well as a change in the diameter itself. The adopted standard value of solar diameter is 959.63" (Auwers, 1891) with an irradiation correction of -1.55".

### 0.1.1 Chapter 1: Solar variability

The title is inspired by "The Sun as a Variable Star" the proceedings of the IAU colloquium n. 143.[1]

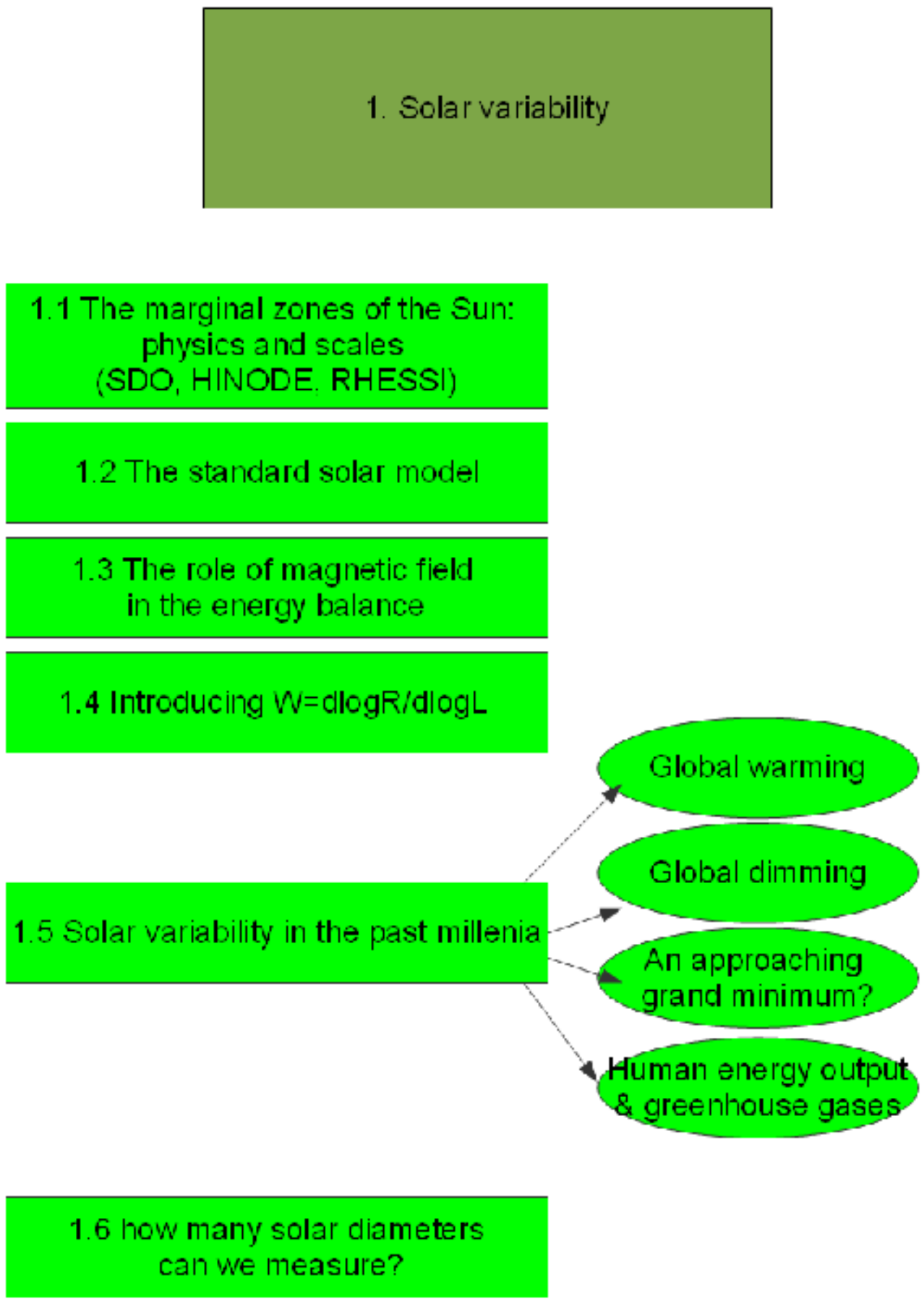

The variation of the luminosity L with radius R is described by a parameter W=dlogR/dlogL. Picard mission will measure W accurately, and its value will be used to recover the past values of L_sun from historical measurements of R_sun in order to feed Earth's climate models.

#### 1.1 The marginal zones of the Sun: physics and scale

This paragraph is dedicated to the marginal zones of the Sun, i.e. the solar surface and the atmosphere nearby when it is seen from a grazing view. Images, scales and physics are here summarized. The solar mesosphere is here defined.

#### 1.2 The standard solar model

The standard solar model has two free parameters: the mixing length scale and the helium abundance, and after 4.52 billion year it should return the present radius of the Sun, its luminosity and the observed metal abundance. During the main sequence phase the solar diameter shrunk of 30%.

[1] The Sun as a Variable Star, Solar and Stellar Irradiance Variations, J. M. Pap, C. Fröhlich, H. S. Hudson & S. K. Solanki, Cambridge University Press 1994.

### 1.3 The role of the magnetic field in the energy balance

The energy stored in the magnetic field plays a fundamental role in the energy balance of the Sun, as well as the temperature of the photosphere and the diameter of the Sun.

### 1.4 Introducing W=dlogR/dlogL

The luminosity and the radius of the Sun are involved in the Stefan-Boltzmann equation, but their logarithmic derivative W is not ½ as for uncorrelated quantities, because of the magnetic field.

### 1.5 Solar variability in the past millennia

The existence of ice ages either in the last million year and in the pre-cambrian age has been explained also with astronomical causes (e.g. Milankovitch cycles), but the periods of global warming and cooling in the past millennia have indeed a solar origin. Nowadays the global warming seems to be anthropogenic through the greenhouse effect. But this is still an open question with political, economical and social implications. Conversely there are emerging topics as the global dimming and the debate about the approaching new solar grand minimum.

### 1.6 Several definition of solar radius

There is the seismic radius, there is a different radius for each wavelength, also for radio. How accurate are these values, and how big are their variations?

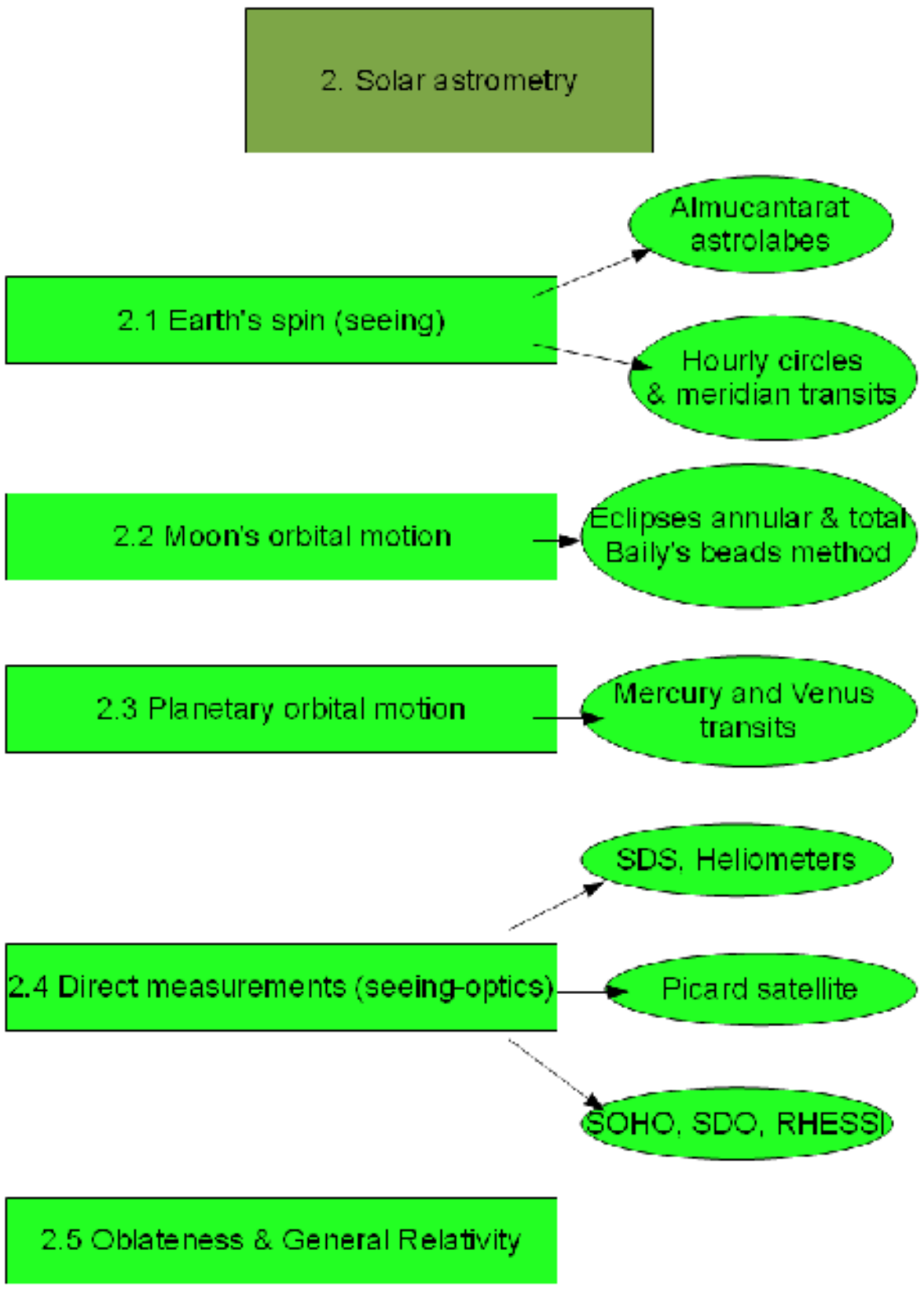

## 0.1.2 Chapter 2: Solar astrometry

The measurement of the diameter of the Sun, even if assumed as a perfect sphere, consists into a definite field of the classical spherical astrometry.

The methods for solar diameter measurements are those exploiting Earth rotation (drift-scan; Danjon solar-type astrolabes); those using the Moon, Earth and planetary orbital motion (eclipses and planetary transits) and finally those making direct angular measurements (heliometers of Fraunhofer, heliometers with objective prisms, SDS, Picard SODISM and MISOLFA, the ground-based telescope dedicated to the direct measurement of the solar diameter.) The principal error sources are discussed in this chapter.

### 2.1 Earth's spin

The rotation of the Earth has an extremely constant rate. In the drift-scan mode the telescope is pointed toward the Sun; the Sun sweeps along the fixed field of view. The time of the drift is proportional to the solar diameter. There are two variants of this method.

The solar astrolabes are used for transits across a circle of constant altitude above the horizon. The last generation of Danjon astrolabes are 10 cm aperture instruments and its data analysis is still controversial. Such instruments have been considered too much influenced by atmospheric turbulences so that measurements made at Calern (OCA) or at Rio de Janeiro were significantly in disagreement.

Other telescopes (not necessarily meridian) are used to observe the transits across the meridian line, or across whatever hourly circle. Because the two limbs and the center of Sun sweep on a fixed field of view, its optical defects act as systematic errors.

### 2.2 Lunar orbital motion: eclipses and Baily's beads

The Moon along its orbit covers approximately its diameter in one hour. The duration of a total eclipse is depending on the angular diameter of the Moon and of the Sun. The circumstances of an eclipse are depending on the shapes and orbits of celestial bodies involved, and no influence arises from atmospheric seeing. During a solar eclipse the onset of totality was considered as a sudden event, because while the Moon progresses above the solar surface, the light coming from the photosphere drops rapidly to zero. The method of Baily's beads was introduced by D. Dunham in 1973. The strategy consists to observe the maximum number of Baily's beads, visible at the shadow's edge of the eclipse. The beads are videorecorded, timed and associated to the corresponding valleys of the lunar limb which produce the phenomenon. Instead of just using the second and third contacts, many Baily's beads improve the statistical error of the measured solar diameter. Baily's beads occurs both during total and annular eclipses and are observable in the whole shadow's path, with their maximum number visible at the edges. With the exploitation of the Baily's beads the timing of disappearance and reappearance of N beads increases of a factor $\sqrt{N}$ the statistical accuracy of the method.

### 2.3 Planetary orbital motion

The transits of Mercury (about 13 per century) and of Venus (two each 120 years) have an angular speed which covers the solar diameter (generally a chord) in about 6 hours. If the time of the contacts are defined with an accuracy of one second an optimal precision on solar diameter is attained. It is explained the example of Venus' transits of 8 June 2004 and 6 June 2012.

### 2.4 Direct angular measurements

The heliometers deal with the whole figure of the Sun, or with the images of the limbs projected through prisms on the focal plane. The optical defects are crucial up to the milliarcsecond level. The Picard astrometric mission is presented here, and its philosophy is compared with SDS balloon-borne telescope and SOHO, SDO and RHESSI.

### 2.5 Oblateness and General Relativity connection

Finally what is the connection between the subject of this thesis and relativistic astrophysics? a slightly oblate Sun was suggested to explain the precession of the perihelion of Mercury. Therefore accurate measurements of solar oblateness were carried out by Dicke (1960-70s), Sofia (SDS 1990s) and now using the RHESSI satellite, to assess classical contributions to this anomalous precession. The required accuracy of these measurements is below one part over 10000, the same order of magnitude of expected solar diameter variability.

## 0.1.3 Chapter 3: Eclipses

Eclipse data are scattered and rare: about one each year is available; in the past 40 years we have some data useful to measure the solar diameter, mainly based on video tape and digital video.

### 3.1 Lunar ephemerides accuracy

The method of Baily's beads originally required data from both North and South limits of the shadow. This was to avoid possible errors on the lunar ephemerides. Here they are described.

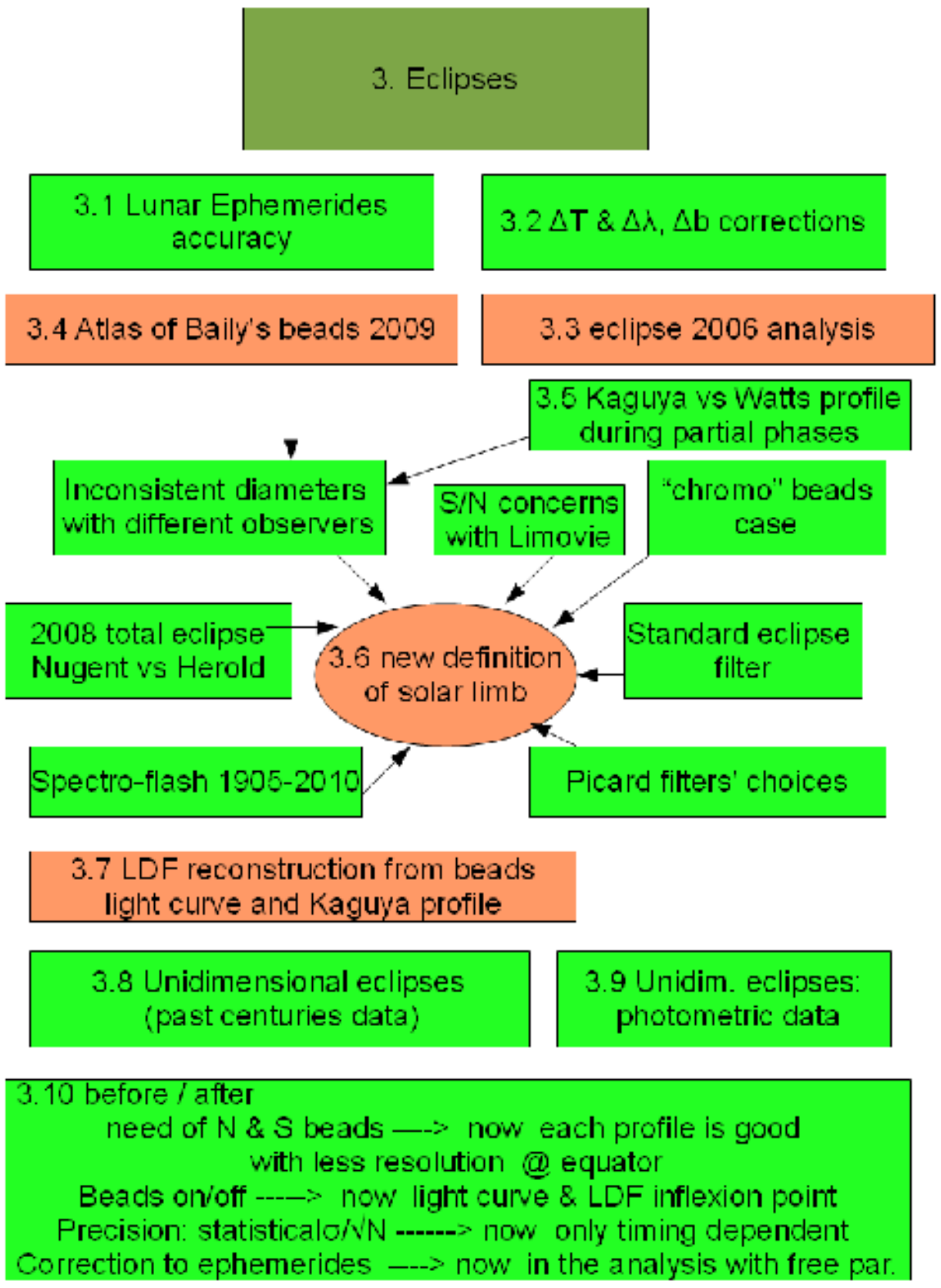

### 3.2 ΔT & Δλ, Δβ corrections

The correction to the lunar ephemerides has been introduced in the data analysis with three free parameters.

### 3.3 2006 eclipse analysis

The total eclipse of 2006 has been analyzed with this method.

### 3.4 Atlas of Baily's beads 2005-2008

The data of the observational campaigns of annular and total eclipses of 2005-2008 made by IOTA members have been published in Solar Physics 2009. Later IOTA members have monitored the 15 January 2010 annular eclipse from some successful unclouded stations.

Few data exists on 2010 total eclipse, mainly from photometers (see unidimensional eclipses).

### 3.5 Kaguya vs Watts profile

To process the data concerning eclipse Baily's beads it is crucial to know the lunar limb profile. Until November 2009 the profile published by C. Watts in 1962, sometimes locally upgraded with stellar occultation data, was the only one available. The precision of Watts atlas, estimated to 0.20", was considered having random uncertainties. With N beads observed the influence of this uncertainty is reduced by the statistical factor √N. Now the profile obtained by the Laser Altimeter LALT of the Japanese lunar probe

KAGUYA is available, with a sampling each 1.5 Km (about 1 arcsecond at the lunar distance) and an height's accuracy of ±1 m. Kaguya data are expected to be error-free.
In the partial eclipse of 4 Jan 2011 we attempted the imaging of the lunar profile (data from Bialkow coronograph) with the purpose to compare it with the profile calculated by Kaguya. Moreover the fit of "Kaguya lunoid" with the observed limb is still to be fully verified (Rocher, 2010).

### 3.6 New definition of solar limb

The observations of Baily's beads with different level of signal to noise show clearly the effect of the emission lines, which extends the measured solar photosphere depending on the instrumental parameters.
Even if the telescopes and the filters have been standardized, the diameters obtained from different observers are inconsistent. The problem of signal to noise ratio is discussed, as well as the (false) case of "chromospheric beads".
The spectro-flash studies of the total eclipses show the evidence of many small emission lines occurring right above the photosphere, and larger telescopes better gather their light, inducing a false perception of a larger photosphere (2008 case). The classic definition of solar limb is the location of the maximum of luminosity profile derivative of the continuum spectrum, already adopted in the oblateness studies. The effect of the blend of tiny emission lines just above the solar limb visible during eclipses is better considered by this definition operative also for eclipses.

### 3.7 LDF reconstruction from Baily's beads light curve and Kaguya profile

Limb Darkening Function studies from single Baily beads (with Andrea Raponi) are presented as a possible new method of diameter's measurement.

### 3.8 Unidimensional eclipses: solar diameter in the past centuries

Reliable past values of the solar radius are believed to be obtained from the durations of ancient total eclipes.
The values of solar diameter calculated from historical (1567 on), and recent (1966 on) eclipses (and planetary transits) using the Watts' lunar profile are discussed and compared with the solar activity.
Before 1966 there are only edge data for eclipses of 1567, 1715, 1869 and 1925. The precision on the solar diameter with these naked eye data is discussed.
The history of past diameter is obtained mainly from Mercury's transits. How reliable are these data? Once we will know the W parameter the past eclipses would give us the luminosity of the Sun in these times, especially right after the Maunder minimum in 1715.

### 3.9 Unidimensional eclipses: photometric data

Eclipse data from photometers have very high time accuracy and almost no spatial information. Their utility in accurate diameter's measurements is discussed. The eclipse mission of 11 July 2010 in French Polynesia is also dedicated to measure the solar diameter from the ground.

### 3.10 Before and after this work

The situation of this filed of research before and after this work is described, putting into evidence the new contributions.

### 0.3.4 Chapter 4: Daily Transits

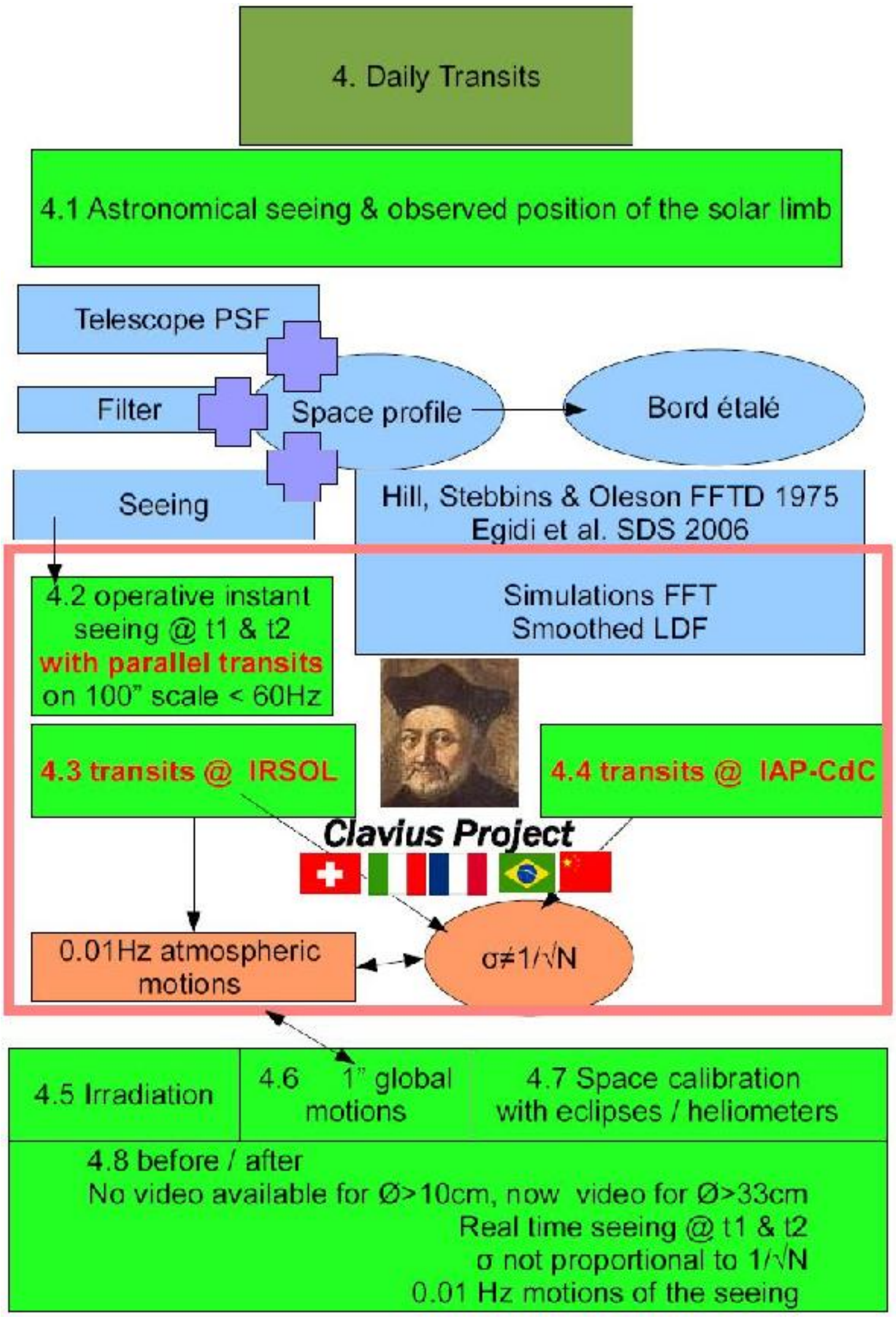

**4.1 Seeing and observed position of the solar limb**

The influence of the seeing on the observed position of the maximum of the derivative of the luminosity profile of the Sun is here described.

**4.2 Instantaneous measurement of the seeing**

The different components of the seeing act upon different angular scales. We have images of a 100" scale. From each image of the solar limb obtained during a drift-scan observation we reconstruct a regular arc of a circle from the distorted solar limb. We measure this medium scale effects of the atmospheric turbulence (seeing) by the irregularity of the motion of such arc. Values of the seeing as low as 0.6" have been measured during summer in Locarno. This determination of seeing is strictly related with the measurement of the solar diameter.

The method of parallel transits is presented.

**4.3 Transits at the IRSOL 45 cm Gregory-Coudé telescope**

The tradition of solar diameter's measurements in Locarno goes back to the late years 1970s with visual observations. Two twin Gregory-Coudé telescopes of 45 cm aperture, designed for solar observations, operated simultaneously in Switzerland and Canary Islands with drift-scan methods between 80's and early 90's. The opening diameter of the telescope is larger than any turbulent atmospheric cell, allowing a better stability in the observed solar diameters.

References: Wittmann and Bianda visual and CCD; Damé private communication.

From 2008 this method has been upgraded in the framework of the Project Clavius, among Italian and Swiss scientific Institutions, for the development of fast detector for physics and astronomy. The drift-scan project at Locarno is being upgraded with fast imaging detectors, exploiting the idea of simultaneous measurement of the seeing. We started with a commercial CMOS at 60 frames per second, reaching a resolution of 0.25" for a single diameter measurement. We discuss how the number of measurements improve the statistics, since we can obtain much more than 30 consecutives diameters (as in the 1990s) per day using the grid method.

The eventuality to perform such an experiment in Anctartica is outlined.

References: Sigismondi 2006 (multiple transits in st. Maria degli Angeli); Sigismondi 2008 AIPC-CLAVIUS; Caccia 2009 internal notes specifications of CMOS detectors; Koutchmy 2009-10; Damé; Sigismondi 2006 IAUC II-drift-scan on the Moon.

### 4.4 Transits at the Carte du Ciel 33 cm refracting telescope

This historical telescope (Institute d'Astrophysique and Observatoire de Paris) used for astrophotography since 1885, has been equipped with a CCD camera Lumenera and screened with a panchromatic filter in astrosolar. The Carte du Ciel 33 cm refracting telescope in Paris, used for our transit measurements, has a PSF particularly clean because of no obstructions in its optics, and a very small scattered light effect.

### 4.5 Operative definition of the irradiation effect

The irradiation is the combined effect of on-axis and off-axis point spread function and stray light. It is relevant for small telescopes, without field stop, especially the ones used for eclipses field observations.

### 4.6 Evidence of 1" slow image motion of the whole solar image

The diameters obtained in consecutive measurements are not consistent within all the experimental errors. The first results of the new measurements tests for both telescopes (2008-2010) are discussed: the effect of low frequency waves (0.01 Hz) in the atmosphere could explain the difference in the successive hourly circle transits' measurements. This finding can cast some light also on some ancient puzzling data (Rome - Campidoglio & Greenwich observations).

### 4.7 Space calibration of transit's measurements

Combined observations space-ground: eclipses and drift-scan. In the occasions of two total eclipses of 1$^{st}$ August 2008 and 22$^{nd}$ July 2009 we have made drift-scan observations at Locarno observatory, and a third session was planned for 11 July 2010 eclipse. A single day of observations made with DORAYSOL on 23 September 2006 can be considered as overlapped with the annular eclipse of 22 September, observed in French Guyana.

The same strategy is planned between SDS and Picard. A new flight of SDS (after the one still unpublished occurred in october 2009), shall be done during the Picard mission to compare the space measurements with the balloon-borne ones.

### 4.8 Before and after this work

The situation of this filed of research before and after this work is described, putting into evidence the new contributions.

### 0.3.5 Chapter 5: Didactic outreach

**5.1 Required accuracy < 1/10000**

The statement on human energy output versus solar input is "1 year of humankind corresponds to 1 hour of Sun". This is the starting discussion of a series of units on solar astrometry, in order to motivate this study in a framework of climate studies.

**5.2 Pinhole studies of real time seeing, PSF and ΔUT1**

A pinhole is a "lens-less telescope", completely free from spherical aberration and perfectly suitable for astrometric studies of the Sun. It has been exploited since 1475 (Ulugh Begh, 1437, measured only solar declination and not diameter) by Paolo Toscanelli for giant pinhole telescope built in Florence's Cathedral as pinhole camera. With the one in Roma, Santa Maria degli Angeli (the Clementine gnomon of 1702) I have demonstrated its extraordinary accuracy down to one arcsecond for all the range of solar declinations, and the possibility to be used for monitoring the ΔUT1 evolution.
Using an evenly spaced grid at the focal plane in a drift-scan method gives the opportunity to have several diameters in a single transit and it can be used for measuring the seeing in real time. Students already have videocameras, and they are ready to start solar astronomy experiments.

**5.3 Pinhole model of Solar Disk Sextant**

A two-pinholes heliometer can be designed to simulate the situation of having two images at the focal plane. Two scientific museums (Coimbra Museum of Science and Robert Hooke Institute in Nice) have contacted me for implementing this instrument.

**5.4 Sunsets and solar diameter**

The measurement of the solar diameter is possible also from timing the sunset over the sea horizon. The arcsecond level of accuracy is reached with video.

**5.5 LDF studies from a simple projected image of the Sun**

Using an ordinary pair of binoculars or a normal telescope, and even a pinhole, it is possible to study the Limb Darkening Function of the Sun by a single digital photo of the image of the Sun projected on a white paper. It is to remember that every commercial videocamera is like a non-linear detector, because it is programmed in order to simulate the eye response.

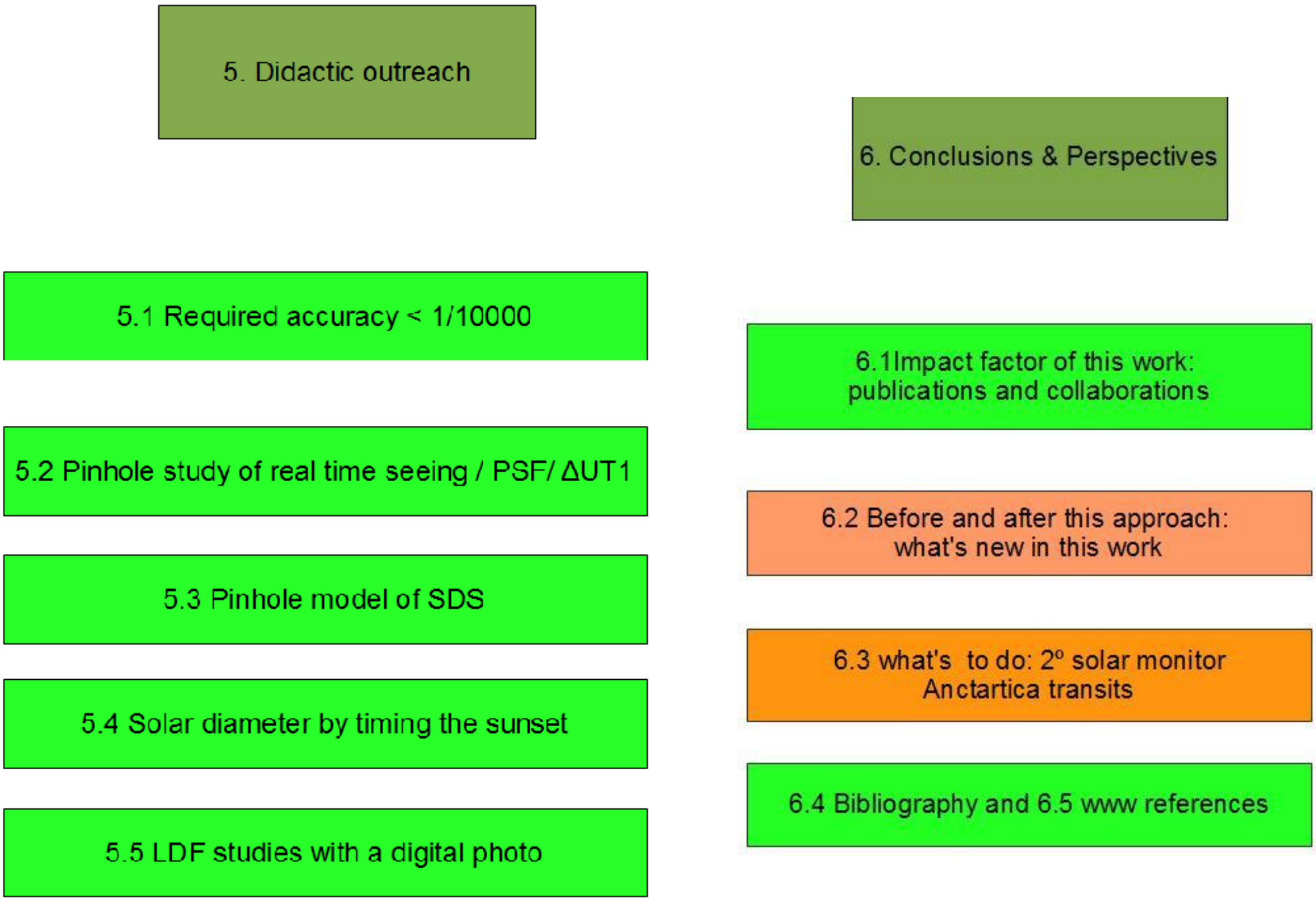

### 0.3.6 Chapter 6: Conclusions and Perspectives

This thesis will be discussed when the first data from Picard will be available and when the calibration of MISOLFA will start.

My contribution to this field has been

1. to have shown the need of a more precise definition of solar limb during eclipses, taking into account the phenomena in the flash spectrum region.
2. to have individuated in the low frequency motion of the atmosphere one of the causes of the inconsistency within seeing errorbars of the daily transit's measurements of diameter.

This is a starting point for many relevant scientific issues on the physics of the Sun. The task to bridge together eclipses observations and direct solar astrometry methods has been tempted, but the process to automatize the procedures was still in progress in Locarno and we could only do preliminary tests at the same time of the eclipses of 2008, 2009 and 2010.

Qualified observers belonging to IOTA, International Occultation Timing Association have firstly published, under my guidance, their valuable data of recent eclipses. Historical eclipses have been re-analized with the latest lunar satellite data. The problem of filters adopted in eclipse observations, risen by me, found a partial solution with a standardization of the IOTA filter. The first atlas of Baily's beads has been published.

The measurements of solar diameter using different sets of beads give different results: a new definition of solar limb is crucial, to take into account the solar mesosphere emitting lines.

**6.1 Impact factor of this work**

The publications and the collaboration activated during this work are here evaluated.

**6.2 Before and after this approach: what's new in this work**

The original achievements of this work are here drafted. They are the new results in this field, and they bring new understandings and points of view.

### 6.3 What's to do: 2° solar monitor and Antarctica

The evidence of slow image motions in the atmospheric seeing, involving the whole figure of the Sun, suggests to realize a 2° solar monitor in order to correct the transits' data for this random-like motion.

From Antarctica a very good and steady seeing conditions are expected, and it should be the ideal place of an experience of solar diameter's monitoring with small scale telescopes (10 cm of objective).

### 6.4 Bibliography and www references

A list of useful books, papers and web links.

## 0.2 Acknowledgments

I am greatly indebted with many people who helped me to carry on this project. A list of people is necessarily incomplete, also because it is difficult to recognize the contributions of other persons in the formation of the ideas: after their genesis they seem to belong to us since ever. Other persons have sustained me with their example, their intellectual honesty and their enthusiasm. Other persons gave to me spiritual support, helping me to focus on my personal growth, motivations and my responsibility toward the society. It is like a night sky full of stars, many of first magnitude and all of different colors, owing to their different roles: masters, mentors, models, inspirers, colleagues, friends, familiars. They supported me, indeed with some criticism, during all situations.[2] Moreover in the last 10 years I made part of my activity in foreign countries, and I found persons who let me experience warm hospitality and sincere friendship so that "wherever I lay my hat that's my home".[3]

My parents Camillo (+2011) and Erminia, my syster Irene, Lisa and Christopher Hoffer, John Sullivan, Bernard Confer, Sabatino Sofia, Priya Natarajan, William Van Altena, Terry Girard, Linghuai Lee, Federico Spada, Dorrit Hoffleit (+2007), Sue Delong, Ann Giangarra, Laura Artusio, David and Joan Dunham, Martha (+2010) and Wayne Warren, Alan Fiala (+2010), Jean Arnaud (+2010), Pierre Assus, Thierry Corbard, Abdenour Irbah, Wassila Dali-Ali, Marianne Faurobert, Jocelyne Bettini, Valérie Chéron,Yan Fantei-Caujolle, Jean-Luc Beaumont, Richard Beaud, Michel Mathieu, Ting Lee, Marie-Thérèse et Solange, Serge Koutchmy, Cyril Bazin, Patrick Rocher, Barbara Obrist, Michele Bianda, Anna Soldati, Aline Bianda, Renzo Ramelli, Flavio Nuvolone, Massimo Caccia, Carlo Monti, Paolo Zanna, Rossana di Gennaro, Paolo Rossi, Runa Briguglio, Andrea Soddu, Giuseppe Blanda, Renzo Giuliano, Mario Catamo, Massimo Fofi, Paolo Gillet, Alberto Egidi, Remo Ruffini, Alessandro Cacciani (+2007), Giovanni Moreno, Paolo de Bernardis, Simonetta Filippi, Cosimo Palagiano, Raffaele Ciambrone, Daniela Velestino, Flora Parisi, Emanuela Celeste Donna, Rita Fioravanti, Pietro Alessandro Giustini (+2007), Sabina Fiorenzi, Cetta Petrollo, Gabriella d'Amore, Orazio Converso, Laura Paladino, Romano Penna, Gian Matteo Botto, Alberto Orlando, Rafael Pascual, Melchor Sanchez de Toca, Paul Poupard, Gianfranco Ravasi, Sergio Pagano, Benedetto XVI, Davide Troise, Danilo Montagnese, Emilio Sassone Corsi, Sabrina Picchi, Laura Comerci, Veronica d'Angelo, Federica di Berardino, Cristina Adamo, Silvia Latorre, Carles Schnabel, Hans-Joachim Bode, Konrad Guhl, Pawel and Kasia Maxim, Pawel Rudawy, Richard Nugent, Gerhard Dangl, Sven Andersson, Antonino Tata, Lee Hyung Won, Fady Morcos, Ahmed Hady, Cristina Mandrini, Alexandre Humberto Andrei, Jucira Lousada Penna, Sergio Calderari Boscardin, Victor d'Avila, Eugenio Reis, Albert Picciocchi, Xiaofan Wang, Jingxiu Wang, Jie Jiang.

I could benefit of a very long "permanent formation", topped with a second PhD being 42 years old. In terms of salaries of teachers, professors and researchers... involved with me[4] during about 39 years of cultural formation, it has been an investment larger than one million of euro, sustained by the Italian State, and by my family.

The cost of "producing a researcher" is probably underestimated, but the order of magnitude is rather correct, and it has to be taken into account in the analysis of the "brain drain" phenomenon.

For some sociological reason, the Italian researcher has to develop a special ability to be actually what he became, i. e. to work in the field in which he or she attained the highest specialization remaining in his own country. This further energy threshold tends to reduce the sense of gratitude towards Italy, which is paradoxically capable of generate big scientists and to avoid them to work for Italy itself. This is not a novelty, it is in the genoma of Italians: already Dante reported "Tu proverai come sa di sale lo pane altrui, e com'è duro calle lo scender e 'l salir per l'altrui scale"[5]. Similarly, on a smaller scale, but relevant for Rome there is the "roman area paradox"[6] with the largest concentration of research institutions without any development for the territory. It is necessary to promote the dialogue between institutions, enterprises and research.
But it is also necessary to develop the culture of responsibility, especially in the Italian academic context, because "the product of academy are persons"[7], avoiding to "produce" persons used to survive in human environments with despotic leadership who will reproduce the same victim-villain relationship when they will be in power.
Positive environments are where both rational and emotional intelligence can develop.

[2] Vanistendael S. et J. Lecomte, Le bonheur est toujours possible, construire la résilience, Paris, Bayard (2000). See the "casita" personality model developed at the BICE/ICCB (Bureau International de L'Enfance di Ginevra/ International Catholic Child Bureau) www.crin.org/docs/Resiliance.doc

[3] Song by Paul Young (1984) lyrics by Marvin Gaye, Norman Whitfield & Barrett Strong.

[4] Salaries divided by the average number of students per teacher.

[5] Dante, Paradiso, XVII - vv 58-66

## 0.3 Preface

*Man go West!* (and I came back to East)
(quoted to me by Alberto Righini)

At the end of 1990s there was the LEST telescope project for a large international solar telescope based on next-generation technology, to be established on the Canary Islands. 6 European and 3 non-European (Australia, China, USA) countries were participating in this enterprise. A prime objective of LEST was to investigate the subarcsec fine structure of solar magnetic fields. The design of the 2.4m aperture telescope was "polarization-free". Helium-filling and adaptive optics were to be used to achieve 0.1 sec of arc spatial resolution.[8] I attended a presentation of this project by Alberto Righini, during a PhD school in 1998 in Monteporzio Observatory. About solar physics and contemporary trends in science he quoted the American refrain "Man go West"[9] explaining that also in astrophysics there is a West, a unknown territory filled by opportunity, and this field is cosmology. After more than 10 years of this presentation a lot of economical resources have been invested into cosmology, and more and more solar physics loses chairs in the most important universities, without immediate replacement. Righini explained to the audience also that the West of cosmology was a region supportable with some exotic physics based upon data with accuracy yet $\Delta X/X \sim 1$ or even larger. Solar physics, on the other hand, was already dealing with much better accuracies $\Delta X/X \sim 10^{-3}$ and the field was always generous of new discoveries.
Cosmology is the West, Solar Physics the East... and I decided to quit the West[10,11] after the wondrous spectacle of the eclipse of 1999.
The observation of 1999 solar eclipse, total over central Europe, triggered my interest in solar astrometry, as well as the following visits in the Basilica of St. Maria degli Angeli e dei Martiri in Rome, where the Clementine Gnomon (1702) is located.
The Clementine Gnomon is a giant pinhole telescope, with maximum focal length of 50 m, capable of a few arcsecond accuracy on the determination of the apparent position of the Sun at the meridian transit.
The atmospheric refraction and its differential effect on the superior and inferior limb of the Sun can be compared with computed ephemerides, and also the secular drift of solstices, due to Earth's obliquity change, and phenomenon of UT1-UTC drift are measurable at this lens-less[12] solar instrument.

---

[6] See the brochure of Parco Scientifico Romano, University of Rome "Tor Vergata".
http://www.parcoscientifico.eu/IMG/pdf/romesciencepark.pdf
[7] William C. Rando, director of the Academy for the Art of Teaching Yale University.
[8] Stenflo, J. O.Vistas in Astronomy, **28**, po. 571-576 (1985).
[9] http://en.wikipedia.org/wiki/Go_West,_young_man
[10] Sigismondi, C., Perspectives on the observation of clusters of galaxies in X-ray band with SAX (X-ray Astronomy Satellite), Nuovo Cimento B, Vol. **112**B, p. 501 – 515 (1997).
[11] Sigismondi, C., et al., Damping Time and Stability of Density Fermion Perturbations in the Expanding Universe, Int. J. Mod. Physics D 10, 663-679 (2001).
[12] See http://www.pinholecamera.com/

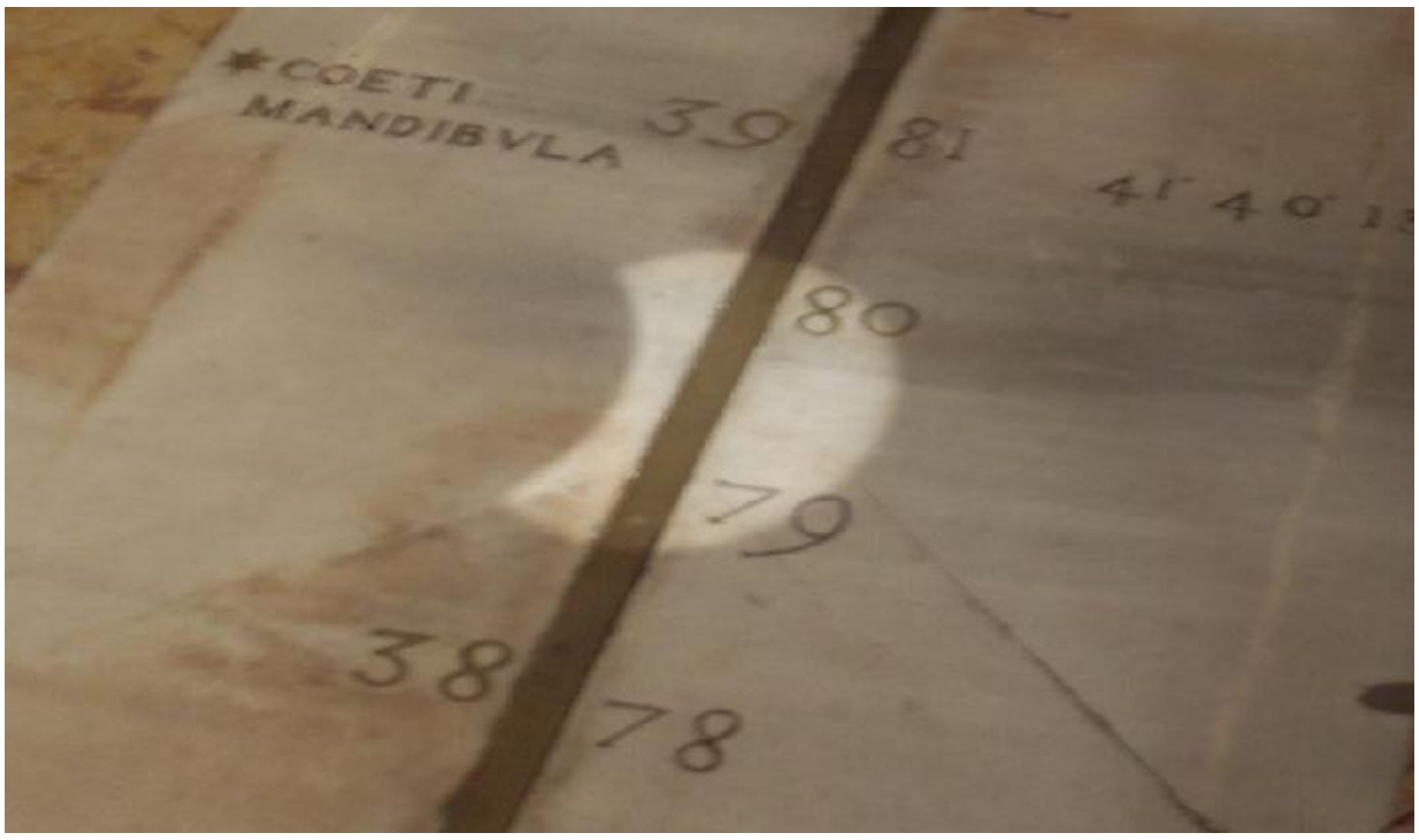

**Fig. 0.1. The eclipse of March 29, 2006 as observed on the meridian line of St. Maria degli Angeli in Rome, at 38.5° from zenith.**

The introduction of video-recording, up to 60 frames per second, in order to recover the maximum accuracy on the meridian transit timing, transformed for me and for my fellow students this historical instrument into a scientific laboratory to test our knowledge on solar astrometry.
In parallel, the cold marbles of the Basilica taught us the story of geodesy and technology at the beginning of XVIII century, and so another field of research on history of science was opened.

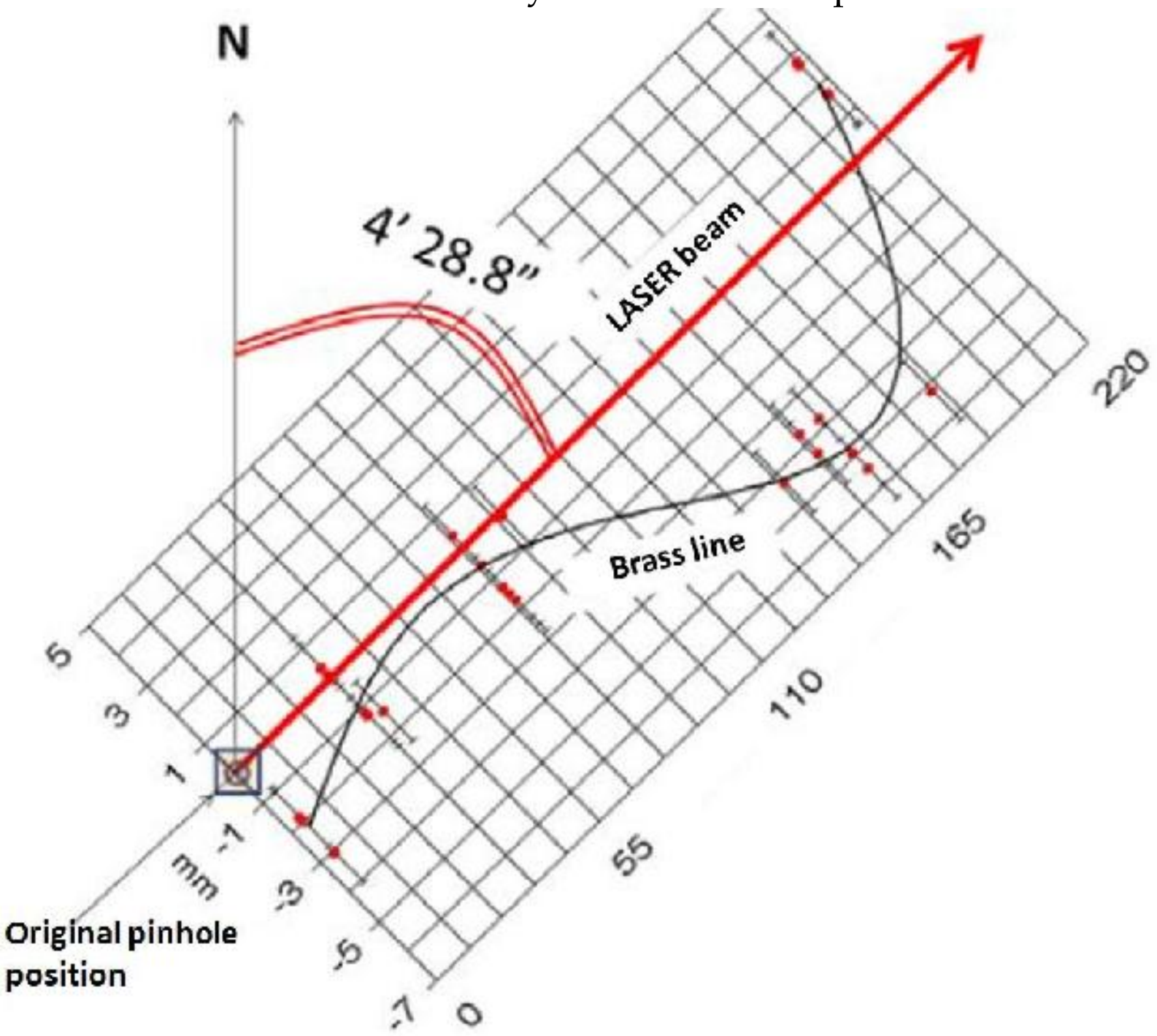

**Fig. 0.2 The azimuth of the Clementine gnomon (1702).**

In this research, started to understand the small deviation between the meridian line and the geographical North, ephemerides and stellar astrometry, proper motions and diffraction optics led us to know our colleagues of three century ago with their sake of experimental accuracy and their strategies, always much cleverer than our initial belief.
The scholarship at Yale University from 2000 to 2002 allowed me to be in contact with one of the most important traditions in stellar and solar astrometry.
At the Astronomy department Prof. Sabatino Sofia and Prof. William Van Altena were joining their efforts

and their teams to reduce the data of the Solar Disk Sextant (SDS), a balloon borne telescope conceived to measure the solar diameter from stratosphere at 37 km of height, with a residual barometric pressure of 3 millibar.

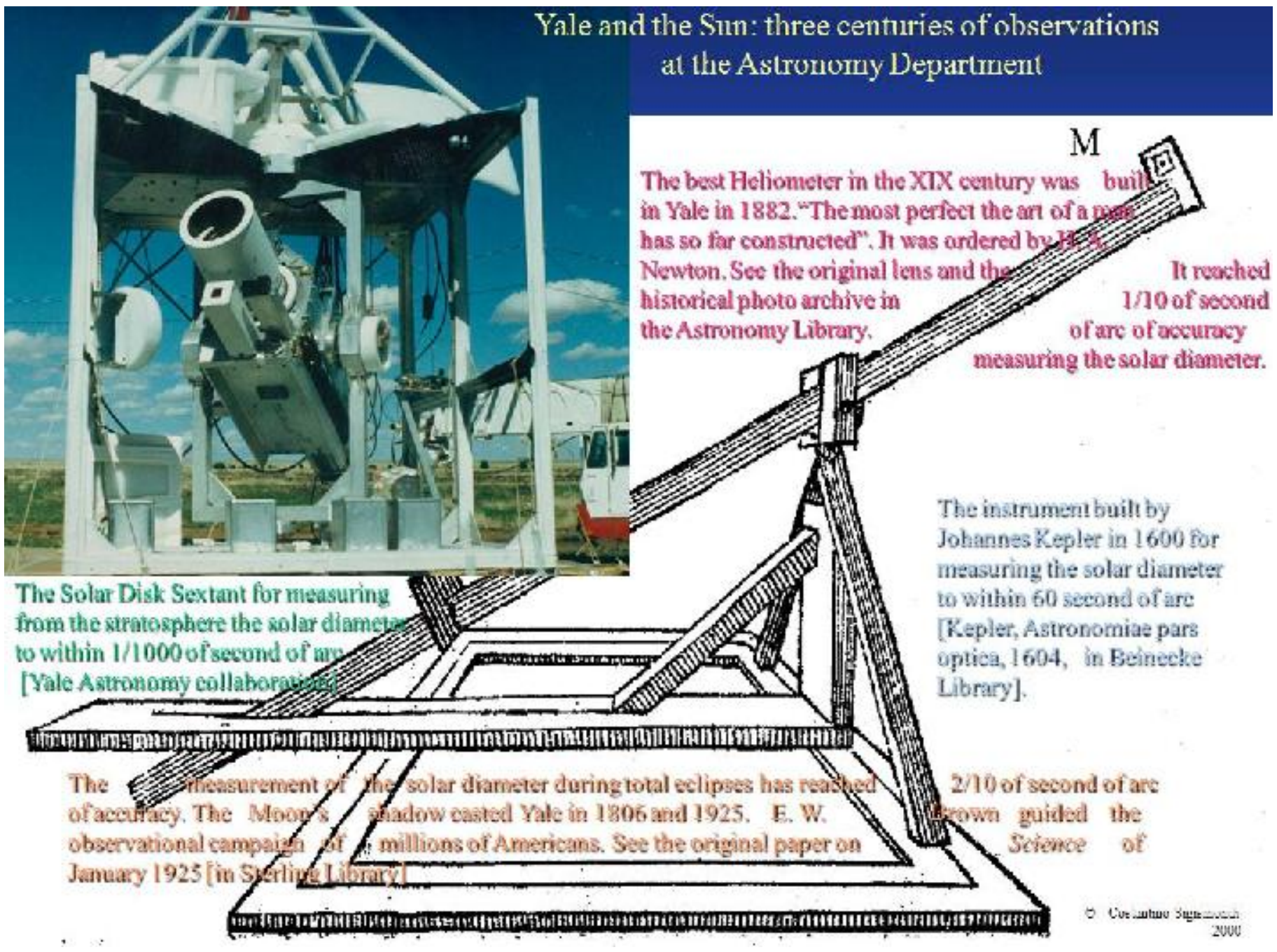

**Fig. 0.3 Yale and the Sun, poster realized by me for the tercentennial of the Yale University. The great astrometric tradition is sketched from H. A. Newton and his Heliometer (used in Venus' transit of 1882) to E. W. Brown lunar theory who changed forever the approach to solar eclipses, to SDS with an accuracy to the milliarcsecond level.**

Meeting and the collaborating also with the late prof. Dorrit Hoffleit (1907-2007), expert in variable stars and witness of the history of astronomy of XX century in America, made this time in Yale unforgettable.
The following acquisition (2002) of the chair of Physics & Laboratory at the institute in Rome dedicated to Giuseppe Armellini [former director of Rome Astronomical Observatory (1924-1958) in the time of translation from Capitol to Monte Mario hill occurred in 1937] forced to dedicate myself to the art of teaching with all my strengths.
A collaboration, in the spare time, with the late prof. Alessandro Cacciani (1938-2007) started at the Sapienza University in his solar physics laboratory and since 2002 several students interested in solar physics could perform experiments in the framework of Astrophysics Lab course of Prof. Paolo de Bernardis.
The work started at Yale on solar astrometry with eclipses was continued in this contest, until a new eclipse casted its shadow on the European continent. It was the time to get our own data after having examined several video and data from other observers. On October 3rd, 2005 an annular eclipse crossed Spain and with Pietro Oliva and Paolo Colona we organized -at our own expenses- an observing expedition.
The original idea was to go to Santiago de Compostela, where the eclipse was central, and move to Lugo at the antumbral limit of the eclipse, where the Baily's beads were visible.
This program would have been perfect under every point of view, but it come out unrealizable. And after landing in Santander we have chosen to intercept the eclipse path in Leon region, south of Burgos and north of Valladolid.

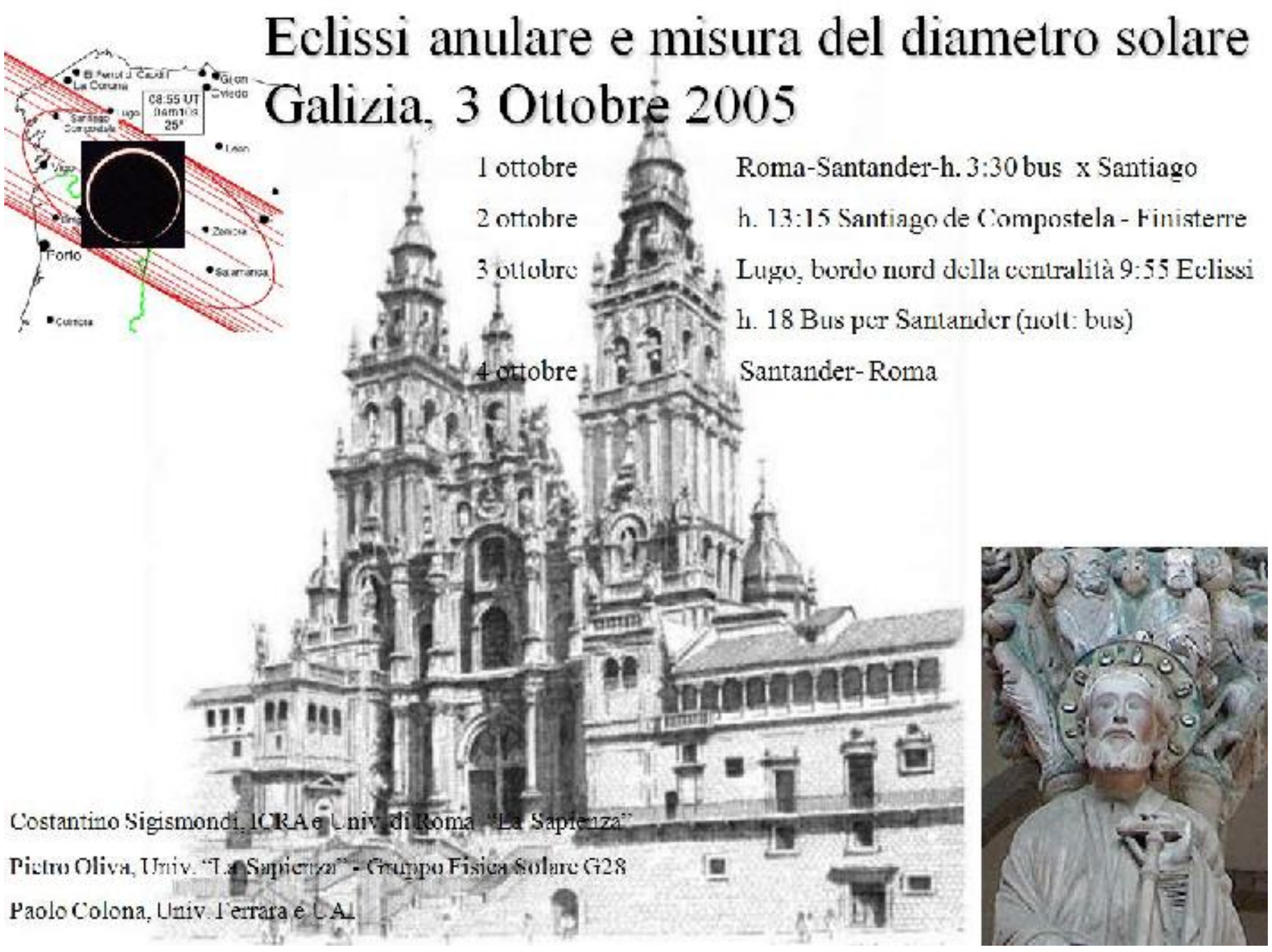

**Fig. 0.4 Poster for the project of eclipse expedition in Galizia**.

Later in 2008 while I was participating to Scientific Instrument Committee meeting in Lisboa, I could land in Santiago de Compostela and visit the local observatory who provided me with the images of that eclipse.[13] But since Santiago was in the centerline the duration of Baily's beads was too short to be useful for our measurements of solar diameter.

So we could display our instruments perpendicularly to the calculated eclipse path's limit. It was the first mission with an array of observers over 1 km. Unfortunately the weather changed over the Spanish meseta half hour around the centrality, and we were clouded out.

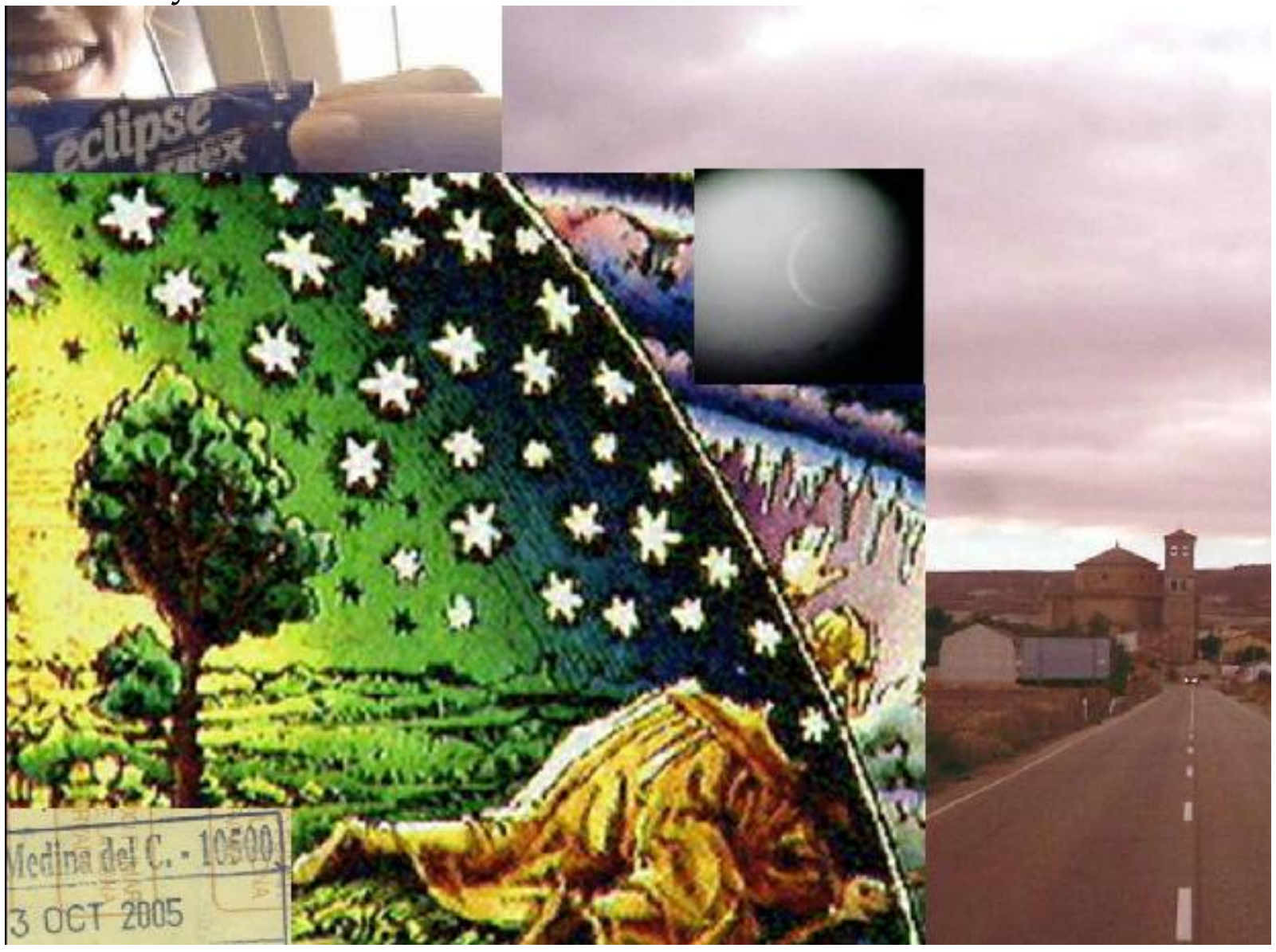

**Fig. 0.5 The annular eclipse of 3 Oct. 2005 at its maximum from Valoria la Buena. No video were possible, for the clouds. Photocomposition.**

This great opportunity to define with a great precision the Northern limit of this eclipse, lost for the clouds, urged us to enter in contact with other observers and Dr. David Dunham, the president of IOTA

[13] http://www.fomento.es/MFOM/LANG_CASTELLANO/DIRECCIONES_GENERALES/INSTITUTO_GEOGRAFICO/ Astronomia/ECLIPSE/

International Occultation Timing Association, put ourselves in contact with Dr. Carles Schnabel of the Astronomical Association of Sabadell[14] (Catalunya), where several video observations were performed according to the metrological standards used in the solar diameter analyses.
We got several DVD of this eclipse at the Northern limit, and Dr. Wolfgang Strickling provided to us the Southern limit's observation needed to complete the data set to recover the whole antumbral path, independently on the ephemerides.
The first opportunity to observe an eclipse with data occurred in 2006, when on March 29 it was total over the western part of northern Egypt. An international IAU colloquium was organized in Cairo just after the eclipse and the LOC prof. Ahmed Hady sent to us Dr. Abdel Fady Morcos who helped us to get to the Southern limit of the eclipse east of Sidi Barrani at Zawyet al Mahtallah. Three observers were positioned over a line perpendicular to the path, and I was the more external one. The eclipse in my location never attained the totality, while the corona was visible to the naked eye, once the remaining beads were covered with the thumb.

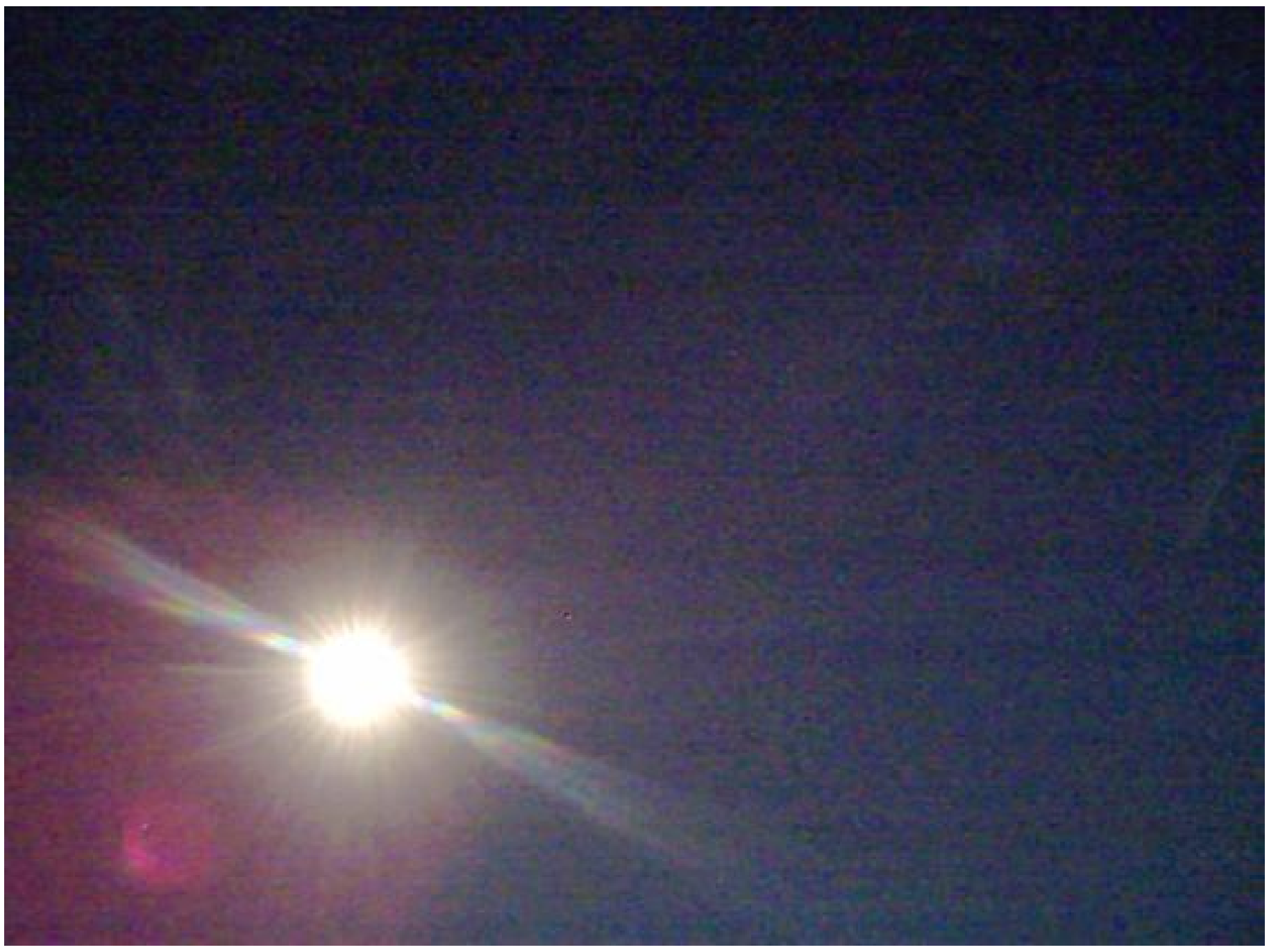

**Fig. 0.6 The eclipse at its maximum phase at Zawyet al Mahtallah: the ghost image lower left shows the position of the beads with respect to the whole disk. Photo with Philips KeyRing 008.**

The luminosity of the two remaining beads at maximum eclipse was dazzling, even if they were barely visible on the white screen over which the solar image was projected through the 7 cm Meade ETX F/5 refracting telescope.

[14] http://www.astrosabadell.org/

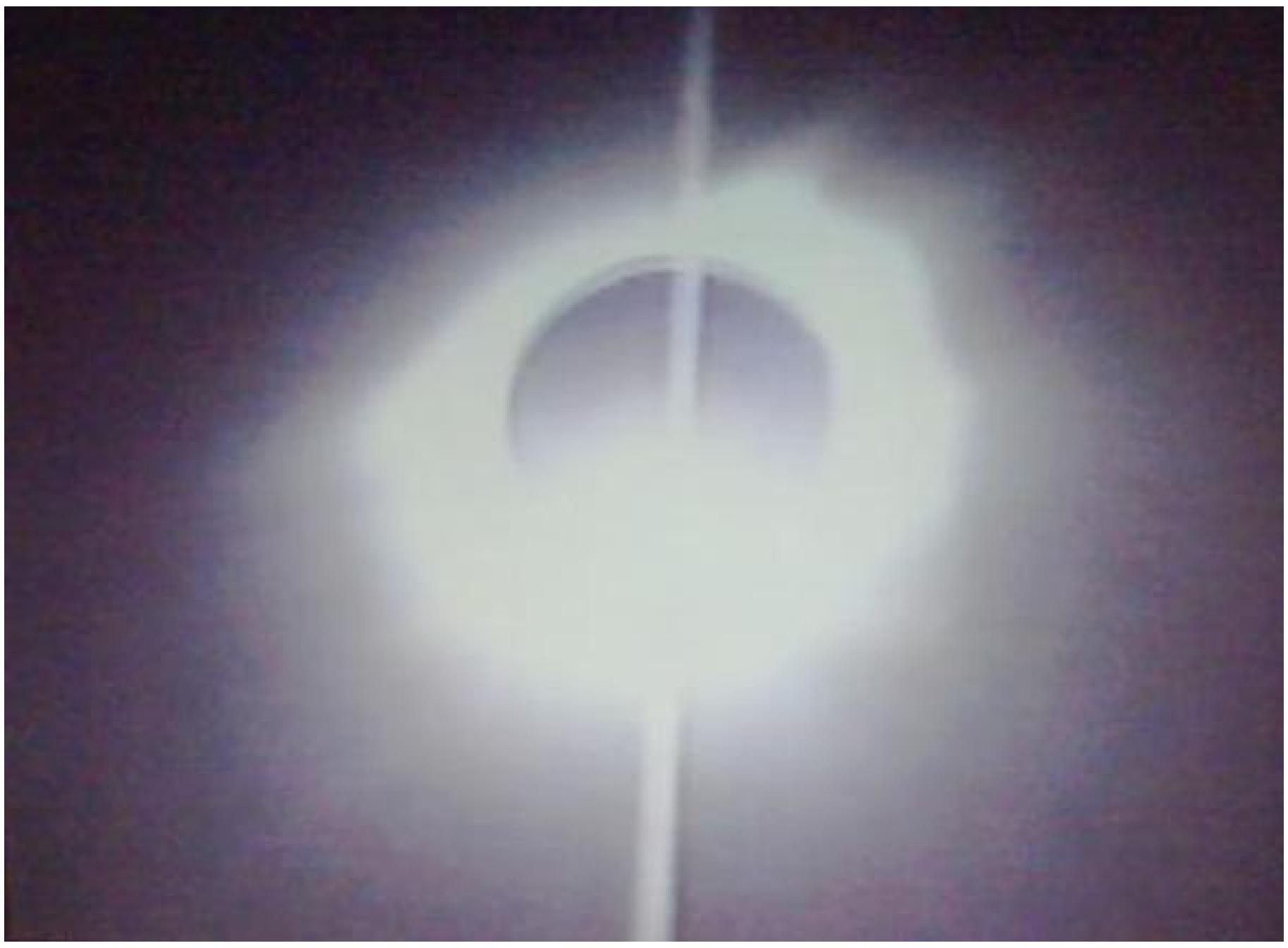

Fig. 0.7 A video taken by Dr. Fady Morcos (250 m inward the umbra) shows the effect of pixel saturation due to the bead's dazzling light. At maximum eclipse its brightness was equivalent to the corona. (Sony miniDV videocamera).

Afterwards, on 22nd September 2006 the whole Draconitic year (354 days) of my directly observed eclipses was completed with the annular eclipse in French Guyana.
At this time I adopted the method learned from Fred Espenak (NASA) the world famous eclipse expert. At Goddard Space Flight Center during my stay in 2001 he presented a talk on the organization of things to do during and near totality in order to maximize the scientific results achievable in such a short time. He registered on a tape in real time the actions to do and he let the tape instructing him during the phenomenon.
I could achieve all the observations thanks to Prof. Albert Picciocchi of the French Ministry of Education who helped me in all logistic needs in French Guyana. So I recorded an audio file, using Occult Baily's beads software for calculating the beads' appearances, with all the operation to do with the videocamera during the phase of annularity.

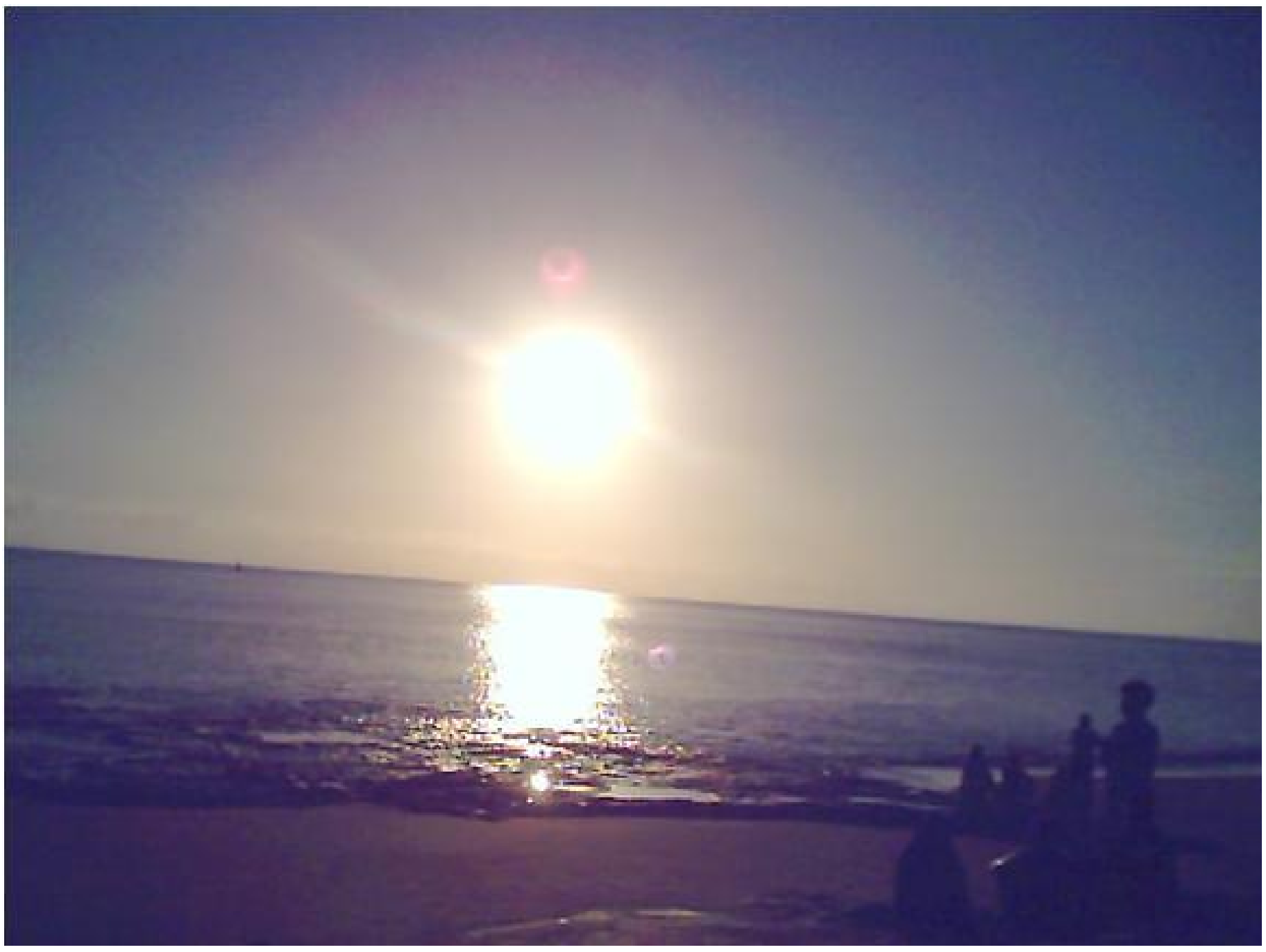

**Fig. 0.8 Annular eclipse of French Guyana at its maximum (92.5%) from Kourou beach. The ghost images up and down (in the water) show the ring of photosphere. (Photo with Philips Keyring 008.)**

At the end of that mission 13 beads were identified in the video, but they were very fast since the observations were possible only near the centerline, because the antumbral limits were on the ocean or in the jungle. No other observers worldwide recorded Baily's beads during this eclipse.

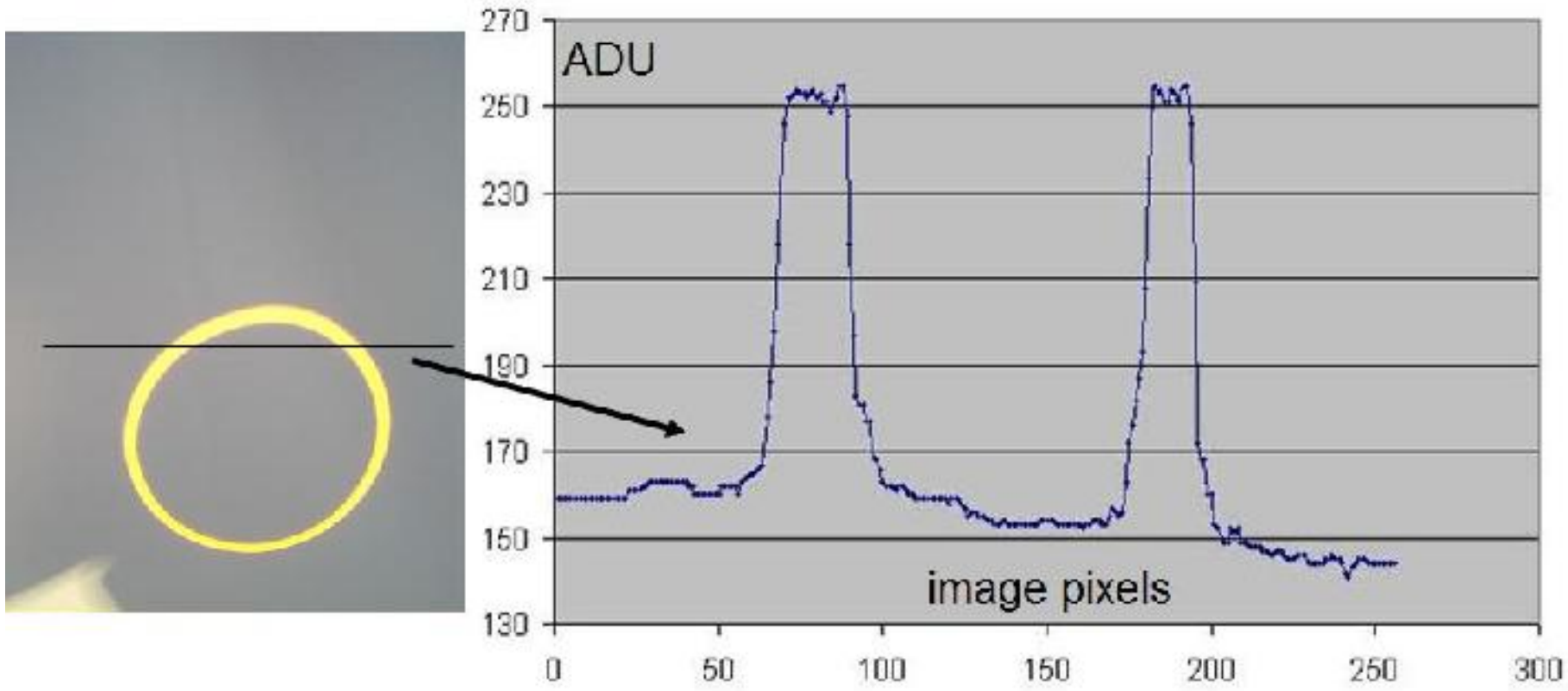

**Fig. 0.9 The analysis of background level in the projected image in Arbitrary Digital Units, at maximum eclipse of 22 September 2006. This photo is made with Philips Keyring008 with 255 levels of intensity (8 bits) for each of R,G and B channels.**

After Spain, Egypt and French Guyana I had a direct, personal experience of observational conditions during a total and an annular eclipse. For the following eclipses I could rely into the observations made by IOTA (International Occultation Timing Association) members, who were eager to publish their data in view of a research on solar diameter variation.

Planetary transits are also sources of data for measuring the solar diameter: in this decade Mercury crossed the Sun in 2003 and 2006 and Venus in 2004.

While in the 2003 transit I made the largest public observation (about 1000 students during the whole transit) with a Newton telescope and with a pinhole device in Armellini Institute (Rome), for the transit of Venus I organized, also in Rome, a public conference and book's exhibit in the library where the 1761 transit was observed.[15] For 2006 Mercury transit I could contact the observatories of Hawaij through Prof. Jay Pasachoff

[15] http://www.casanatense.it/index.php/en/gli-eventi-in-biblioteca/le-mostre/159-saros.html

and Sacramento Peak through Dr. Kevin Reardon. The transit of Mercury was carefully observed in Sac Peak, but not perfectly for our purposes, [16] while it was very windy on Haleakala and the dome could not be opened until about a half hour after first contact.[17] The data of MEES telescope of the Hawaiian solar physics group are published in the web page of November 8 2006 data.[18]
We get very close to a big result in solar diameter determination…
One of the reasons why this was not achieved in the last 3 transits of 2003-2004 and 2006 can be the fame of unreliability of the planetary transits for solar diameter studies, due to the black drop phenomenon. Conversely I had a simple idea to overcome black drop and seeing effects by analyzing a sequence of chrono-dated images near the internal contacts.
A very promising result was obtained with Venus transit in 2004, from 50 photo made each minute with 1 second accuracy, made by Anthony Ayomamitis[19] in Athens with a 160 mm Apochromatic telescope in Alpha line: the instants of the internal contacts were determined by extrapolating to zero the chord draft by the planetary disk and the solar limb with ±8s and ±1 s for different seeing conditions at sunrise and near the local meridian.
With these experiences in mind, and with a paper published on Nuovo Cimento on the oblateness of the Sun[20] in order to put in evidence the links between Celestial Mechanics, General Relativity and Solar Physics, I have participated to the IRAP-PhD competition in 2007, being selected among the winners.
This international PhD program involved also the Université de Nice-Sophia Antipolis, institution participating to the satellite mission Picard through the Fizeau department, and in end January 2008 the *cotutèle* between Rome and Nice Universities was stated. The astronomer Jean Arnaud, former director of Themis telescope in Canary Islands, recently arrived in Fizeau department, and prof. Alberto Egidi former director of the physics department of Tor Vergata University (Rome 2) accepted to be my co-directors. After

[16] http://www.nso.edu/press/merc_trans06/20061108hl.mpg a Video in 656.28 nm; the web site for all the proceedings is http://www.nso.edu/press/merc_trans06/ "Unfortunately, we did not get good observations of the first contact. We do have observations of the second contact, but the seeing was not very good and the adaptive optics system was not yet locked on resulting in a rather distorted image. I am attaching one of the better images we obtained during second contact, but clearly it is hard to define exactly the time of second contact due to image motion and blurring. Third contact occurred right near sunset and in the minutes leading up to the contact it was hard to even identify Mercury in the highly distorted images." [K. Reardon, Arcetri Astrophysical Observatory, private communication 10 Nov.2006].

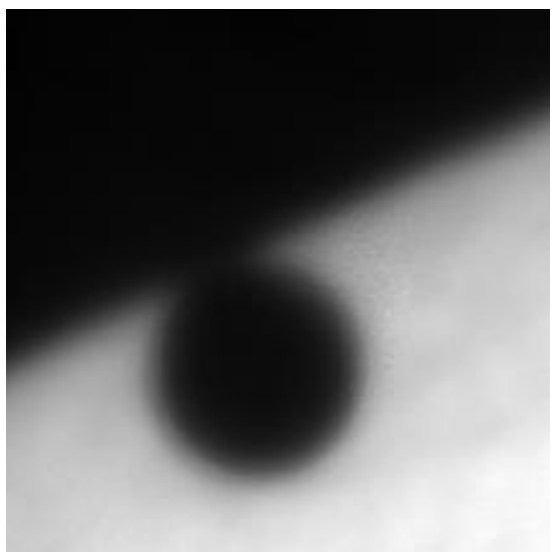

**Fig. 0.10 Mercury transit of 2006 second contact.**

Observations were carried out at the Dunne Solar Telescope using the Interferometric Bidimensional Spectrometer (IBIS) to observe the exosphere of Mercury, and the Williams College Portable Occultation, Eclipse, and Transit System (POETS) camera system to observe the solar photosphere.

[17] Jay Pasachoff [private communication 9 Nov. 2006]
[18] http://www.solar.ifa.hawaii.edu/cgi-bin/MeesLogSearch?date=20061108
[19] "I just checked my CD backups and I do have both ingress and egress at one-minute intervals. More specifically, I have 08:16:00-08:42:00 UT+3 and 14:00:00-14:27:00 UT+3 and which cover completely both ingress and egress since I included a three-to four-minute window outside the official contact 1-4 times." [A. Ayomamitis, private communication 20 Nov. 2006] I answered "Let me quote Horatius (Epistulae, II, 1.156) *Graecia capta ferum victorem cepit*. [free translation: Once the Greece was kept (from Rome), it prevailed (culturally) on the rude conqueror] Thanks and regards from Rome." Anthony saved his original data with an admirable foreseeing and after some years they were useful to measure the solar diameter in H alpha line, with an accuracy of 8 parts over 20000, i.e. about 0.7".
[20] Sigismondi, C.; Oliva, P., *Solar oblateness from Archimedes to Dicke*, Il Nuovo Cimento B, vol. 120, 1181 (2005).

the death of Jean Arnaud in September 2010, Marianne Faurobert has taken his duty in November 2010, for accompanying the last months of this thesis work. In 2008 the VINCI fellowship of the Université Franco-Italienne was granted to this thesis, covering all the local expenses in France and the travels between Italy and France.
During the three years of PhD, according to the current Italian law for public school tenures, I could benefit of the same economical treatment as if I were teaching at school, which was 20% better than the PhD scholarship. Therefore it is important to show also the didactic aspect of this work.
I will try to be clear and simple, rather than verbose and complicate, as it should be for whatever scientific work, in whatever language.
At the time of the "siècle des lumières" Antoine de Rivarol (1753-1801)[21] said "Ce qui n'est pas clair n'est pas Français" and he was proposing the French language as the universal one. This should be the aim of every scientific contribution, in whatever language, and I hope not to be so far from this goal.

## 0.4. Introduction

The theme of solar diameter is introduced using an historical approach.
This thesis deals with the accuracy of ground-based measurements of solar diameter and their reference values obtained during solar eclipses observations. Thanks to the scientific Institutions which supported this thesis and the observational missions this thesis finalizes a 12 years work on the field of high precision solar astrometry, focusing on methods of measurements, identifying the range of solar diameter variability and rising interest in this subject to young generations. Father Angelo Secchi, who published in 1877 the book "Le Soleil", started the tradition in Rome on solar diameter measurements. Later a 60 years series of measurements on solar diameter made at the Capitol's Observatory become of international interest. The Picard satellite mission will monitor the solar diameter at least in the next 3 years. Daily data from space are expected to be at milliarcsecond level of accuracy. It has been planned that the SODISM data will be exploited to calibrate a ground-based method (SODISM II-MISOLFA now in the Nice/Calern Observatory) for solar diameter measurements to continue monitoring its tiny variations in the years to come.

There are two other methods which can achieve an accuracy comparable with the one of space missions in solar astrometry: they are the eclipses and the transits.

These two methods have been extensively analyzed in this book in order to ascertain their accuracies.

[21] http://www.tlfq.ulaval.ca/axl/francophonie/HIST_FR_s7_Lumieres.htm see chapter 5.1.

# Chapter 1: Solar variability

The title is inspired by "The Sun as a Variable Star" the proceedings of the IAU colloquium n. 143.[22]

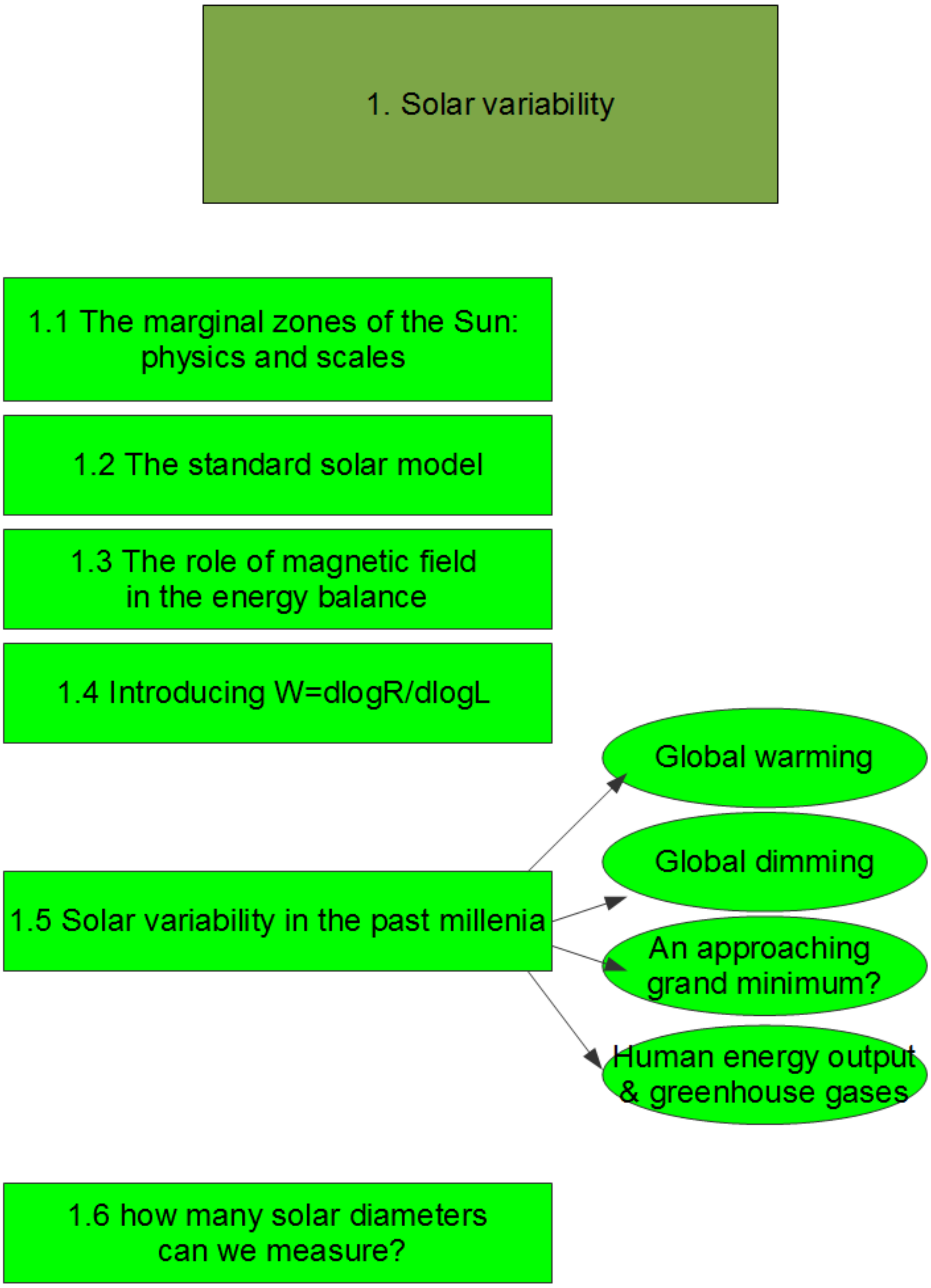

The variation of the luminosity L with radius R is described by a parameter W=dlogR/dlogL. The Picard mission is conceived to measure W accurately, and its value will be used to recover the past values of L_sun from historical measurements of R_sun in order to feed Earth's climate models.

[22] Pap, J. M., C. Fröhlich, H. S. Hudson & S. K. Solanki eds., *The Sun as a Variable Star, Solar and Stellar Irradiance Variations*, Cambridge University Press (1994).

## 1.1 The marginal zones of the Sun: physics and scale

This paragraph is dedicated to the marginal zones of the Sun, i.e. the solar surface and the atmosphere nearby when it is seen from a grazing view. Images, scales and physics are here summarized.

The marginal lines-emission region is the solar mesosphere.
In the following I will describe the reasons of this definition.

The flash spectrum (see fig. 1.1) has been obtained with a spectrograph during total eclipses, and the continuum of the spectrum corresponds to the light coming from the Baily's beads (the photosphere limb with Fraunhofer (F) lines seen in absorption). Also a myriad of faint emission lines are seen simultaneously superposed on this continuum, in the upper part of the figure. The thin layer associated with these faint emission lines between the photosphere (F-lines) and the low chromosphere, corresponds to the region which can be seen only during the internal contacts of a total eclipse.
This particular region has to be taken into account while defining the solar edge, and the continuum spectrum should be measured between these faint emission lines for accurately measuring the true solar diameter.
The spectrum of figure 1.1 has been obtained as the following. A moving plate spectrograph has been used in 1905 eclipse by W. W. Campbell [Campbell, 1906], then director of the Lick Observatory, in order to photograph the evolution of the spectrum with the reduction of the area of the visible Sun.
It is visible the transition between the absorption dark lines (photosphere with Fraunhofer spectrum) and the tiny emission lines. More up appear the lines of the chromosphere (few and brilliant), and finally the corona continuum (not well visible in the image, which has been scanned from the original).

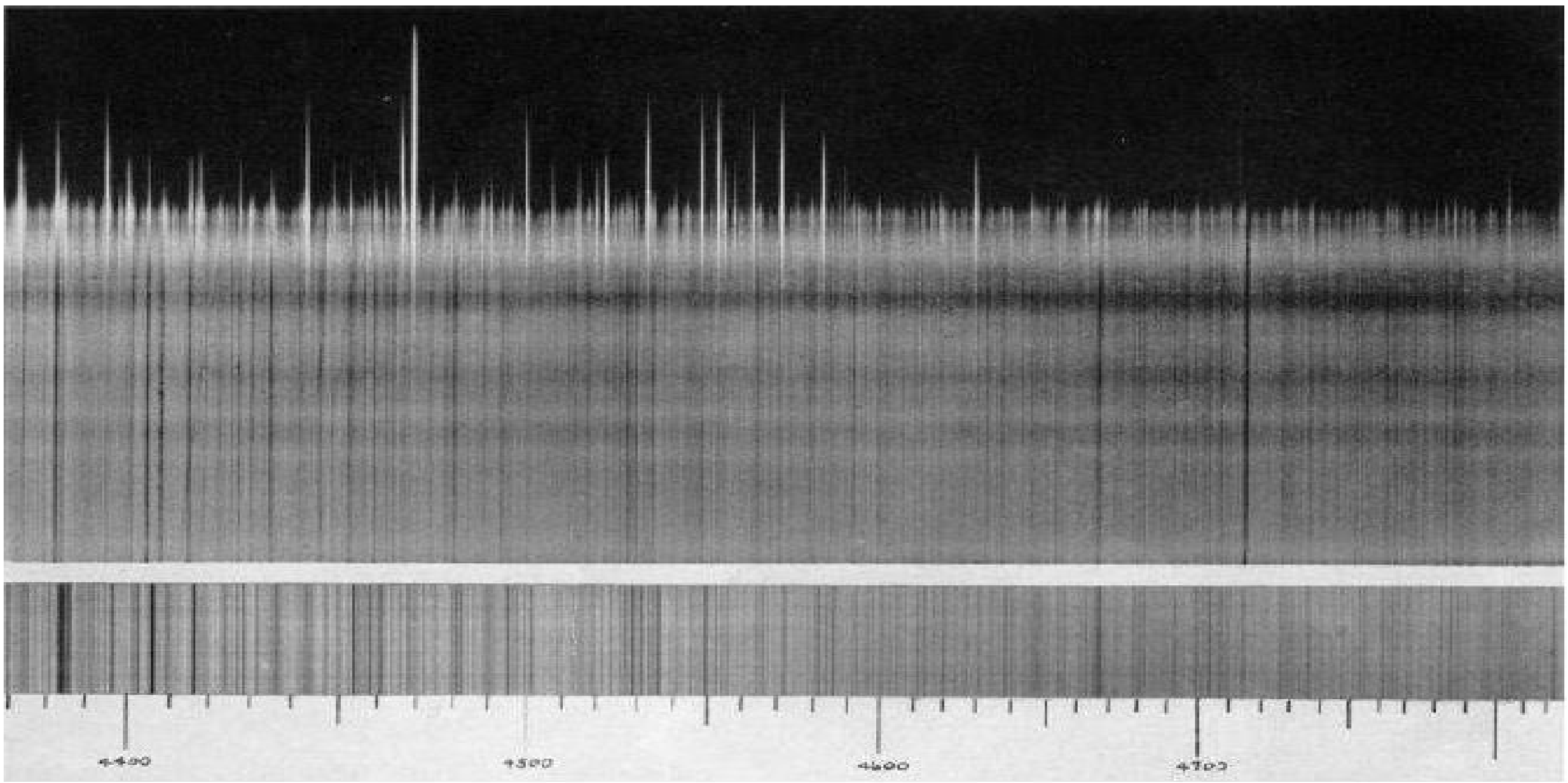

**Fig. 1.1 Flash spectrum of 1905 total eclipse.** The width of the slit was w =1.3mm and the plate moved of  w each second. The lower part of the figure is the spectrum before the totality, where the continuous of the photosphere and a lot of absorption lines. In the upper part there are only the lines of chromosphere; their blend is $10^{-4}$ times the intensity

of the continuum. After the beginnig of totaly only the continuum of the corona can be detected. The blend of the small emitting lines, just below the bright and few lines of the chromosphere, is perceived as white light and it is $10^{-3}$ times the intensity of the photosphere's continuum.

The blend of these tiny emission lines just above the solar limb is perceived as white light, and even if it is about 1000 times dimmer than the photosphere's brightness it can be confused with the photosphere itself, when the photosphere is seen through an area 1000 times smaller and without spectrograph, as it occurs during the last stages of a total eclipse. This successful 1905 eclipse expedition, sponsored by the billionaire Crocker, remained a cornerstone for the following decades.

Nowadays the definition of reversing layer, used in the original paper of Campbell[23] is no longer physically satisfying.

This region, in analogy with the situation in the Earth's atmosphere, can be defined as the **solar mesosphere**.

There the temperature has a local minimum of about 4500 K instead of 5800 K of the photosphere, after the temperature rises at the beginning of the chromosphere.
This region, over the photosphere, starts with a layer where the contrast of the granules changes.
The vertical magnetic field between granules and supergranules emerges and it spreads horizontally.
The propagating waves along the magnetic field become shock waves, and they release the energy in these layers.

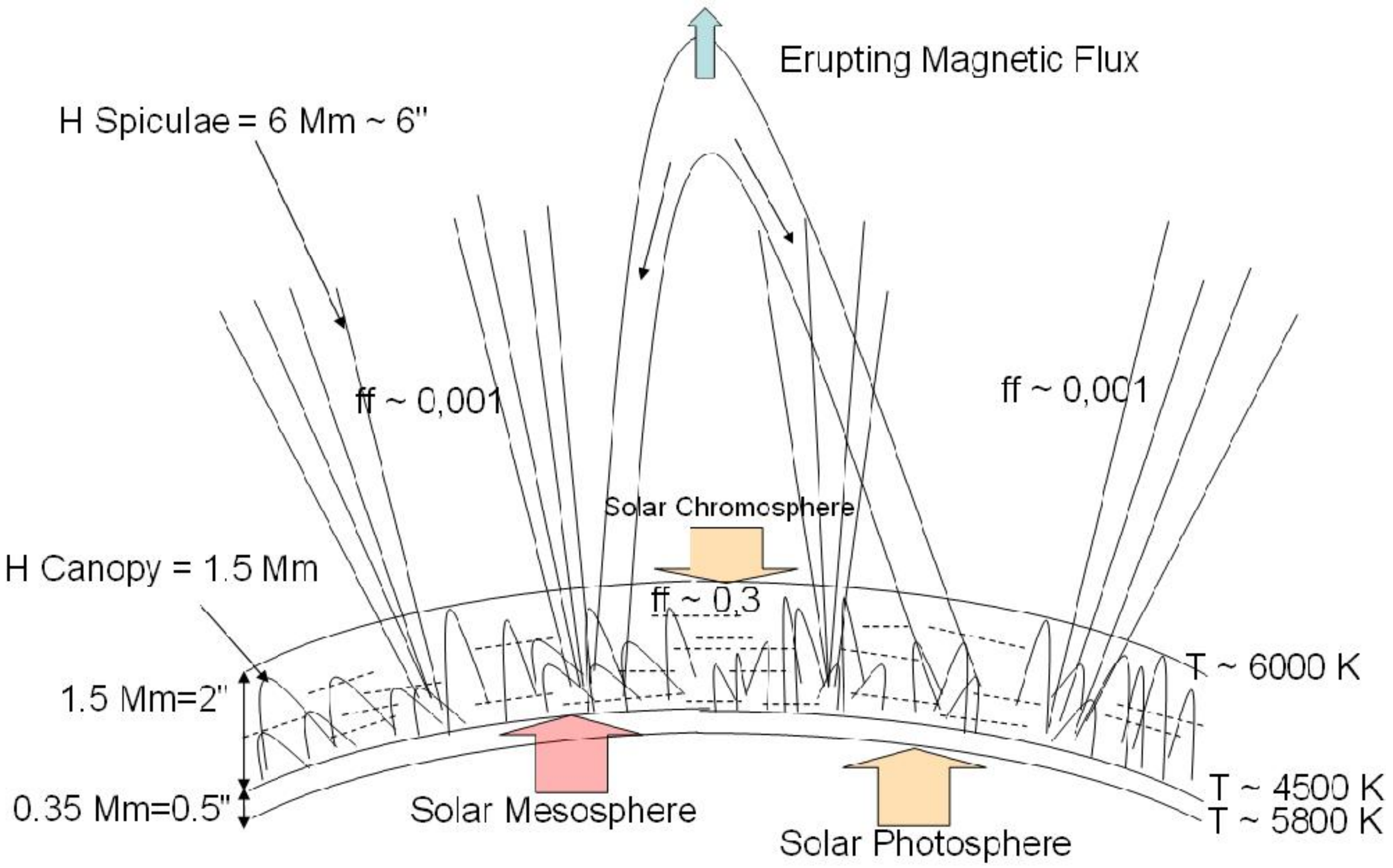

**Fig. 1.2 Diagram of the Solar Mesosphere with length scales.** The Filling Factor (ff) or

[23]Campbell, W.W. and C. D. Perrine, Proc. Astron. Soc. Pacific **18** (1906) 13.

heterogeneity factor, is the ratio between the volume occupied by the plasma over the total volume, because the plasma is magnetically confined. The inversion of the temperature occurs around 350 Km, or 0.50 arcsec above the photosphere.

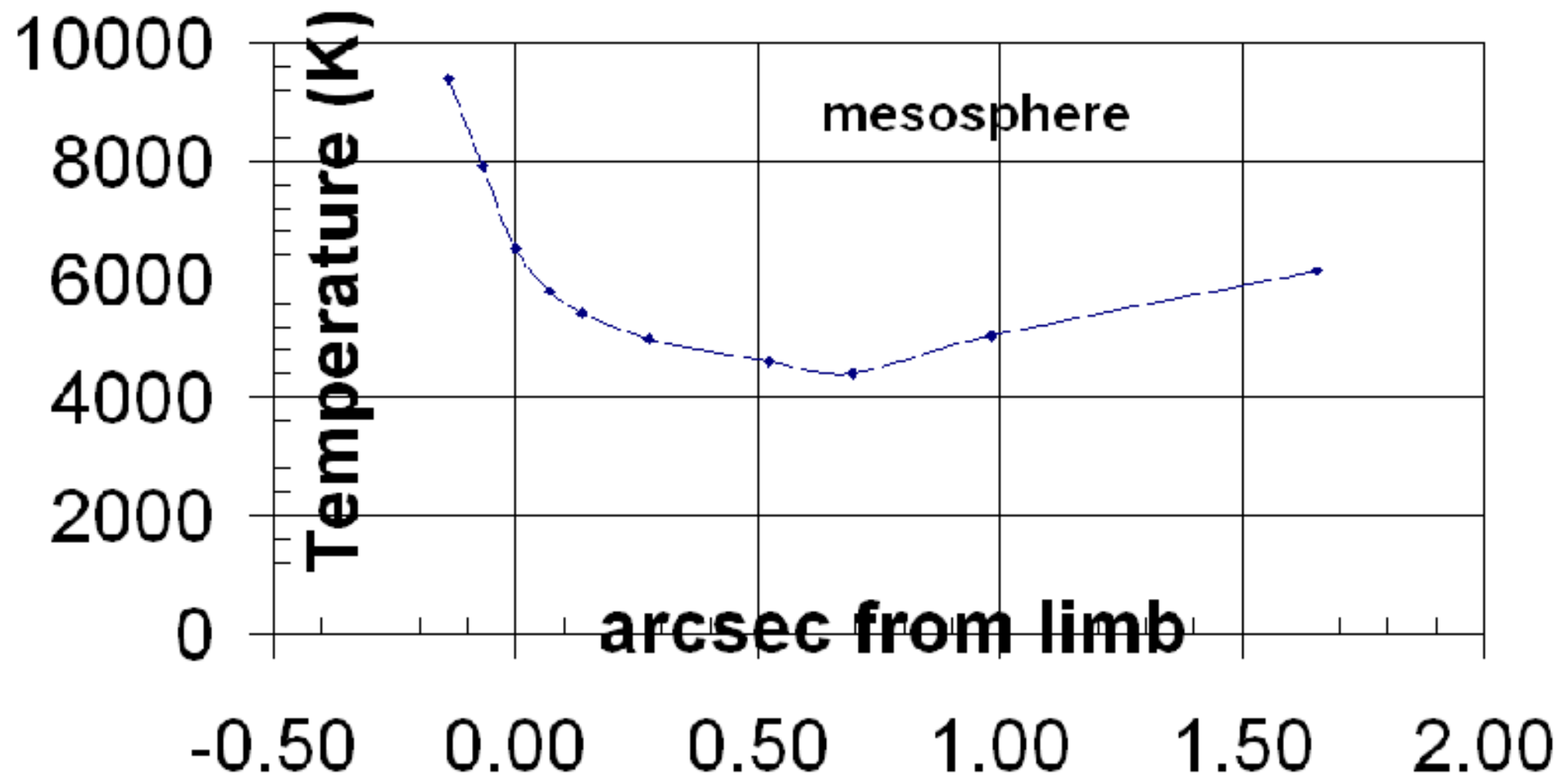

**Fig. 1.3 The temperature's profile at the solar limb.** In an hydrostatic one-dimensional model (data plotted from Foukal[24] tab. 5.2 ) the inversion of the temperature occurs around 350 Km, or 0.50 arcsec above the photosphere.

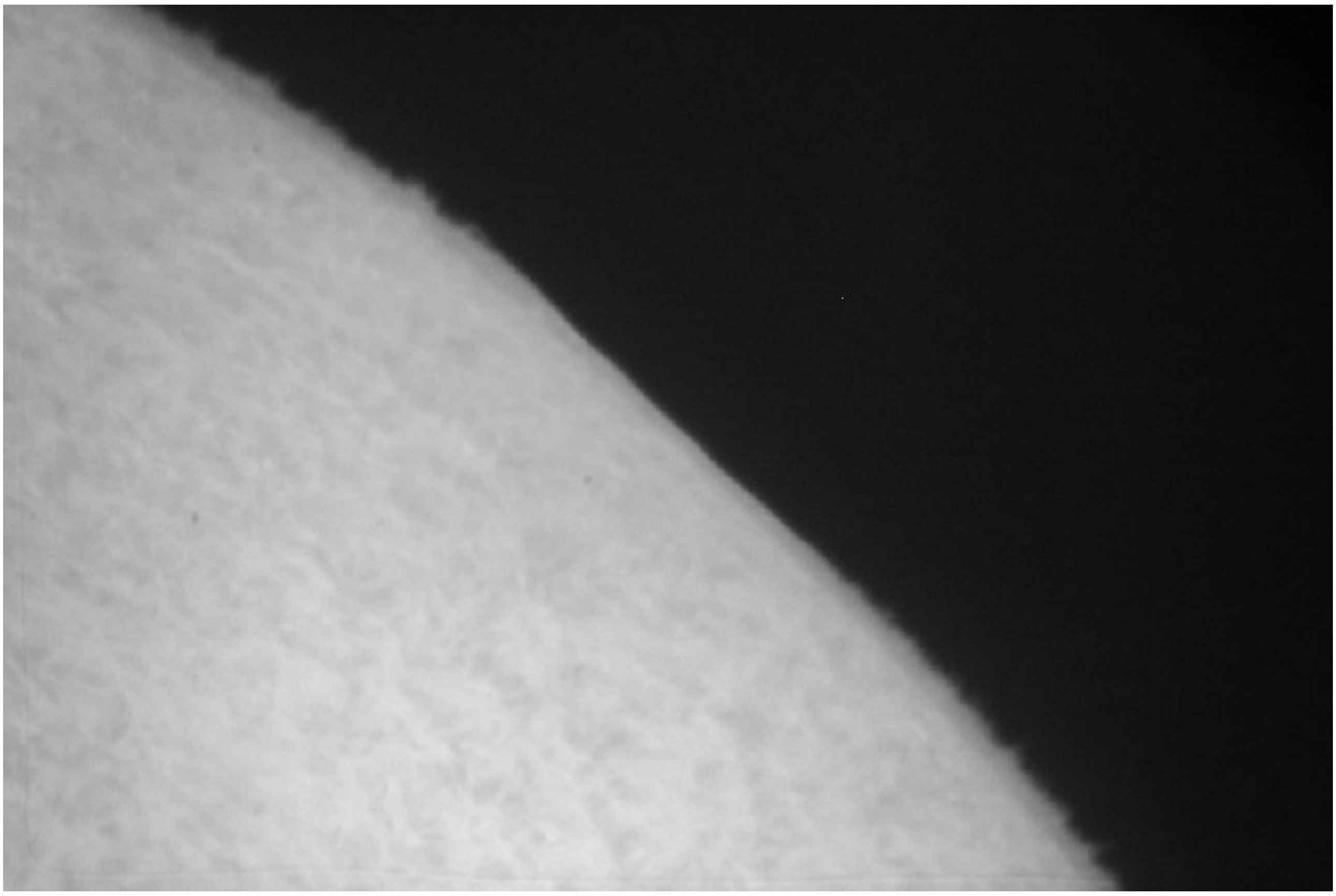

**Fig. 1.4 The solar and the lunar limb: end of January 4th 2011 partial eclipse, in H alpha.**

[24] P. V. Foukal, Solar Astrophysics, (Wiley-WCH, Weinheim, 2004).

The region under study is here imaged with the Bialkow Corograph (Poland) of 53 cm, the largest in the World, during the partial solar eclipse of January 4, 2011 at 10:57:01 UT in the red channel (H alpha).

The end of the eclipse has been calculated (with Occult 4 software) for a solar standard photospherical radius at 10:56:53 UT, which corresponds to the time in which the lunar limb is tangent to the solar limb.
In the following 15 seconds the Moon sweeped on the solar mesosphere and the spiculae.
The lunar limb is leaving the solar disk at a relative velocity of 0.343 arcsec/s, and it is visible in the central part of the figure, cutting partially the spiculae.
By the duration of the transit of the lunar limb (16 s) above the spicules their length is measured with an accuracy of ±1 s, i.e. ±0.3 arcsecs: 5.8±0.3 arcsecs, as presented in the figure 1.2, where 6 arcsec are indicated as the height of the spiculae.

## 1.2 The standard solar model

The standard solar model has two free parameters: the mixing length scale and the helium abundance, and after 4.52 billion year it should return the present radius of the Sun, its luminosity and the observed metal abundance. During the main sequence phase the solar diameter shrunk of 30%.[25]
The Sun is a star which is experiencing the main sequence phase of its evolution, therefore it is stable over a very long time span.

The standard solar model is used as a test case for the stellar evolution calculation because the luminosity, radius, age and composition of the Sun are well determined.[26]

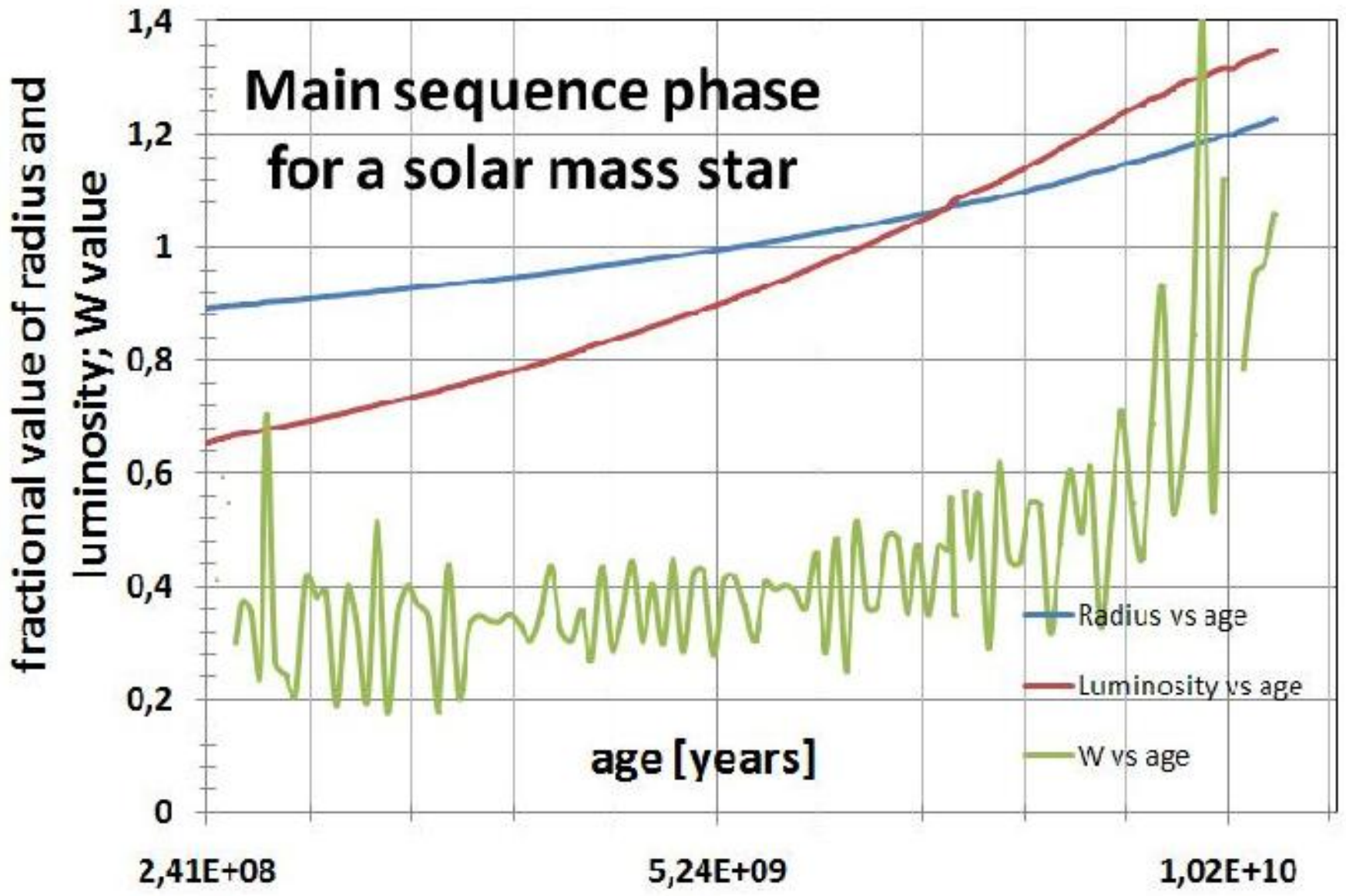

[25] The paradox of faint young Sun has been outilned firstly by Carl Sagan in 1972 is due to the fact that the solar luminosity grows up gradually during the main sequence phase. Early in the Earth's history, the Sun's output would have been only 70% as intense during that epoch as it is during the modern epoch. Different combinations of greenhouse effects have been claimed to solve this paradox. Moreover in the proterozoic epoch the Earth would have experienced periods of complete glaciation: the phenomenon is known as *Snowball Earth* at it is end with volcanic massive immission of carbon dioxyde, which led to an heavy greenhouse effect (A Neoproterozoic Snowball Earth, P. F. Hoffman, A. J. Kaufman, G. P. Halverson and D. P. Schrag, Science **281**, 1342, (1998)).
[26] Guenter, D., *What is a solar model?* http://www.ap.stmarys.ca/~guenther/evolution/what_is_ssm.html (2010)

**Fig. 1.5 A reference evolutionary model for a solar mass star** calculated by L. Siess (1999) on his website of Liège University.[27] I have added the plot of W=dlogR/dlogL as computed from the same data.

Nevertheless certain properties of the Sun are observed to vary during the course of a sunspot cycle.

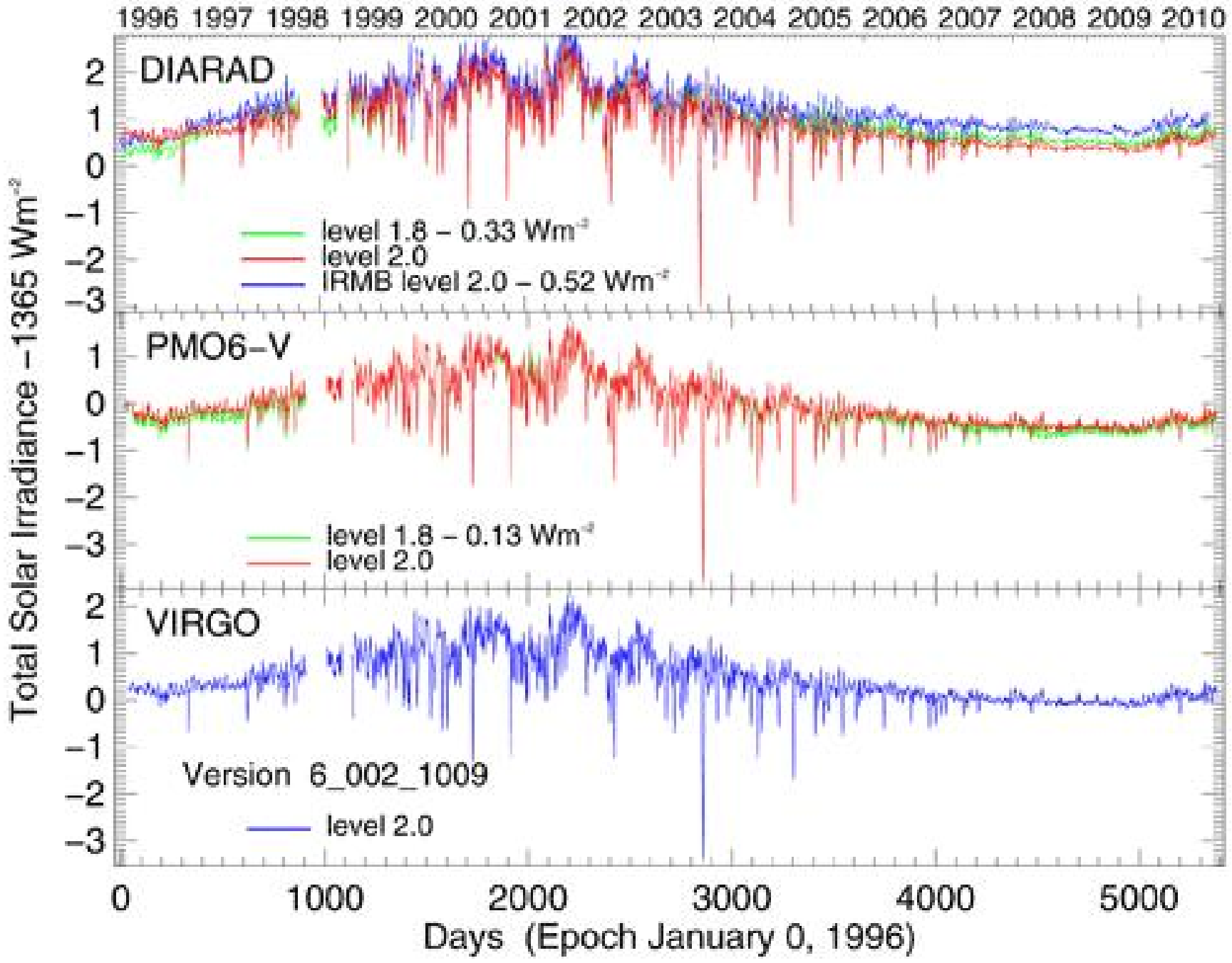

**Fig. 1.6 The total solar irradiance** measured in the last 14 years with the VIRGO experiment onboard of SOHO satellite.[28]

The daily averages of the solar irradiance have excursions between a minimum of 1362 W/m² and a maximum of 1368 W/m². The irradiance is affected by the occurrence of individual sunspots on the disk. Taking into account yearly averages, the peak to peak amplitudes of luminosity[29] are found to be about $\Delta L/L \approx 0.001$.

Moreover the variations of surface temperature are limited at 1.5 ± 0.2 K and they are in phase with other indicators of the cycle[30].
Using these constraints on the equation of Stefan-Boltzmann of a radiating sphere we can set an upper limit for the variations of the solar radius.

---

[27] http://www-astro.ulb.ac.be/~siess/MODELS/PMS/OV02/m1.0z02d02.hrd
it is the evolution of a solar mass star with z=0.02 and overshooting parameter d=2 pressure scale heights. The Sun has Z/X = 0.02497, and this explains why radius and luminosity do not cross each other at 1 value for present age (4.52 billion years).

[28] Source: http://www.ias.u-psud.fr/virgo/

[29] The difference between irradiance and luminosity is that the first can be affected by solar spots or active regions, while the second is independent on the local properties of the solar disk. Irradiance, hereinafter, is the instant value of the energetic output of the Sun while Luminosity is its reference, global, value.

[30] Gray, D. and W. Livingston, *Monitoring the Solar Temperature: Spectroscopic Temperature Variations of the Sun*, Astrophysical Journal v.474, p.802 (1997). Also a secular trend of 0.014 K per year was presented in that paper. No further mention of that trend in Livingston, W., Gray, D., Wallace, L., & White, O. R., Quiet Sun unaffected by Activity Cycle, Large-scale Structures and their Role in Solar Activity ASP Conference Series, Vol. 346, Proceedings of the Conference held 18-22 October, 2004 in Sunspot, New Mexico, USA. Edited by K. Sankarasubramanian, M. Penn, and A. Pevtsov, p.353- 355.

$$\Delta L/L=2\Delta R/R+4\Delta T/T \quad (1.0)$$

So with $\Delta L/L \leq 0.001$ and $4\Delta T/T \leq 0.001$, where T=5777 K for the photosphere, and L=1365 W/m². Assuming both L and T in phase, the amplitude of $2\Delta R/R$ is $\Delta R/R \ll 0.0001$ i.e. $\Delta R \ll \pm 0.1''$; while assuming them in anti-phase $\Delta R/R \approx 0.001$, and the maximum value for radius oscillation is $\Delta R = \pm 1''$where R=959.26" at 1 AU of average distance.
We can consider these constraints on the possible variations of the diameter of astrophysical origin.

Nevertheless if the variation of the radius is due only to an outer layer, the whole variation can be larger, with a reasonable energy balance.

Variations of the order of 0.1 arcsec per solar cycle can be explained by various models, with different energy involved.
A variation of the solar radius implies a variation in the gravitational energy according to the following equation (1.1) where E is the accumulated energy, R is the variation of the solar radius, G the universal constant of gravitation and M the solar mass. A variation ΔR=0.1% requires an energy variation 75 times larger than the energy irradiated.[31]

$$\Delta E=\Delta R\cdot(3GM^2/5R^2) \quad (1.1)$$

Considering the radiant surface and the energy flux per unit area as constant, the radius variation produces a variation of the irradiance I according to the formula (1.2) where R is the solar radius

$$\Delta I=-\Delta R\cdot(2I/R) \quad (1.2)$$

Therefore with a ΔR/R=0.1% half of the variation of irradiance (the total is ΔI/I=0.1%) during a solar cycle is explained, the other half requires other explanations. If the radius variation is restricted only to the convective zone of the Sun, the observed variation of 0.1% requires a negligible gravitational energy of about 0.01% during a whole solar cycle.[32]

## 1.3 The role of the magnetic field

The energy stored in the magnetic field plays a fundamental role in the energy balance of the Sun, as well as the temperature of the photosphere and the diameter of the Sun.

Nevertheless the magnetic field is not taken into account in the standard solar model.
Moreover understanding the reasons of the cyclic variation of the total solar irradiance is one of the most challenging targets of modern astrophysics. These studies are essential also for climatologic issues, associated to the global warming discussion.
All attempts to determine the solar contribution to this phenomenon must include the effects of the magnetic field, whose strength and shape in the solar interior are far from being completely known.

Modelling the presence and the effects of a magnetic field requires a two-dimensional approach, since the assumption of radial symmetry is too limiting for this topic. A 2D evolution code has been introduced by

---

31 Boscardin,S., A. Andrei, E. Reis Neto, J. Penna, V. d'Ávila, W. Duarte, P. Oliveira, XXXIII Reunião da SAB, resumo 362 (2007).

32 Boscardin, S. C., *UM CICLO DE MEDIDAS DO SEMIDIÂMETRO SOLAR COM ASTROLÁBIO*, Ph D Thesis Observatorio Nacional, Rio de Janeiro, Brasil, 2011.

Sabatino Sofia and his team at Yale University: rotation, magnetic field and turbulence are taken into account in such a model.[33]

There are other stars of colour index B-V close to the solar one B-V = 0.66, which show similar behavior in the chromospheric flux, with periodicities ranging from 7 to 16 years, very close to the solar cycle of 11 years.
The study of these stars will help to place the solar activity in a general perspective.

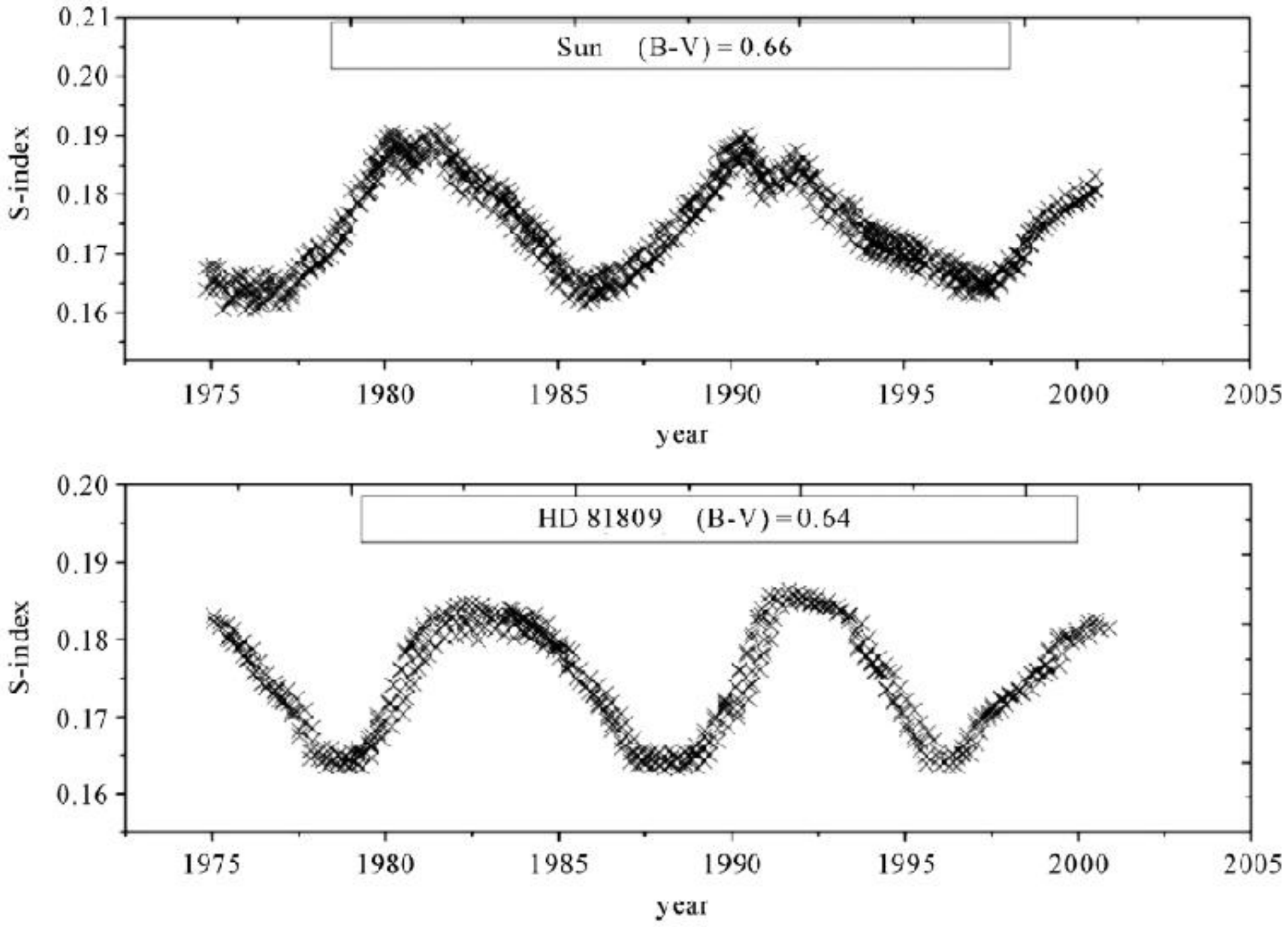

**Fig. 1.7** The relative intensity of Ca II H (396.8 nm) and K (393.4 nm) emission lines for the Sun and the star HD 81809 (HR 3750) of Mv=5.4, Mount Wilson Observatory.[34]

## 1.4 Introducing W=dlogR/dlogL

The luminosity and the radius of the Sun are involved in the Stefan-Boltzmann equation, but their logarithmic derivative W is not ½ as for uncorrelated quantities, because of the magnetic field.

It is introduced the parameter W=dlogR/dlogL in order to describe the connection between luminosity and radius. For an isothermal Sun W=0.5.
From the model plotted in fig. 1.5 W ranges from 0.2 to 0.6 during the main sequence phase of a solar-type star, but it is from the Picard satellite mission that this parameter is expected to be determined observationally with an unprecedented accuracy.

The W parameter can be considered as a characteristic of the Sun during the present phase of its evolution, and it will help to know the solar luminosity in the past centuries, if some measurements of the diameter are available.

Data on the past solar diameter are expected to be obtained by total eclipses timing, with the required

[33] Li, L., S. Sofia, P. Ventura, V. Penza, S. Bi, S. Basu and P. Demarque, *Two-Dimensional Stellar Evolution Code Including Arbitrary Magnetic Fields. II. Precision Improvement and Inclusion of Turbulence and Rotation*, ApJS **182** 584 (2009).

[34] Bruevich, E., *On chromospheric variations modeling for main-sequence stars of G and K spectral classes*, Natural Science **3** 641 (2011).

accuracy to feed opportune climate models. Eclipses are treated in chapter 3.

For a daily monitoring it is necessary a direct measurement or the timing of a transit on a meridian, an hourly circle or an almucantarat.
The methods based on timing are the other subject of this thesis in chapter 2.

## 1.5 Solar variability in the past millennia

The existence of ice ages either in the last million year and in the pre-cambrian age has been explained also with astronomical causes (Milankovitch cycles), but the periods of global warming and cooling in the past millennia have indeed a solar origin. Nowadays the global warming seems to be anthropogenic through the greenhouse effect. But this is still an open question with political, economical and social implications. Conversely there are emerging topics as the global dimming and the debate about the approaching new solar grand minimum.
The study of solar activity along centuries has permitted to identify a number of phenomena which are reliable indicators of the solar activity, as it is the solar spot number, available with continuity only since 1700.
These indicators, or proxies, show that the Sun undergoes some variability experiencing periods of grand maxima (10 to 15% of the time), normal activity and grand minima (15 to 20% of the time).[35]

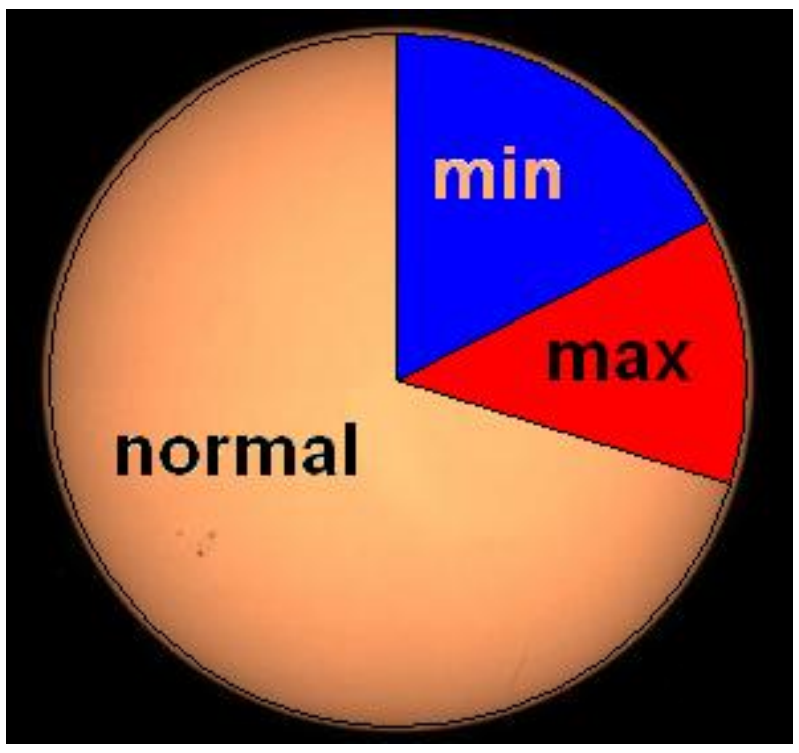

**Fig. 1.8** The ratio between grand minina, grand maxima and normal activity.

The Maunder minimum, occurred between 1645 and 1715, is one of these period of grand minima, the only one witnessed in the telescope age; another relevant minimum has been named Dalton minimum and occurred around 1810.[36]

[35] Usoskin, I., *A History of solar activity over millennia*, Living Rev. Solar Phys. **5**, 3 (2008). also available on http://cc.oulu.fi/~usoskin/personal/lrsp-2008-3Color.pdf

[36] Axel D. Wittmann, German astronomer and historian of science answered to my question about the names of solar minima in the following way. “If I remember correctly, some of the extended solar activity minima were named by John "Jack" Eddy in the 1970-ies after persons who investigated sunspot number records - in particular Wolf [Zürich], Spoerer [Potsdam], Maunder [Greenwich], and Gleissberg [Istanbul]. I knew Gleissberg in person, but I think the "Gleissberg cycle" is, like most other sunspot cycles except the 11-year cycle, a transient phenomenon which disappears after several hundred or thousand or hundred thousand years... I do not know much about the chemist and "atomic physicist" John Dalton (ca. 1766-1844) and his possible work in solar physics. But he flourished (lived) around the time of the "Dalton Minimum" (i.e. around 1810) which may be the reason for giving it his name. What remains is the Oort Minimum (around 1050) named after the Dutch astronomer Jan Hendrik Oort, who in 1927 investigated the rotation of the Milky Way (see "Oort's constants"). I am not aware of any "solar cycle research" by Oort, except that he was a member of the allied commission that investigated solar research in Germany after World War II. But more recently some evidence has accumulated that variations (fluctuations) of cosmic ray intensity - due to the motion of the Sun around the center of the galaxy - may cause climatic changes by influencing the formation of clouds in the Earth's

The Dalton minimum is the most similar to the recent exceptionally low solar minimum, occurring at the end of cycle XXIII and the rumor of an approaching grand minimum is gaining consensus in the science community.[37]

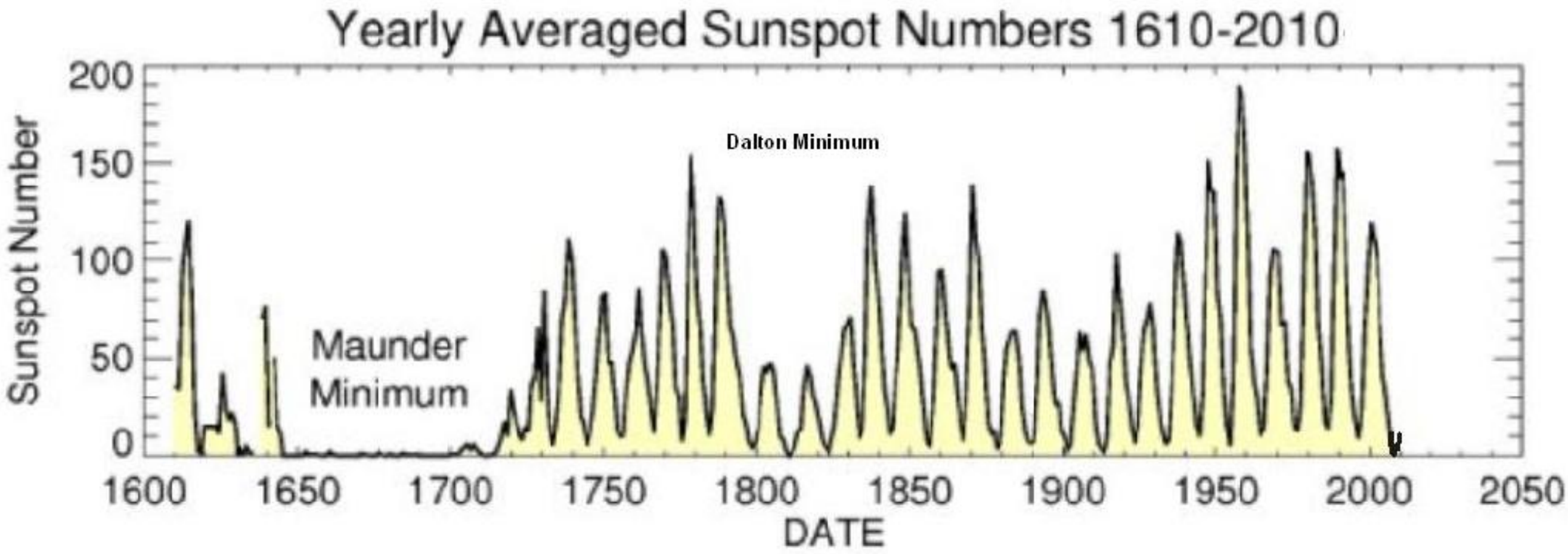

**Fig. 1.9 The yearly averaged sunspot numbers from 1610 to 2010 from SIDC website.[38]**

According to the observations from space of a large change in solar UV emission in the same phase of previous solar cycles there is evidence for centennial-scale change.[39]
Other confirmations come from the open solar magnetic flux derived from geomagnetic observations.[40]

atmosphere. In one of my publications (around 1980) I have tentatively suggested to name the most recent maximum (approx. around 1980/1990) "Eddy maximum" in honor of Jack Eddy's work on the solar cycle, but I do not know whether this has found a wider acceptance.”

[37] Lockwood, M., *Solar change and climate: an update in the light of the current exceptional solar minimum*, Proc. of the Royal Society A, **466**, 303-329 (2010).

[38] http://www.sidc.be/

[39] Lockwood, M. , et al., *Top-down solar modulation of climate: evidence for centennial-scale change*, Environmental Research Letters **5** 034008 (9pp) 2010, IOP Publishing Ltd.

[40] Lockwood, M. , et al., Are cold winters in Europe associated with low solar activity?, Environmental Research Letters **5** 024001 (7pp) 2010, IOP Publishing Ltd.

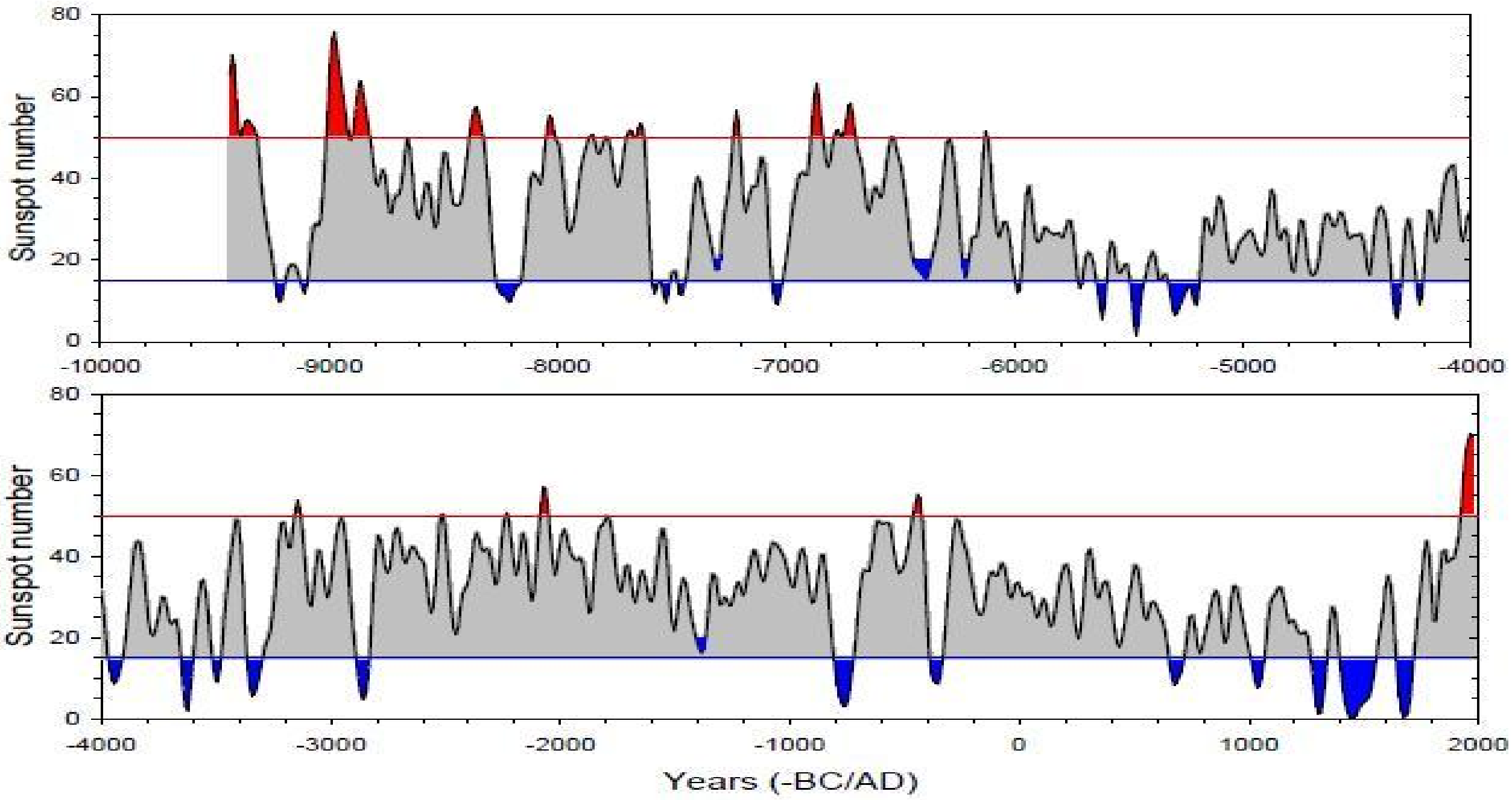

**Fig. 1.10 The solar activity in the past millenia**, with grand minina and grand maxima evidenced in blue and red (from Usoskin, 2008).[41]

### 1.5.1 Global Warming

There is a long debate with this title, peaked with a Nobel Peace Prize assigned to a political personality.
The basic question is if this is anthropogenic, or if its origin is solar. Looking at the history of solar activity in the last few millennia, the role of mankind seems marginal.
Nevertheless the debate nowadays gained momentum because the anthropogenic energy on a yearly base corresponds to the % variation of the solar irradiance observed in the last three cycles.
In other terms the annual input of energy from human activities on the Earth system corresponds to one single hour of the solar input, i.e. $1/365/24=1.1\cdot 10^{-4}$ of the total annual input.
The variations of the solar irradiance (see fig. 1.6) range up to 1 part over 1000, during a solar cycle, so the human component starts to be significant with respect to these solar variations.
Here is not the place where to present all the issues related to the Global Warming, there are two books written by solar physicist which I recommend: "The role of the Sun in climate change"[42] and "The Maunder minimum and the variable Sun-Earth connection".[43]
For the variability of the interpretations of data it is very interesting the website dedicated to the "infamous hockey stick"[44] which would have paved the way to that Nobel Prize...

---

[41] Usoskin, I., *A History of solar activity over millennia*, Living Rev. Solar Phys. **5**, 3 (2008).
[42] Hoyt, D.V. and Schatten, K.H., *The role of the Sun in climate change*, New York Oxford, Oxford University Press, (1997).
[43] Wei-Hock Soon, W. and S. H. Yaskell, The Maunder minimum and the variable sun-earth connection, River Edge, NJ: World Scientific (2003).
[44] http://www.john-daly.com/hockey/hockey.htm (2004).
The author is J. L. Daly, he is not a professional scientist, but this field is now belonging to politcs, and all opinions have to be considered. Nevertheless even the Nobel Prize for peace are not necessarily chosen by a scientific board, and moreover he publishes books on the Global Warming sold in the science section of the libraries from US to South Korea, like Kyobo Bookstore in Seoul visited twice in 2005 and 2009 by me.

What it is certain is that a global vision of the climate on the Earth is a very challenging task.
The dendrochronology helps to see locally the alternance between dry and wet years. Hot and cold seasons can be detected by the time of blossoms of some particular trees, but all these data can be valid in a restricted area. The lack of general information on the climate over the whole planet makes the interpretation of such data rather arbitrary.
Another example of problems in data analysis is the record of temperature.[45] The meteorological stations which in the XIX century were out of the cities now are included in the urban area. All cities behave like *heat islands*, storing the solar heat during the day and releasing it during the night.
Therefore the correctness of measurements on Global Warming on the Earth is a very delicate topic.
Moreover the models dealing with greenhouses gases are very delicate for what concern the servomechanism which would let the temperature exponentially grow once the raising started.
The fact that the atmosphere of the Earth contained more than 30 times the present amount of $CO_2$ during Jurassic,[46] or just before Cambrian, shows that the planet could bear such gaseous composition without losing its equilibrium, and also that the planet undergo greenhouse conditions without the contribution of mankind.
The influence of solar minima on the climate on Earth has been evidence by the authors above mentioned.
Frost fairs on the river Thames in London during 1814 at the end of the "Dalton minimum" is one example.
The "small ice age" during the Maunder minimum (1645-1715) has been represented by various artists.
Also the consequences of the explosion of Krakatoa volcano (1887) were represented by painters, and a study on the color of sunsets has been conducted in Greece. Sunsets appeared redder, caused by enhanced absorption. The paintings were analyzed with astronomical techniques as for looking for a color index. [47] Another effect has been a more pronounced and prolonged twilight, similar to what is sometimes produced by high clouds in the stratosphere: polar stratospheric clouds (PSC's). In that case twilight might also be spread wider along the horizon than usual.[48]

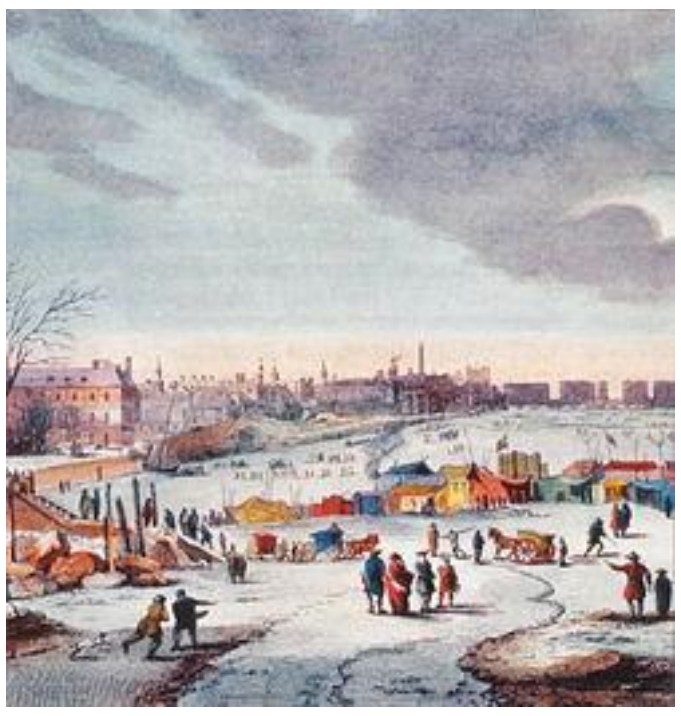
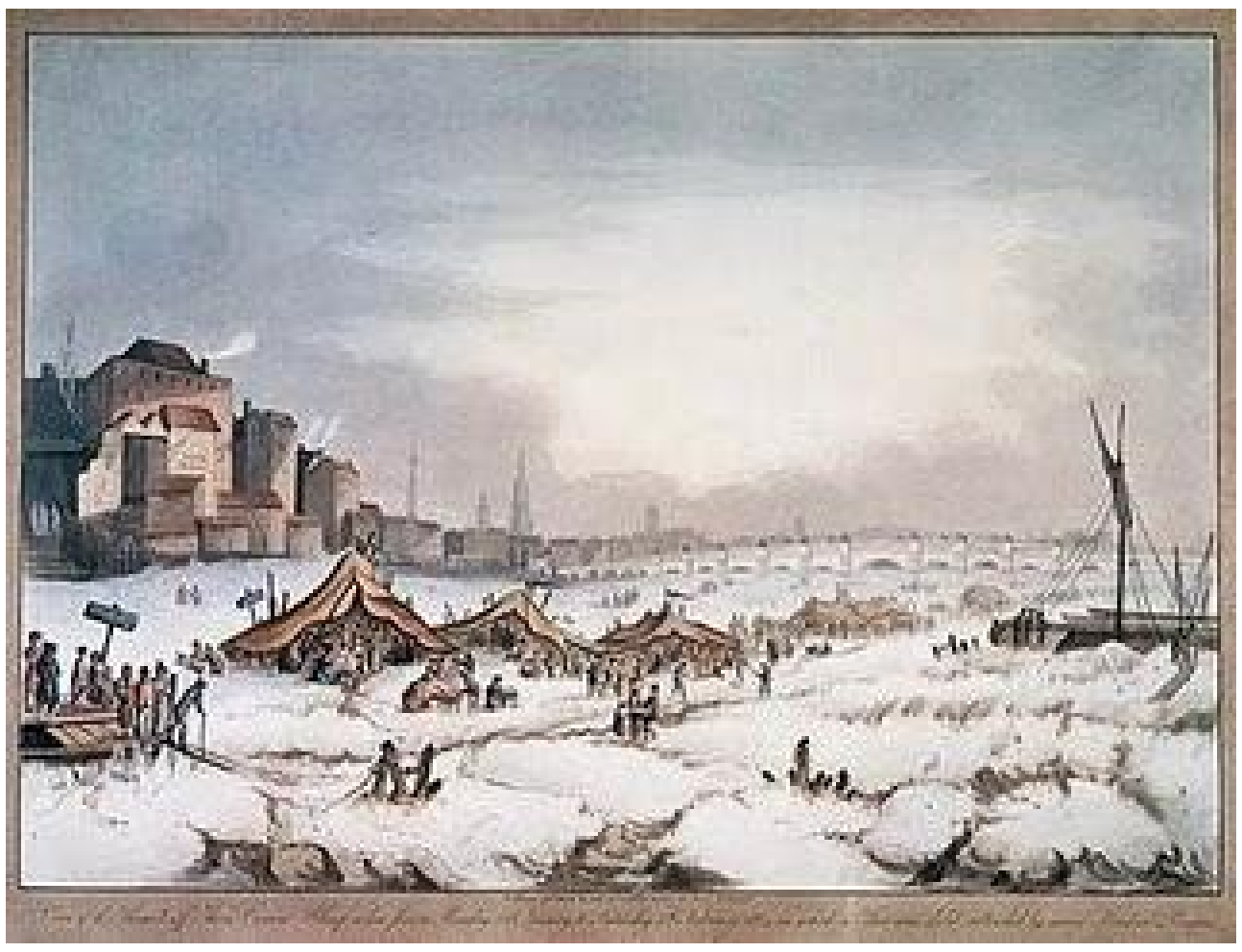

**Fig. 1.11 Frost fairs on the river Thames in 1814.**

[45] On this topic there are the books of Leroy Ladurie, E., Histoire humaine et comparée du climat. Canicules et glaciers, XIIIe-XVIIIe siècles, Paris, Fayard (2004). Alexandre, P., *Le climat au Moyen Age*, Paris, EHESS, 1987.
Titov, J., *Evidence of Weather in the Account Rolls of the Bishopric of Winchester*, 1209-1350, in Economic History Review, 1960
[46] That news is in a diorama exhibited in the Science Museum of Sydney, accessible to the kids, as well as the information that the oceans can absorb large amounts of $CO_2$.
[47] Zerefos., C., V. Gerogiannis, D. Balis, S. Zerefos, A. Kazantzidis, *Atmospheric effects of volcanic eruptions as seen by famous artists and depicted in their paintings*, Atmos. Chem. Phys., **7**, 4027-4042 (2007).
[48] Van der Verfe, S. Private communication (2010), in the occasion of the volcanic eruption in Iceland, which blocked the fly in the northern europe for a few weeks in April 2010.

## 1.5.2 Global Dimming

The reversal of the Global Warming is also claimed. The condensation trails (“contrails”) of the airplanes act like a filter for the solar radiation coming toward the Earth’s surface.[49] The contrails reduce the daily temperature range.[50] During the period 11–14 September 2001, without commercial airplanes flying on the US it was registered an anomalous increase in the average diurnal temperature range (that is, the difference between the daytime maximum and night-time minimum temperatures) . Because persisting contrails can reduce the transfer of both incoming solar and outgoing infrared radiation, and so reduce the daily temperature range, at least a portion of this anomaly has been attributed to the absence of contrails over this period.

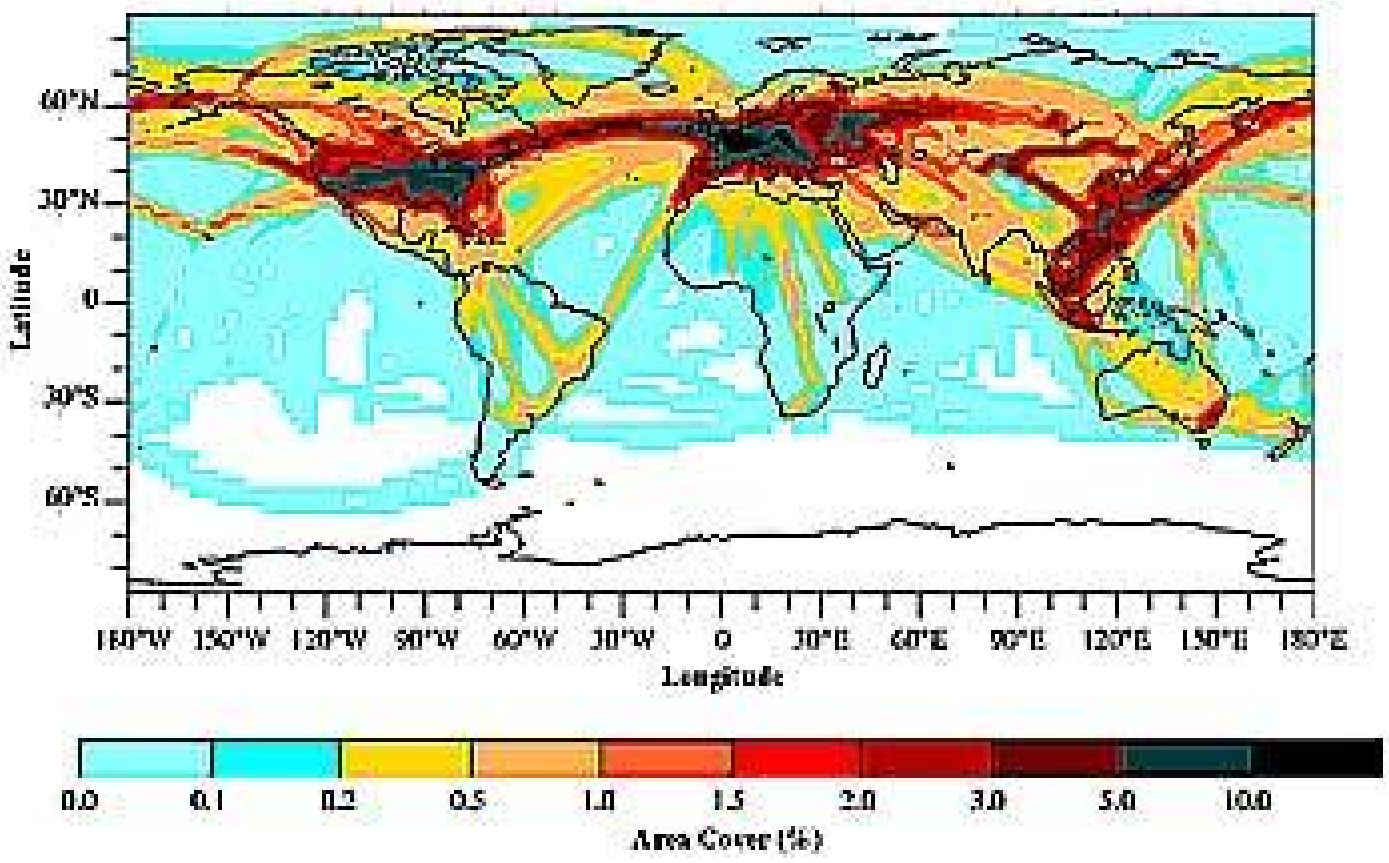

**Fig. 1.12 Persistent contrail coverage** (in % area cover) based on meteorological analysis data and on fuel emission database for 2050 (fuel consumption scenario for 2050), assuming linear dependence on fuel consumption and overall efficiency of propulsion h of 0.5; global mean cover is 0.5%.[51]

More in general also smog is a source for the Global Dimming, which would amount to 4% of solar radiation arrived to the Earth since 1960.[52]

## 1.5.3 An approaching Grand Minimum?

After an unusual period of no spot activity, the XXIV cycle of the Sun has started to resume a rather normal activity in the month of July 2011.[53] Groups of spots appeared on the solar surface, leading the international solar community to switch the theme of solar meetings from "Solar Astrometry and Grand Minima of Activity"[54] to "The Sun: from quiet to active – 2011".[55]

---

[49] http://www.ipcc.ch/ipccreports/sres/aviation/index.php?idp=40

[50] Travis, D. J., A.M. Carleton and R. G. Lauritsen, Nature **418** 601 (2002).

[51] Source: IPPC, Aviation and the Global Atmosphere, http://www.ipcc.ch/ipccreports/sres/aviation/index.php?idp=40 fig 3.23. According to these authors the increasement of the coverage by contrails will correspond to an increasement of the heat through greenhouse effect. Just the contrary of the previous article of Travis et al. on Nature. The IPPC Intergovernmental Panel on Climate Change is also the recipient of the aforesaid Nobel Peace Prize. Because of its scientific and intergovernmental nature, the IPCC embodies a unique opportunity to provide rigorous and balanced scientific information to decision makers. By endorsing the IPCC reports, governments acknowledge the authority of their scientific content. *The work of the organization is therefore policy-relevant and yet policy-neutral, never policy-prescriptive.* That Nobel Price recipient was indeed not neutral.

[52] http://en.wikipedia.org/wiki/Global_dimming

The long period of time with the Sun spotless, almost 3 years, has been compared with the "Dalton minimum" occurred in 1810-14.

The trend of diminution of surface magnetic field of the Sun, would led to zero within the present decade. Livingstone and Penn have observed spectroscopic changes in temperature sensitive molecular lines, in the magnetic splitting of an Fe I line, and in the continuum brightness of over 1000 sunspot umbrae from 1990-2005. All three measurements show consistent trends in which the darkest parts of the sunspot umbra have become warmer (45K per year) and their magnetic field strengths have decreased (77 Gauss per year), independently of the normal 11-year sunspot cycle. A linear extrapolation of these trends suggests that few sunspots will be visible after 2015.

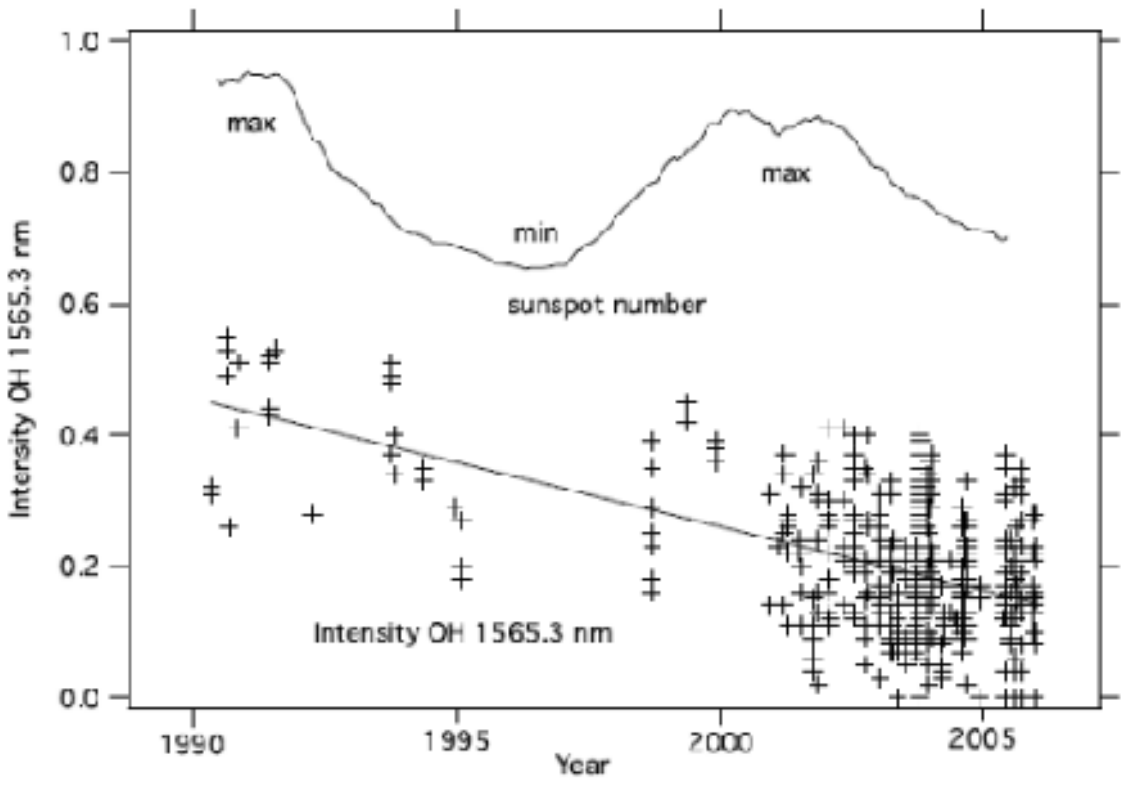

**Fig. 1.13 The line depth of OH 1565.3 nm for individual spots.** The upper trace is the smoothed sunspot number showing the past and current sunspot cycles; the OH line depth change seems to smoothly decrease independently of the sunspot cycle.[56]

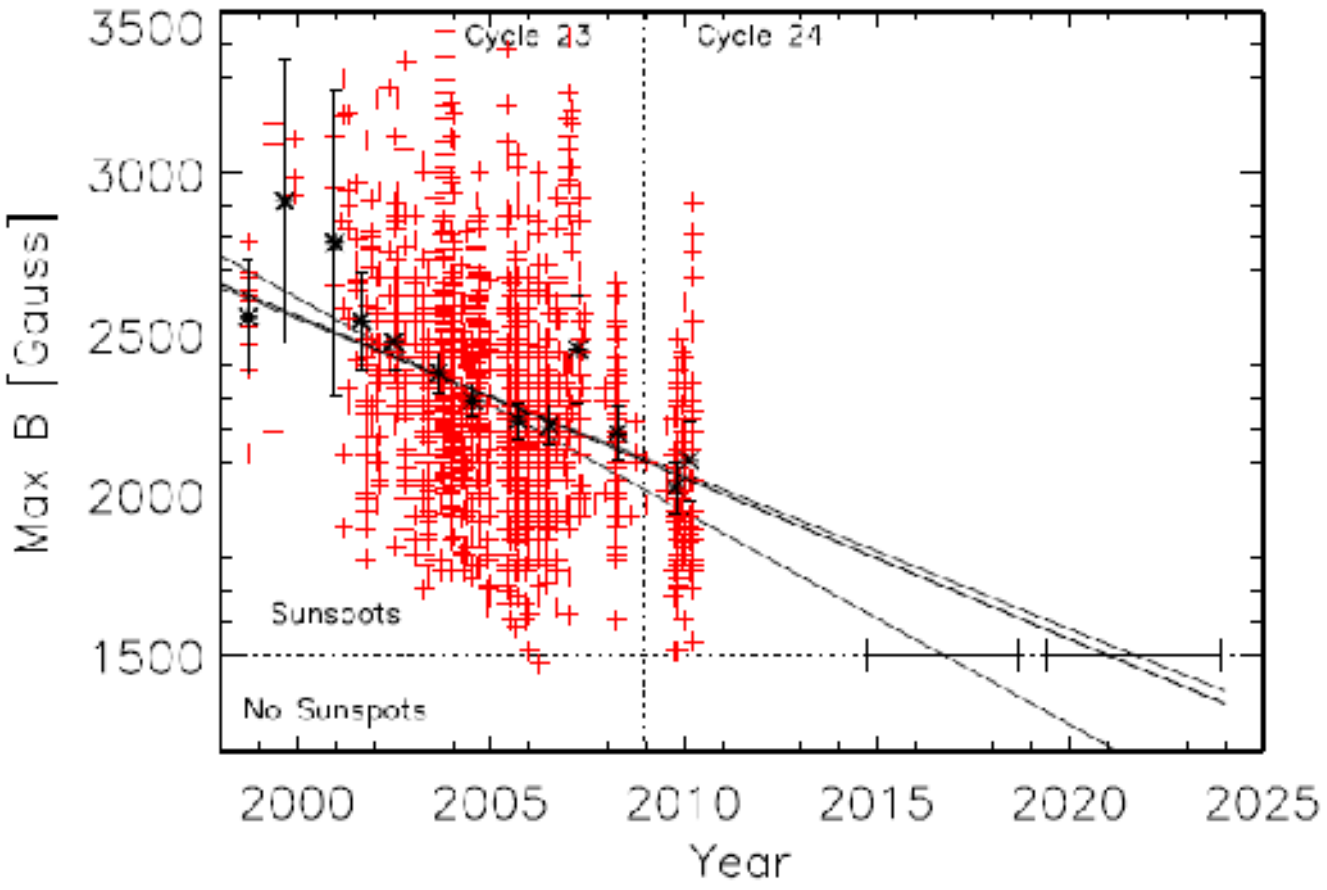

**Fig. 1.14 Measurements of the total magnetic field strength at the darkest location in umbrae and pores as a function of time.** The crosses show the individual measurements, the asterisks show annual bins. Three linear fits are shown: the bottom fit line fits data

[53] http://www.sidc.be/products/quieta/

[54] Galileo – Xu III meeting, Beijing, October 11-15, 2011, parallel session, conceived in March 2011.

[55] International Workshop on Solar Physics, Lebedev Institute of the Russian Academy of Sciences, August 29 – September 02, 2011.

[56] Penn, M.J. and Livingston, W., ApJ (Letters), **649**, L45 (2006).

from 1998-2006 as done in their 2006 paper. The top line fits all the data from Cycle 23, and the middle line fits all of the data.[57]

The absence of spot would mean a new grand minimum, like the minimum of Maunder.
According to other theories there is a cycle of climatic changes lasting from 80 to 90 years, determined by the movements of the center of mass of the solar system.

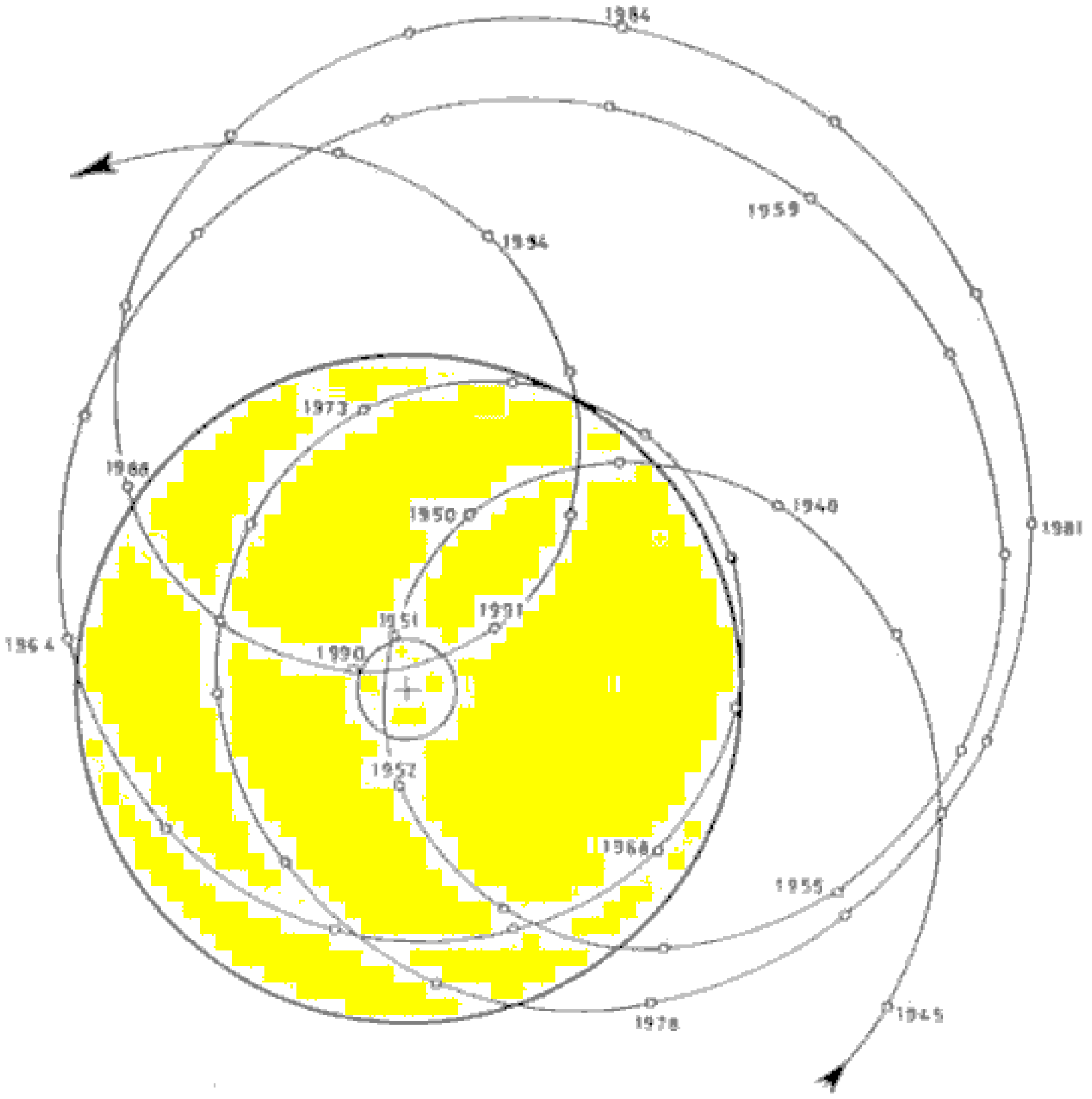

**Fig. 1.15 Motion of the baricenter of the solar system in and around the Sun.[58] The baricenter departs up to 2.2 solar radii from the center of the Sun.**

Predictions of maxima of single solar cycle based upon this indication[59] failed for predicting the time of the present 24th Gleissberg cycle, whose maximum was predicted on 2011.8 while the Sun remained in unusual very low activity and only during 2012 started to show an increasement of suspots activity. Nevertheless the expected strenght of the cycle (R<80) was correct, and the influence of the motion of the center of mass of the solar system upon the sunspot cycle cannot be excluded.
Other theories associate the angular momentum of the solar system with the sunspot cycles, and they involves in particular 179 years-cycle of Uranus and Neptune alignements.[60]

---

[57] Penn, M.J. and Livingston,W. http://arxiv.org/abs/1009.0784 (2011).
[58] Landscheit, T., *Swinging Sun, 79-Year Cycle, and Climatic Change*, J. Interdiscipl. Cycle Res., 1981, vol. 12, number 1, pp. 3-19 (1981).
[59] Landscheit, T., *Extrema in sunspot cycle linked to Sun's motion*, Solar Physics 189, 413 – 424, (1999).
[60] Sharp, G., *Are Uranus & Neptune responsible for Solar Grand Minima and Solar Cycle Modulation?* http://arxiv.org/pdf/1005.5303v3 (2010).

## 1.6 Several solar radii

There is a different radius for each wavelength, also for radio. There is also the seismic radius. How accurate are these values, and how big are their variations?

The solar diameter is defined as the position of the inflexion point of the limb darkening function. This function is the intensity variation of the Sun a
long its radius.

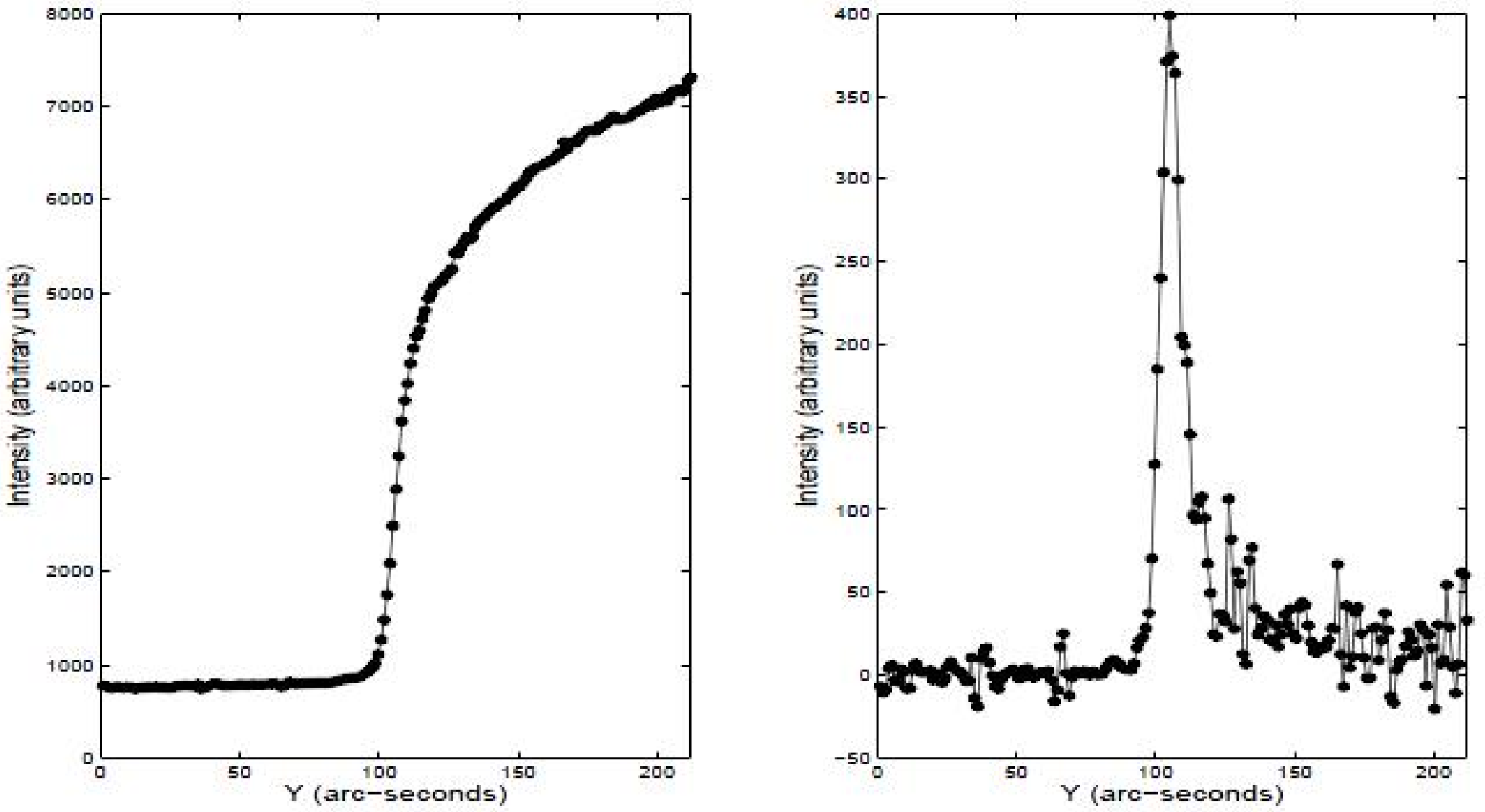

**Fig. 1.16 Solar limb profile and its derivative of the image of the Sun at 535.75 nm made with SODISM II on August 11, 2011.**[61]

The different solar diameter measured at different wavebands is due to the differet height sampled within different wavebands.

Neckel and Slabs[62] have published accurate measurements of the solar diameter at the Kitt Peak observatory around the visible range of wavelengths, from 303 to 1099 nm.

These measurements still remain a fundamental reference for all models of solar emission.

The recent measurements made with the instruments of Picard-Sol show very well the difference between the wawelenghts.

[61] Irbah, A., M. Meftah , T. Corbard, R. Ikhlef, F. Morand, P. Assus, M. Fodil, M. Lin, E. Ducourt, P. Lesueur, G. Poiet, C. Renaud and M. Rouze, *Ground-based solar astrometric measurements during the PICARD mission*, submitted to SPIE (2011), fig. 12.

[62] Neckel H. and D. Slabs, Solar Phys. 153 (1994) 91.

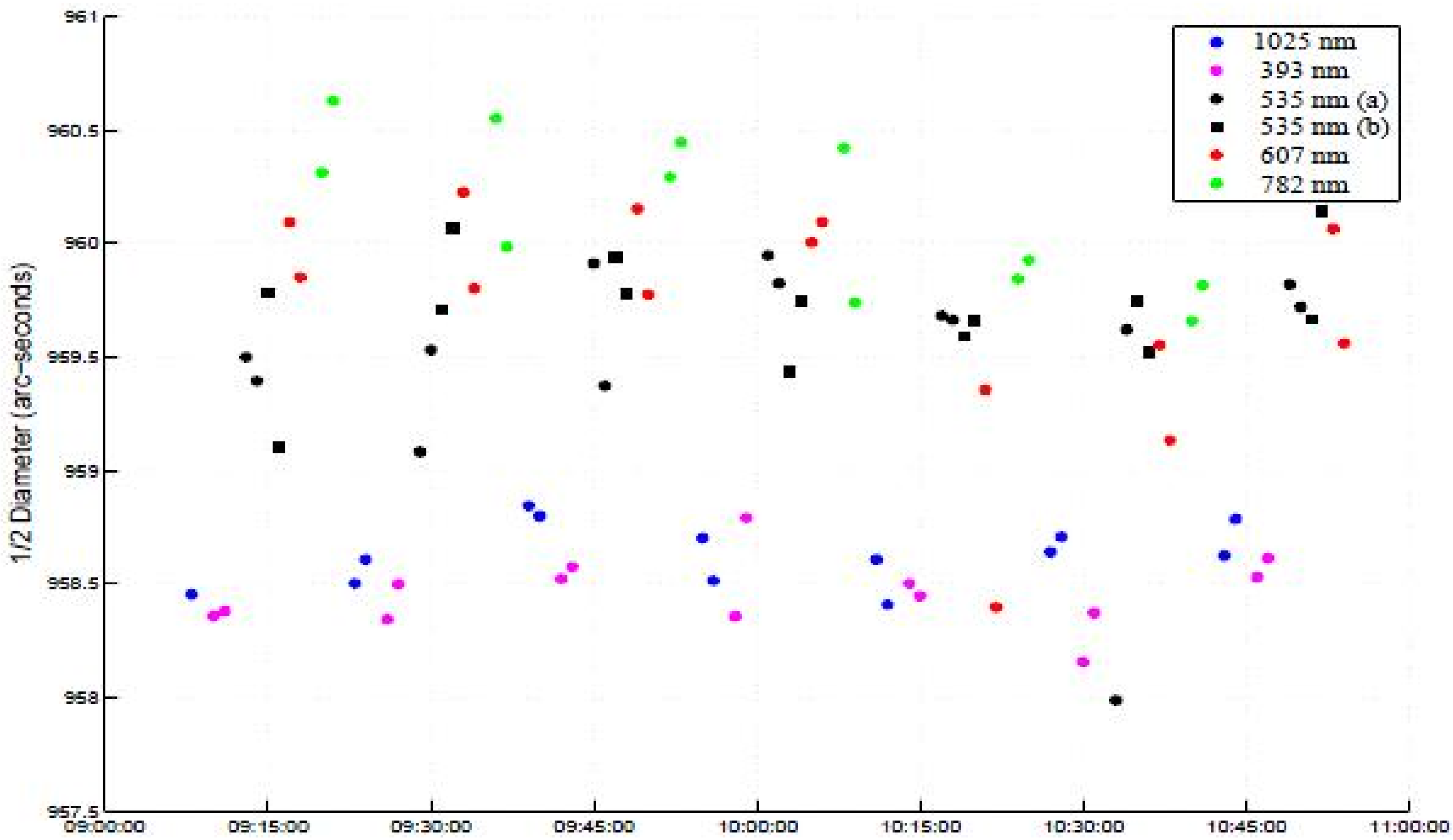

**Fig. 1.17 The diameters measured with SODISM II on August 11, 2011 in different wavebands.** SODISM (Solar Diameter Imager and Surface Mapper) is an 11-cm diameter Cassegrain telescope associated with a 2048x2048 pixels CCD detector where the whole SUN is formed. Wavelengths are selected by mean of interference filters placed on 2 wheels. Wavelength domains have been chosen free of Fraunhofer lines (535.75, 607.1 and 782.2 nm). Active regions are detected in the 215 nm domain and the CaII (393.37 nm) line. Helioseismologic observations are performed at 535.75 nm. From Irbah et al. (2011).[63]

[63] Irbah, A., M. Meftah , T. Corbard, R. Ikhlef, F. Morand, P. Assus, M. Fodil, M. Lin, E. Ducourt, P. Lesueur, G. Poiet, C. Renaud and M. Rouze, *Ground-based solar astrometric measurements during the PICARD mission*, submitted to SPIE (2011), fig. 13.

## Chapter 2: Solar astrometry

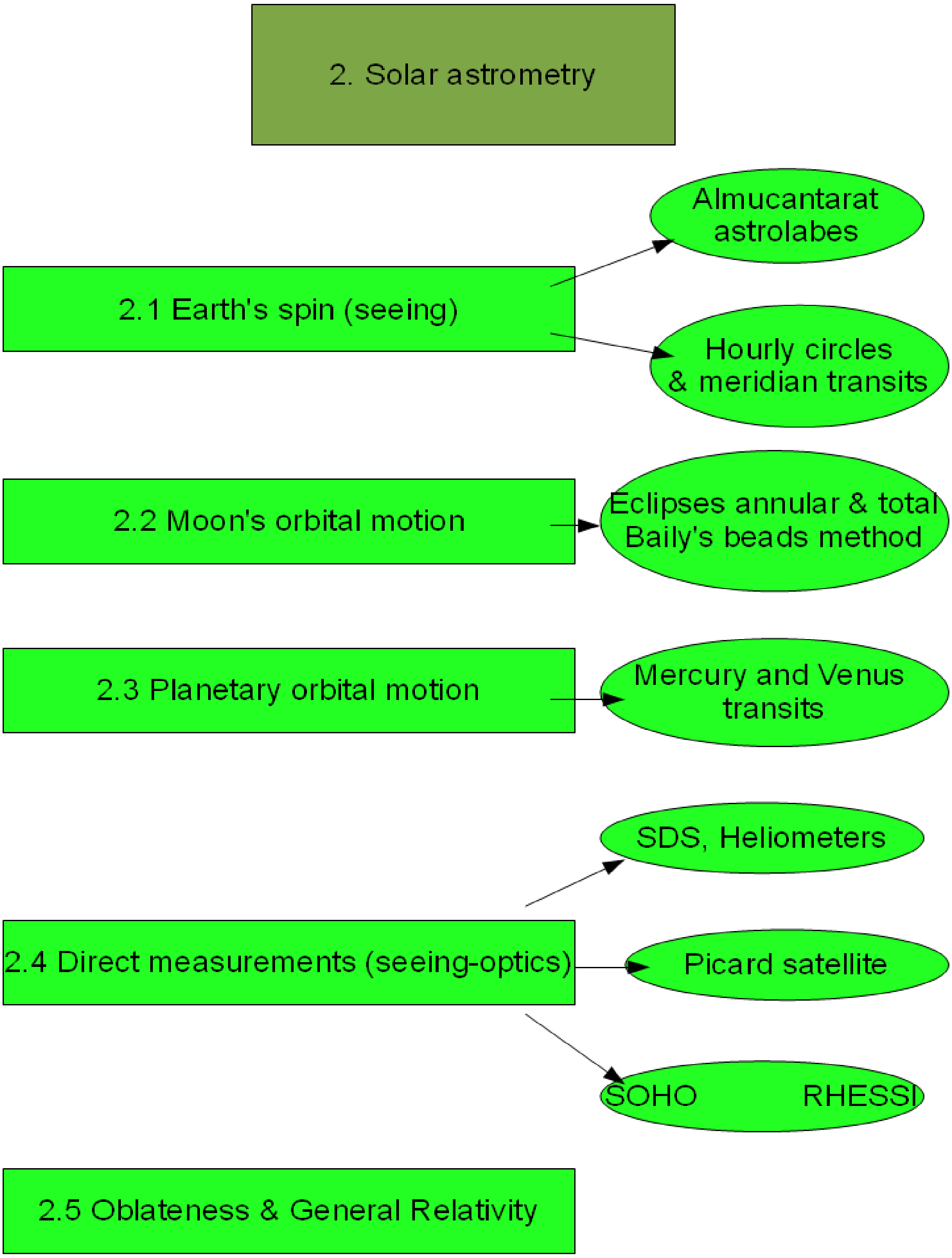

The word "astrometry" is usually related to very accurate (down to the milliarcsecond level) measurement of the positions of the stars. By adding "solar" it is to show that the milliarcsecond level is also a goal for the measurement of the variations of the solar diameter. Solar astrometry is a field of classical astrometry.
The measurement of the diameter of the Sun was firstly started quantitatively by Archimedes before 212 b.C. he published his method in the "Arenarius", the *Sand reckoner*.[64]
Afterwards several methods have been developed, and here I have classified them. We start with the method using a fixed telescope and measuring the drift time of the Sun, due to the diurnal motion, i.e. the rotation of the Earth.

## 2.1 Earth's spin and seeing problems

These methods exploit Earth's rotation (drift-scan over meridian circles; Danjon solar-type astrolabes).

### 2.1.1 The solar astrolabes

André Danjon, former director of the Observatoire de Paris, invented an istrument which was called astrolabe,[65] and become later the "impersonal astrolabe". These instruments measure the transits of stars over an almucantarat, an arabic name standing for equal height circle.
The history of the astrolabes and solar diameter has is craddle in France, at the Calern Observatory, where since 1978 Francis Laclare and his colleagues did work with them to measure the solar diameter.
Originally the astrolabe was conceived to recover the position of the stars, the exact timing when a star reached a given altitude.[66] Laclare started in Calern a long series of observations with impersonal astrolabes, and later in 1999, a new type of astrolabe, the "DORaySol", was created.[67] Also in Brasil, at the Observatorio Nacional, in 1983 the astrolabes started to be used for measuring the solar diameter, after their classical use to correct the reference systems.[68]
Due to the inclination of the daily orbit of the Sun, the duration of these transits range from the minimum 1887/15=125.8 s at the equator during equinoxes, to infinty near the time of the meridian transit.
Operational durations of such transits range only up to 6 minutes.
While the standard astrolabes were limited to a few measurements in the morning and in the afternoon, because of the number N of prisms available, DORaySol using a prism movable with springs, had a continuous range of measurements potentially available. DORaySol operated from 1999 to 2006, with the last documented observation in 2008.[69]
At the Observatorio Nacional in Rio de Janeiro, an astrolabe was modified with a "continuous prism" of the same type of DORaySol, and it is still performing measurements according to the project in which there is DORaySol: the R3S2 Réseau de Suivi au Sol du Rayon Solaire, ground-based network of solar radius monitoring.
DORaySol, which is an acronym standing for Definition et Observation du RAYon SOLaire, made up to 2006 about 20000 observations. A similar number, 21640, was attained at the Rio astrolabe from 1998 to 2009.[70]

---

[64] Sigismondi, C., and P. Oliva, Solar oblateness from Archimedes to Dicke, Il Nuovo Cimento **B 120**, 1181 (2005).
[65] Danjon, A., Astronomie Générale, J. & R. Sennac, Paris (1952) ; seconde édition, revue et corrigée, 1959.
[66] Débarbat, S., Sur l'extension des applications de l'astrolabe de Danjon, Thèse, Université de Paris (1969).
[67] Morand, F., Ch. Delmas, T. Corbard, B. Chauvineau , A. Irbah, M. Fodil and F. Laclare, Comptes Rendus Physique **11** 660 (2010).
[68] Penna, J. L., private communication.
[69] The observation of october 14 to which I could assist. The experiment was interrupted because of lack of personnel, engaged with priority on Picard-sol set up (Morand, et al, Comptes R. Phys. **11** 660 2010).
[70] Boscardin, S. C., UM CICLO DE MEDIDAS DO SEMIDIÂMETRO SOLAR COM ASTROLÁBIO, Ph. D. Thesis, Observatorio Nacional, Rio de Janeiro (2011).

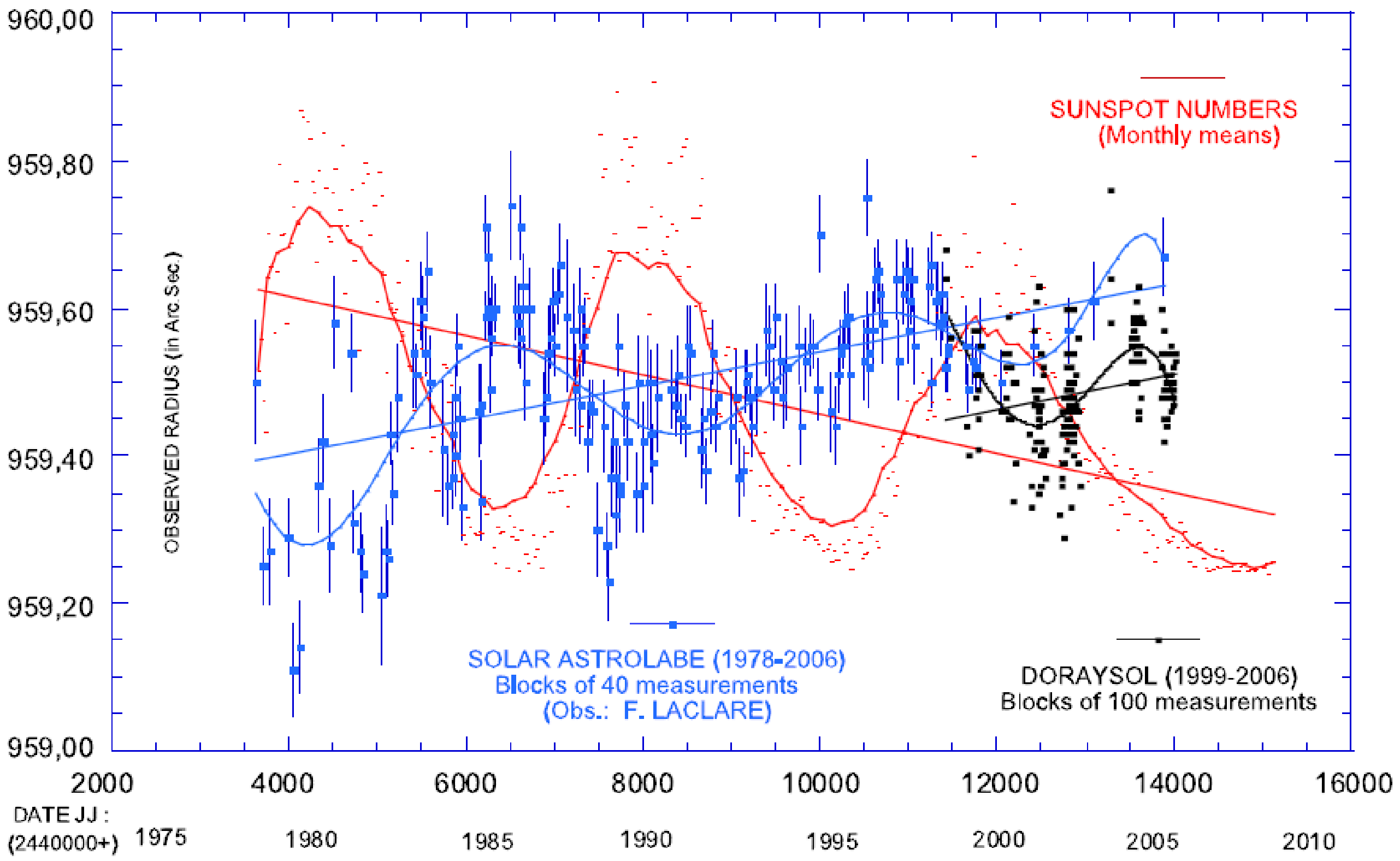

**Fig. 2.1 Solar Radius at Calern and magnetic activity (1978-2006) from Morand, et al (2010).**

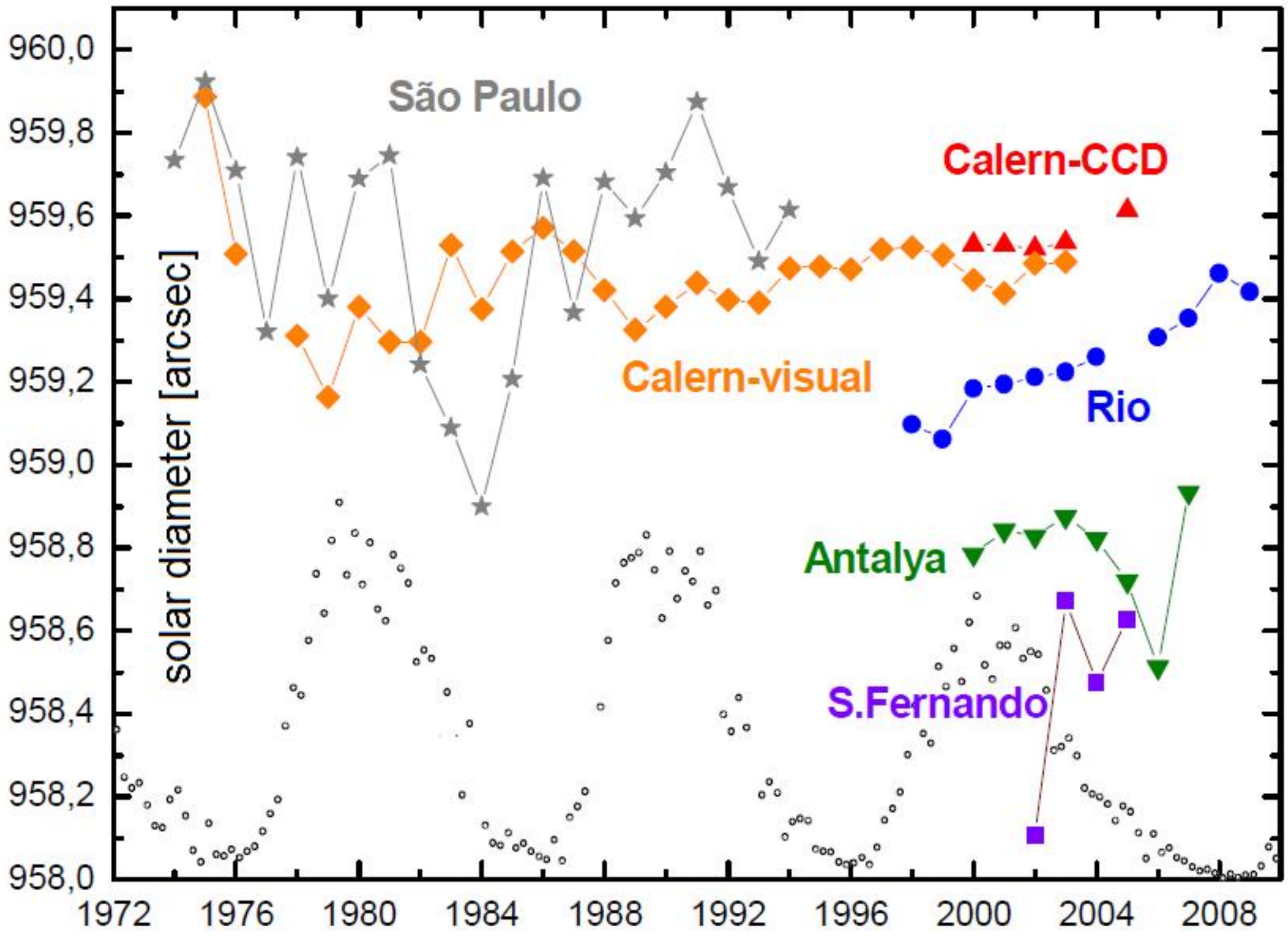

**Fig. 2.2 Solar Radius changes as from the R3S2 network. CERGA CCD stands for Calern CCD "impersonal astrolabe"**, from S. Boscardin (2011).

The difference between the various instruments of the R3S2 network is due essentially to a difference between the filters adopted. The trend in anticorrelation with the solar activity of the variation of the diameter is confirmed.

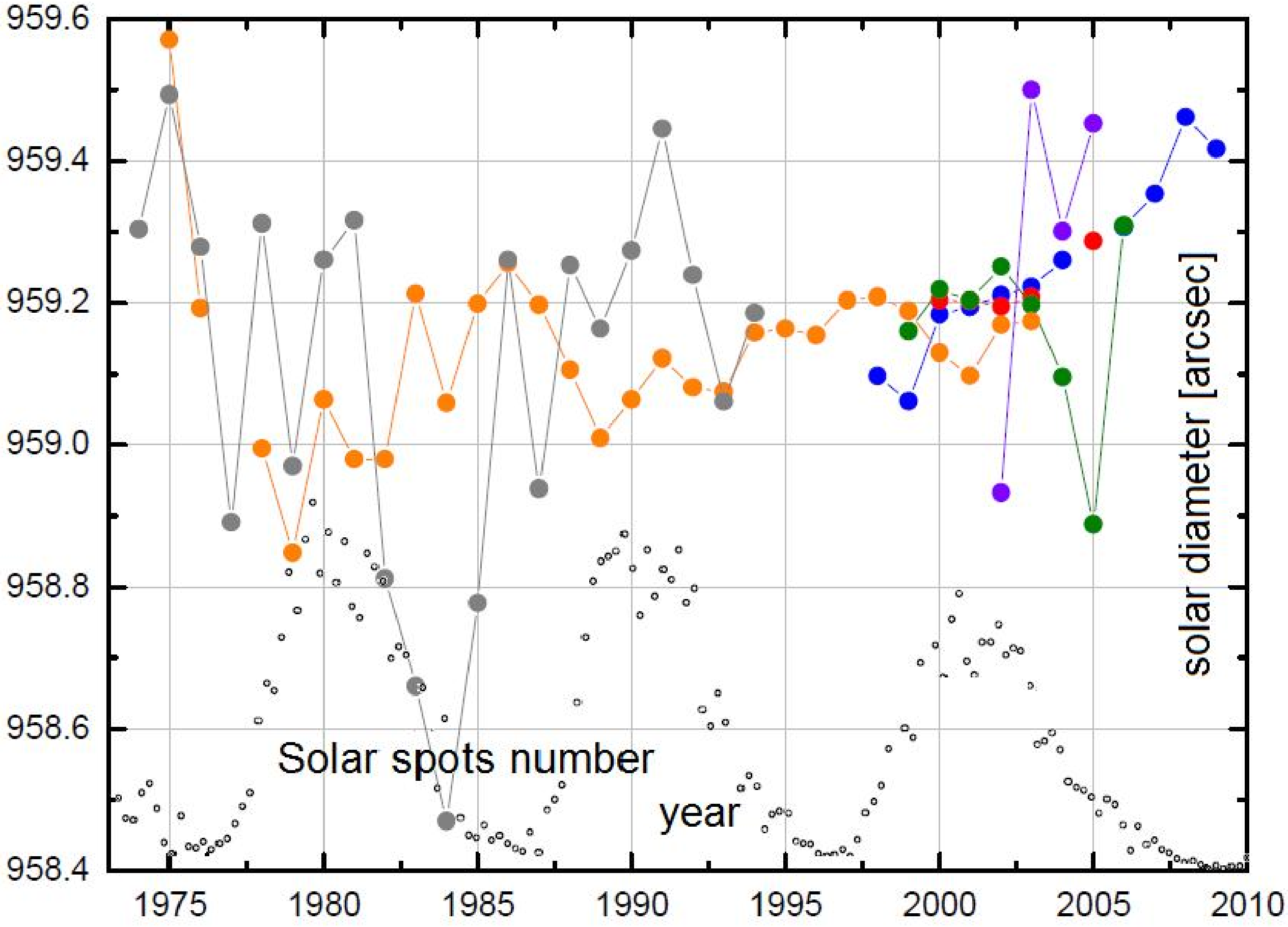

**Fig. 2.3 Solar Radius changes of fig 2.2 traslated to the same reference of Rio de Janeiro,** from S. Boscardin (2011).

The optical and mechanical scheme of DORaySol is shown in the following figure.[71] The main issue about this instrument is the diameter of the telescope, 11 cm, but since half of the objective is used at time, its point spread function is aymmetric, with the vertical axis more affected than the horizontal one.

The average Fried parameter for DORaySol[72] has been measured as $r_0$=4 cm, and therefore the instrument

[71] – A Cassegrain reflector (110 mm diameter; 3450 mm focal length) is horizontally disposed on a rotating plate, in order to ensure azimuth pointing;

– A varying prism, whose edge has to remain horizontal and perpendicular to the optical axis, ensures the altitude pointing. Associated with the mercury surface materializing the horizon, it allows to image in the focal plane of the telescope the two symmetric components of the Solar edge;

– A CCD camera and its acquisition system reconstruct the Solar edge; they also time its transit through the parallel of altitude. A spectral filter limits the wavelength range at a bandwidth of 60 nm around a central wavelength of 548 nm. A rotating shutter, in front of the telescope, alternately triggers the acquisition of the direct and reflected images of the solar edge;

– A 4.5 density filter and a shield (not sketched here) protect the whole instrument;

– Five computer-driven motors pilot the instrument: rotating plate (azimuth), angle of the prism (altitude) and inclination of the density filter. Accurate controls of the horizontality of the edge of the prism and of the optical axis are achieved.

[72] Fried, D. L. Optical resolution through a randomly inhomogeneous medium for very long an very short exposures, J. Opt. Soc. Am. **56** (1966) 1372–1379. See also: Irbah, A., F. Laclare, J. Borgnino, G. Merlin, Solar diameter

aperture is fully exploited with respect to the seeing.
Nevertheless the choice to use big solar telescope, with lenses larger than 33 cm, gives another perspective to these measurements. That happened in the Clavius project, that we will examine later in chapter 4.

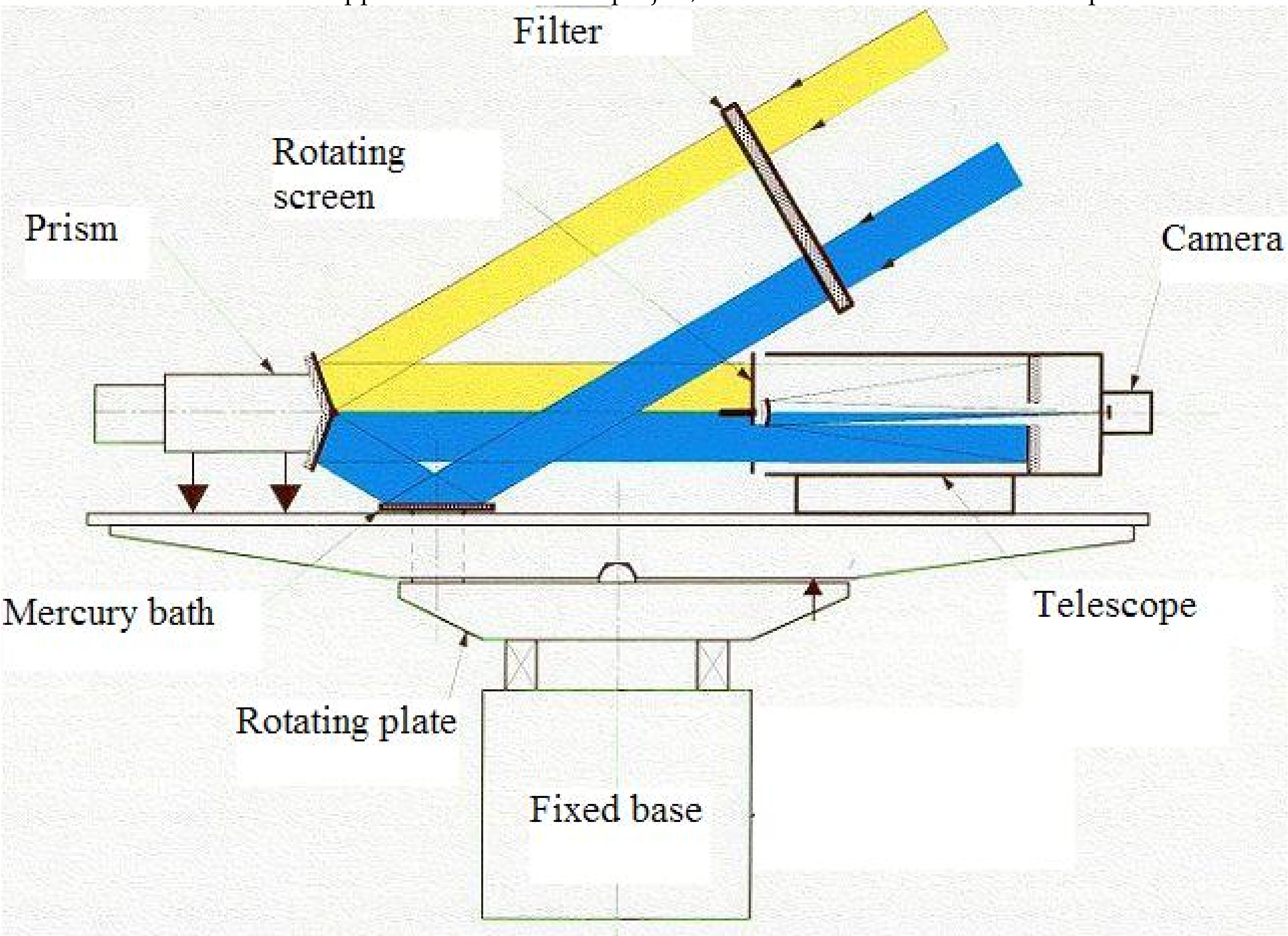

**Fig. 2.4 DORaySol optical scheme.**The yellow ray is direct the blue one reflected.[73]

The measure of the solar diameter, on different heliolatitudes, made with the astrolabes is not an instantaneous measurement. Up to the first order the refraction at the same altitude is the same, but some density waves in the atmosphere can modify the relative position of the two observed limbs of the Sun. Another advantage of having fixed optics is that the optical defects act like systematic effects, the same for the first and the second limb (admitting that the paths are symmetrical).
A single measurement made with the astrolabe is not significant, only statistical errors can be associated to a series of measurements. Thermal effects and seeing play a crucial role in the interpretation of the data, if the accuracy required is better than one part over 10.000.

The influence of the solar spots on the measurements made at the astrolabe has been debated.
A statistical study shows that the presence of a solar spot at limb may have affected 4.13% of the observations made at the astrolabe of Rio de Janeiro. The influence of a solar spot in the measurement of the solar diameter has been evaluated, with simulations of the software of analysis, less than 0.002 arcsec, even if the probability to have a spot on the solar limb is relevant.[74]

---

measurements with CALERN Observatory Astrolabe and atmospheric turbulence, Solar Phys. **149** (1994) 213–230.
[73] Source: http://www.oca.eu/gemini/equipes/ams/doraysol/index.html
[74] Boscardin, S. C., UM CICLO DE MEDIDAS DO SEMIDIÂMETRO SOLAR COM ASTROLÁBIO, Ph. D. Thesis, Observatorio Nacional, Rio de Janeiro (2011), chapter 6.

While the effect of the faculae is not relevant, for other scholars[75] the solar spots at the limb can have a significant effect, therefore this aspect requires more attention in the future research.

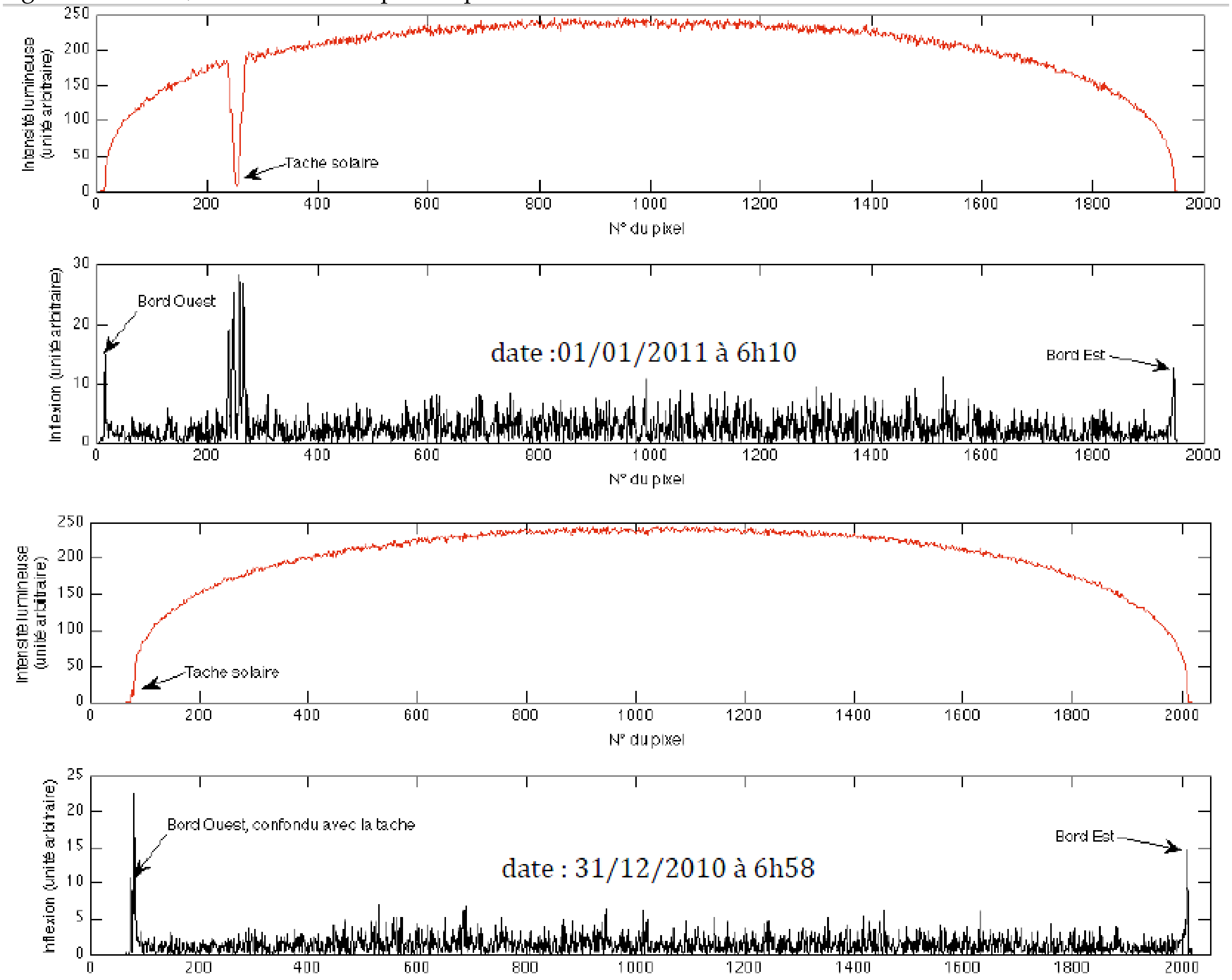

**Fig. 2.5 The algorithm of edge finiding in the case of a big spot on the limb between December 31, 2010 and January 1st 2011, gave a difference of 25 arcsec, i. e. 1.7% of variation on the diameter's evaluation.[76] Original images from SDO–HMI[77] experiment.**

[75] Courcol, B. et S. Koutchmy, Mesures et variations du diamètre solaire, IAP, Paris (2011).

[76] Courcol, B. et S. Koutchmy, Mesures et variations du diamètre solaire, IAP, Paris (2011). Figure 2 and 3.

[77] http://hmi.stanford.edu/

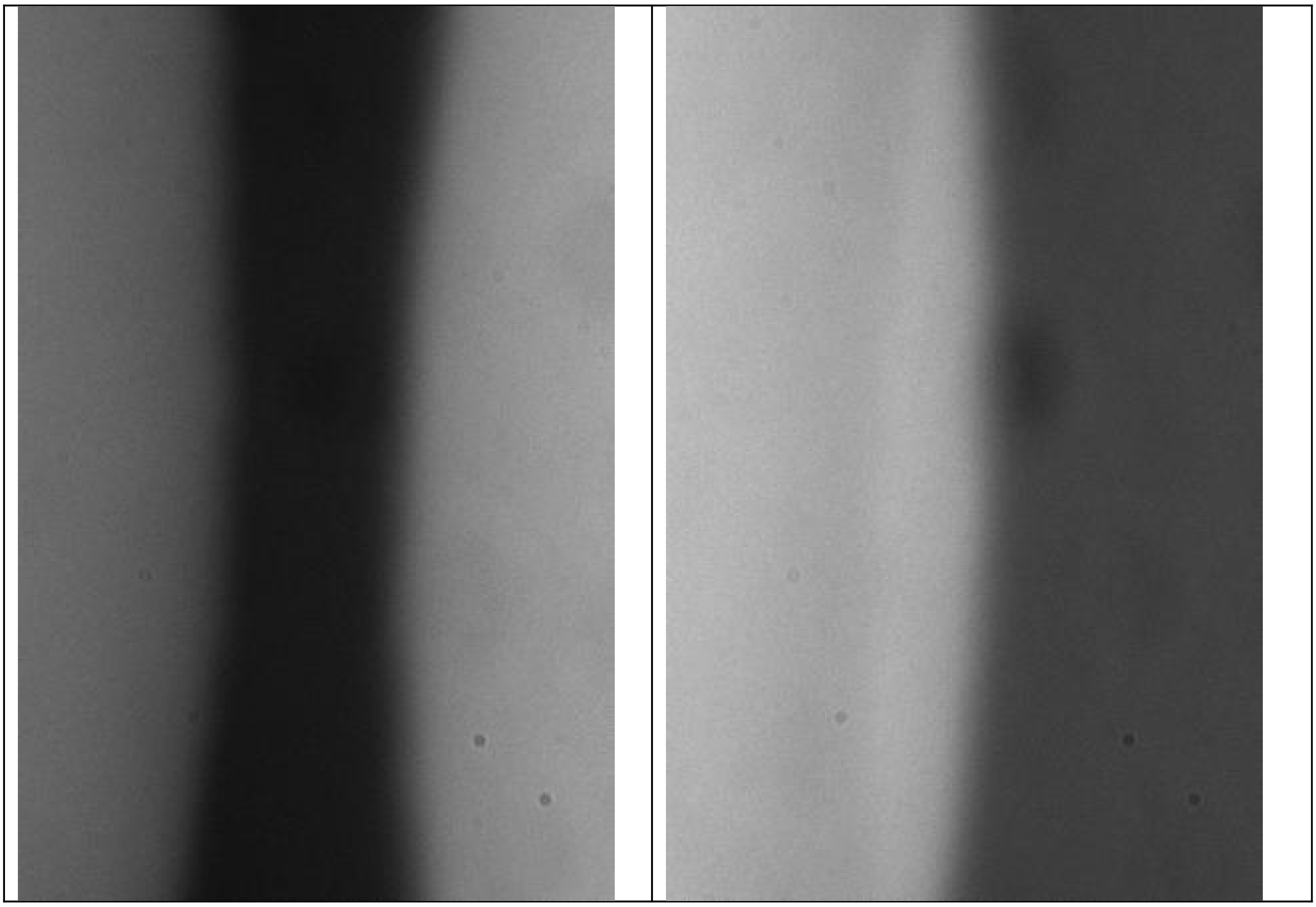

**Fig. 2.6 The two images of the solar limb at the Astrolabe of Rio de Janeiro.** Observations of March 18, 2011. The image reflected by the bain of mercury is on the left. While in DORaySol there is a shutter which shows alternatively the direct and the reflected image, in Rio the two images superpose, like in this right side image.

Presently only the Astrolabe of Rio is working, and its acquisition system is being renovated, thanks also to my contribution (see paragraph 2.4 on the Heliometer of Rio). It is important that the activity of this instrument could be prolonged during the Picard mission, in order to overlap its data with the space ones, and also with the new heliometers and the eclipses. In this way all the four decades of solar diameters recovered with astrolabes can be used in understanding better the relationship solar radius-activity.

### 2.1.2 Meridian and hourly circle transits

The measurement of the diameter of the Sun using meridian transits, and later the hourly circles transits, is the method used since the invention of the telescope. The french abbot Jean Picard measured the solar diameter before 1680 with this method. The duration of a transit is typically about 2 minutes.
The largest angular diameter of the Sun is 1952 arcsec on Jan 4, corresponding with the largest declination 23°26′, therefore the maximum duration is about 1952/(15·cos(±23°26′))= 141.8 s.
Chapter 4 will be dedicated to this topic more wider, when presenting the CLAVIUS project.

## 2.2 Lunar orbital motion: eclipses and Baily's beads

The Moon along its orbit covers approximately its diameter in one hour. The duration of a total eclipse is depending on the difference between the angular diameters of the Moon and of the Sun. By timing the totality near the centerline an error of ±1 s corresponds approximately to ±0.5 arcsec in the uncertainty of solar diameter. If the timing is made near both the edges of the umbral path a greater accuracy can be achieved. A variation of Δt=1 in the duration of the totality phase corresponds to a few meters on the ground, and the dimension of the umbra, and then of the solar diameter, can be measured better than one part over 10.000.[78] More details concerning the Baily's beads method are explained in the chapter 3, on the eclipses.

## 2.3 Planetary orbital motion

The transits of Mercury (about 13 per century) and of Venus (two each 120 years) have an angular speed which covers the solar diameter (generally a chord) in about 6 hours. If the times of the contacts are defined with an accuracy of one second an optimal precision on solar diameter is attained.
Shapiro[79] in 1980 used the historical data on Mercury transit to check the hypothesis of variations of the solar diameter claimed by Eddy and Boornazian[80] and Dunham, et al.[81]
Other works reconsidered the question of the transits of Mercury.[82],[83]
The problem of the ancient data is due to the difference between the instruments utilized. I have presented the comparison between the observations of the transit of Mercury of 1832 made by Bessel and Gambart respectively in Königsberg with the Fraunhofer heliometers of 16 cm and in Marseille with a 7 cm refractor.[84] The younger and skilled observer Gambart, under perfect meteorological conditions, observed a transit shorter (and therefore a shorter solar diameter) than Bessel. This occurred because the instrument of Gambart was smaller and its point spread function poured many photons into the disk of Mercury. The measurement of the diameter of Mercury, also published by the two observers on the Astronomische Nachricthen in 1833, showed clearly this effect. Gambart measured the diameter of Mercury one arcsecond less than Bessel (6.91", very close to the ephemerides value).
In this direction is also the modern explanation of the famous black drop effect: an interplay between the solar limb darkening function, rapidly varying near the limb, and the point spread function of the telescope.[85]

The method of chords has been applied by me already in 2006, to the data of the Venus' transit of 8 June 2004.[86] The influence of the seeing on the determination of the internal contacts of Venus with the solar photosphere (observed in Halpha by Anthony Ayomamitis in Athens) was evidenced. An uncertainty of ±7 s arose from the morning measurements with the Sun still low in the sky, while it decreased to ±1 s with the Sun at the local noon. The solar diameter calculated from this observation was consistent with Halpha diameter.
The errorbar of 7 seconds could have been reduced if the number of photo would be larger. Ayomamitis made a photo each minute, and only 25 photo per contact were available. Nevertheless this is the only observation of Venus' transit of metrological quality I have found after many years of research on the internet and in various solar observatories in Europe and worldwide. This is rather unbelievable!
I tried to coordinate some observatories (Big bear and Mauna Kea) for the transit of Mercury in 2006, but the meteo conditions and the geometry prevented the measurements of both internal contacts.
In the following transit of Venus of June 6, 2012, we gather data in Huairou Solar Observing Station and we joined the Venus Twilight Experiment.[87] The reason is that the aureole formed around the disk of Venus is reducing its apparent diameter and affects the timing of the internal contacts. Therefore the knowledge of the aureole is complementary to the one of the contact timings.
Shapiro in 1980[88] and Parkinson et al., 1980[89] calculated the solar diameter after the historical data on Mercury transits. In the following figure are reported these measurements (triangles) along with two squares representing the last two determination of the solar diameter made with SOHO satellite with the transits of 2003 and 2006 by Emilio et al., 2012.[90]

---

[78] Sigismondi, C., Measures of Solar Diameter with Eclipses: Data Analysis, Problems and Perspectives, AIPC **1059**, 183 (2008).
[79] Shapiro, I. I., Science, **208**, 51 (1980).
[80] Eddy, J. A. and Boornazian, A.A., Bulletin of the American Astronomical Society, **12**, 437 (1979).
[81] Dunham, D. W., S. Sofia, A. D. Fiala, P. M. Muller and D. Herald, Science **210,** 1243 (1980).
[82] Gilliland, R. L., Astrophysical Journal, **248**, 1144 (1981).
[83] Parkinson, J. H., L.V. Morrison and F. R. Stephenson, Nature **288**, 548 (1980).
[84] Sigismondi, C., AIPC **1059**, 183 (2008).
[85] Pasachoff, J. M., G. Schneider, L. Golub, The black-drop effect explained, in Transits of Venus: New Views of the Solar System and Galaxy, Proceedings of IAU Colloquium #196, held 7-11 June, 2004 in Preston, U.K.. Edited by D.W. Kurtz. Cambridge: Cambridge University Press, 242-253 (2004).
[86] Sigismondi, C. and P. Oliva, Astronomia UAI, **3**, 14 (2006).

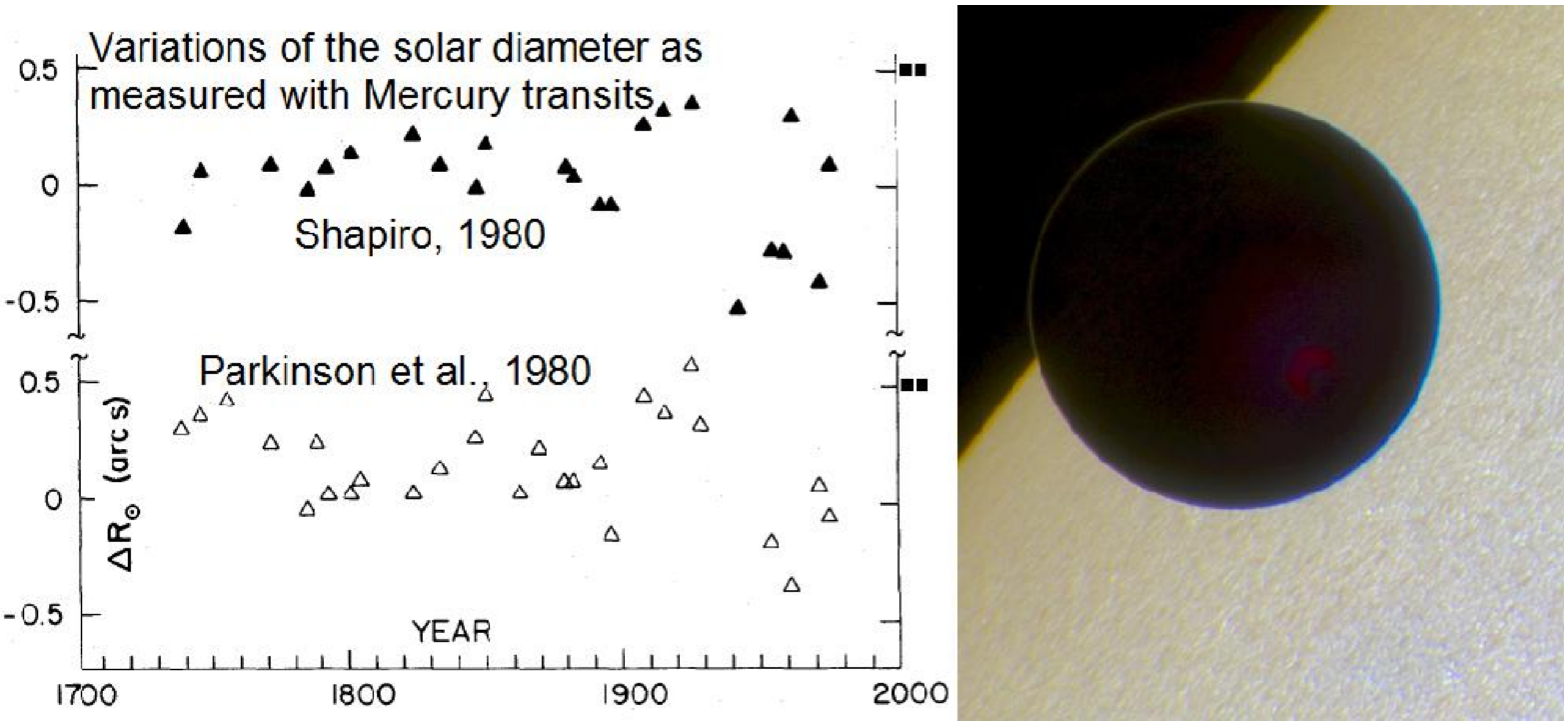

**Fig. 2.7 Solar diameter measured with past transits of Mercury and (right) the transit of Venus as seen with HINODE satellite.** The two squares at right, after year 2000, are the data of SOHO reduced by Emilio et al. 2012.[27]

The different length of the solar diameter measured with the transits of Mercury can be done either to intrinsic variations of the solar diameter either to instrumental errors which changed the contact timings.
The scatter of the measurements may be due both to an intrinsic variation of the solar diameter and to the difference between the instruments adopted for the observations, as it was in the case of 1832.
Bessel (1832)[91] observed the transit of Mercury of 5 May 1832 with a 16-cm Fraunhofer heliometer and perceived the diameter of Mercury as 6.7 arcsec and

[87]Xie, W., C. Sigismondi, X. Wang and P. Tanga, Venus transit, aureole and solar diameter, in Solar and Astrophysical Dynamos and Magnetic Activity (A.G. Kosovichev, E.M. de Gouveia Dal Pino and Y. Yan, editors) Proceedings IAU Symposium No. 294 (2013). http://arxiv.org/pdf/1210.8451v1
[88]Shapiro, I. I., Science **208**, 51 (1980).
[89]Parkinson, J. H., Morrison, L. V. and F. R. Stephenson, Nature **288**, 548 (1980).
[90]Emilio, M., Kuhn, J. R., Bush, R. I. and I. F. Scholl, I. F., Astrophys. J. **750,** 135 (2012).
[91]Gambart, J. , Astronomische Nachrichten **10** 257 (1832).

Gambart (1832)[92] observed with a 6.7-cm Dollond refracting telescope, measuring the diameter of the planet as 9.3 arcsec; from the ephemerides the diameter of Mercury was 6.9 arcsec, and both the observations were done under optimal weather conditions. The Point-Spread Function of the telescope ruled a major role on the determination of the contact timings, more affecting the observations made with smaller instruments.

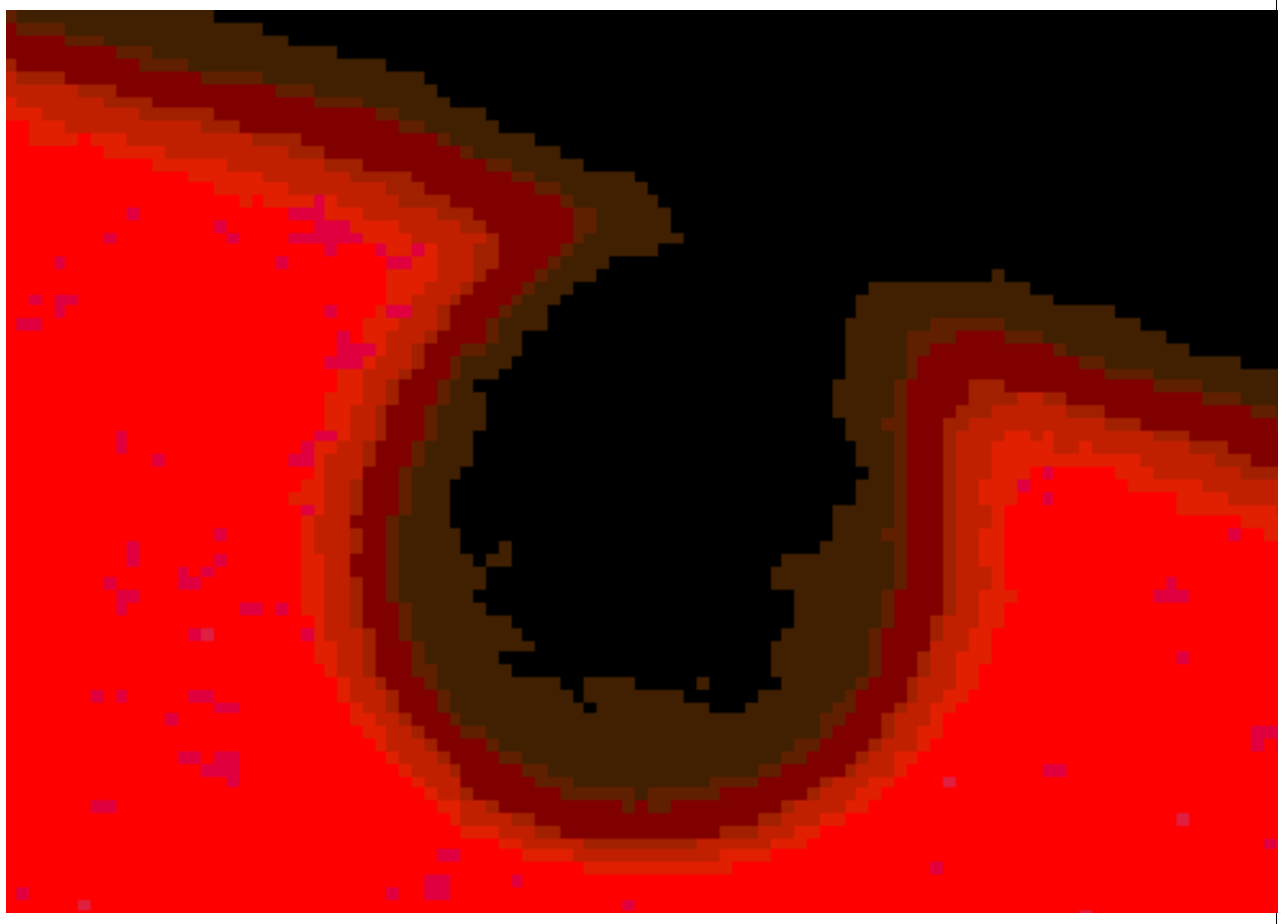

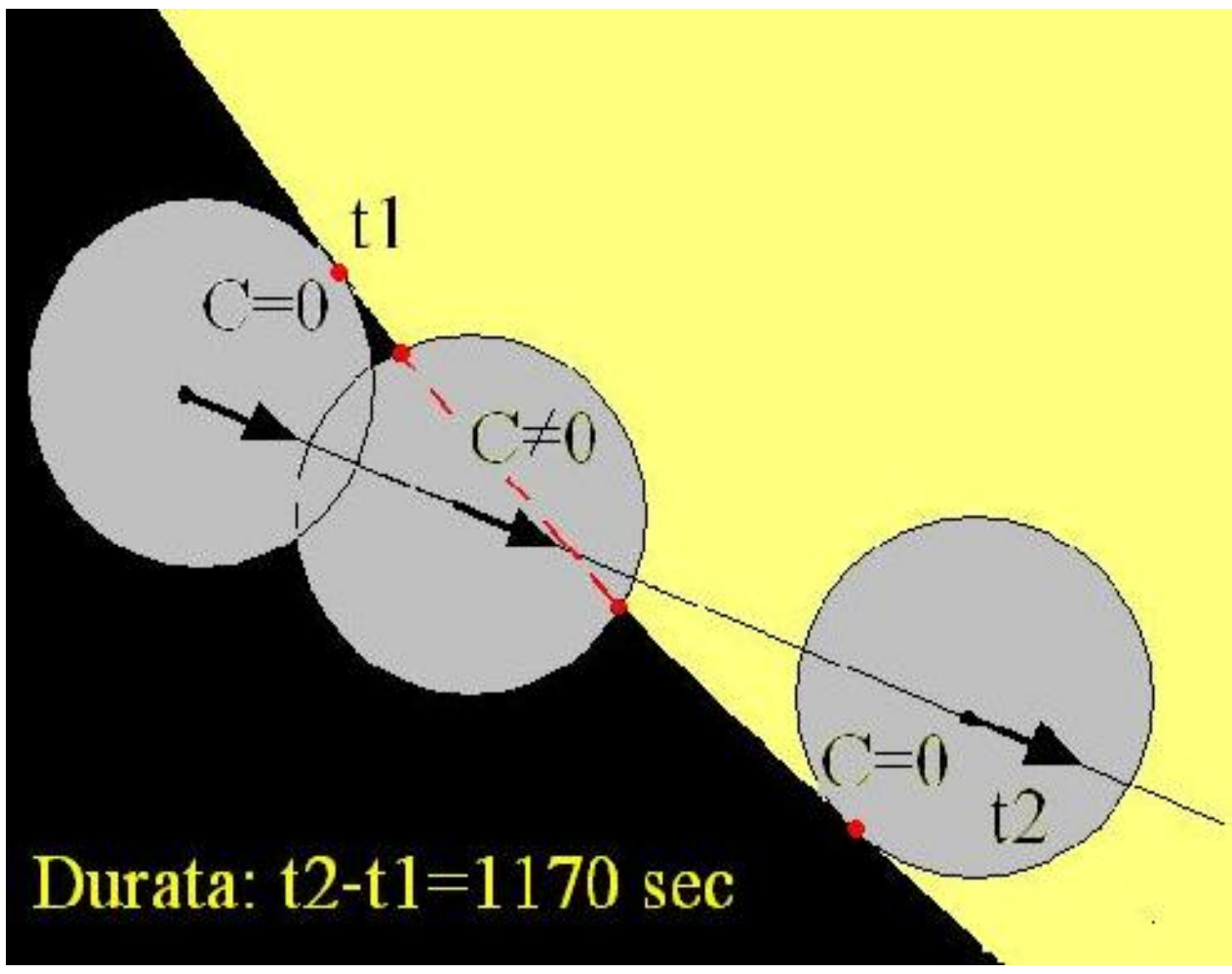

**Fig. 2.8 Venus' transit of 2004 in H$\alpha$** (A. Ayomamitis,[93] 60 mm refractor and 0.7Å H$\alpha$ bandpass) ingress phase at 7:37:00 Athen's time. The image of Venus is deformated because a) it is offaxis, since the center of the Sun was centered in the image b) it is probably slightly defocussed.

**Fig. 2.9 (right side) The sketch of chord's method, on the right side.** Figure from G. Di Giovanni.[94]

---

[92]Bessel, F. W., Astronomische Nachrichten **10** 185 (1832).
[93]http://www.perseus.gr/Astro-Planet-Ven-Tr2004.htm
[94]Di Giovanni, G., Astronomia UAI **2**, 15 (2005).

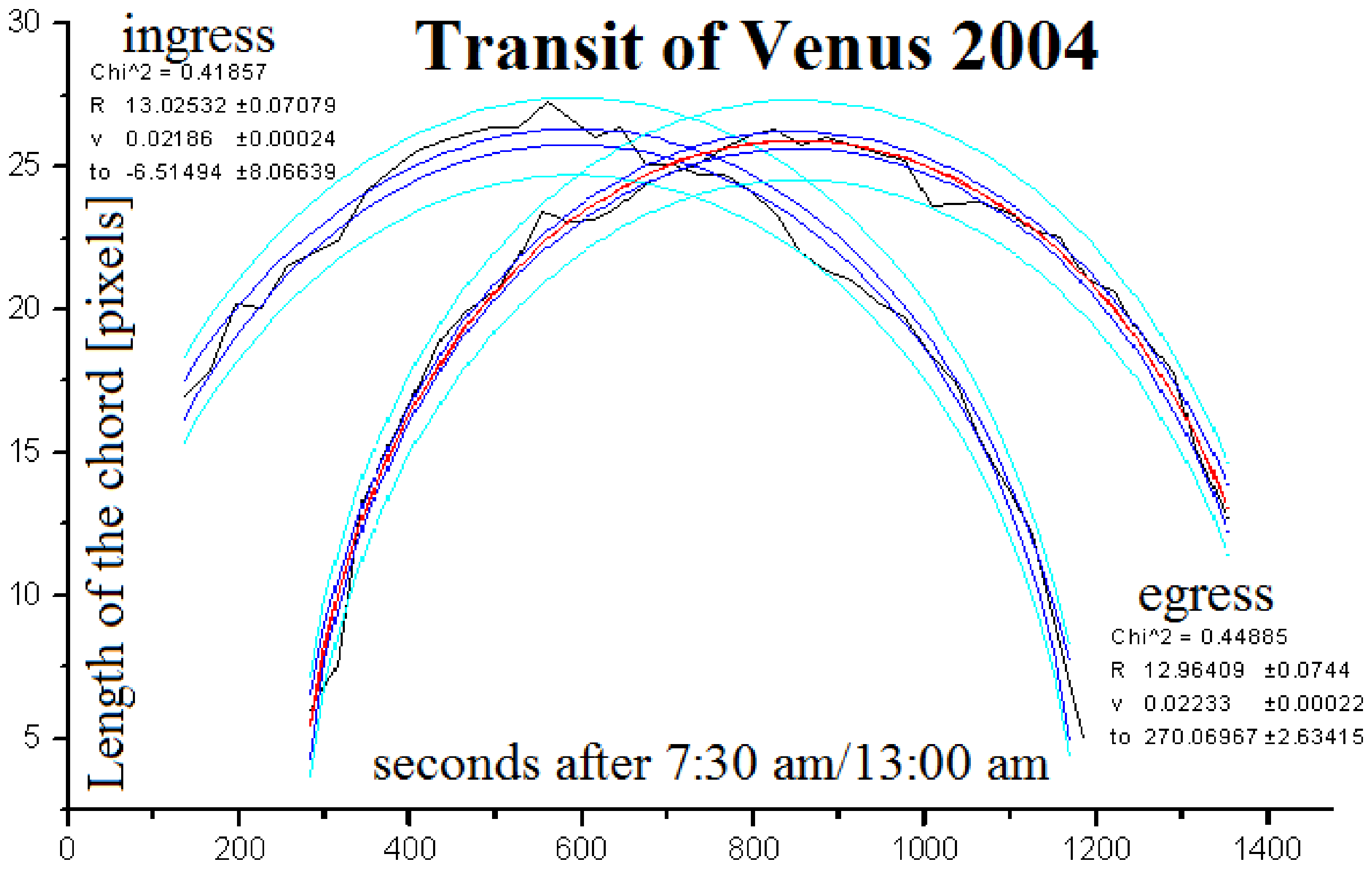

**Fig. 2.10** A parabolic fit applied to the length of the chords vs time, is required to find t1 and t2. The ingress is on the left (±8 s of uncertainty because of the seeing), the egress on the right (±2.6 s). The angular velocity of Venus above the Sun was about 0.08 arcsec/s; this corresponds to 0.64 arcsec of uncertainty in the ingress and 0.20 arcsec at the egress. The velocity in the fit is measured in pixels/s and the time $t_0$ in seconds after 7:30:00 am for the ingress and after 13:00:00 am for the egress. The calculus of the ingress/egress timings has individuated the first contact of the Venus limb with the solar one t1 [see fig. 2.9] and t4, the last contact. The seeing through all the layers of the atmosphere near sunrise is more than three times larger than the seeing when the Sun was much higher over the horizon and transiting on the local meridian.

If we consider these values of t1 and t4, with their relative accurate timing (the same along the 6 hours of the whole phenomenon) we find an extra-duration of the transit corresponding to a $\Delta R=0.38\pm0.67$ arcsec for H$\alpha$ line. This value is not definitive because of two reasons:

a) the influence of the Point Spread Function over the perception of the diameter of Venus

b) the effect of the aureole which reduces the diameter of Venus especially on the side immersed in the photosphere and

c) the effect of a defocussed image on the measurement of the chord's length.

The example of the transit of Mercury in 1832 observed by Bessel and Gambart is typical for this kind of analysis: the smaller instrument (6.7 cm Gambart) perceived a larger planet, while Bessel with the Heliometer of Fraunhofer (16 cm) measured a diameter of the planet in very good agreement with modern ephemerides. In our case we can assume that all the three effects a), b), c) are negligible and the final result of the solar diameter during Venus transit of 2004 was $R_0+\Delta R=959.63+0.38\pm0.67$ arcsec, leading to a radius R(H$\alpha$ line)=960.01±0.67 arcsec. The solar radius in H$\alpha$ line can be considered as 0.07 arcseconds larger than in white light according to Neckel and Labs.[95] The H$\alpha$ line filter used for these observations has a nominal bandwidth of ±0.7Å and so the corresponding measured radius is very similar to the corresponding radius measured in the photospheric continuum.[96] Finally the value in white light for 2004 transit of Venus is 959.94±0.67 arcsec in good agreement with the measurements made by Emilio et al. (2012) on the transits of Mercury with SOHO in white light and with our measurements of SDO/HMI[97] (see par. 2.4.6). The assumption that effects a) b) and c) are negligible is only valid for a first-order analysis, but nevertheless the seeing and the low number of images available (25 for the ingress and 25 for the egress) heavily limited this analysis.
The transit of Venus of 2012 has been observed from China, Huairou Solar Observing Station, with the possibility to have

1) ingress and egress phases visible from the same location

2) full disk images with 1 second cadence (1800 for the ingress and 1800 for the egress phases)

3) 4096 level of intensity images in order to avoid saturation effects and individuate the inflection points.

The analysis of these data, along with the reprocessing of 2004 images is still ongoing and it will appear on the web repository of Cornell University Library arxiv.org as soon as possible.

## 2.4 Direct angular measurements

The heliometers deal with the whole figure of the Sun, or with the images of the limbs projected through prisms on the focal plane. The optical defects are crucial up to the milliarcsecond level.
The advantage of these measurements is that they are instantaneous.
At the same time the two opposite limb of the Sun are observed one in front of the other, and from the distance between them the diameter is recovered.
The first application of the divided object-glass and the employment of double images in astronomical measures is due to Servington Savary from Exeter in 1743. Pierre Bouguer, in 1748, originated the true conception of measurement by double image without the auxiliary aid of a filar micrometer, that is by changing the distance between two object-glasses of equal focus. John Dollond, in 1754, combined Savary's idea of the divided object-glass with Bouguer's method of measurement, resulting in the construction of the first really practical heliometers. As far as we can ascertain, Joseph von Fraunhofer, some time not long before 1820, constructed the first heliometer with an achromatic divided object-glass, i.e. the first heliometer of the modern type.[98]

### 2.4.1 Heliometers

There are various type of heliometers, starting from the lenses splitted in two halves, the classical one, to the Göttingen's type (1890s) with a prism in front of the objective used also in the balloon-borne solar telescope

---

[95] Neckel, H. and D. Labs, *Solar limb darkening (1986 to 1990) λ=330 to 1099 nm*, Solar Phys. **151**, 93 (1994) p. 95.
[96] If the Hα filter has a bandwidth of ±0.3Å therefore the structures belong to the chromophere, and they are 7 arcseconds higher than the photosphere, but as we move±2Å the measured radius becomes similar to the continuum.
[97] Wang, X. and C. Sigismondi, Geometrical information on the solar shape: high precision results with SDO/HMI, in Solar and Astrophysical Dynamos and Magnetic Activity (A.G. Kosovichev, E.M. de Gouveia Dal Pino and Y. Yan, editors) Proceedings IAU Symposium No. 294 (2013). http://arxiv.org/pdf/1210.8286v1
[98] Chisholm, Hugh, ed. (1911). Encyclopædia Britannica (11th ed.). Cambridge University Press. Volume 13, pp. 224-230.

SDS Solar Disk Sextant, to the mirror heliometer and to its recent variant called annular heliometer both invented by V. d'Ávila and developed at the Observatorio Nacional in Rio de Janeiro (2011).
The last two instruments being based on reflecting splitted mirrors have the advantage of no chromatic aberrations, as well as of no optical aberrations.

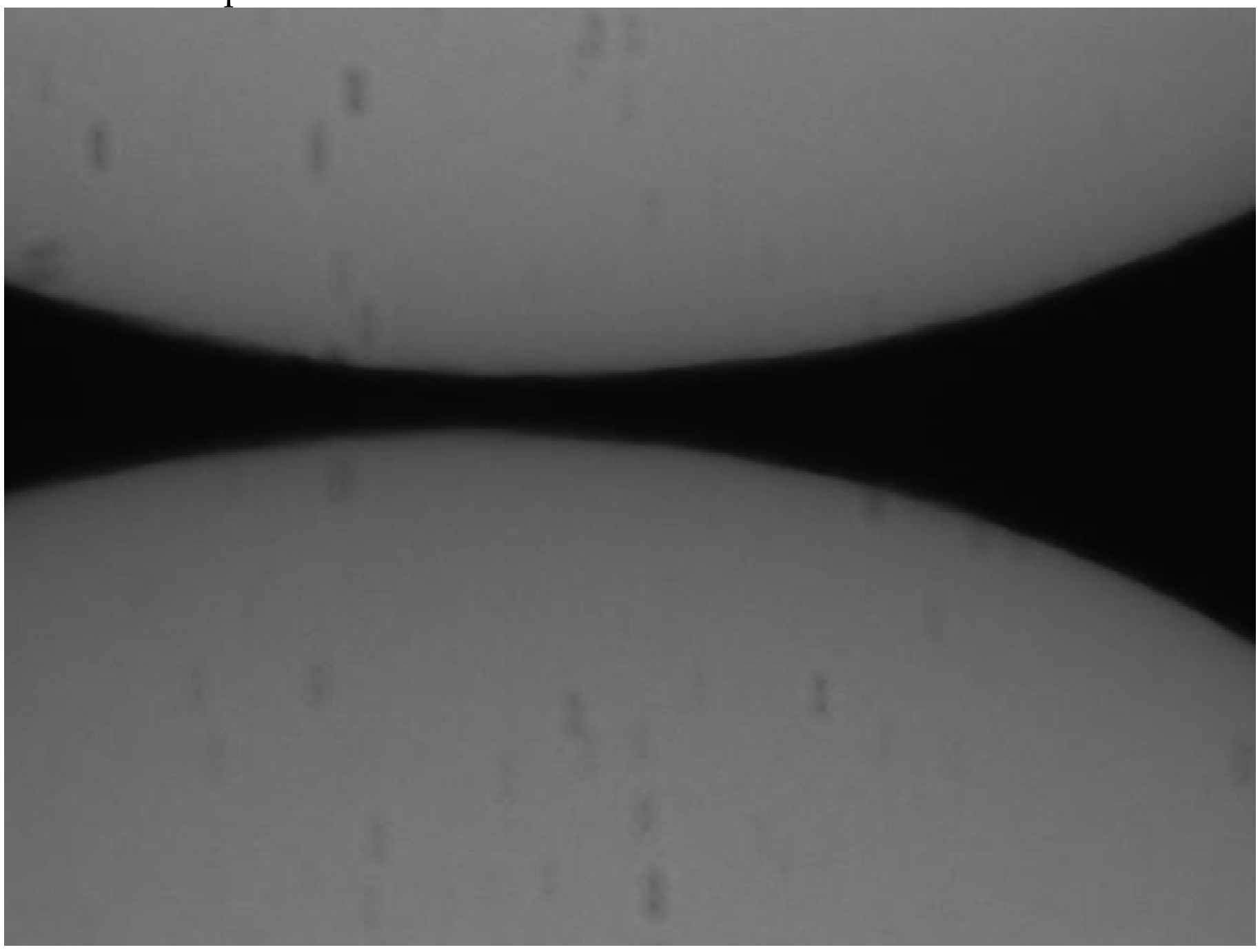

**Fig. 2.11 The two limbs of the Sun at the Heliometer of Rio de Janeiro on March 28, 2011.**

The heliometer of Rio de Janeiro now performs a series of 100 measures each minute, this increase the number of measurements of solar diameter well beyond the DORaySol-type instruments.
The heliometer of Rio de Janeiro is being calibrated also with the results of diameter's measurements from eclipses. For this purpose it has been transported to the Easter's Island in occasion of the total eclipse of July 11, 2010, and it will observe also the eclipse of May 20, 2012 in southern China.
The software and camera used for the heliometer recognizes the two limbs of the Sun, and calculates the position of the centers, and the radii. The distance d between the two limbs depends on the focal length F of the telescope, on the angle θ between the mirrors and on the angular solar diameter ds". $d=F\cdot \theta - F\cdot ds''$.
F and θ are constant (the temperature is under control in order to keep F constant) and therefore the angular diameter of the Sun is $ds''=d/F-\theta$.
The same software has been adapted to the astrolabe, where the point of contact of the two limbs is crucial.
Therefore the measure of d, and its extrapolation to zero, is the strategy of the new acquisition system of the the Rio Astrolabe. The possibility is offered by the fact that the Rio Astrolabe yields both images, reflected and direct, at the time. I implemented this idea during my stay in the Observatorio Nacional in Rio de Janeiro from February to April and June 2011. In this way the astrolabe will work together with the heliometer during the Picard mission.

### 2.4.2 MISOLFA

On the same principle, but with the purpose of measuring the parameters of the atmosphere, is the solar monitor MISOLFA[99] (*Moniteur d'Images SOLaires Franco-Algerien*). It is placed in the same dome of SODISM

[99]Irbah, A., M. Meftah , T. Corbard, R. Ikhlef, F. Morand, P. Assus, M. Fodil, M. Lin, E. Ducourt, P. Lesueur, G. Poiet, C. Renaud and M. Rouze, Ground-based solar astrometric measurements during the PICARD mission, submitted to SPIE (2011).

II is devoted to measure the caracteristics of the atmospheric turbulence during the measurements of SODISM II, made in parallel with SODISM instrument onboard of Picard. At the end of the space mission a procedure of accurate measurement from ground with SODISM II and MISOLFA will work routinely.

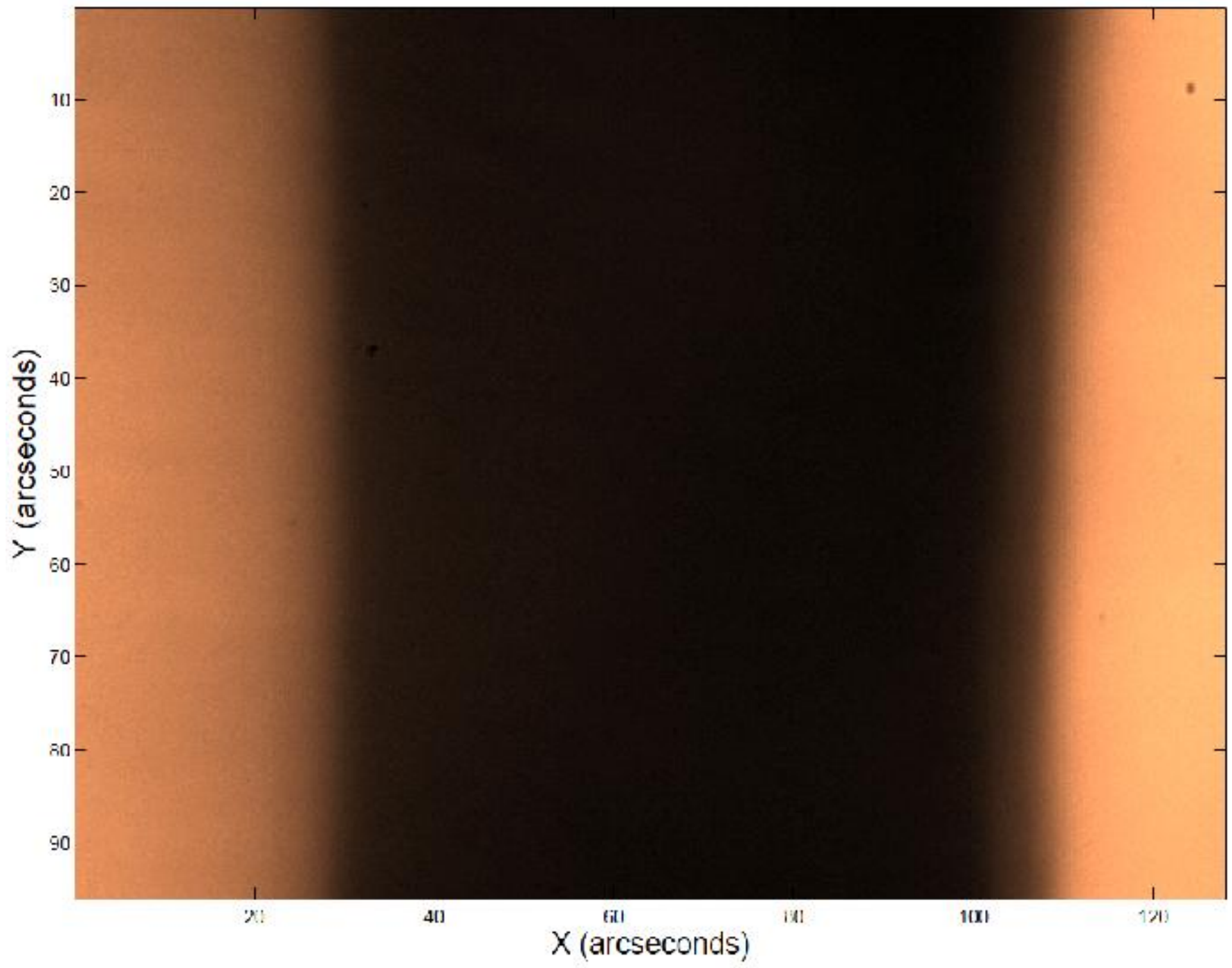

**Fig. 2.12 The two limbs of the Sun observed with MISOLFA on August 11, 2011.**

### 2.4.3 SDS, the Solar Disk Sextant

The third instrument of this type, and the first to produce results, has been SDS. The two images of the Sun are projected on the focal plane, where are 7 linear CCDs. These CCDs detect the luminosity profile, and an algorithm individuates the inflexion point.

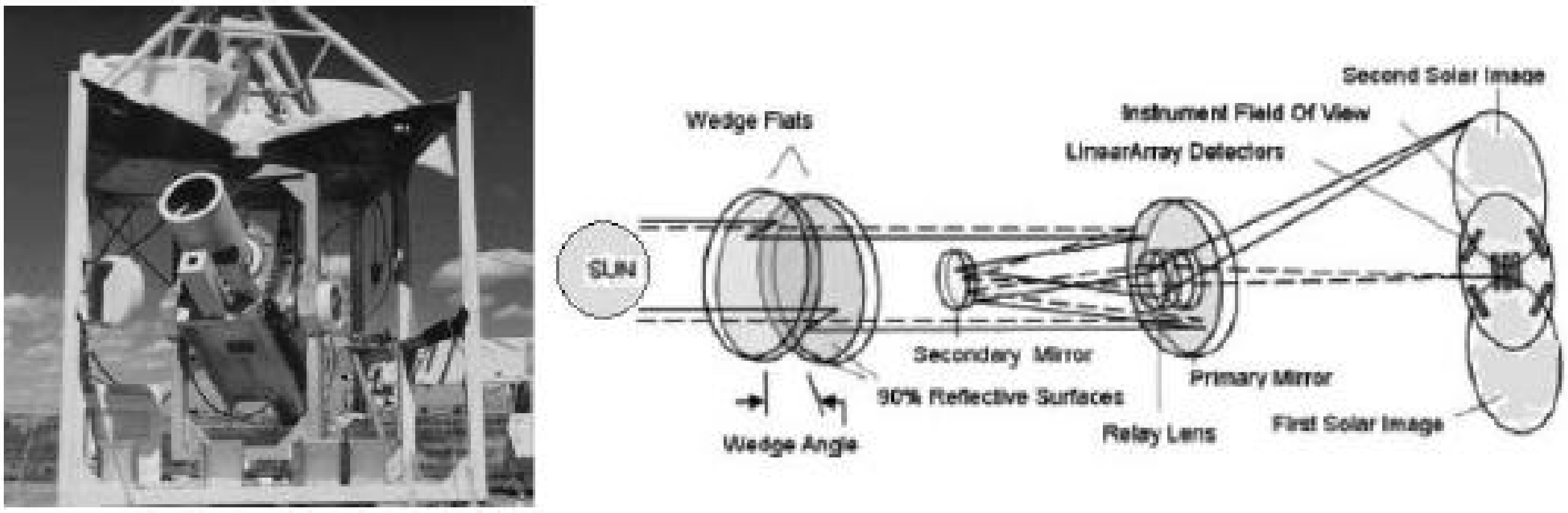

**Fig. 2.13 SDS: focal plane scheme.**[100]

SDS has made 5 useful flights, 4 of them have been published, with two different analyses,[101] leading to similar trends with a systematic difference between them. The solar diameter increased since 1992, inversely to the magnetic activity. This result is in agreement with the result of the astrolabes.

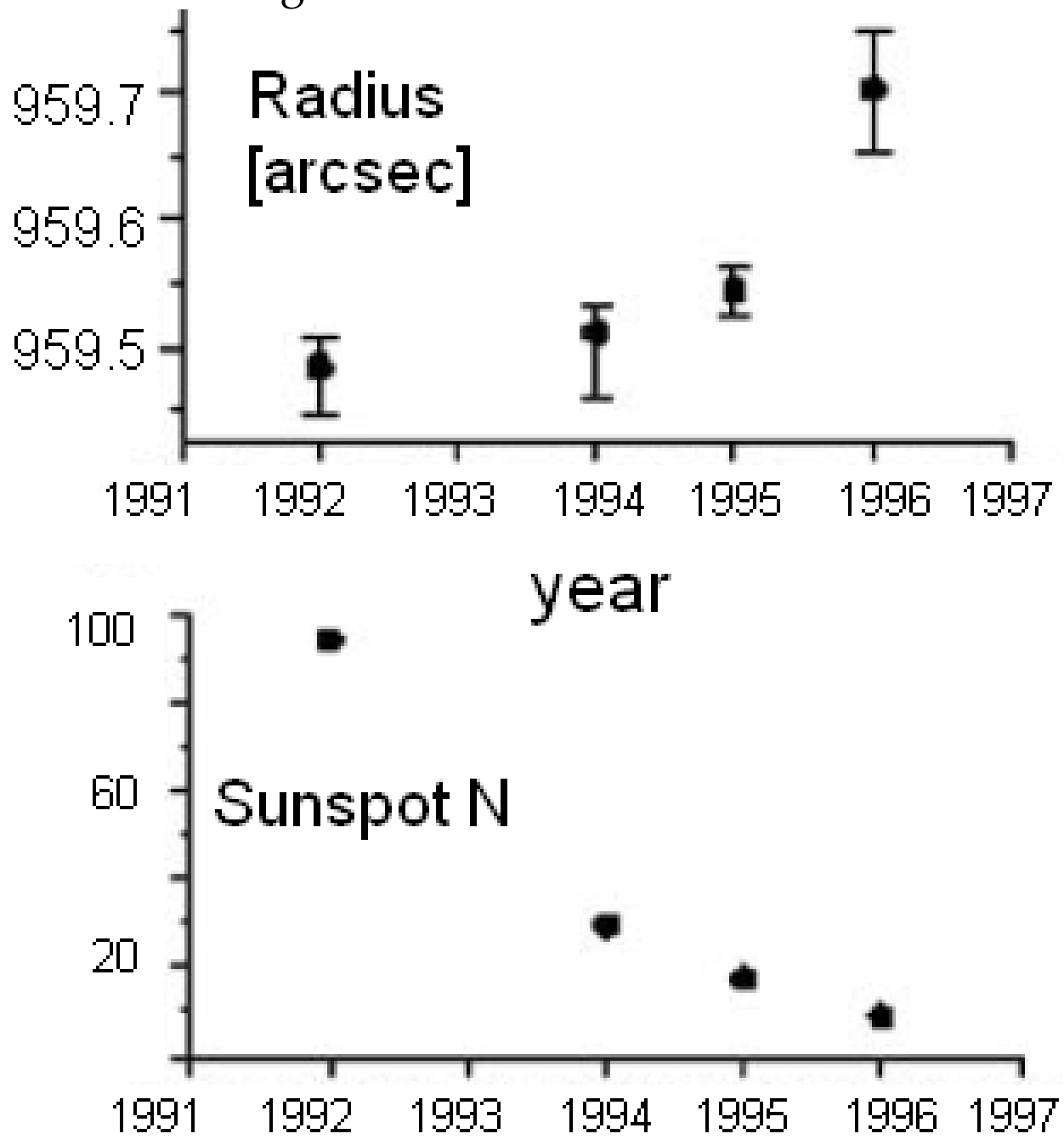

**Fig. 2.14 SDS: measurements of solar diameter**, from Egidi et al. (2006).

[100] Chiu, H.Y., Maier, E.,0 Schatten, K. H., Sofia, S. Solar disk sextant optical configuration, Applied Optics, **23**, 1230 (1984).

[101] Egidi, A., B. Caccin, S. Sofia, W. Heaps, W. Hoegy, L. Twigg, High-Precision Measurements of the Solar Diameter and Oblateness by the Solar Disk Sextant (SDS) Experiment, Sol. Phys. 235, 407 (2006).

Djafer, D., G. Thuillier, S. Sofia, A. Egidi, Processing Method Effects on Solar Diameter Measurements: Use of Data Gathered by the Solar Disk Sextant , Sol. Phys. 247, 225 (2008).

### 2.4.4 Picard satellite

Picard is a satellite dedicated to the simultaneous measurement of the solar diameter, the solar shape, the solar irradiance and the solar interior. It has been launched on June 15, 2010. The Picard payload consists in absolute radiometers and photometers measuring the total solar irradiance and in the SODISM instrument which is an imaging telescope developed to determine the diameter, the limb shape and the asphericity of the Sun. The Picard mission has also a ground segment consisting of several instruments based at the Calern observatory. The ground segment is composed with the qualification model of the space instrument, MISOLFA the solar seeing monitor and some others ground-based instruments giving useful data such as the solar irradiance, air temperatures, the wind velocity and directions, the nebulosity etc. The Picard ground-based instruments are installed at the Calern observatory.[102]

SODISM (Solar Diameter Imager and Surface Mapper) is an 11-cm diameter Cassegrain telescope associated with a 2048x2048 pixels CCD detector where the whole SUN is formed. Wavelengths are selected by mean of interference filters placed on 2 wheels. Wavelength domains have been chosen free of Fraunhofer lines (535.7, 607.1 and 782.2 nm). Active regions are detected in the 215 nm domain and the CaII (393.37 nm) line. Helioseismologic observations are performed at 535.7 nm. The satellite platform is stabilized within 36 arcseconds field. The telescope primary mirror stabilizes then the Sun image within 0.2 arc-second using piezoelectric actuators. An internal calibration system composed with 4 prisms, allows to follow scale factor variations induced by instrument deformations resulting from temperature fluctuations in orbit or others causes.[103] This is obtained by processing the 4 corner images formed at 535 nm by the prisms. The diameter measurements are referred to star angular distances by rotating the spacecraft towards some doublet stars several times per year. The instrument stability is assured by use of stable materials (Zerodur for mirrors, Carbon-Carbon and Invar for structure). The whole instrument is temperature stabilized within 1°C. The CCD is also temperature stabilized around -7°C within 0.2°C. In order to limit the solar energy, a window is set at the telescope entrance limiting the input to 5% of the total solar irradiance. No significant ageing has been measured in laboratory for all the duration of the mission.[104]

### 2.4.5 SOHO

SOHO MDI has been used for determining the solar diameter,[105] even if its configuration is not optimized for this purpose. The value of the solar diameter has been constant within 0.05 arcsec according to the last analysis, around the standard value of 959.63 arcsec,[106] but the calibration of the instrument was made using the Mercury transit of 2003,[107] and the solar diameter was consistent, 959.25 arcsec[108] ($\Delta R$=-0.38 arcsec) with the one measured during 2006 total eclipse where $\Delta R$=-0.41 arcsec.[109]

A new analysis on the two transits of Mercury of 2003 and 2006 led Emilio et al.[110] to claim a diameter of the Sun rather constant between these two transits and of 960.12 arcsec with a positive $\Delta R$=+0.49 arcsec

---

[102] Sigismondi, C., Picard satellite for solar astrometry, Proc. 2nd Galileo-Xu Guangqi Meeting, Ventimiglia - Villa Hanbury, Italy, 11-16 July 2010, also on arXiv:1106.2198 (2011).

[103] Assus, P., A. Irbah, P. Bourget, T. Corbard and the PICARD team, Astron. Nachr. **329**, 517 – 520 (2008).

[104] Irbah, A., et al., *Ground-based solar astrometric measurements during the PICARD mission*, submitted to SPIE (2011).

[105] Bush, R. I., M. Emilio and J. R. Kuhn, On the Constancy of the Solar Radius. III, Astrophys. J. **716**, 1381 (2010).

[106] Auwers, A., *Der Sonnendurchmesser und der Venusdurchmesser nach den Beobachtungen an den Heliometern der deutschen Venus-Expeditionen*, Astronomische Nachrichten **128**, 361 (1891).

[107] Kuhn, J. R., R. I Bush, M. Emilio, and P. H. Scherrer, On the Constancy of the Solar Radius. II Astrophys. J. **613**, 1241 (2004).

[108] i.e. $\Delta R$=-0.38 arcsec with respect to the standard angular solar radius at 1 AU 959.63 arcsec.

[109] Kilcik, A., C. Sigismondi, J. P. Rozelot and K. Guhl, Solar Phys. **257**, 237 (2009).

[110] Emilio, M., Kuhn, J. R., Bush, R. I. and I. F. Scholl, I. F., Astrophys. J. **750,** 135 (2012).

Since the first publication[111] the authors claim to have fully under control the systematic errors as ageing processes of the detector and temperature forcing, but the accuracy of their result (0.01 arcsec) is two order of magnitude below the systematic errors. For this reason and for their calibration made with the Mercury transit (which disconfirm the thesis of Sun radius close to the standard one) I take these conclusions with care.

SOHO did not observe the last two transits of Venus, because from its orbit it was unvisible.

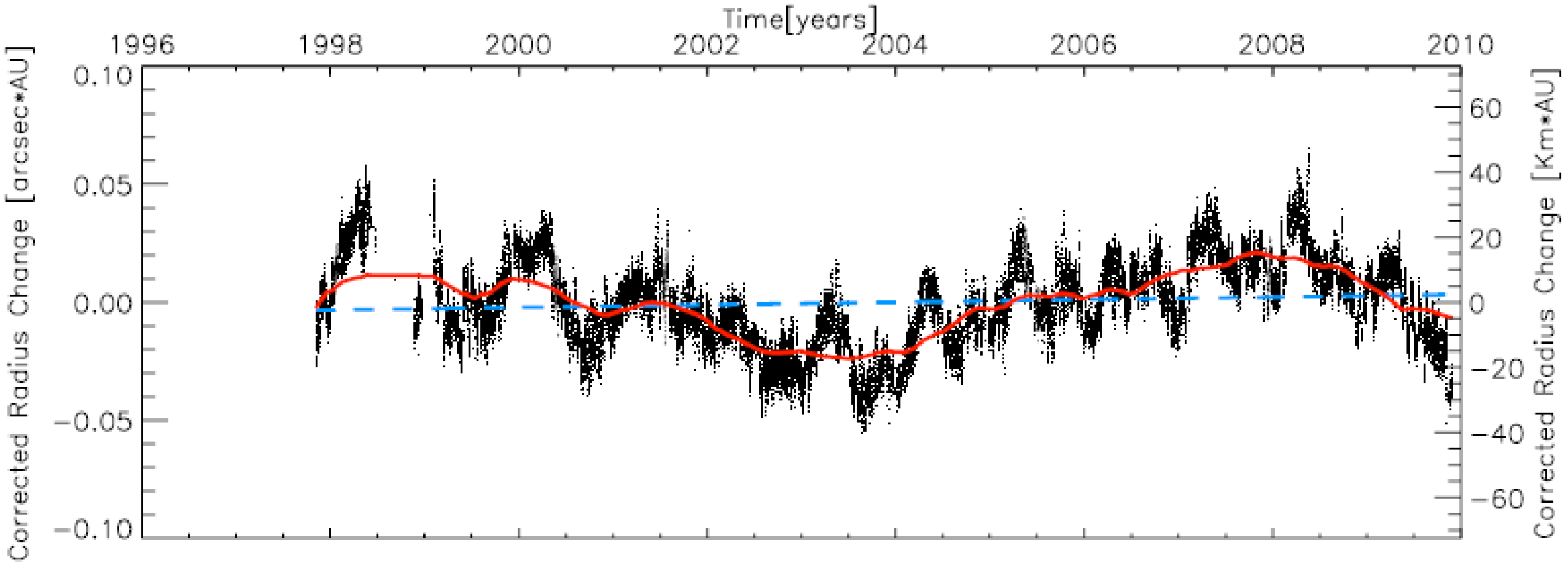

**Fig. 2.15 SOHO/MDI measurements of solar diameter** (version 2010). Here the anticorrelation seen with the astrolabes between solar spots and radius would disappear.

### 2.4.6 SDO/HMI

The debate around the variations of the solar diameter is still open and also we have presented some measurements made with SDO/HMI (Solar Dynamics Observatory/ Helioseismic Magnetic Imager) satellite during 2011.[112] In the process of calculation of the solar diameter as seen from 1AU there are still some residual oscillations of the same timescale of the orbital periods of the satellite around the Earth. This may be done also to other instrumental effects, than the orbital ones.

[111] Kuhn, J. R., R. I Bush, M. Emilio, and P. H. Scherrer, On the Constancy of the Solar Radius. Astrophys. J. **543**, 1007 (2000).

[112] Wang, X. and C. Sigismondi, *Geometrical information on the solar shape: high precision results with SDO/HMI*, in *Solar and Astrophysical Dynamos and Magnetic Activity* (A.G. Kosovichev, E.M. de Gouveia Dal Pino and Y. Yan, editors) Proceedings IAU Symposium No. 294 (2013). http://arxiv.org/pdf/1210.8286v1

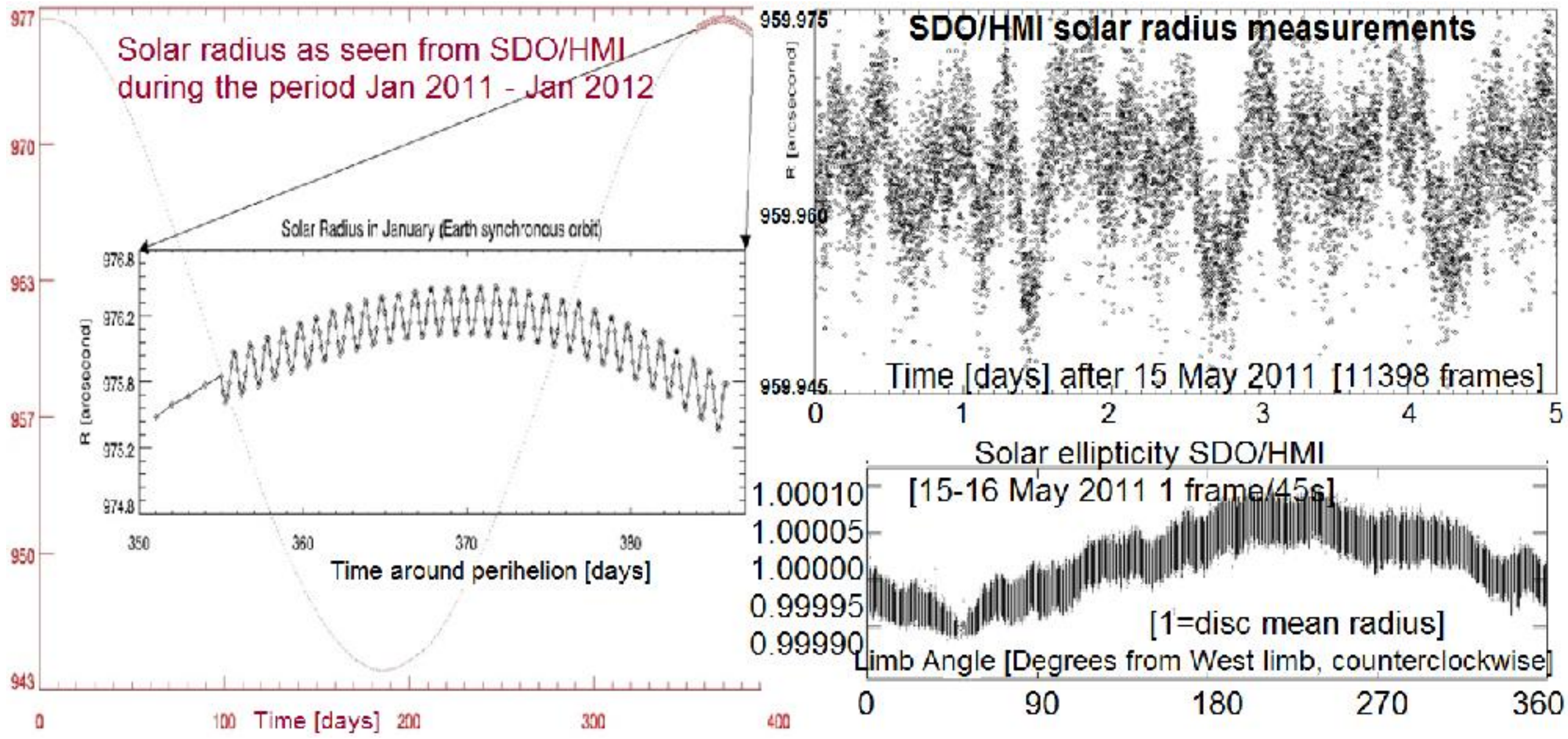

**Fig. 2.16 SDO/HMI measurements of solar diameter** The apparent solar radius affected by the Earth orbital motion and the satellite orbital motion (left image with enlargement). The solar radius appears as a fluctuating signal when the orbital motions are removed (upper-right image), and this shows that there is still some orbital motion residuals. The apparent shape of solar disc calculated from 1920 full-disc images on very quiet activity period (15-16 May 2011) shows small departures from the sphere (lower-right image).

The solar radius averaged over one year in this analysis of SDO/HMI data is 959.963±0.005 arcsec (1 σ).
It is 0.333 arcsec larger than the standard value (959.63 arcsec) and 2 σ smaller than 960.12±0.12 arcsec of Emilio et al. (2012).
Higher accuracy could depend on better orbit information, better data set (single wavelength), and absolute pixel scale calibration using Mercury and Venus transits (Sigismondi and Wang, 2013); Averages over shorter timescales can be done. The pixel scale determination is crucial in this analysis and only the transit of Venus can help to know this scale with the utmost precision.
The oblateness is just a preliminary result and the rough estimate is 0.048±0.03 arcsec (1 σ) since the deformation of the focal plane is unknown. The solar oblateness (ellipticity) is a difficult topic, more related to the dynamics and physics of the solar interior. The absolute pixel calibration, orbit information, and focus step are no longer essential, but flat field, focus plane deformation, and CCD small tip-tilt can affect seriously this measure.

### 2.4.7 RHESSI

RHESSI is a satellite devoted to the study of solar flares. RHESSI is a solar X-ray/$\gamma$-ray observatory, and the astrometric data come serendipitously from the Solar Aspect Sensor. The RHESSI measurement essentially follows Dicke's[113] method of using a rapidly rotating telescope to control systematic errors. The Solar Aspect Sensor consists of three independent optical systems, each with a simple lens (4 cm diameter) mounted on the front tray of the RHESSI modulation collimators, and a linear CCD sensor mounted on the rear tray at a separation of 1.55 m. The 2048-element CCD pixels are 1.73 inches square, and the observing wavelength is a 12-nm bandpass at 670 nm wavelength. The telemetry provides frequent samples (16/s) of each of the six limb intercepts in nominal pointing conditions, plus full CCD images at a slower cadence (1/min).

The solar oblateness (equator-pole radius difference) are determined from the axisymmetric quadrupole term of the Fourier components of the limb position given by the RHESSI data. The radius measurements are numerous, telemetered at about 100 samples/s, and are distributed approximately uniformly in azimuth around the limb.

Sunspots (negative excursions) and active-region faculae (positive excursions) produce obvious effects.

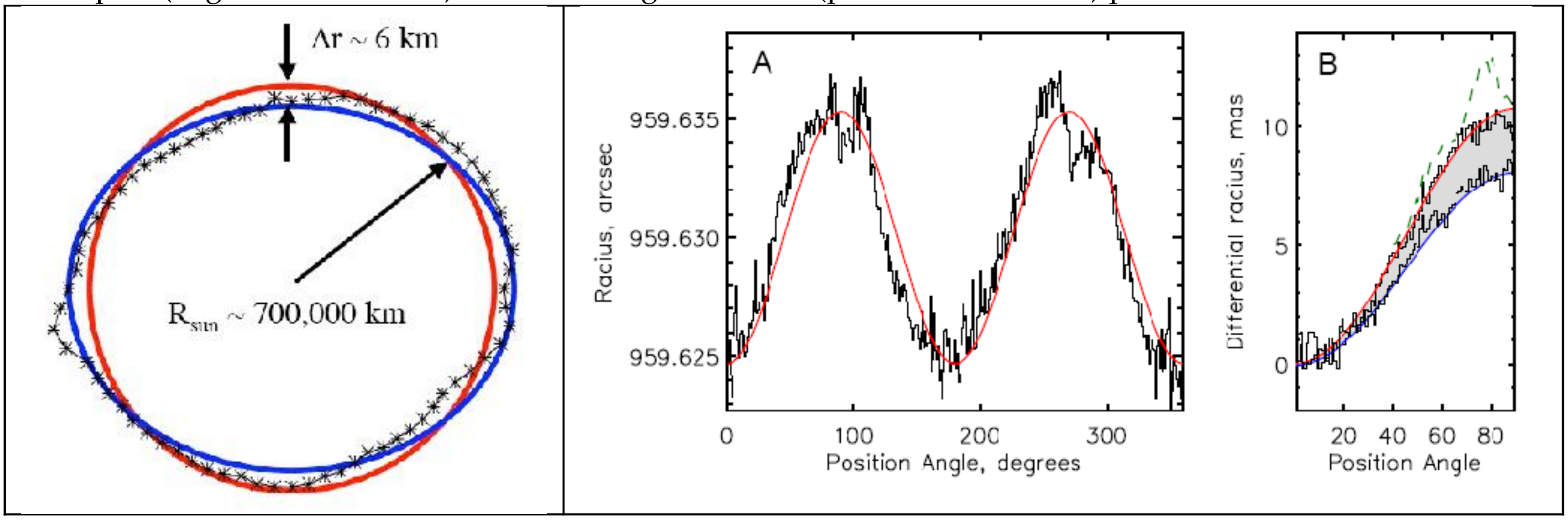

**Fig. 2.17 RHESSI measurement of the solar figure**: the normalization has been done with respect to the standard solar radius. From 2002 to 2007 the oblateness of the Sun is slightly larger (10.74 ± 0.44 mas) than the one (8.7 mas) due to the rotation of the Sun, because of magnetic activity.[114]

[113] Dicke, R. H. and M. Goldenberg, Phys. Rev. Lett. **18**, 313 (1967).

[114] Fivian, M. D., H. S. Hudson, R. P. Lin, H. J. Zahid, Science **322**, 560 (2008).

## 2.5 Oblateness and General Relativity connection

A slightly oblate Sun was suggested to explain the precession of the perihelion of Mercury. Therefore accurate measurements of solar oblateness were carried out by Dicke (1960-70s), Sofia (SDS 1990s) and now using the RHESSI satellite, to assess classical contributions to this anomalous precession. The required accuracy of these measurements is below one part over 10000, the same order of magnitude of expected solar diameter variability. On this topic I have published three papers: two on Precession in General Relativity[115], [116], where among the various possible precessions there is the one determined by the oblateness of the central body which is the source of the gravitational filed; and a third one on the direct measurement of the oblateness from SDO/HMI data.[117] In particular the measurement made on SDO data has shown an oblateness of 0.048±0.03 arcsec, which means that the ratio between polar and equatorial diameter of the Sun is $(1-a/b)=2.5\cdot10^{-5}$.

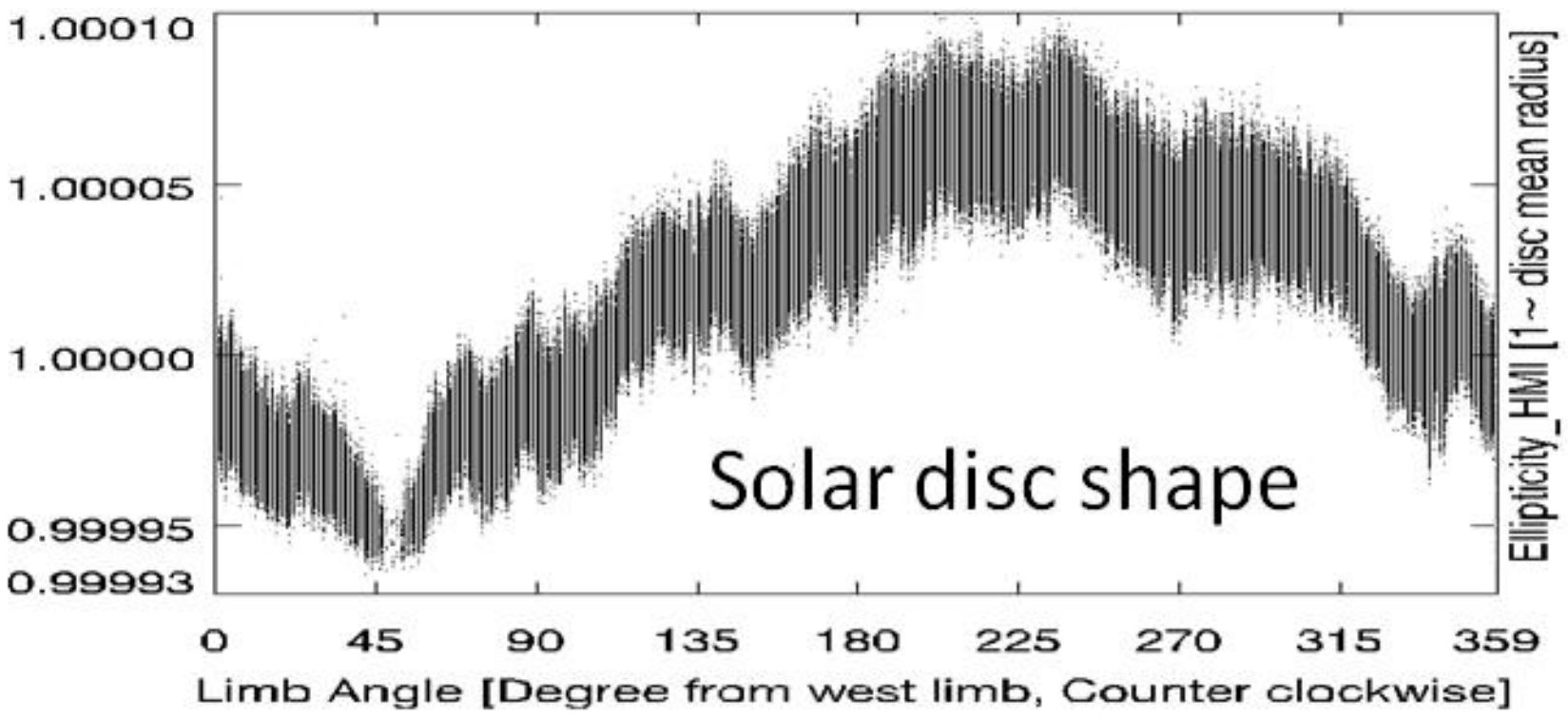

**Fig. 2.18 SDO/HMI measurements of the solar oblateness.** Data of 15-16 may 2011.

115 Sigismondi, C., *Astrometry and Relativity*, Il Nuovo Cimento B, **120**, 1169 (2005).
116 Sigismondi, C., *Relativistic Implications of Solar Astrometry*, http://arxiv.org/abs/1106.2202v1 (2011).
117 Sigismondi, C., *Geometrical information on the solar shape: high precision results with SDO/HMI* , http://arxiv.org/pdf/1210.8286v1 (2012).

## Chapter 3: Eclipses

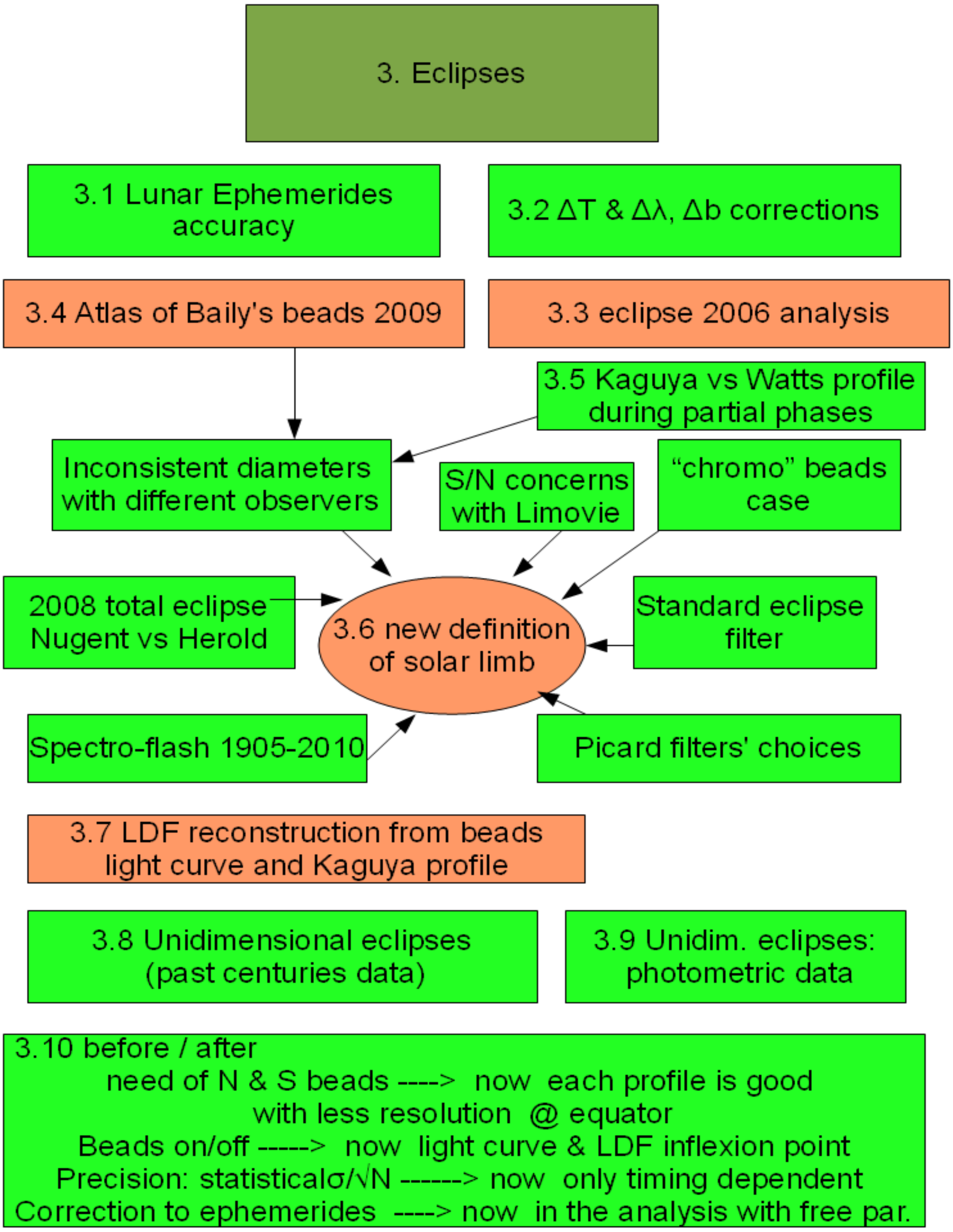

The method of measurement the solar diameter with the total eclipses is based on the fact that the last glimpse of photospheric light changes sudden the aspect of the sky, the darkness appear and some stars are visibile. The transition between partial and total phase occurs rather instantaneously.

Several eclipses have been studied, historical and recent, both with naked eye and electronic detectors and the information of the light during the transition before totality becomes crucial to achieve resolutions down to 0.01 arcsec in the solar diameter measurement.
The eclipse data are scattered and rare, about one each year is visible; I have inspected the past 40 years of eclipses in video tape and digital video.

### 3.1 Lunar ephemerides accuracy

The ephemerides of the Sun are much better known with respect to the lunar ones. This is since the times of Tycho Brahe, who gathered several observations of the Sun and produced the first accurate ephemerides of his time. Later Giandomenico Cassini, using the great meridian line of Bologna, recovered further precision, such that he was able to distinguish between Ptolemaic ephemeris and Keplerian, in the famous problem of Bissection of eccentricity.[118]
The two body problem of the Sun – Earth has much less perturbations from the other planets, than the three body one including the Moon. Newton produced a "theory" of the lunar motion with several parameters in 1702.[119] Later the "theory" of the Moon was constantly studied, because the method of the lunar distances was promising for the solution of longitude's problem in the sea.
E. W. Brown[120] at Yale after studying the works of G. W. Hill (the one of the restricted three body problem) completed the lunar theory entirely based upon gravitation, publishing tables with precision up to 0.001 arcsec. Later a discrepancy of 10.71 arcsec and the secular acceleration of the Moon, already discovered by Halley studying the occultations published in the Almagest of Ptolemy, drove Brown to attribute to the Earth's rotation variations the cause.
This is to understand why the "lunar theory" is much more complicated than the solar one. Nowadays more than 400 terms are used to compute the polynomials for the numerical lunar theories (of JPL and of IMCCE, the former Bureau de Longitudes) which have names like DE200 (DE are ephemerides numerically integrated) or Chapront (the French astronomers prefer to use series of analytical polynomials, usually of Chebitchev type).

The inclusion of laser-ranging data allows getting very high precisions in the lunar ephemerides.
Nevertheless the method of Baily's beads for measuring the solar diameter, originally, required data from both North and South limits of the shadow. This was to avoid possible errors on the lunar ephemerides.
Nowadays the problem of the position of the center of the Moon, which is the issue of a lunar theory, has two main problems: the difference between center of mass and center of figure, and the radius of the Moon.
For the first problem there is a maximum distance between the center of mass and the center of the figure of 0.5 arcsec in longitude and 0.24 in latitude. This space vector rotates in the space with the Moon, such that there is an instantaneous value each time.[121] Generally the ephemerides are referred to the center of mass, because they are calculated with the gravity laws.
The second problem appeared when the lunar profile changed from Watts to Kaguya (see chapter 3.5).
The Japanese team of the Kaguya lunar probe, calculated the altimetry of the Moon with respect to a sphere, centered on the center of mass. The radius of this reference sphere is different from the one used in the classical Watts profiles published in 1963 and tested with more than 60000 occultations.
The Watts profile has fewer points than Kaguya, and therefore less resolution, but the Kaguya profile still lacks of the test of occultations.

---

[118] Heilbron, J. L., *The Sun in the Church*, Harvard University Press, Boston (1999). Kollerstrom, N., *Newton's Forgotten Lunar Theory: His Contribution to the Quest for Longitude*, Green Lion Press, Santa Fe, (2000).
[119] Cohen, B., *Isaac Newton, Theory of the Moon's Motion (1702). With a bibliographical and historical introduction*, Dawson, London (1975).
[120] Brown, E.W., *Tables of the Motion of the Moon,* Yale University Press, New Haven CT, (1919).
[121] Sigismondi, C., *Relativistic Corrections to Lunar Occultations*, Journal of the Korean Physical Society, **56**, 1694 (2010).

| Source | Lunar radius |
|---|---|
| **Van Flandern[122] (1970)** | 1738,11 ± 0,03 km |
| **Morrison[123] (1979)** | 1738,23 ± 0,02 km |
| **Morrison & Appleby[124] (1981)** | 1738,05 ± 0,02 km |
| **Rosseló & Jordi[125] (1982)** | 1738,07 ± 0,02 km |
| **Newhall et al.[126] (1983)** | 1738,09 km |
| **Sôma[127] (1985)** | 1738,107 ± 0,004 km |
| **Rosseló & Jordi[128] (1991)** | 1738,103 ± 0,002 km |
| **Kaguya[129] (2009)** | 1737,400 km |

Table 3.1 Different lunar radii for the Watts profile, in the last row the Kaguya radius.[130] As it is evident, the Moon of Kaguya is smaller than the others.

Accurate tests have still to be done in view of clarifying the compatibility between the figure of the Moon of Kaguya and the Watts profile. The Watts profile was adjusted by occultations, while Kaguya is produced by the LALT Laster ALTimeter of that Japanese lunar probe.
There are several cases which demonstrate that the profile of Kaguya is more accurate, but the times of contacts suffer of systematic differences due to the different radius and the position of the center of the figure different from the center of mass.
At the moment, as an example, the duration of totality in 2010 July 11 eclipse, in the Hao atoll, French Polynesia, has been calculated with two programs based on the two profiles Watts and Kaguya, and the result is rather different.

| Phenomenon | Watts (Occult4) | Watts IMCCE | Kaguya Occult4 | Kaguya IMCCE Optical center | Kaguya IMCCE Center of mass |
|---|---|---|---|---|---|
| Second contact | 18h 37m 43.8s | 18h 37m 43.8s | 18h 37m 41.6s | 18h37m 41.459s | 18h 37m 43.367s |
| Third contact | 18h 38m 8.6 s | 18h 38m 8.6 s | 18h 38m 12.2s | 18h 38m 8.279s | 18h 38m 11.349s |
| Duration of totality | 24.8s | 24.780s | 30.6s | 26.82s | 27.98s |

Table 3.2 Different prediction for the duration of the totality of July 11, 2010 at Hao atoll.[131]

The reason of these discrepancies is not yet clear. Nevertheless the calculation with Kaguya profile made with the center of mass is to be preferred.

[122] Van Flandern, T., Astron. J. **75**, 744 (1970)
[123] Morrison, L. V., *An analysis of lunar occultations in the years 1943-1974 for corrections to the constants in Brown's theory, the right ascension system of the FK4, and Watts' lunar-profile datum*, Monthly Notices of the Royal Astronomical Society **187**, 41 (1979).
[124] Morrison, L. V. and G. M. Appleby, Monthly Notices of the Royal Astronomical Society **196**, 1013 (1981).
[125] Rosseló, G. and C. Jordi, The Moon and the Planets **27**, 131 (1982).
[126] Newhall, X. X., E. M. Standish, J. G. Williams, *DE 102 - A numerically integrated ephemeris of the moon and planets spanning forty-four centuries*, Astron. Astrophys. **125**, 150 (1983).
[127] Sôma, M., Celes. Mech. **35**, 45 (1985).
[128] Rosseló, G. , C. Jordi and A. Salazar, Astrophysics and Space Science **177**, 331 (1991).
[129] Araki, H., et al., *Lunar global shape and polar topography derived from KAGUYA-LALT laser altimetry*, Science **323**,897 (2009).
[130] This table has been compiled by Patrick Rocher of IMCCE (Paris), to whom I am particularly indebted for the fruitful discussions and for the research material he provided to me.
[131] Also this table has been computed by Patrick Rocher.

| Source | Ellipticity correction |
|---|---|
| **Morrison (1979)** | –(0,09" ± 0,01") cos(2p – 146°±1°) |
| **Morrison & Appleby (1981)** | –(0,09" ± 0,01") cos(2p – 146°±1°) |
| **Rosseló & Jordi (1982)** | –(0,087" ± 0,007") cos(2p – 146°±4°) |
| **Sôma (1985)** | –(0,128" ± 0,003") cos(2p – 135°±1°) |
| **Rosseló & Jordi (1991)** | –(0,091" ± 0,003") cos(2p – 145°±4°) |

Table 3.3 Different ellipticity corrections to the Watts lunar profile.[132]

Moreover the North pole of the Watts profiles (WA=Watts Angle) is displaced from the real North pole (AA=Axis Angle) of 0.241°. The formula WA=AA + 0.241° is valid for conversions.[133]
The profile of Kaguya is not corrected as happened with the Watts profile. It is centered on the center of Mass of the Moon, while Watts was centered on the figure's center in continuous rotation around the center of mass.

## 3.2 ΔT & Δλ, Δβ corrections

Despite of the claimed accuracy of the ephemerides, dealing with hundredths of arcsecond, there are differences between analytical and numerical ephemerides.
In the analysis of the total eclipse of 2006 I have assumed that the profiles of Watts (the only one available before November 2009) with and without corrections of Morrison and Appleby, were the better approximations of the real lunar limbs, or the lunar figure.
The rigid figure of the Moon has been left free to move around the first solution for Baily's beads timing with a modified solar diameter, in order to minimize the residuals.
In other words the differences between Observed and Calculated timings $\Sigma(O\text{-}C)^2$ for each bead event has been minimized first with respect to a ΔR for the solar radius; after, starting from this solution, other minimizations have been calculated numerically with respect to

4) ΔT which changes slightly the phase of the Earth rotation; it is the ΔUT1 parameter which is available daily from the IERS[134] service "International Earth Rotation System". The accuracy of the published values ΔUT1 is better than 0.001 s, but its use here is like a freedom parameter to include all possible uncertainties on ephemerides and their implementation in the softwares.[135]
5) Δλ moves the Moon in longitude, along its orbit, leaving the Earth rotation fixed. The effect is similar to the previous correction, but there are small differences.
6) Δβ is a change in lunar latitude (zero corresponds to the Moon's center of mass exactly on its orbit). This correction was originally[136] computed by arithmetical average between the solar radius correction calculated using the beads observed at the Southern edge of the eclipse, and the solution coming from the Northern ones.

[132] Also this table has been computed by Patrick Rocher.
[133] Sôma, M., Celes. Mech. **35**, 45 (1985).
[134] www.iers.org
[135] The software utilized by me for the analysis of the eclipses is Occult 4. Since the year 2000, when I started these analyses many versions of the software have been modified by D. Herald. I never get access to the code. That's why I assume as fairly good the software, but I always concern with a possible small error. The same consideration stands valid for other ephemerides softwares.
[136] The method of the Baily's beads during total eclipses has been developped since 1973.The prescription to cover both Southern and Northern edges of the umbral shadow was made for being able to correct for the Δβ.

| Variations due to the librations | Motion of the origine of selenographic coordinates | Visibility |
|---|---|---|
| $\beta > 0$ | The center of the Moon is moved down | The North of the Moon is more visible |
| $\beta < 0$ | The center of the Moon is moved up | The Sud is more visible |
| $\lambda > 0$ | The center of the Moon is moved toward geocentric East | The East is more visible at the geocentric West |
| $\lambda < 0$ | The center of the Moon is moved toward geocentric West | The West is more visible at the geocentric East |

Table 3.4 The variations of the appearances of the Moon when β and λ change.[137]

According to the table 3.4 there is a small change also in the profile of the lunar limb, if β and λ change. These variations are of the second order, and I neglected them in my analysis, leaving the profile unperturbed.

The complete procedure has been described by me in 2009,[138] and applied to the analysis of 2006 eclipse.[139]

The advantage of this method is to reduce the errors of the ephemerides or of the software which implement them. The disadvantage is that, if the figure of the Moon departs significantly from the real one, this method converges on a solution affected by this systematic error. But this would occur even with correct ephemerides and wrong profile.

The case of Kaguya profile is still under examination: this profile is much detailed than Watts, but the radius of the "lunoid" is 0.7 km smaller, which may introduce significant systematic errors, if the altimetry of the Moon made with Kaguya does not include automatically all past corrections made upon more than 60.000 occultations. More tests have still to be done on this respect.

## 3.3 2006 eclipse analysis

The total eclipse of 2006 has been analyzed following the method described above in paragraph 3.2.

According to this analysis the correction to the diameter of the Sun on March, 29 2006 was ΔR=-0.41 ± 0.04 arcsec.

The event of appearance or disappearance of a Baily's bead has been inspected on the video, visually and using the LIMOVIE software.[140] The first or the last glimpse of light has been considered as the first or the last appearance of the photosphere.

## 3.4 Atlas of Baily's beads 2005-2008 and 2010 eclipses

The data of the observational campaigns of annular and total eclipses of 2005-2008 made by IOTA members have been published in Solar Physics in 2009.[141] Later IOTA members[142] have monitored the 15 January 2010 annular eclipse, and also other colleagues in Sri Lanka.[143]

[137] Also this table has been computed by Patrick Rocher.

[138] Sigismondi, C., *Guidelines for measuring solar radius with Baily beads analysis*, Science in China, **52 G**,1773 (2009).

[139] Kilcik, A., C. Sigismondi, J. P. Rozelot and K. Guhl, Solar Phys. **257**, 237 (2009).

[140] http://www005.upp.so-net.ne.jp/k_miyash/occ02/limovie_en.html

[141] Sigismondi, C., et al., Solar Phys. **258**, 191 (2009).

[142] Raponi, A., C. Sigismondi, K. Guhl, R. Nugent, A. Tegtmeier, *The Measurement of Solar Diameter and Limb Darkening Function with the Eclipse Observations*, Solar Physics **278**, 269 (2012).

[143] Adassuriya, J., S. Gunasekera, N. Samarasinha, *Determination of the Solar Radius based on the Annular Solar*

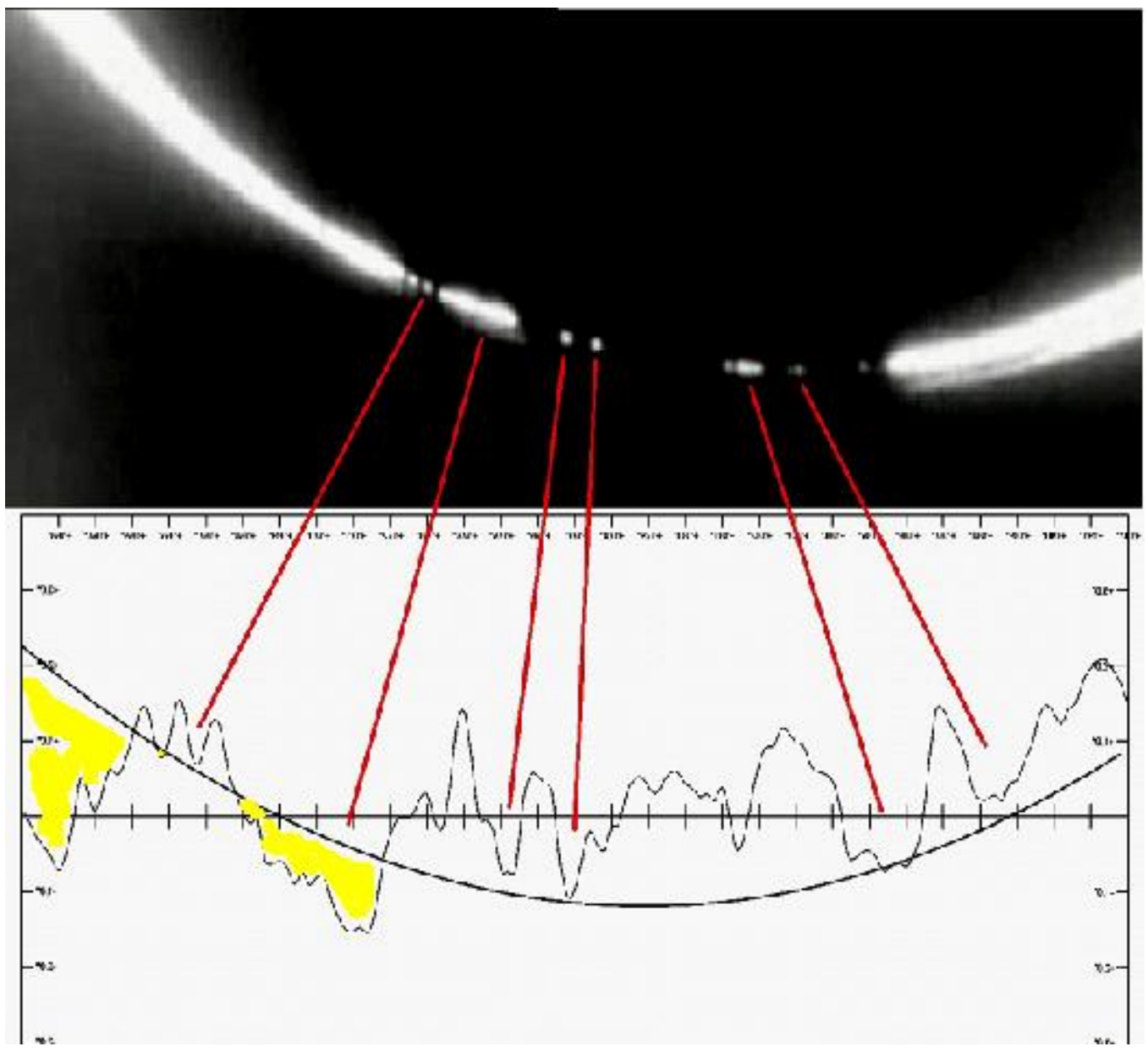

**Fig. 3.1 Baily's beads from the annular eclipse of January 15, 2010.**[144]

The total eclipse of 2010 on the Easter Island is being used at the Observatorio Nacional of Rio de Janeiro to calibrate the new heliometer, and the mission of CNRS in French Polynesia with 12 photometers is also under analysis.

We gather all the observations of Baily's beads on a single paper, and therefore we create a reference atlas.

The following analyses of these beads on these data did not drive us to clear conclusions: the diameter found by the various observers at different locations did not match together.

We looked for a reason, and this fact proved the need of changing the solar edge definition.

No more the ON/OFF treatment of considering the limb darkening function as a "step function", but the search of the inflexion point of the same function. This was needed because a different combination of telescope + filter + detector has different limiting magnitude for the beads, and so the outer part of the limb darkening function can or cannot be detected. In the "step function" view the limiting function of the instrument determines the position of the observed limb, yielding to inconsistencies between the diameters determined by different instruments. The following figures explain better this concept.

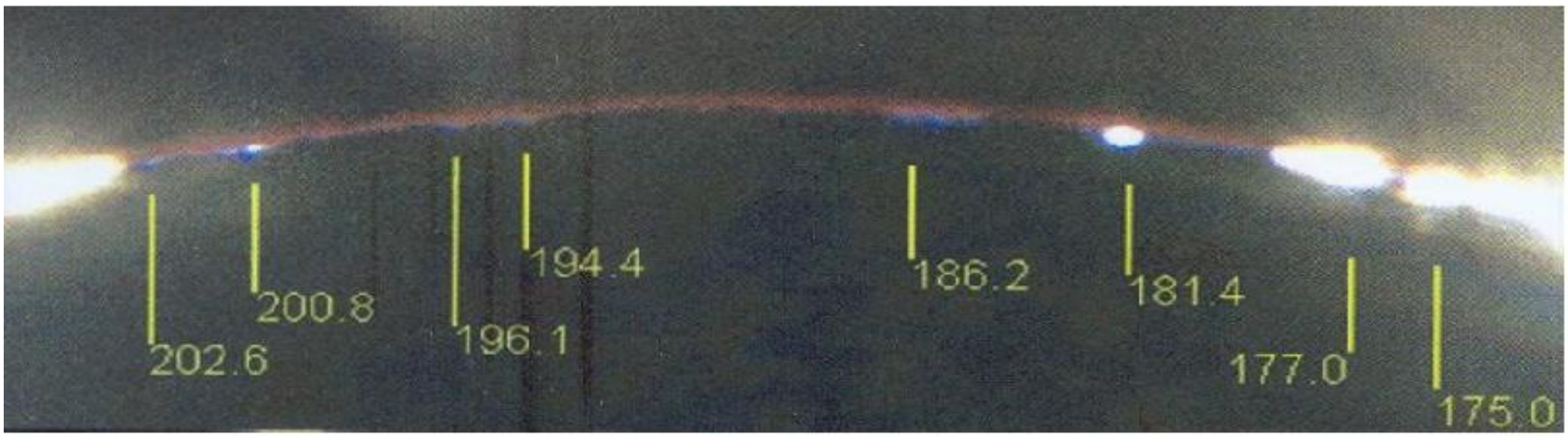

**Fig. 3.2 Baily's beads and light from the solar mesosphere**, as observed in the annular eclipse of October 3rd, 2005.[145]

---

*Eclipse of 15 January 2010*, Sun&Geosphere **6**, 17 (2011).

[144] Raponi, A., C. Sigismondi, K. Guhl, R. Nugent, A. Tegtmeier, *The Measurement of Solar Diameter and Limb Darkening Function with the Eclipse Observations*, Solar Physics **278**, 269 (2012).

[145] Schnabel, C., in *Trabajos de investigacion II* (Agrupacion Astronomica de Sabadell) p. 46 (2009).

The thin light has firstly been interpreted as chromospheric light, observable during total eclipse even with filters at density 4 to 5. After the presence of chromatism suggested that it was white light. Later it become clear that it was the light from the region of emitting lines which is above the solar photosphere and presented in the chapter 1.1 as solar mesosphere.

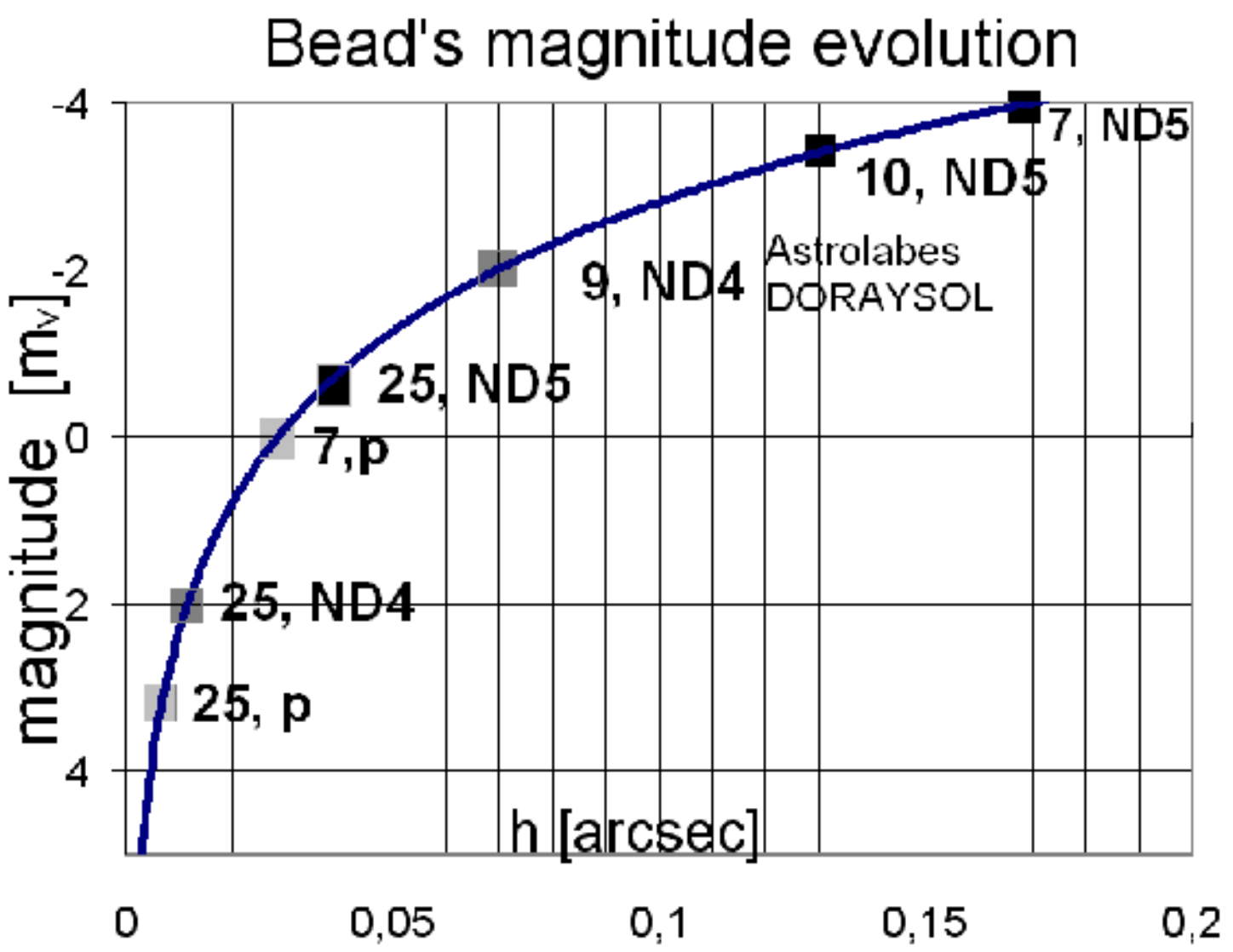

**Fig 3.3 Bead´s magnitude evolution**: the height of the solar limb above the valley is on the abscissa. The various square dots represent different type of telescopes. [25, p] means 25 cm opening with projection of the image. [25, ND4] the 25 cm telescope with a filter of Neutral Density of transmittance 1/10000; ND5 stands for 1/100000. **For this reason IOTA/ES decided in 2009 to standardize the filters and the telescopes for the eclipse missions**.

The visibility of the light from the solar mesosphere, which is usually difficult to be seen when observing the full disk for background reasons, and the limiting magnitude of a bead which depends on the telescope+filter+detector show that considering as a step function the true Limb Darkening Function (LDF) drives to estimate different photospheric diameters, as it is sketched in the following figure.

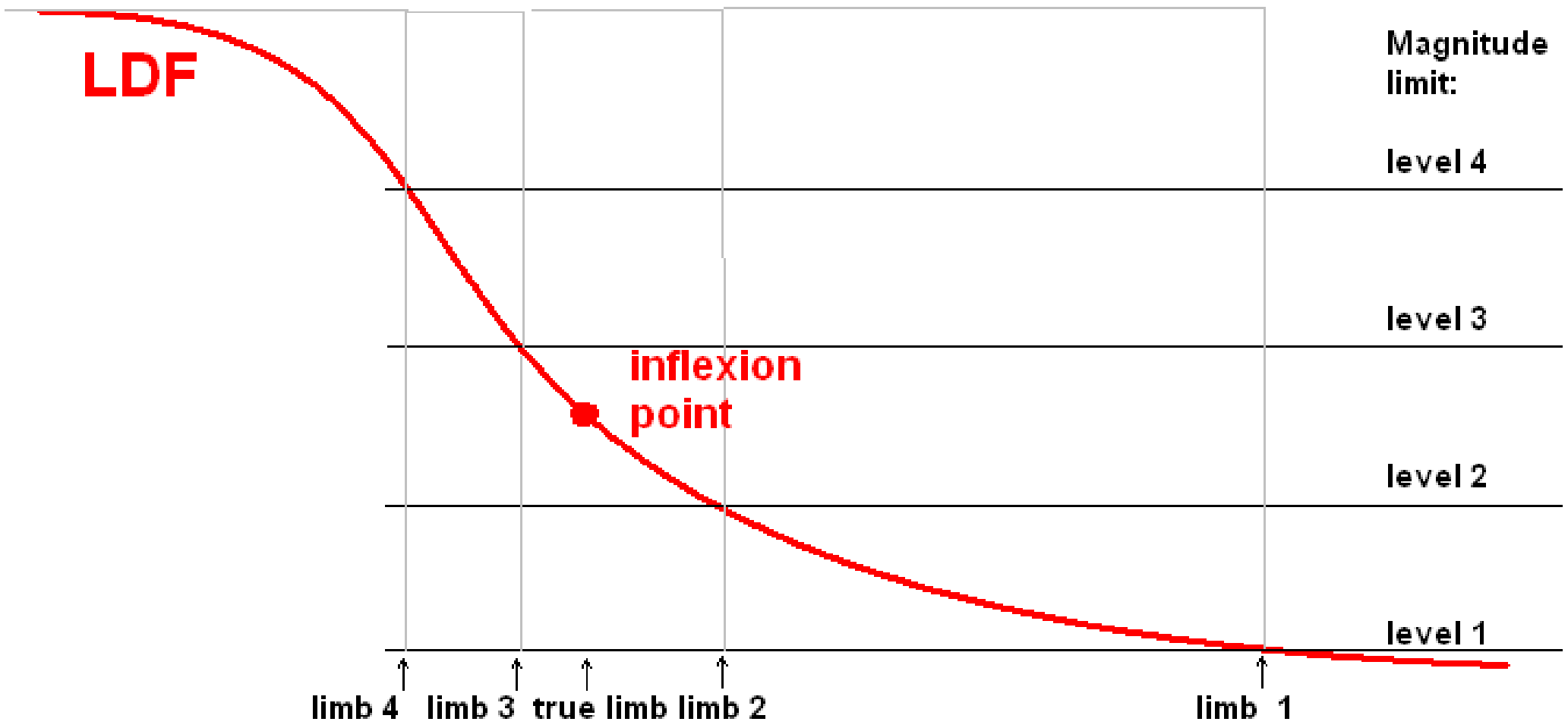

**Fig. 3.4 Variation of the estimated limb with different limiting magnitudes**.

The choice of Picard satellite filters has been done in order to select either regions of the spectrum with emission lines and without (see paragraph 2.4.4).

## 3.5 Kaguya vs Watts profile

To process the data concerning eclipse Baily's beads it is crucial to know the lunar limb profile. Until November 2009 the profile published by C. B. Watts in 1963,[146] sometimes locally upgraded with stellar occultation data, was the only one available. The precision of Watts atlas has been estimated to 0.20 arcsec. With N beads observed the influence of this uncertainty is reduced by the statistical factor √N.

During the occultation of Pleiades occurred on August 7-8, 2007 with a Schmidt Cassegrain telescope of 20 cm I recorded 9 stars occulted, spread over 90° of axis angles.

A correction of 0.7±0.8 arcsec was required to the lunar longitude, to minimize the residuals (O-C).[147]

The occultations of Venus (June 18) and Saturn (May 22, 2007) were also used to control the lunar longitude correction to the Occult 3 software.

After that correction a residual of 0.093 arcsec represented the departure of Watts profile from the real one in the case of Venus, the better measured.

Now the profile obtained by the Laser Altimeter LALT of the Japanese lunar probe KAGUYA is available,[148] with a sampling each 1.5 Km (about 1 arcsecond at the lunar distance) and an height's accuracy of ±1 m. Kaguya data are expected to be error-free.

In the partial eclipse of 4 Jan 2011 we attempted the imaging of the lunar profile (data from Bialkow coronograph)[149] with the purpose to compare it with the profile calculated by Kaguya. Moreover the fit of "Kaguya lunoid" with the observed limb is still to be fully verified.

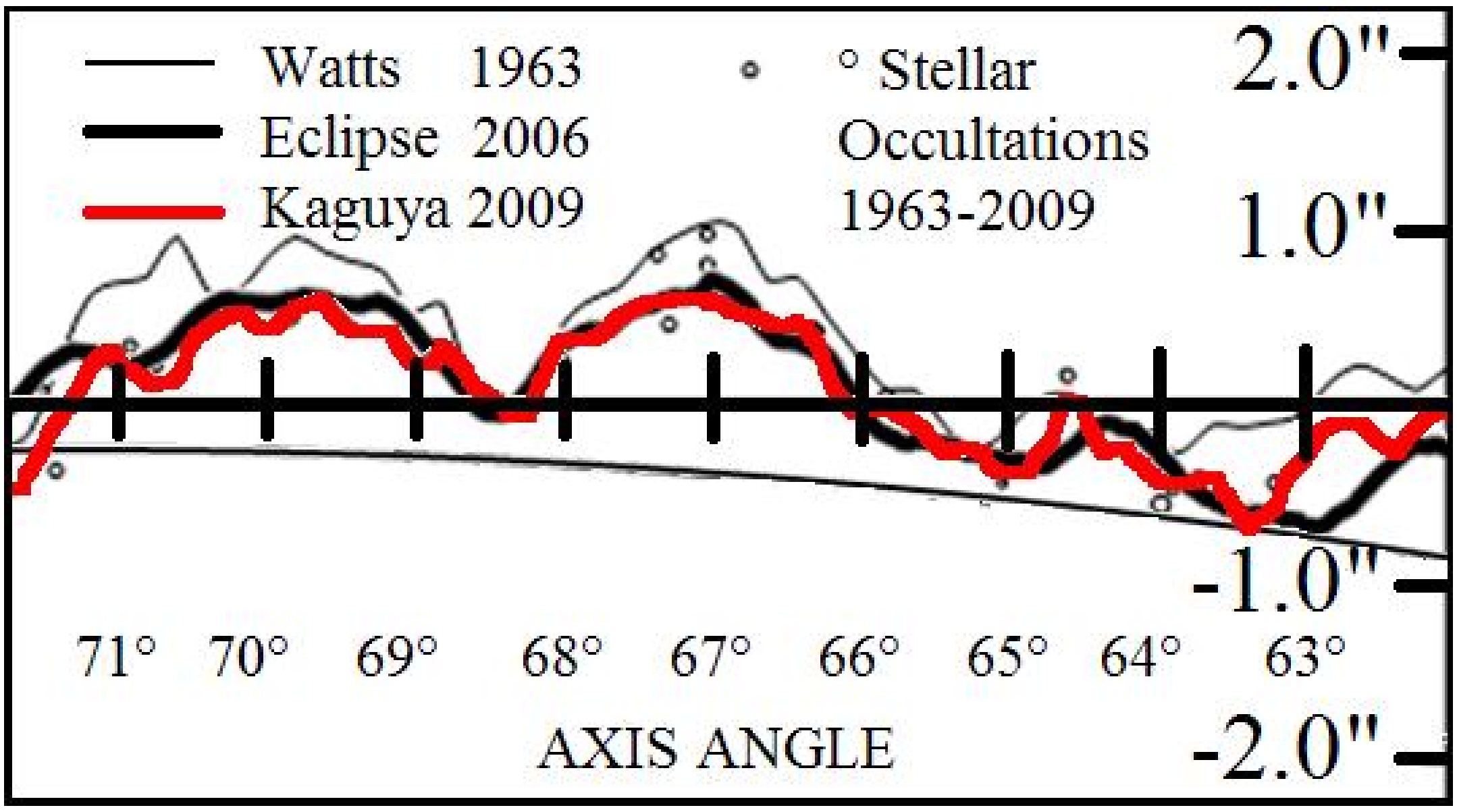

**Fig 3.5** The Watts and Kaguya profiles versus the real lunar profile.

This image has been obtained superposing to the Watts profile the lunar profile as obtained with an edge finding algorithm applied to the total eclipse of 2006 observed from centerline with a 20 cm Schmidt-Cassegrain telescope near Antalya, Turkey. The small dots are data from the archive of stellar occultations occurred near the same Watts angle (in abscissa) and near the same libration.
The thick (eclipse) line is the real profile, obtained with the algorithm to find the edge.[150]
The thin line is the Watts profile and red thick one the Kaguya profile. Departure as big as 0.4 arcsec occur between Watts and real.
A similar situation occurs when comparing Watts and Kaguya.
The great sampling of the Laser Altimeter of the Japanese probe Kaguya (Selene) gives a point each 1.5 km all over the Moon, with an accuracy of 4 m in height.

[146] Watts, C. B., *The Marginal Zone of the Moon, Astronomical Papers prepared for the use of the American Ephemeris and Nautical Almanac*, Vol. XVII (U. S. Government Printing Office, Washington, 1963).

[147] Sigismondi, C., *Relativistic Corrections to Lunar Occultations*, J. of Korean Physical Society **56** , 1694 (2010).

[148] Araki, H., et al., *Lunar global shape and polar topography derived from KAGUYA-LALT laser altimetry,* Science **323** 897 (2009).

[149] The image of this eclipse has been presented for the first time in the first chapter of this book, and used to measure the height of the spiculae.

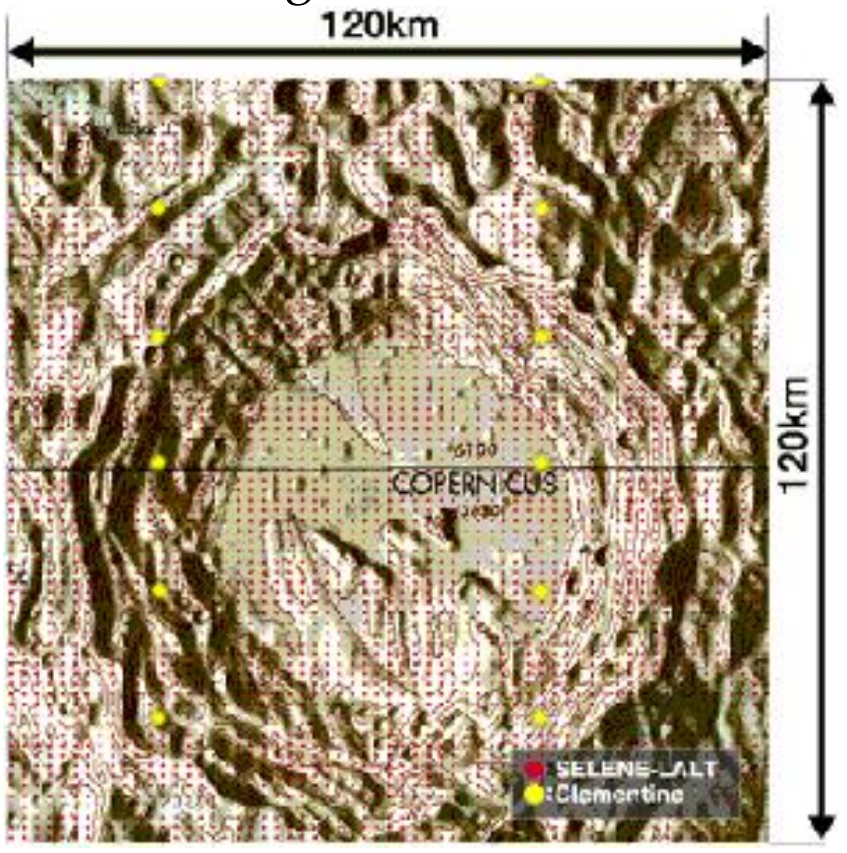

**Fig 3.6 The sampling rate of Kaguya in red dots, compared with Clementine (yellow).**

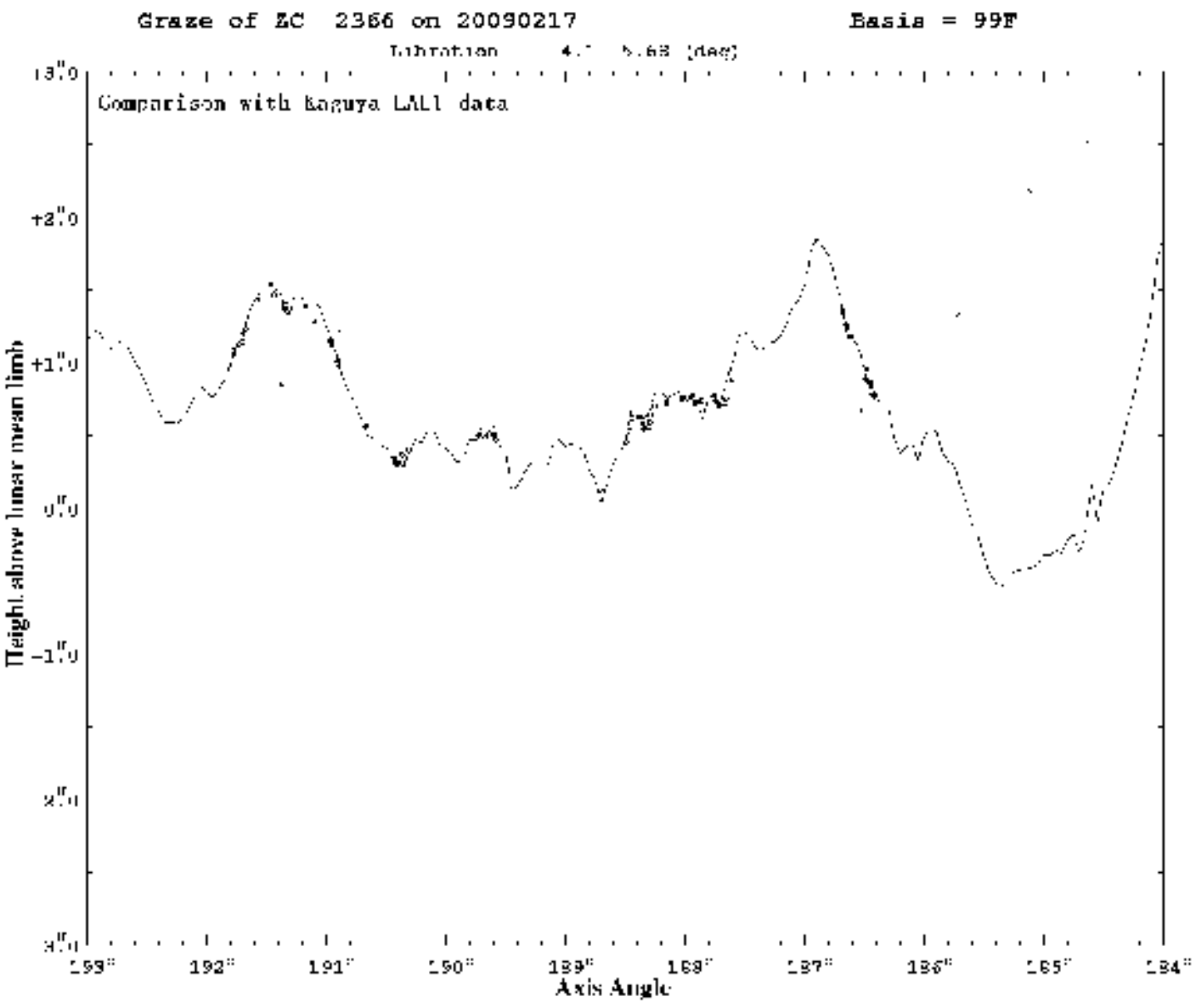

**Fig 3.7 The comparison between Kaguya and a grazing occultation of the star ZC2366 of February 17, 2009, observed by IOTA/ES.**

From the previous figure the agreement seems to be perfect, but there are other positions in which the departure is significant.

[150] Canny, J., *A computational approach to edge detection*, IEEE Trans. Pattern Analysis and Machine Intelligence **8**, p. 679-714 (1986).

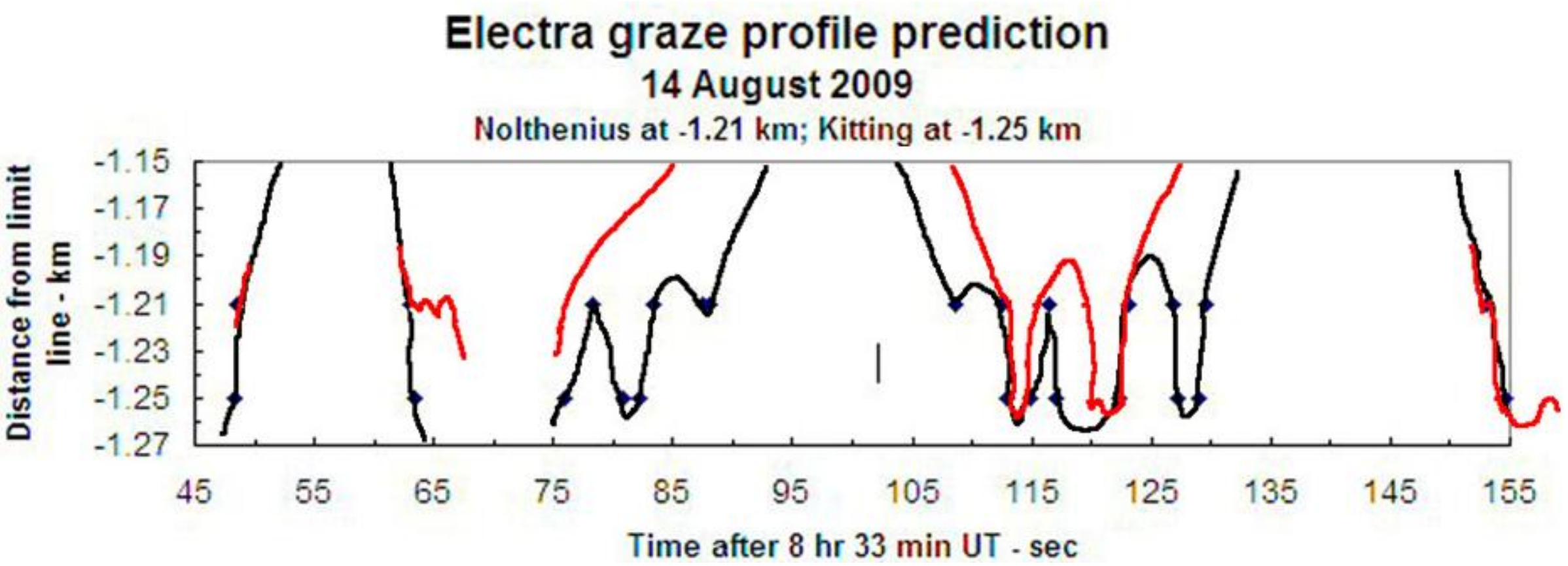

**Fig 3.8 The predictions (red) vs observations (black with dots) of the grazing occultation of Electra (Pleiades) on August 14, 2009.** There is a discrepancy of 0.8 arcsec, due probably to the different radius of Kaguya's lunoid.[151]

The tests on the Kaguya profile are still ongoing, the "lunoid" used by Kaguya with its smaller radius has to be checked again, before using directly the Kaguya profile.

## 3.6 New definition of solar limb

Different extension of the solar photosphere are measured at different wavelengths.
The classic definition of solar limb is the location of the maximum of luminosity profile derivative of the continuum spectrum,[152] already adopted in the oblateness studies. The effect of the blend of tiny emission lines just above the solar limb[153] visible during eclipses is better considered by this definition operative also for eclipses.
Otherwise the observations of Baily's beads with different level of signal to noise show clearly the effect of the emission lines, which extends the measured solar photosphere depending on the instrumental parameters.
With the use of the inflexion point in the eclipses observations we adopt for the first time the same approach as the observations at full disk.

## 3.7 LDF reconstruction from Baily's beads light curve and Kaguya profile

The studies on the Limb Darkening Function made with Baily's beads of the annular eclipse of 2010 is presented in the annexed paper, submitted to Solar Physics.[154]
The inflexion point position is reconstructed from the luminosity profile of a bead and from the geometrical form of the lunar valley at the limb, available from Kaguya data.
The light curve of two beads have been determined and the Limb Darkening Function deconvoluted by a numerical procedure including the area of photospheric light through these Baily's beads varying with the time.

---

[151] Bredner, E., *Moon Limb after Kaguya*, XXVIII ESOP meeting, Niepolomice, Poland 29 August – 3 September 2009.
[152] Hill, H. A., R. T. Stebbins and J. R. Oleson, Astrophys. J. **200** 484 (1975).
[153]Flash spectrum is related to spectroscopy: a better definition is the one of solar mesosphere give in chapter 1.1
[154] Raponi, A., C. Sigismondi, K. Guhl, R. Nugent, A. Tegtmeier, *The Measurement of Solar Diameter and Limb Darkening Function with the Eclipse Observations*, Solar Physics **278**, 269 (2012).

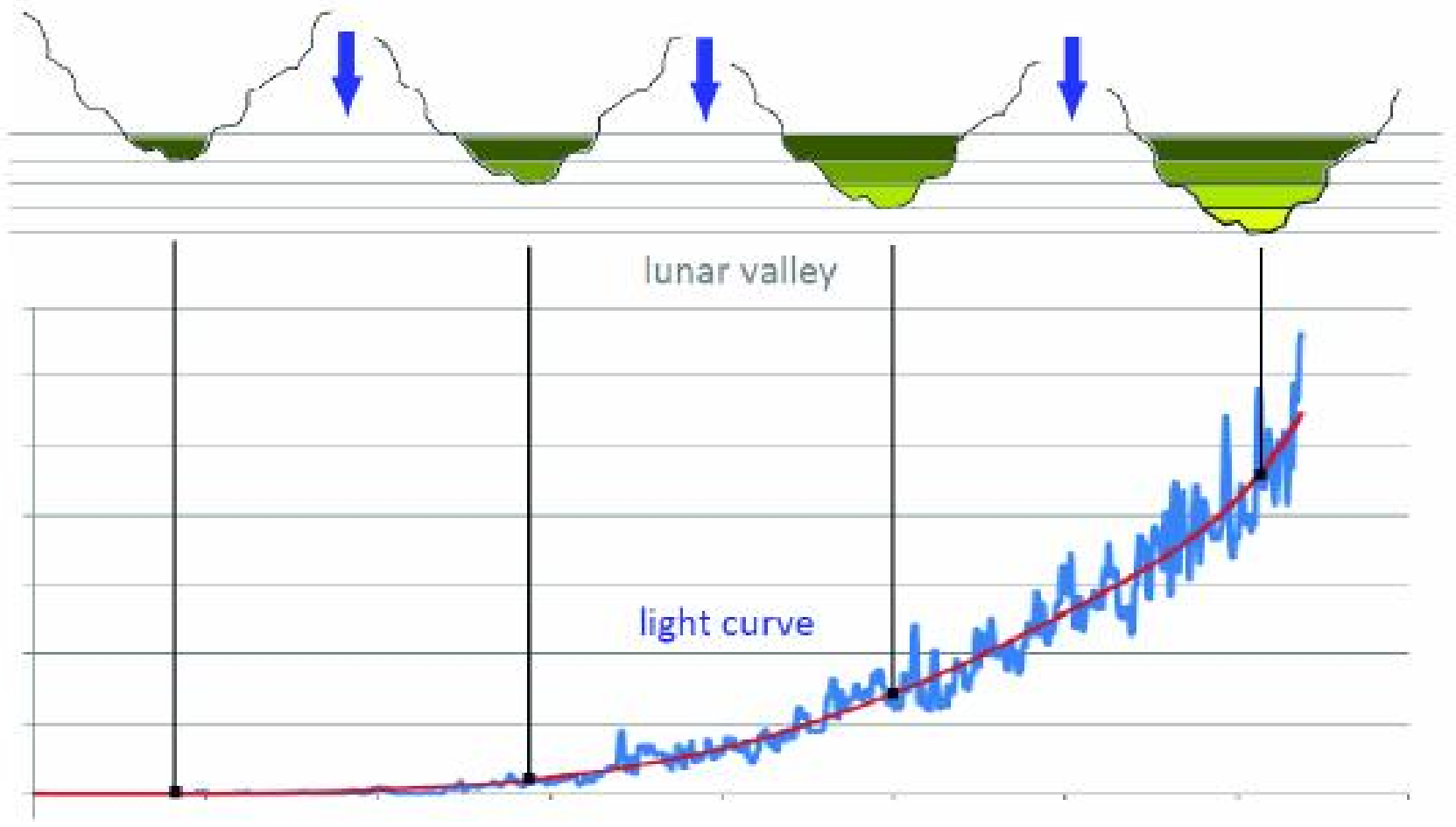

**Fig 3.9 The light curve of a bead is the product of the LDF by the exposed area.**

In this figure the variation of the LDF is represented by layers of different colors, which covers gradually the different strips of the lunar valley. The area of the strips is computed on the Kaguya profile, the light curve is observed and the LDF is obtained with the inverse operation described in the caption. More details of the algorithm adopted are in the paper published in Solar Physics.[37]

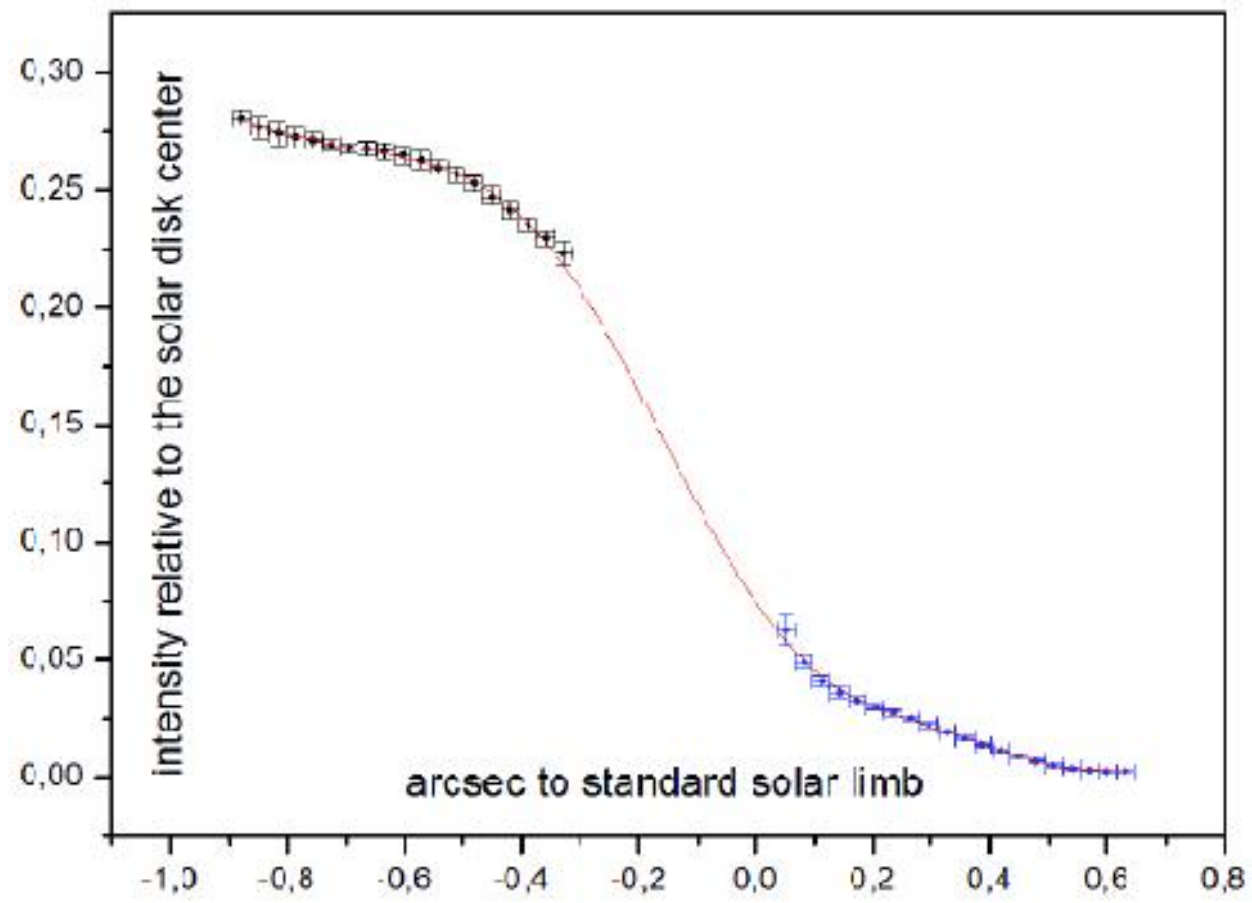

**Fig 3.10 A Limb darkening profile obtained by the inversion technique here described.** The left and right parts of the curve are obtained with two different 8-bits CCDs, the lower right saturated before the inflexion point, the upper left had a limiting magnitude brighter than the inflexion point. A photometric resolution of 12-bits CCD is recommendable, and in more spectral channels, in addition to white light. This is now technologically possible and at affordable costs for the observers.

The luminosity profile is normalized to the center of the solar disk according to Rogerson (1959)[155] for the inner parts, and in arbitrary way for the outer parts. The zero of the abscissa is the position of the standard solar limb with a radius of 959.63 arcsec at 1 AU. The error bars on y axis are the 90% confidence level. The error bars on x axis are the thickness (h) of the lunar layers. The solid line is an interpolation between the profiles and gives a possible scenario on the position of the inflection point.

The results shows light at least 0.85 arcsec beyond the inflection point, and this suggests to reconsider the evaluations of the historical eclipses made with naked eye.

From the two beads examined it is impossible to infer an exact location of the inflection point, but it is

[155] Rogerson, J.B., *The Solar Limb Intensity Profile*, Astrophys. J. **130**, 985 (1959).

possible to deduce an upper and lower limit corresponding to the points that better constrain it:
-0.190 arcsec < ΔR < +0.050 arcsec, for the eclipse of January 15, 2010.

## 3.8 "Unidimensional" eclipses: solar diameter in the past centuries

Reliable past values of the solar radius are believed to be obtained from the durations of ancient total eclipes. The values of solar diameter calculated from historical (1567 on), and recent (1966 on) eclipses (and planetary transits) using the Watts' lunar profile are discussed and compared with the solar activity.
Before 1966 there are only edge data for eclipses of 1567, 1715, 1869 and 1925. The precision on the solar diameter with these naked eye data is discussed.

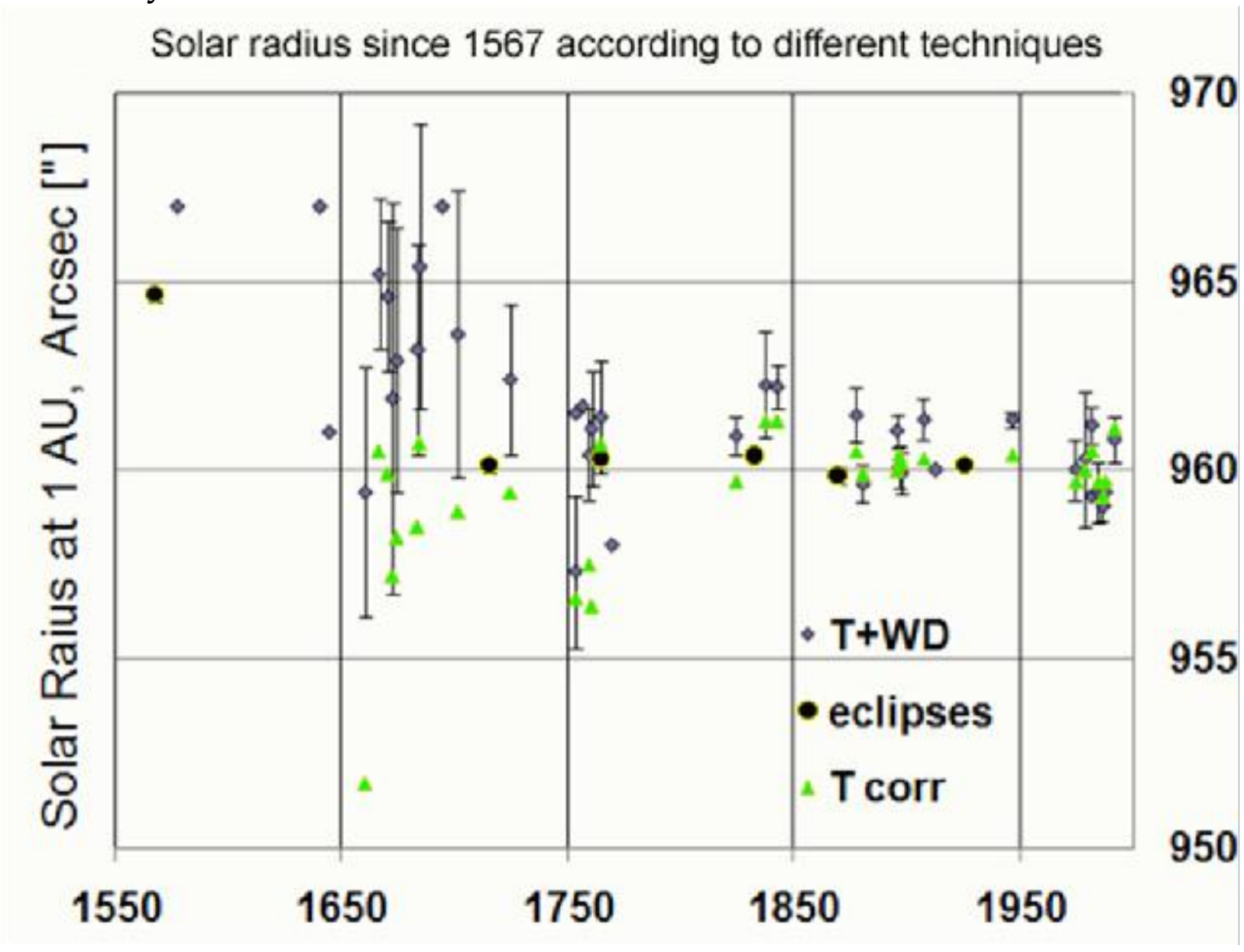

**Fig. 3.11 All historical data available on the solar diameter. T stands for Toulmonde,[156] WD for Wittmann and Débarbat.[157]** The eclipses are much stable and with narrower errors, excepted the eclipse of Clavius.

The history of past diameter is obtained mainly from Mercury's transits. How reliable are these data?
I already discussed this topic in the paragraph 2.3 about the transit of 1832.
Once we will know the W parameter the past eclipses would give us the luminosity of the Sun in these times, especially right after the Maunder minimum in 1715, in order to understand the role of our star in the little ice age.

The studies for the following historical eclipses show a solar radius significantly larger than the standard one, but they must be reconsidered in the light of the results obtained in the previous section: determination with the "classical approach" can be misleading.

[156] Toulmonde, M., *The diameter of the Sun over the past three centuries*, Astronomy and Astrophysics **325**, 1174 (1997).
[157] Wittmann, A. D. and S. Debarbat, *The solar diameter and its variability*, Sterne und Weltraum **29**, 420 (1990). In German.

### 3.8.1 Clavius, 1567, Rome

Stephenson, Jones, and Morrison[158] studied the observation of an annular eclipse made by the Jesuit astronomer Christopher Clavius in April on 9, 1567 in Rome, they derived limits to the Earths spin rate back to that time. They attributed the appearance as a ring of the annular eclipse to the effect of the ″inner corona″ of the Sun. If the ring of that annular eclipse was instead a solar layer inner to the inflection point position, the average angular radius of the Sun would have been some arcsec larger than its standard value of 959.63 arcsec.

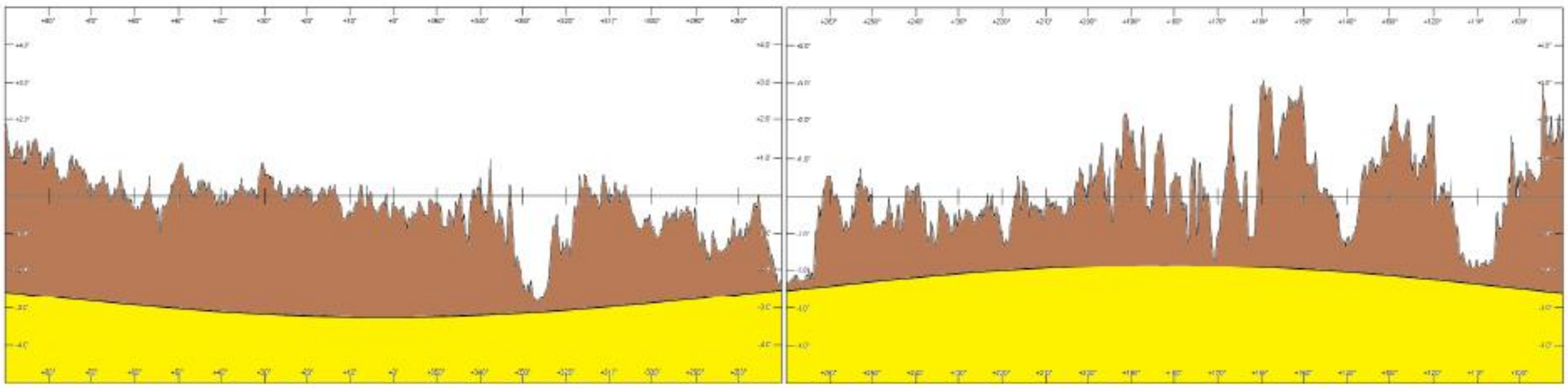

**Fig. 3.12 The lunar limb during the eclipse of Clavius. The solar limb would have been 4.5 arcsec below the standard value.**

The figure 3.12 shows the eclipse of April 9, 1567 simulated with Occult 4.0 software. View from Collegio Romano (lat = 41.90◦ N, long = 12.48◦ E) where probably he made his observation. The lunar limb's mountains are plotted in function of the Axis Angles (the angle around the limb of the Moon, measured Eastward from the Moon's North pole). 1◦ in the Axis Angle (abscissa) corresponds to 16 arcsec at the mean lunar distance. The solid line under the mountains is the standard solar limb. The figures are the Northern (left) and the Southern (right) semicircle.
To have the solar limb higher than all the mountains of that lunar limb it should be $\Delta R \geq +4.5$ arcsec.

According to the formula of Rayleigh for the angular resolution $\varrho=1.22\lambda/D$, and taking into account a pupil diameter D=2 mm (day vision), one gets $\varrho$=50 arcsec.
Details smaller than 50 arcsec are not visible on the Moon profile.
This means that the ring of the annular eclipse could have not been complete but divided by mountains no more than 50 arcsec, that is about 3◦ in Axis Angle. For explain the observation of a complete ring with naked eye, for this eclipse, is thus sufficient $\Delta R$ = +2.5 arcsec, that still remains a great value.
According to this proof of a larger solar diameter Eddy et al. (1980)[159] assumed a secular shrinking of the Sun from the 17th century to the present. The observation of Clavius was the subject of several studies and publications: even Kepler asked Clavius to confirm it was a solar ring, rather than diffuse appearance, that he would attribute to the lunar atmosphere, but Clavius always confirmed what he
already wrote (Clavius, 1581).[160] The interpretation for an annular eclipse is still debated in scientific publications.
Because this observation was made with naked eyes, a more careful study on this eclipse has to take in account the angular resolution of an eye pupil.

[158] Stephenson, F. R., J. E Jones, L. V. Morrison, *The solar eclipse observed by Clavius in A.D. 1567*, Astronomy and Astrophysics **322**, 347 (1997).
[159] Eddy, J., Boornazian, A. A., Clavius, C., Shapiro, I. I., Morrison, L. V., Sofia, S., S*hrinking Sun. Sky and Telescope* **60**, 10 (1980).
[160] Clavius, C., *Commentarius in Sphaeram Iohannis de Sacrobosco*, Venezia (1581).

### 3.8.2 Halley, 1715, England

Edmund Halley attempted to measure the size of the umbral shadow by observing the total eclipse of 1715 in England. Halley as secretary of the Royal Astronomical Society collected the numerous reports of this observation (Halley, 1717).[161]

His idea was to associate the duration of the eclipse with the position for each observer, in order to assess the size of the shadow of the Moon on the Earth.

But from these observations we can obtain also interesting informations about the solar diameter.

In the present work, thanks to the Occult 4 software and the new data on lunar profile we are able to reanalyze the 1715 eclipse data. In particular we consider the observations on the Southern and Northern path of totality, where timings are not required to deduce the diameter of the Sun. We simply require to know where the observers were and if they saw a total or a grazing eclipse to infer an upper or a lower limit of ΔR from each eyewitness.

Table 3.5 shows the observations we consider. The first is located in the Northern limit of the shadow, the second and the third in the Southern one. According to Occult 4 we have from the observers in the Southern limit: +0.85 arcsec <ΔR < +0.94 arcsec. The error, due to the uncertainties on the observation's coordinates is 0.2 arcsec.

| Location of observation | Position coordinates | How appeared the Sun in the instant of maximum occultation | Lunar - Solar limb position by Occult 4 |
|---|---|---|---|
| **Darrington** | 53°, 39', 50.4" N<br>358°, 44', 09.6" E | "Point like Mars" | + 0.34 arcsec |
| **Bocton Kent** | 51°, 17', 16.8" N<br>0°, 56', 16.8" E | "Point like Star" | + 0.85 arcsec |
| **Cranbrook Kent** | 51°, 06', 03.6" N<br>0°, 31', 44.4" E | "Duration instant" | + 0.94 arcsec |

Table 3.5 The Eclipse in 1715, England. Observation in the Northern (Darrington) and Southern (Bocton Kent and Cranbrook Kent) limit of the umbra shadow. Position coordinates by Dunham *et al.* (1980).

Because of the uncertainties on ephemeris, the center of the solar disk could have been in a different position with respect to the simulated standard Sun.

Thanks to the observation in the Northern limit, an eventual error on ephemeris could be bypassed, obtaining a larger gap: +0.59 arcsec < ΔR < +1.28 arcsec.

The lower and upper limit are obtained moving the center of the solar disk in the North-South direction till the positions where the eyewitnesses are still valid.

Both the evaluations of the historical eclipses are made without any reference to the inflection point, and considering the LDF as a step function. This can enlarge the measured ΔR. It is the main concern with naked eye observations.

[161] Halley, E., *Observations of the Late Total Eclipse of the Sun on the 22d of April Last Past, Made before the Royal Society at Their House in Crane-Court in Fleet-Street, London. By Dr. Edmund Halley, Reg. Soc. Secr. with an Account of What Has Been Communicated from Abroad concerning the Same.* Philosophical Transactions (1683-1775) **29**, 245 – 262 (1717).

### 3.8.3 Recent eclipses

Here the plot of the analysis of the recent eclipses published and discussed in the annexed paper.[162]

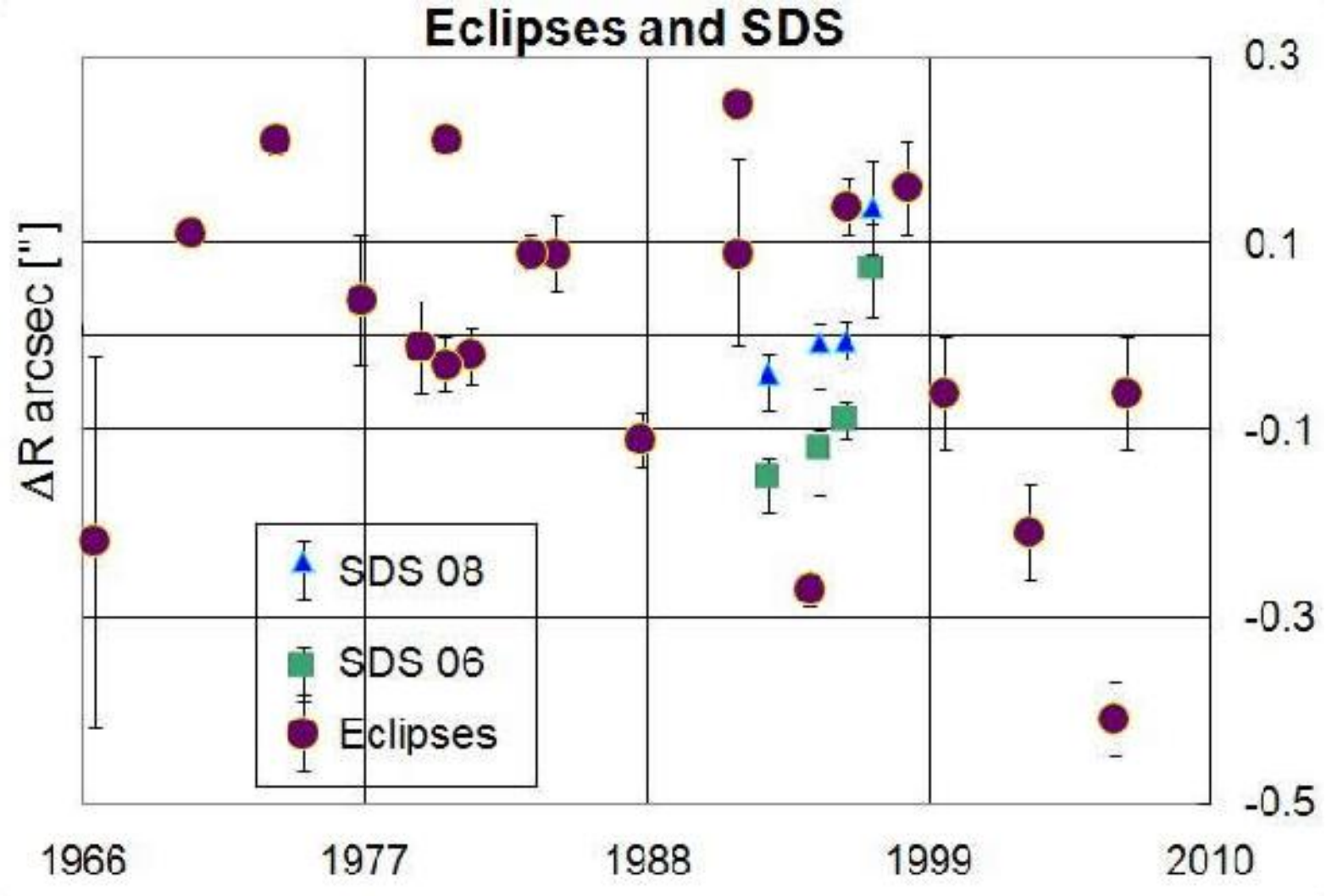

**Fig. 3.13 The eclipses in the last 45 years.** The four eclipses without errorbars are the photometric data of Y. Kubo. The data of SDS are superimposed.

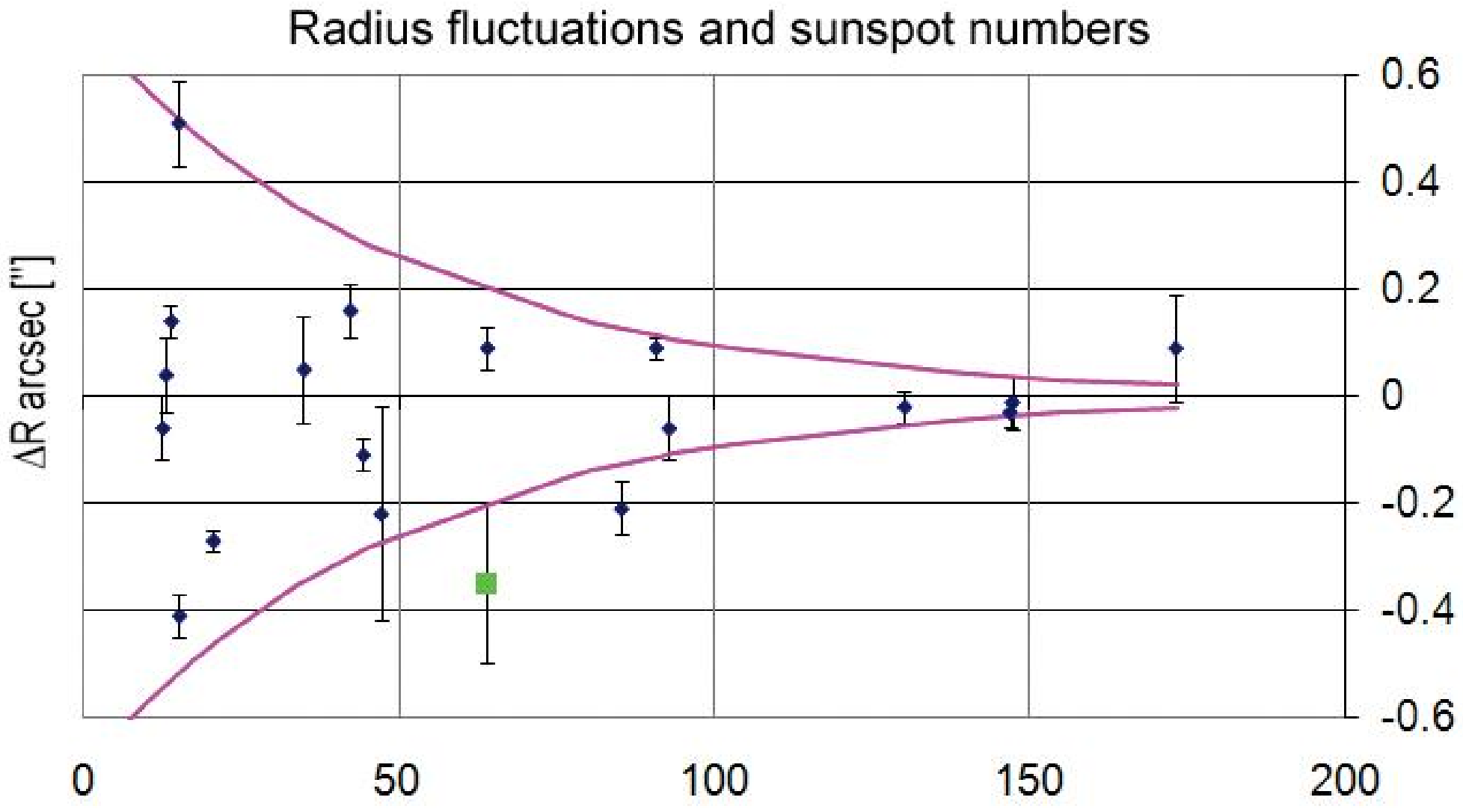

**Fig. 3.14 The eclipses data redrawn as function of the solar activity,** with the sunspot number in abscissa. The green square is the transit of Mercury of 2003 used by SOHO for a calibration of the MDI.

### 3.9 Unidimensional eclipses: photometric data

Eclipse data from photometers have very high time accuracy and almost no spatial information. Their utility in accurate diameter's measurements is discussed.
The first attempts of this kind of measurement have been done by Kubo in 1973.[163]

[162] Kilcik, A., C. Sigismondi, J. P. Rozelot and K. Guhl, Solar Phys. **257**, 237 (2009).
[163] Kubo, Y., Publications of the Astronomical Society of Japan, **45**, 819 (1993).

The eclipse mission of 11 July 2010 in French Polynesia is also dedicated to measure the solar diameter from the ground.
The difference between Baily's beads and photometer's concepts is in the "memory" of the detectors.
The pixels of CCD used in the video for eclipses can have a memory up to 1s after their illumination.
And this occurs also with CMOS detectors, as verified during the occultation of Venus by the Moon of December 1, 2008,[164] where another glimpse of light appeared 1s after the complete decrease of the fitted curve.

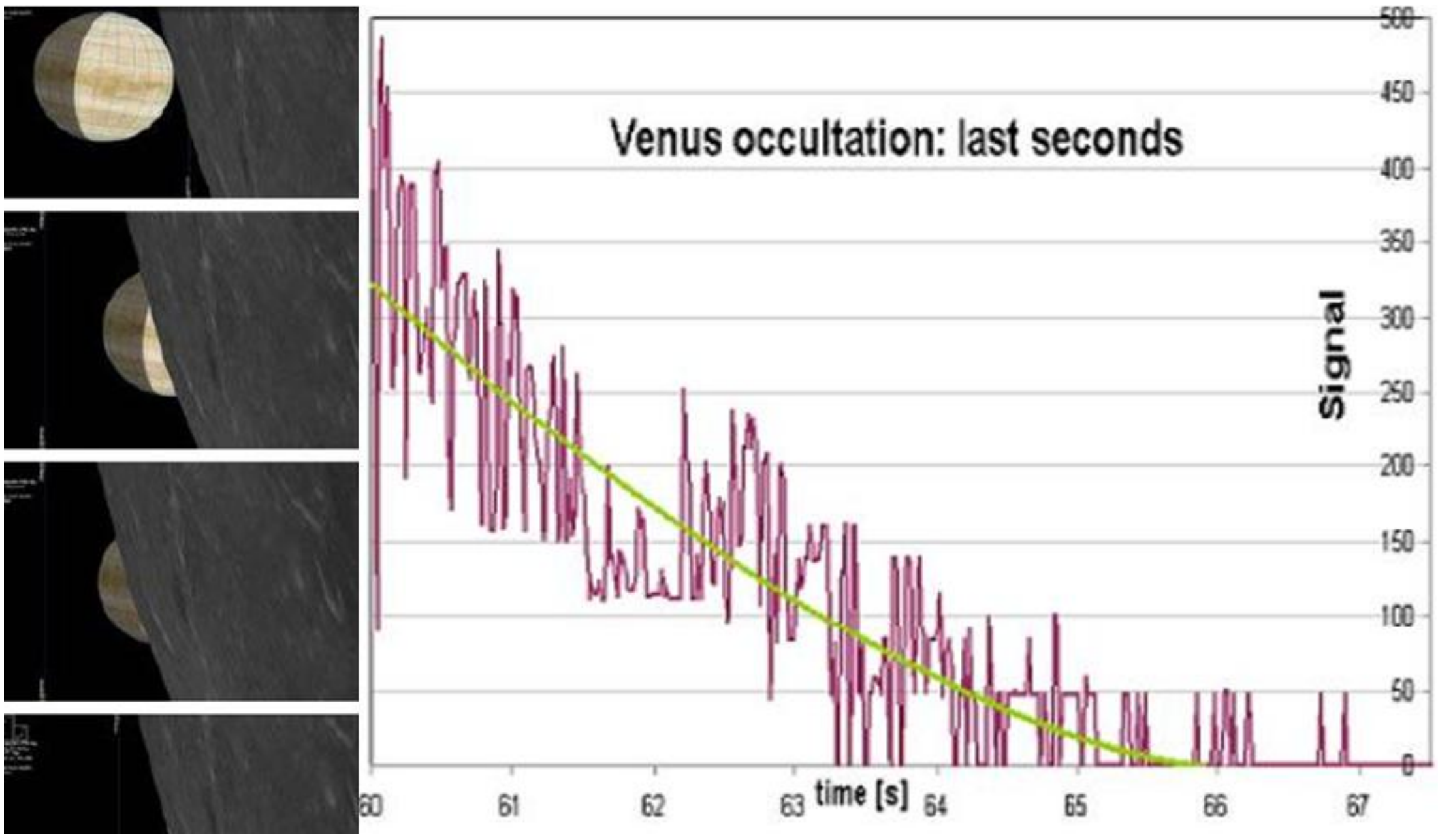

**Fig. 3.15 Venus occultation of December 1st, 2008 observed from Rome, Regina Apostolorum University. See the two signals one second after the zero of the fit.[165]**

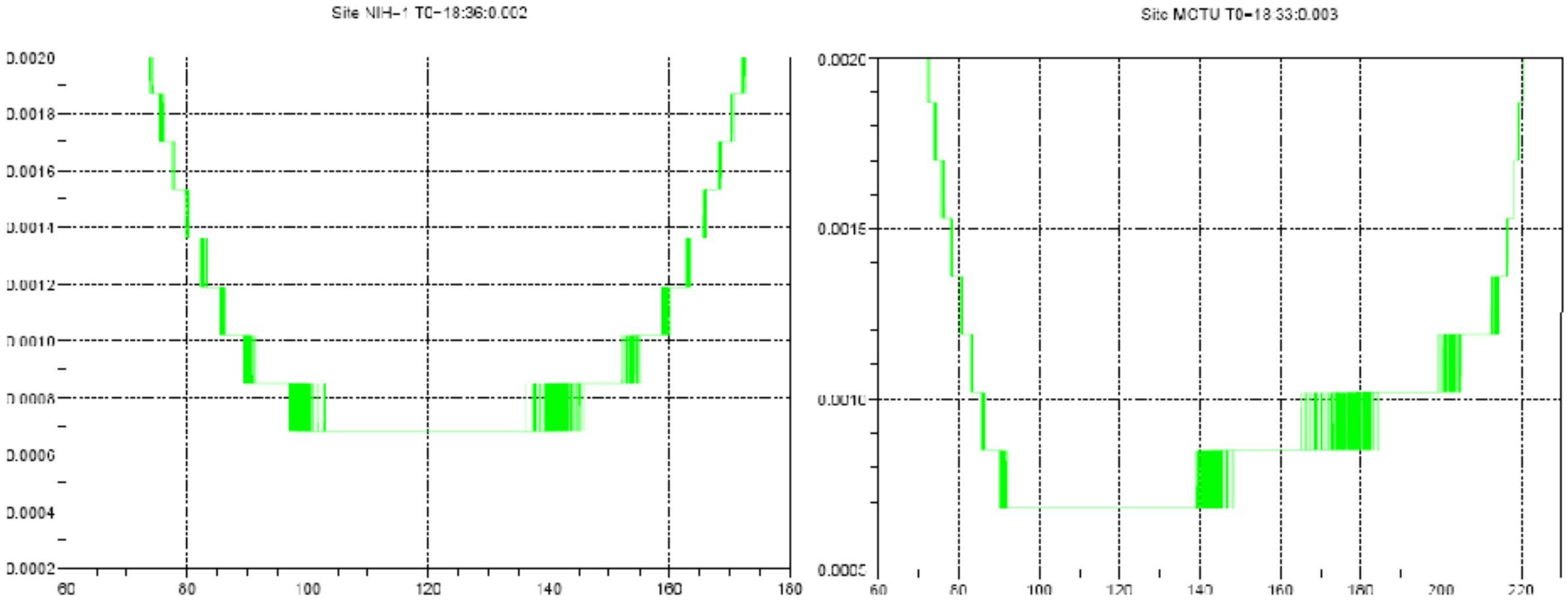

**Fig. 3.16 Signals from two photometers at different sites in the CNRS mission in French Polynesia (2010).[166]**

---

[164] Sigismondi, C., R. Nugent, G. Dangl, *Measuring solar disk shape up to relativistic accuracy: the role of scintillation in ancient naked eye data*, Proceedings of the 3rd Stueckleberg Workshop on Relativistic field Theories, Pescara, Italy, 8-14 July 2008. edited by N. Carlevaro, R. Ruffini and G. V. Vereshchagin, Cambridge Scientific Publishers, p. 303 (2011). arXiv:1106.2451

[165] Figure from Sigismondi, C. R. Nugent and G. Dangl, arXiv:1106.2451(2011).

[166] Courtesy of Jean-Yves Prado, CNES.

The observable is the duration of totality (the lower horizontal line in the figure). With respect to Baily's beads observations here we have the information without spatial resolution.
One chord for each observing station can be recovered on the solar disk, i.e. only one diameter at time.
The passage of clouds can also affect the last stages before totality.
The use of a network of 12 photometers displaced on the ground at several km each from the other, compensates the unidimensionality of each measure.
The same strategy is exploited in Australia, near Cairns, for the November 14, 2012 eclipse.[167]
The oscillation of the signal before totality can also be attributed to the electronic noise and to scintillation.[168]
The photometers are 2.4x2.4 mm without collecting optics, for these reason the scintillation can have important effect.
A formula[169] for estimating scintillation without taking altitude into consideration is:
$\Delta mscint = (0.09\cdot A^{7/4}) / (D^{2/3}\cdot\sqrt{(2\cdot t)})$
where A is the airmass, D is the aperture in cm and t is the integration time in seconds.
For A=1.89, because the Sun altitude was h=32° and A=1/sin(h), D=0.24 cm and t=0.01 s, $\Delta mscint$ = 5 magnitudes, or a factor of 100 in intensity, which is beyond of what observed.
Nevertheless in the case of the eclipsing Sun, the poissonian approach is more correct.
With N photons detected, a temporal variation within $\pm\sqrt{N}$ is expected, and this can be responsible of the noise near the beginning of the totality.
The scintillation of the last crescent of Sun, moreover, produces the flying shadows, observed for the first time during total eclipse of 1870 in Sicily.

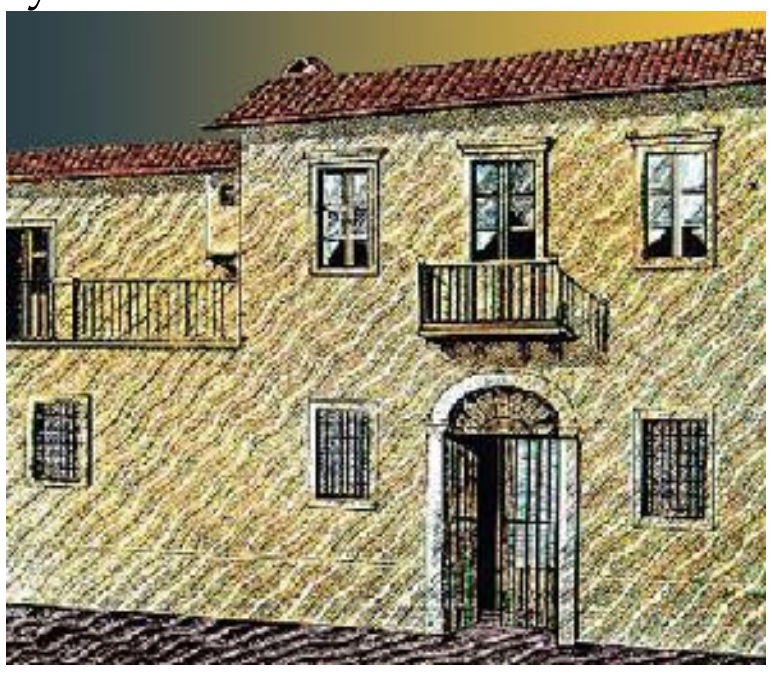

**Fig. 3.17 Flying Shadows in Sicily during 1870 eclipse. Image of the epoch.**

For the experiment of Australia with the network of photometers, I strongly recommend to put a "mighty mini"[170] device near to each photometer, in order to have also spatially resolved information.
The mighty mini devices have been developed in IOTA, for asteroidal occultations, with spectacular results obtained by a single observer controlling several stations.
The cost of the mission would not increase significantly.

[167] Espenak, F., http://eclipse.gsfc.nasa.gov/SEgoogle/SEgoogle2001/SE2012Nov13Tgoogle.html
[168] Sigismondi, C., R. Nugent, G. Dangl, *Measuring solar disk shape up to relativistic accuracy: the role of scintillation in ancient naked eye data*, Proceedings of the 3rd Stueckleberg Workshop on Relativistic field Theories, Pescara, Italy, 8-14 July 2008. edited by N. Carlevaro, R. Ruffini and G. V. Vereshchagin, Cambridge Scientific Publishers, p. 303 (2011). arXiv:1106.245
[169] This formula is adapted by Radu Corlan from a series of papers by Dravins et al. beginning with *"Atmospheric Intensity Scintillation of Stars, I. Statistical Distributions and Temporal Properties"* PASP 109, 173 (1997) and published in The AAVSO CCD Observing Manual, AAVSO, Cambridge MA (2009).
[170] Degenhardt, S., http://scottysmightymini.com/ (2011).

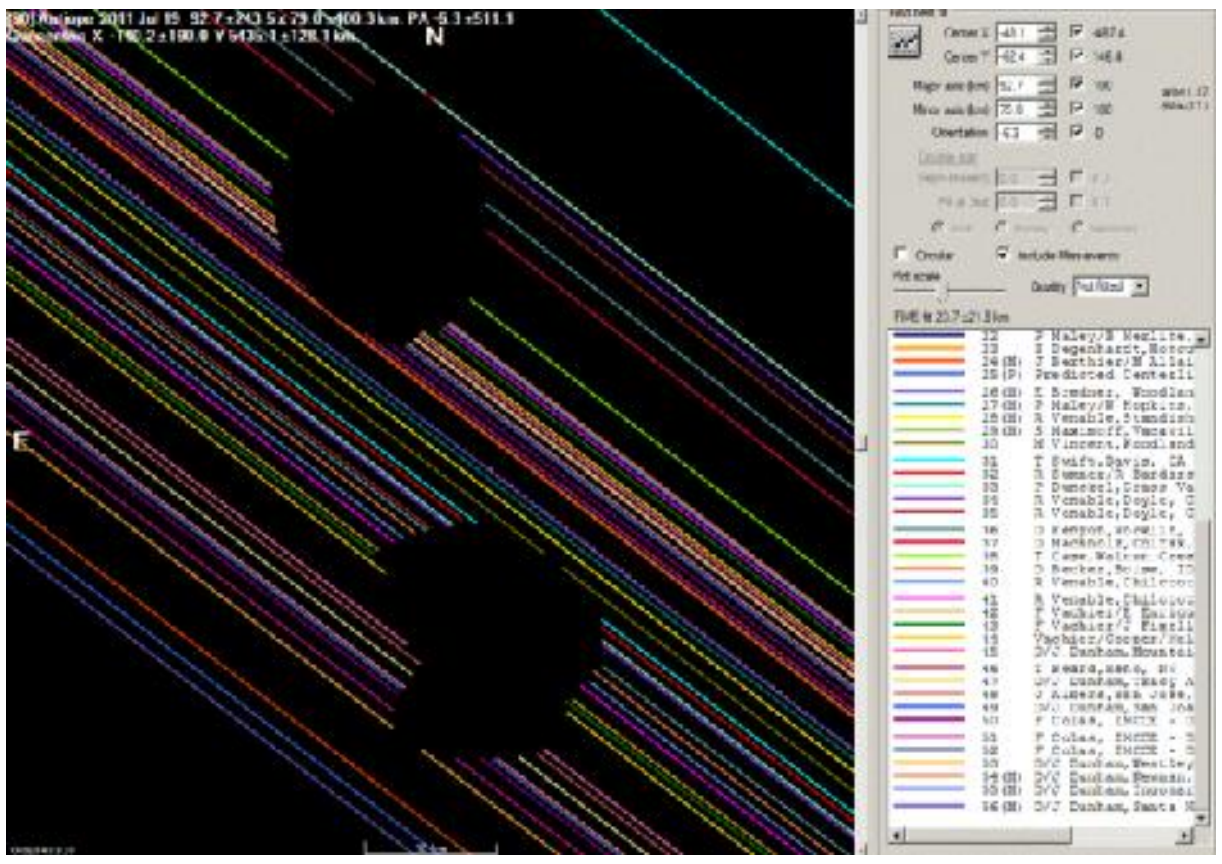

**Fig. 3.18 The asteroidal occultation of the double asteroid 90 Antiope (July 19, 2011) observed with an array of mighty mini devices coordinated by IOTA/US section.**

## 3.10 Before and after this work

The situation of this field of research before and after this work is described, putting into evidence our new contributions.

The unification of the methods to find the solar edge between the full disk observations and the eclipse observations has been the achievement of this work.

This was possible only after gathering all observations of Baily's beads by the European Section of the IOTA members since the year 2005.

The coordination of this activity has permitted to exploit the observational competences of these amateurs and their economical effort (about 30 travels all around the World, i.e. 100-150 K€ of budget) to converge on a more accurate and general procedure for the evaluation of the solar diameter with eclipses.

Now the different limiting magnitudes of the instruments, or the different saturation levels of the detectors are no longer affecting the estimate of the solar diameter.

As in the new procedure adopted with the annular eclipse of 2010, it is possible to see if the CCD are saturated before or after the inflexion point by the inclination of the recovered LDF curve.

CCD at 12 bits are recommended for the next eclipse missions, in order to avoid this saturation problem which prevents to locate correctly the inflexion point of LDF.

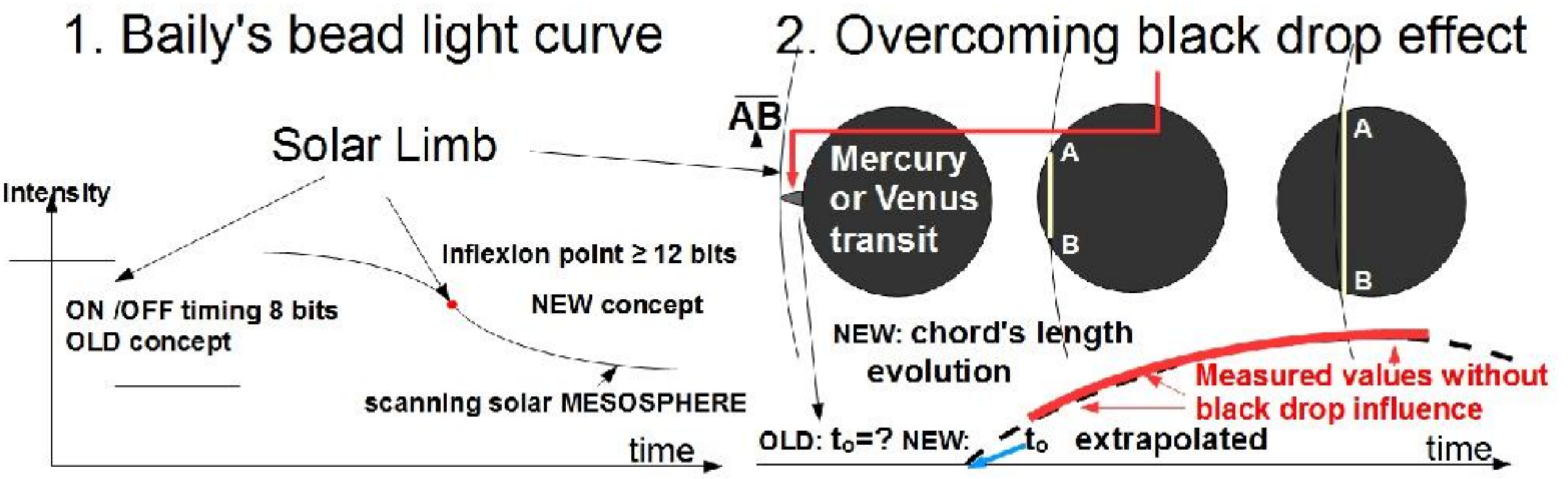

**Fig. 3.19 Scheme representing the novelties in the approach to eclipses and planetary transits measures of solar diameter, as exposed in chapters 2 and 3.**

## Chapter 4: Daily Transits

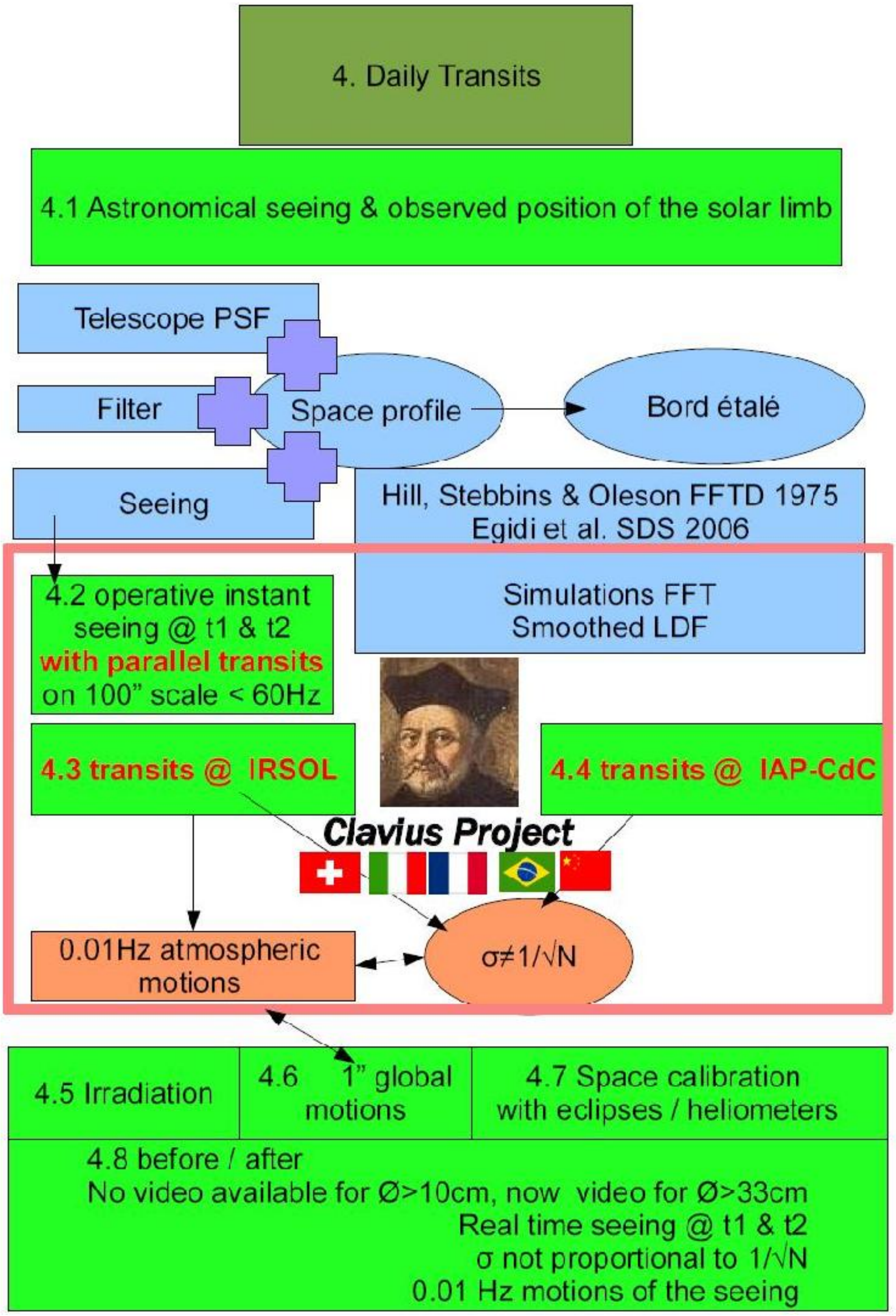

### 4.1 Seeing and observed position of the solar limb

The influence of the seeing on the observed position of the maximum of the maximum of the derivative of the luminosity profile of the Sun is here described.

The Limb Darkening Function used for the evaluation of the effect of the seeing on the inflection point position is from Rogerson (1959)[171] with the analytic formula used in Hill, Stebbins and Oleson (1975).[172]
The seeing effect is represented with a gaussian Point Spread Function with Half Width Half Maximum=0.5", 1" and 1.5".
The analytic formula[173] adopted for the PSF is PSF=exp(-((x-xo)/HWHM)^2).
In the figure 4.1 this is represented.

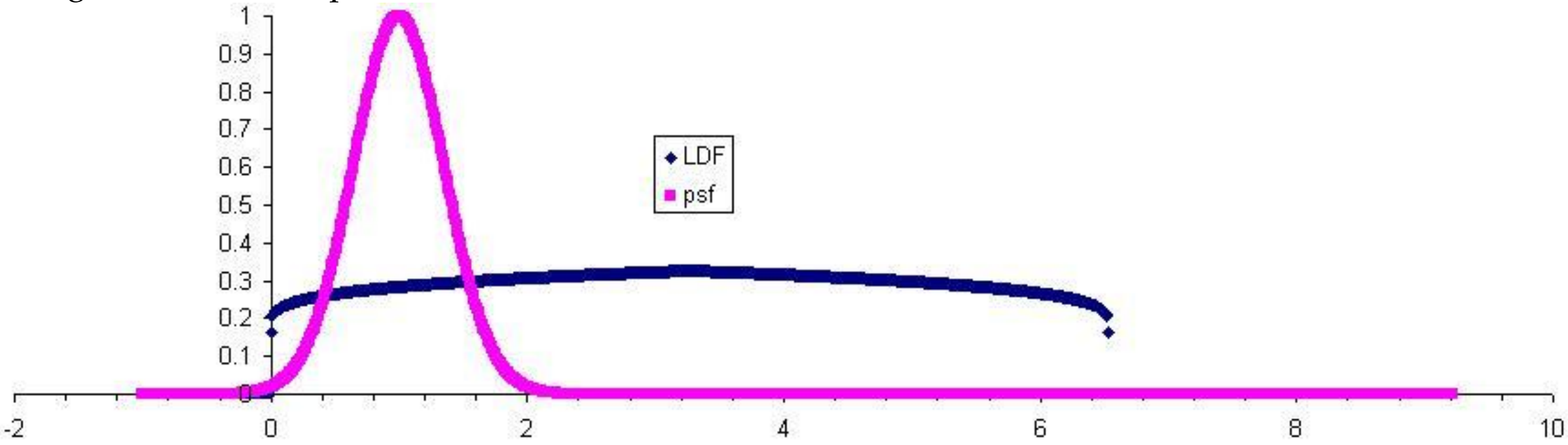

**Fig. 4.1 The Point Spread Function is a gaussian of given HWHM, to represent the seeing. The Limb Darkening Function formula is from Rogerson (1959); it has been adapted to 1024 points in order to perform the FFT, Fast Fourier Transform.**

In order to study the regions near the inflection points, in this numerical example, I have created a Limb Darkening Function composed by two mirrored parts representing the regions around the two inflection points. The LDF in the center of the Sun is not interesting for this study and does not affect the positions of the inflection points. Moreover the points are 1024 and the synthetic LDF is included within some points at zero value. This is in order to avoid the Gibbs' phenomenon and border effects during Fourier transforms. Through Fast Fourier Transform algorithm and the convolution theorem, the LDF is convoluted with the PSF and the new positions of the inflection points are detected through the maxima of the first derivative. The results are normalized at the standard solar diameter.

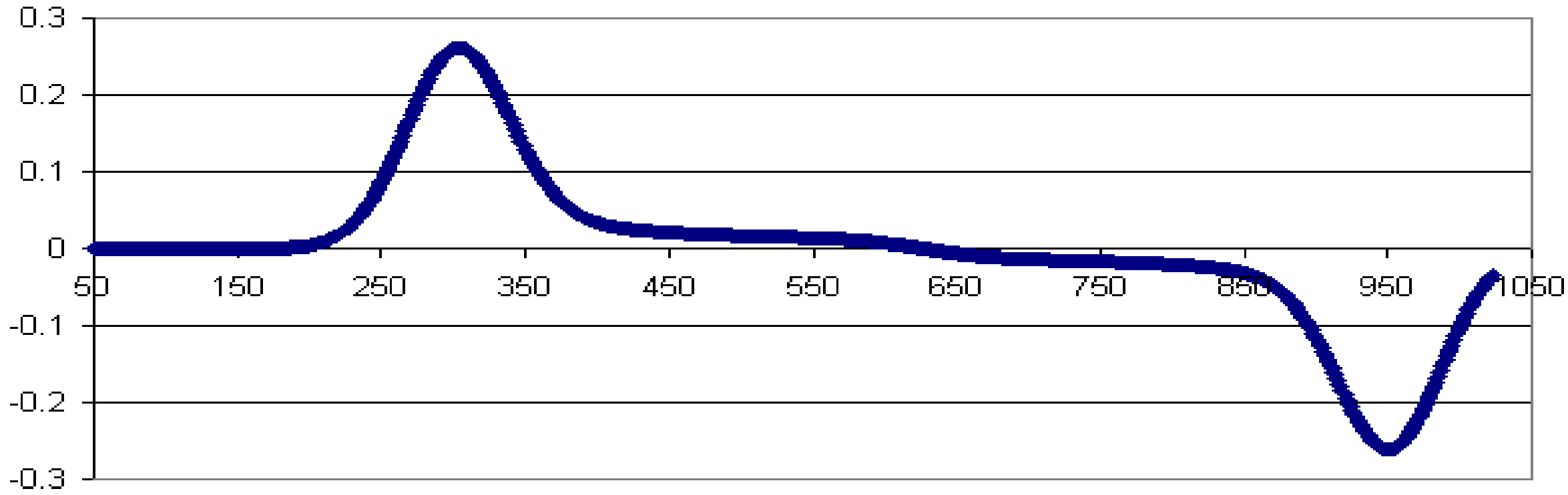

**Fig. 4.2 The first derivative of the LDF convoluted with a gaussian Point Spread Function** (see figure 4.1). The abscissa is in arcsec. The positions of the inflection points shift inwards. **PSF with HWHM respectively of 1; 2 and 3 arcsec give perceived diameters 0.07, 0.17 and 0.28 arcsec smaller than the unperturbed value.**

---

[171]Rogerson, J. B. Jr, *The Solar Limb Intensity Profile*, Astrophysical J. **130** 985 (1959)

[172]Hill, H. A., Stebbins, R. T. and J. R. Oleson, *The finite Fourier transform definition of an edge on the solar disk*, Astrophysical J. **200** 484 (1975)

[173] Instead of the Gaussian there is also the Moffat formula I(r)=Io/[1+(r/ro)^2]^b with the parameter ro measuring the image's amplitude, slightly depending on the wavelength, b~3. I have preferred the Gaussian.

Similar results have been obtained either theoretically[174], [175],[176] and observationally.[177]
The work realized at the Observatoire de la Côte d'Azur, at the site of Calern, with the solar astrolabe and later with DORaySol has shown this phenomenon.

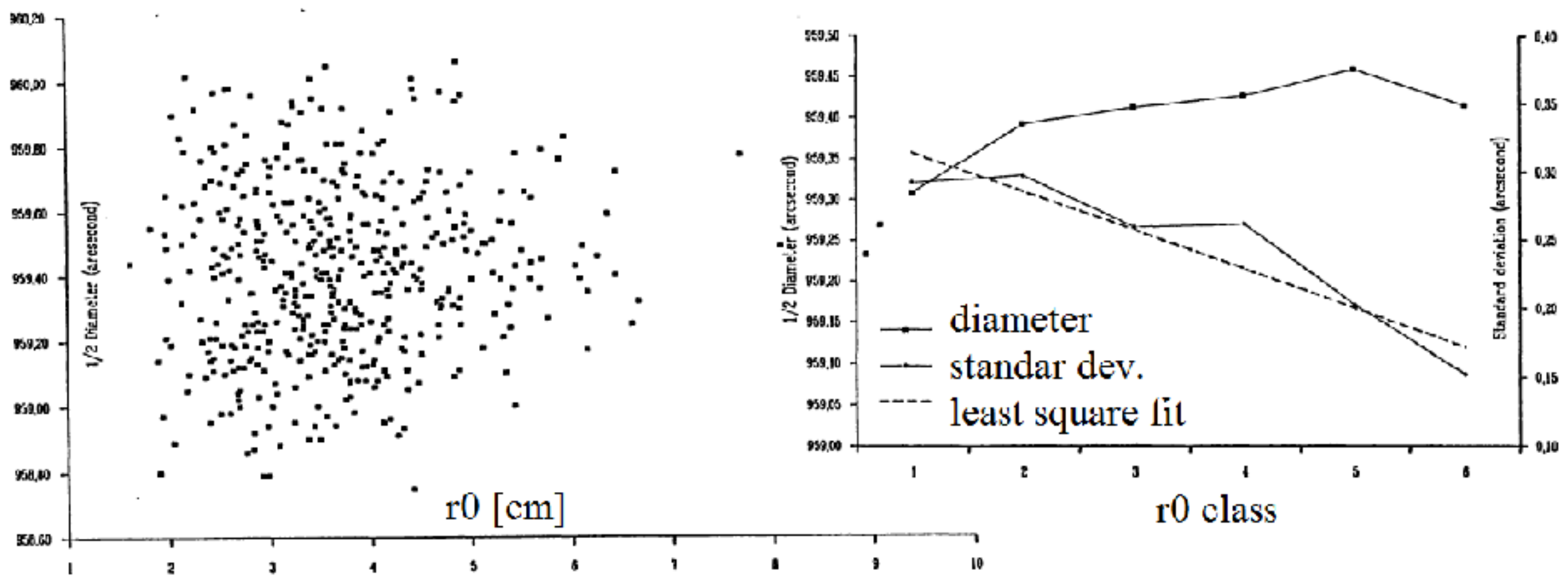

**Fig. 4.3 Semidiameters of the Sun as observed at the Calern Astrolabe.** Left: the distribution as function of the Fried's parameter r0. The vertical scale is 2 arcsec. Right: Average values of the measured diameters (0.5 arcsec of vertical scale) and their standard deviations. The difference between the observed diameters was less than 0.08 arcsec.[178]

The evolution of the error with the seeing parameter shows a decrease of the diameter measurement error with good seeing conditions as predicted by the theory. The figure 4.3 reproduced from fig. 10a and 10b of Irbah et al.[8] (1994) has been plotted with the intent of observing the atmospheric effects on the diameter's value. Solar diameter measurements are represented in the left part, according to the seeing parameter values. The diameter values have been divided into 6 classes with equal width from r0=1 cm to r0=7 cm. A mean value of the solar diameter has been calculated for each r0 class and it is represented in the right part of figure 4.3. In this right part, however, it is shown a slight decrease of the solar diameter with a degradation of seeing conditions. The difference of solar diameter values corresponding to the worst and best seeing at Calern Observatory is of order of 0.08 arcsec. The standard deviation has been calculated for each r0 class. As already observed with the diameter measurement error, the results improve with good seeing. In fact the standard deviation of diameter values is about 0.3 arcsec when the r0 value is 2 cm and falls down linearly to 0.15 arcsec when r0 is equal to 7 cm.
The authors of this study concluded that "it can be noted that variations less than 0.08 arcsec could not be observed in the mean solar diameter. Since the maximum occurrence of the Fried's parameter during the observations at Calern Observatory is 4 cm, variations less than 0.26 arcsec of the solar diameter can not currently be observed".
The more simple approach to demonstrate the phenomenon of the shifting of the inflection point which reduces the observed diameter, expressed in French as **"étalement du bord"**, is the aforementioned numerical method of FFT of the LDF convoluted with the PSF of the seeing. The analytical formula ΔD(A, K, K1, r0, k) presented in the same paper, requires 4 parameters in addition to r0. These parameters have to be adjusted from observations. I prefer a single parameter[179] formula, depending only on r0: the advantage of the whole formula does not seem of any suitable use in practical cases. Moreover the authors declare that 0.08 arcsec is the maximum sensitivity of the Astrolabe at the seeing conditions of the Calern site: this is one more reason to use my linear fit from a numerical approach.
The numerical approach has the advantage to put in clear evidence the physical origin of the phenomenon: the convolution of a PSF with the LDF. With other types of LDF (as a step function) this phenomenon could not appear as well.

[174]Borgnino, J. *Etude de la dégradation des images astronomiques diurnes par analyse statistique des fluctuation d'ange d'arrivée*, PhD Thesis, Université de Nice (1978).
[175] Berdja, A. and J. Borgnino, *Modelling the optical turbulence boiling and its effect on finite-exposure differetial image motion*, Monthly Notices of the R. Astronom. Soc. **378**, 1177 (2007)
[176]Berdja, A., *Effets de la Turbulence Atmosphérique lors de l'Observation du Soleil à Haute Résolution Angulaire*, PhD Thesis, Université de Nice-Sophia Antipolis (2007)
[177]Irbah, A., Laclare, F., J. Borgnino and G. Merlin, *Solar diameter measurement with Calern Observatory astrolabe and atmospheric turbulence*, Solar Physics **149**, 213 (1994)
[178]Irbah, A., Laclare, F., J. Borgnino and G. Merlin, *Solar diameter measurement with Calern Observatory astrolabe and atmospheric turbulence*, Solar Physics **149**, 213 (1994) Fig. 10a and b.

Already in 1975 Hill Stebbins and Oleson did show the effect of a convolution with Gaussians of the Limb Darkening Function, as it is shown in their figure 2.

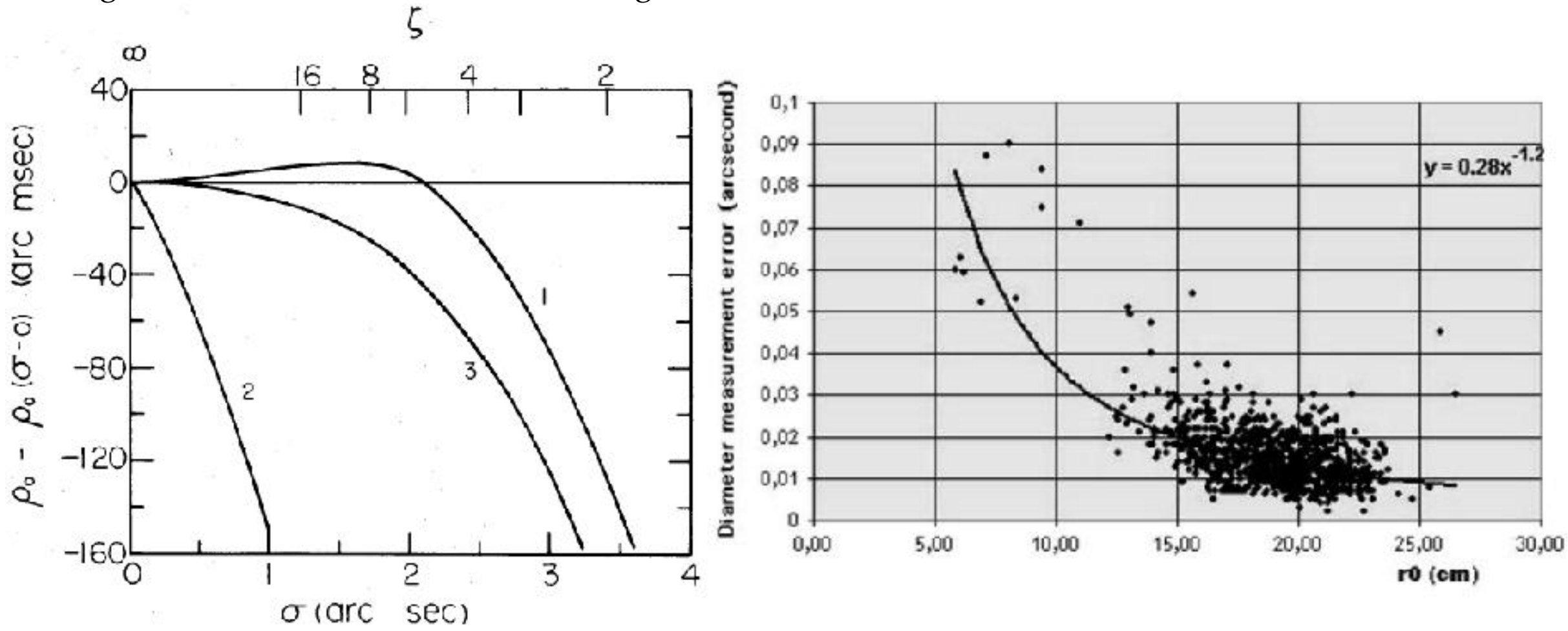

**Fig. 4.4 Other Seeing – étalement relationships.** Left: from Hill Stebbins and Oleson (1975) the reduction of the solar radius (displacement of the inflection point) as function of the seeing σ. The methods of edge definition are three: 1 stands for FFTD, 2 for second derivative technique and 3 for the integral definition. Right: from Rozelot, Lefebvre and Desnoux (2003).[180]

The diameter of the Sun in the paper of Rozelot, Lefebvre and Desnoux (2003) is considered at larger r0 parameter with respect to Calern. This is because the observations have been carried out at the Pic du Midi Observatoire at 2861 m on the Mounts Pyrenées, under exceptional seeing conditions.
Nevertheless, considering that r0=10 cm corresponds to 1 arcsec and r0=5 cm is 2 arcsec, the order of magnitude of the "étalement" is verified.

[179]*Entia non sunt multiplicanda praeter necessitate* (William of Ockham).
[180]Rozelot, J. P., Lefebvre, S. and V. Desnoux, *Observations of the Solar Limb Distortions*, Solar Physics **217** 39 (2003).

My synthetic numerical data are framed in the following synoptic **table 4.1**:

| Seeing | 1 arcsec | 2 arcsec | 3 arcsec |
|---|---|---|---|
| Fried's Parameter R0 | 10 cm | 5 cm | 3.3 cm |
| Δ Diameter FFT (this work) | -0.07 arcsec | -0.17 arcsec | -0.28 arcsec |
| Δ Diameter Hill&al.'75(aver.) | -0.12 arcsec | -0.22 arcsec | -0.32 arcsec |
| Δ Diameter Rozelot & al.'03 | -0.038 arcsec | -0.10 arcsec | -0.17 arcsec |
| Δ Diameter Irbah & al. '94 | -0.07 arcsec | -0.17 arcsec | -0.27 arcsec |

## 4.2 Instantaneous measurement of the seeing

The method of parallel transits[181] is here presented.

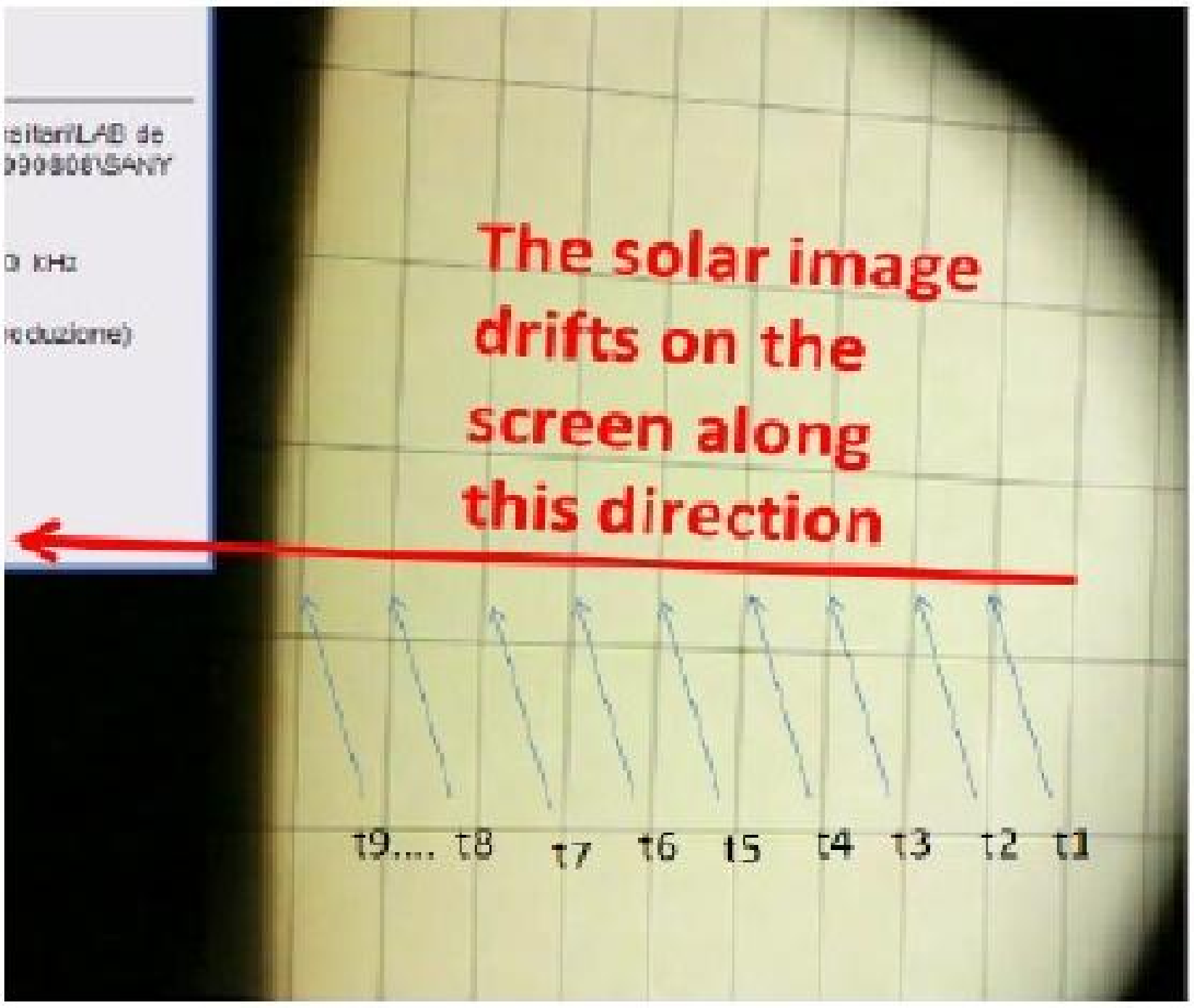

**Fig. 4.5 Parallel transits technique:** A regular grid is on the focal plane, the projected image of the Sun sweeps on the grid.

Without seeing the time intervals tn+1-tn are all equal. The standard deviation σ of time intervals is related to the seeing ϱ by the formula **ϱ″=15″·σ·cos(δ)** where δ is the solar declination

This formula has been applied to calculate the contact errors in arcsec, for the figure 4.5.

The most simple way to measure the seeing is by projection of the solar image on a regular grid during a drift-scan observation. A videocamera records the transit of the solar limbs above the grid, and the time intervals required to cover the evenly spaced intervals of the grid are measured by a frame by frame inspection. The preliminary studies on this subject have been conducted at the Meridian Line of the Clementine Gnomon in Santa Maria degli Angeli in Rome. The advantage of this location is to have a fixed pinhole projecting the solar image on the floor at large distance. The image of the Sun is studied with a video frame by frame.

[181]Sigismondi, C., IAUC **233**, 522 (2006).

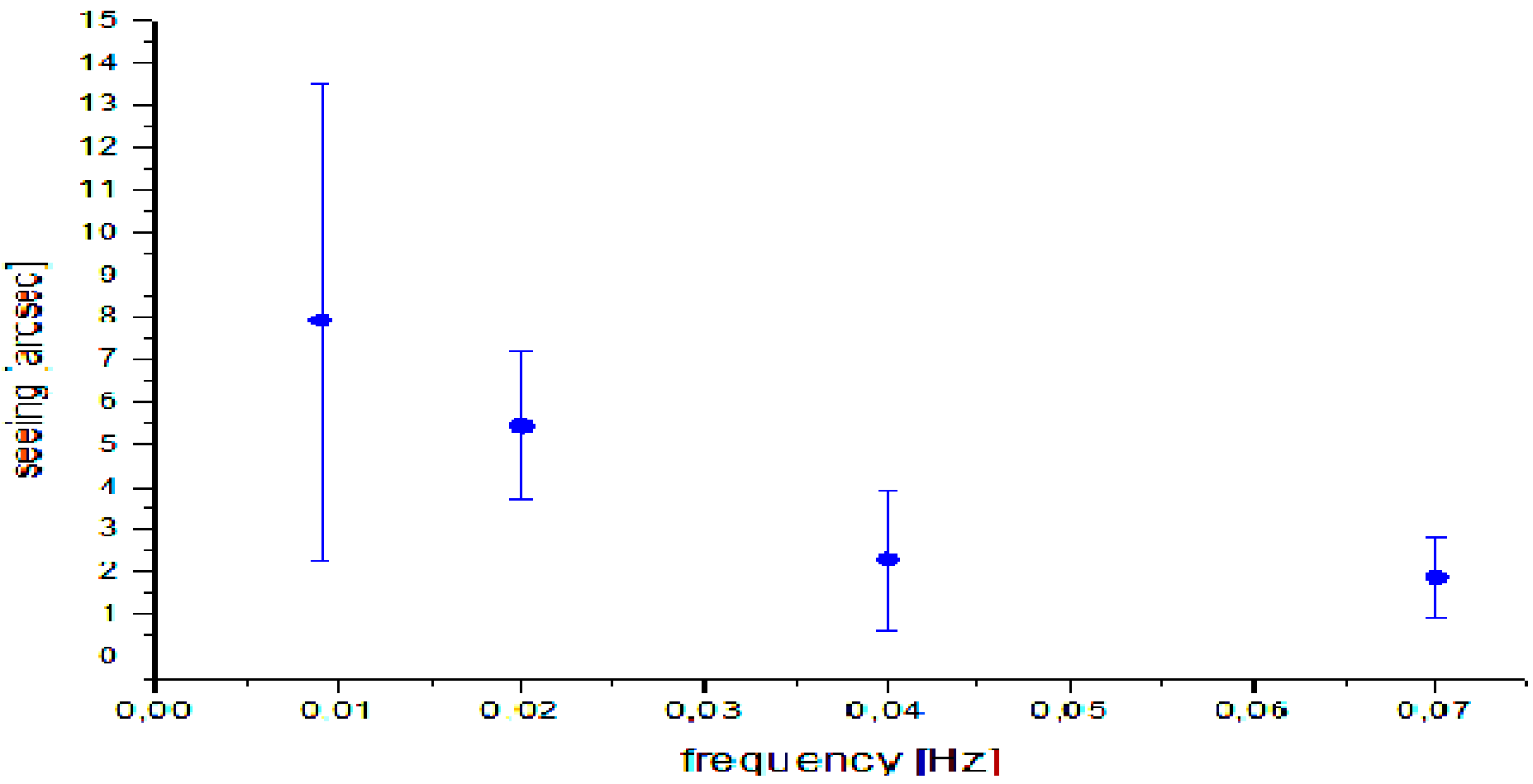

**Fig. 4.6 Parallel transits technique: the power spectrum of the seeing.** As an application of this technique we obtained the power spectrum of the seeing at discrete values of the frequence. The data around 0.07, 0.04, 0.02 and 0.01 Hz correspond to the time intervals measured over 1, 2, 4 and 8 rows respectively.[182]

The diffraction is the lower limit of the detectable amplitude of the seeing.

For our application we used the Lucernaria Dome indoors Basilica Santa Maria degli Angeli e dei Martiri in Rome. The Lucernaria lenses are fixed solar telescopes with opening 6.3 cm and focal lenght 20 m.[183] The resulting diffraction is 2.31 arcsec for $\lambda$ = 550 nm.

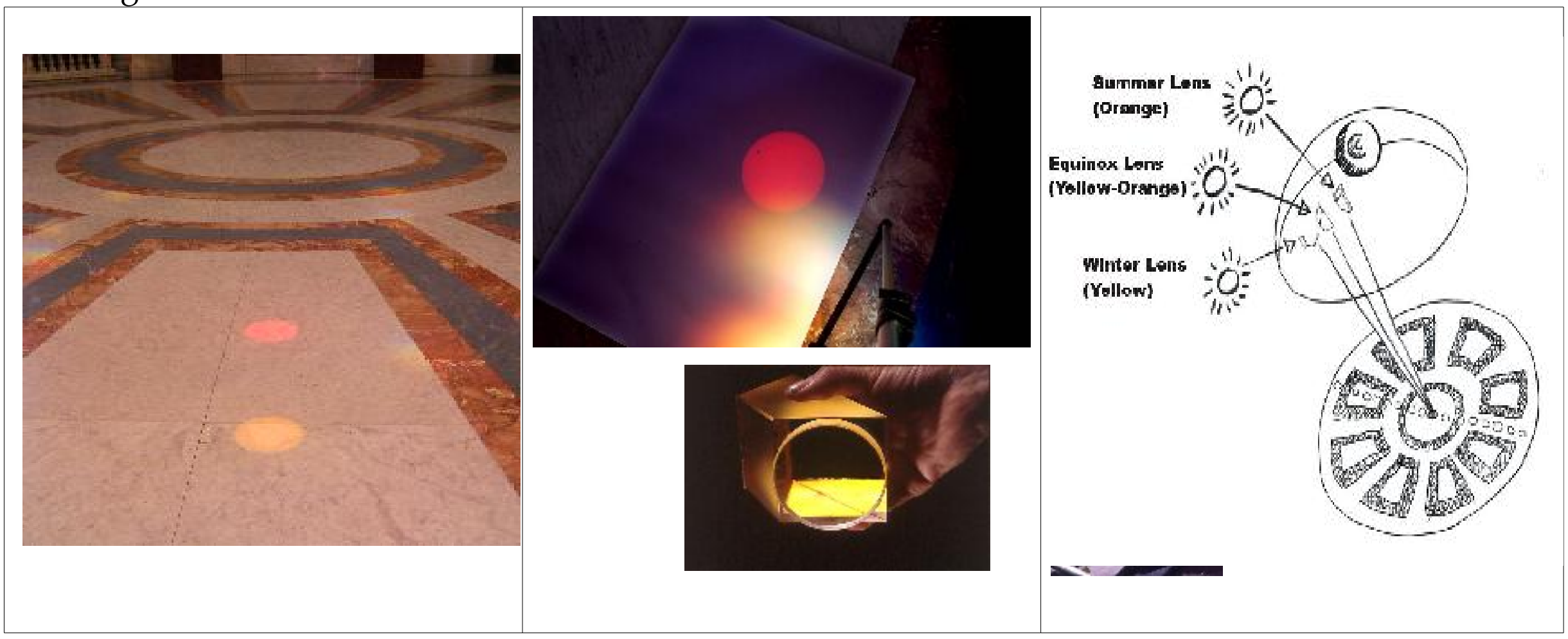

**Fig. 4.7 Lucernaria lenses at Santa Maria degli Angeli in Rome.** The red image of the Sun at the center of this collage is a summer image taken on June 8, 2004 during the transit of Venus. The yellow-orange and red ones are taken on August 21, 2009. The lenses are shown alone and built in the external part of the lucernaria dome. The lenses have been designed and settled by Salvador Cuevas Cardona.[184]

[182] A. Raponi, X. Wang, C. Sigismondi, G. De Rosi, M. Bianda, R. Ramelli, M. Caccia, *The power spectrum of the seeing during solar observations*, 4th French-Chinese meeting, Nice 15-18 November 2011.

[183] Cuevas, S., *Lucernaria Prismatic Lenses*, Il Cigno GG Roma, ISBN: 978-88-7831-213-4 (2009).

[184] It is remarkable that S. Cuevas Cardona has studied at the école de optique in Paris, as the major optic engineers of Nice University. I come in contact with them independently. He is professor of Astronomy at the Autonoma University of Mexico City.

In another paper by Irbah et al. (2003)[185] based on Calern observations with the solar astrolabe gave larger errorbars on the diameter measurements. I report their figures in order to show the correspondance between errorbars in time and arcsec.

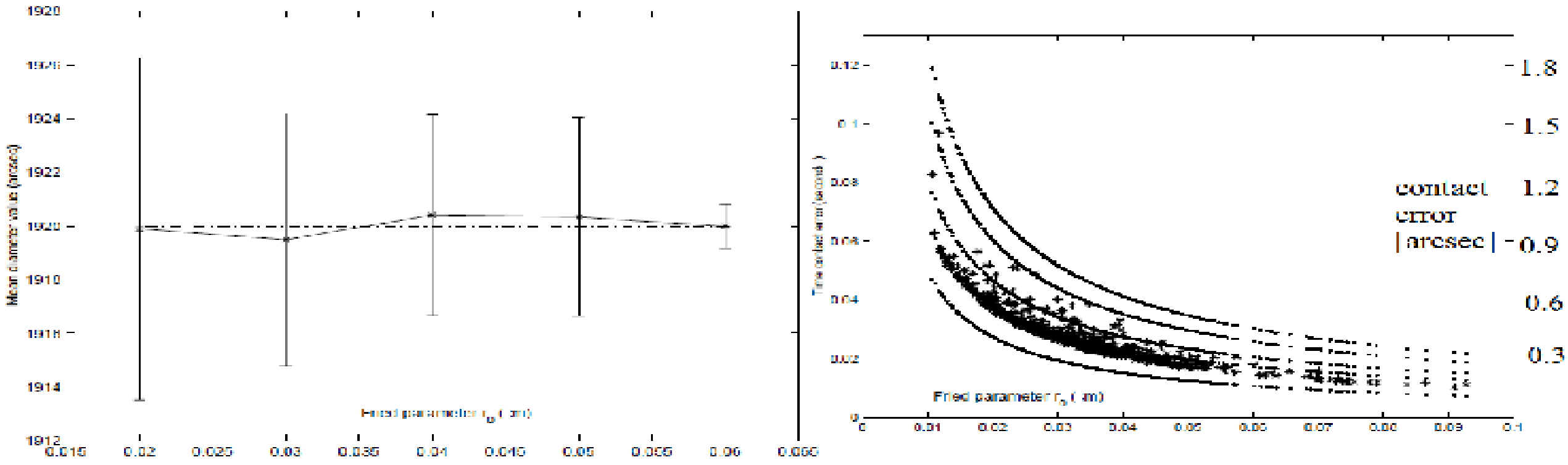

**Fig. 4.8 Seeing – étalement relationships and time contact errors vs seeing.** Left: the measured diameter as function of the Fried's parameter. Right: time contact errors from long to short exposure times (from bottom to top); the stars correspond to experimental values. Adapted from Irbah et al. (2003).

## 4.3 Transits at the IRSOL 45 cm Gregory-Coudé telescope

The tradition of solar diameter's measurements in Locarno goes back to the late years 1970s with visual observations. Two twin Gregory-Coudé telescopes of 45 cm aperture, designed for solar observations, operated simultaneously in Switzerland and Canary Islands with drift-scan methods between 80's and early 90's. The opening diameter of the telescope is larger than any turbulent atmospheric cell, allowing a better stability in the observed solar diameters.

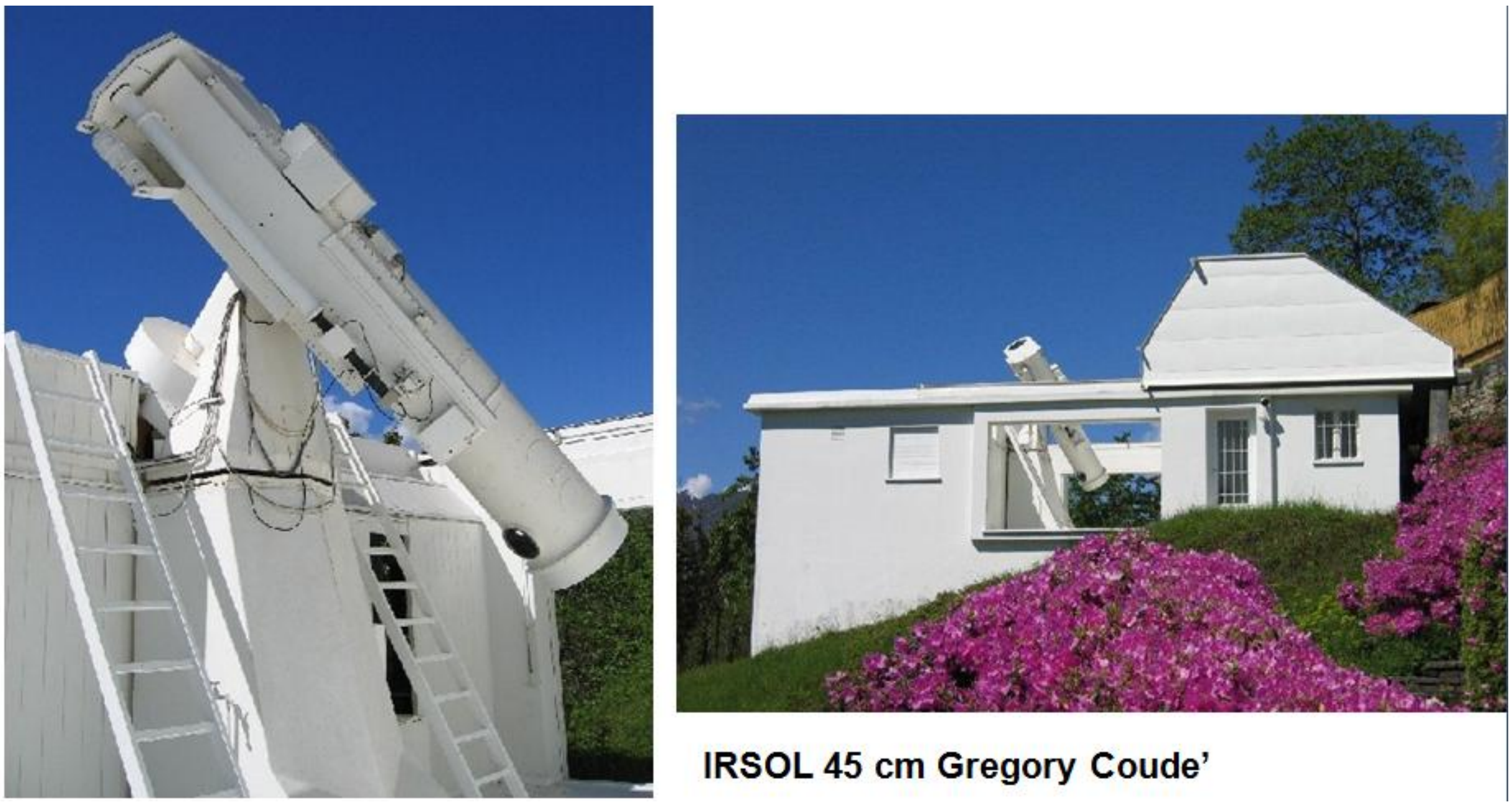

**Fig. 4.9 The Gregory-Coudé solar telescope at IRSOL, 500 m above sea level, and 300 m above the lake Maggiore.** This instrument benefits of very good seeing.

[185]Irbah, A., Bouzid, A., Lakhal, L., Seghouani, N., Borgnino, J., Delmas, C. and F. Laclare, *Atmospheric turbulence and solar diameter measurement*, The Sun's Surface and Subsurface: Investigating Shape. Edited by J.-P. Rozelot., Lecture Notes in Physics, vol. 599, p.159-180 (2003).

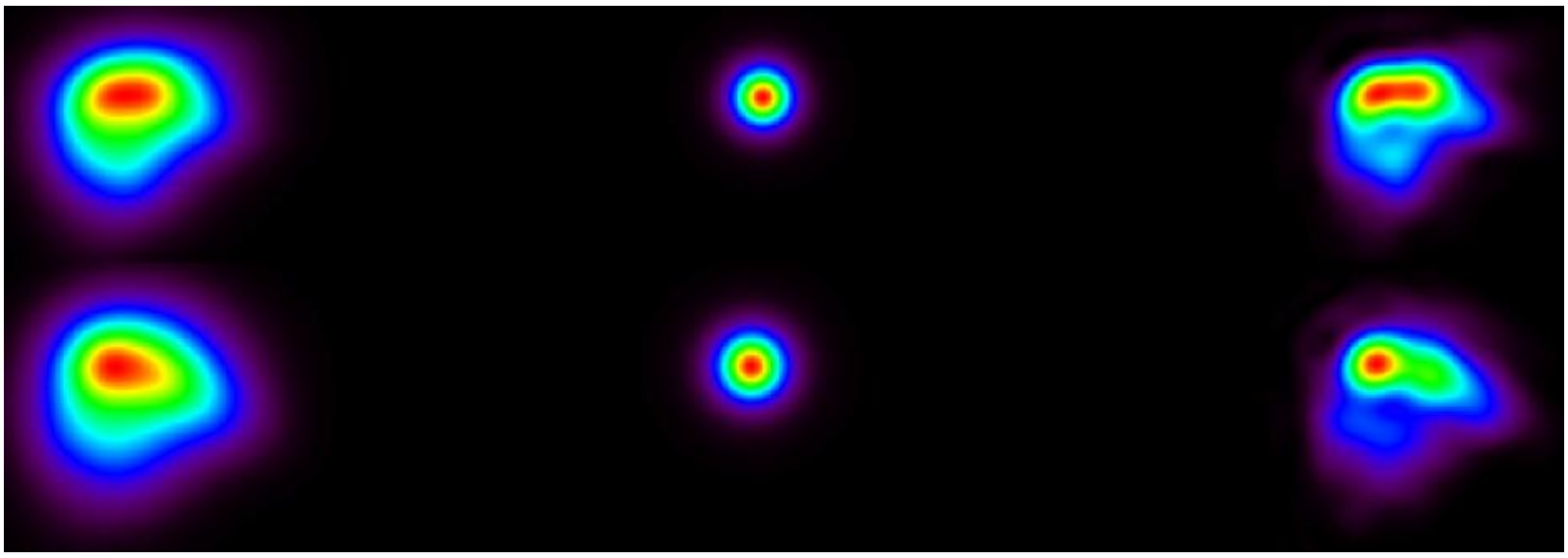

**Fig. 4.10 PSF tests with Markab α Peg (up) and Sadamelik α Aqr (down) at IRSOL Gregory-Coudé solar telescope.** The seeing spreads the instrumental PSF, but the asymmetries are due to misalignments in one secondary mirror.[186] Left and right mages are defocussed[187] images (encoder 530/554 and 624/648). Center: on focus (encoder 589) ideal PSF.[188] An irradiation effect 2.5 times larger than the ideal PSF affects this telescope.

In the section 4.5 there is an operative definition of irradiation. It is due to the PSF wings beyond the diffraction limit.
The study on the PSF is important in order to define the solar limb. Near the limb the PSF is convoluted with a rapidly variating Limb Darkening Function. A symmetrical PSF, and very sharp, is the ideal situation.
The different components of the seeing act upon different angular scales.
There are three main effects of the seeing: blurring, image motion and lensing. The blurring is at the arcsecond scale, the image motion involves all the figure of the Sun at a scale of about 2000 arcseconds, and the lensing corresponds to a deformation comparable with the full disk scale.

We have images of a 100 arcsec scale. From each image of the solar limb obtained during a drift-scan observation we reconstruct a regular arch of a circle from the distorted solar limb. We measure this medium scale effects of the atmospheric turbulence (seeing) by the irregularity of the motion of such arch.
The motion of an arch defined by a fitting curve over 100 arcsec is averaged over the same space scale. Therefore this motion is more regular than the motion of a single point. For this reason values of the seeing as low as 0.6 arcsec have been measured during August 2008 in Locarno.
The afternoon in summer presents usually good observing conditions in Locarno, due to the presence of the lake, but the low value is a cumulative value on the whole observed arch.
A single point-like solar spot would produce a much more scattered motion, as observed with the solar heliometer at the National Observatory in Rio de Janeiro on 14 February 2011 and in Santa Maria degli Angeli Clementine Gnomon, which is a giant pinhole solar telescope, on 1st July 2006.

---

[186]Jolissaint, L., J. Christou, P. Wizinowich, and E. Tolstoy. *Adaptive optics point spread function reconstruction: lessons learned from on-sky experiment on Altair/Gemini and pathway for future systems,* In Astronomical Telescopes and Instrumentation, volume **7736** of SPIE Proceedings, July 2010.
[187]Roddier, C. and F. Roddier, *Wave-front reconstruction from defocused images and the testing of ground-based optical telescopes*, Journal of the Optical Society of America A **10**, 2277-2287 (1993).
[188]Images obtained by Renzo Ramelli and Michele Bianda at IRSOL in November 2010, and elaborated by Laurent Jolissaint (Aquilaoptics.com)

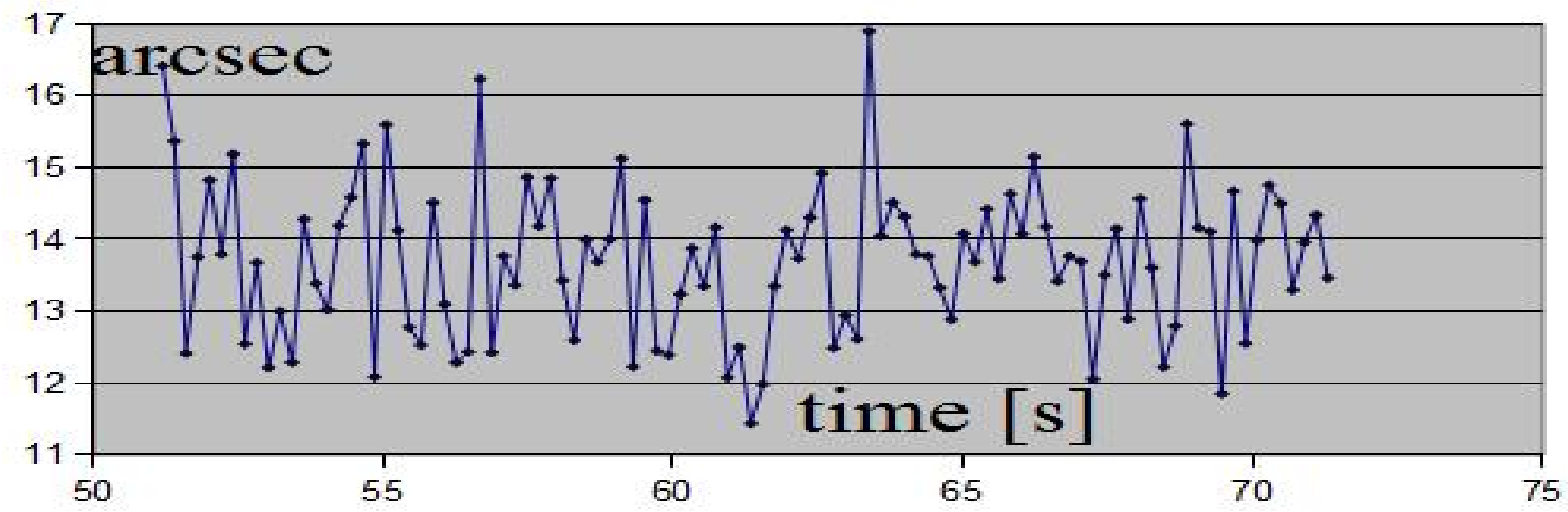

**Fig. 4.11 Position of a small solar spot, observed at the Heliometer of Rio de Janeiro, in drift-scan mode.** The seeing is the standard deviation of such oscillations:1.07 arcsec.[189]

This determination of seeing is strictly related with the measurement of the solar diameter: if the seeing is determined from the scatter of contact times, as in the solar astrolabes, this value has to be associated with the spatial scale of the fitting parabola, and not with a point-like source. Consequently the corresponding seeing ("Fried" like) parameter changes, and in general, it increases.

The observation of the transits of the Sun at Locarno has been made with the Gregory-Coudé vacuum telescope of 45 cm aperture and 25.0 m of focal length. The instrumental field of view gives a portion (about 100 arcsec x 200 arcsec) of the solar image, from which we recover the curvature of the limb and the solar center.

The image of the Sun is projected on the CCD Baumer camera, and it is digitalized. The figures 4.10 hereafter represent the motion of the center of the Sun as recovered from the Northern limb, tracked for 1000 s, and the corresponding FFT power spectrum over frequencies, in abscissa, from 1 to 1/100 of such timespan.

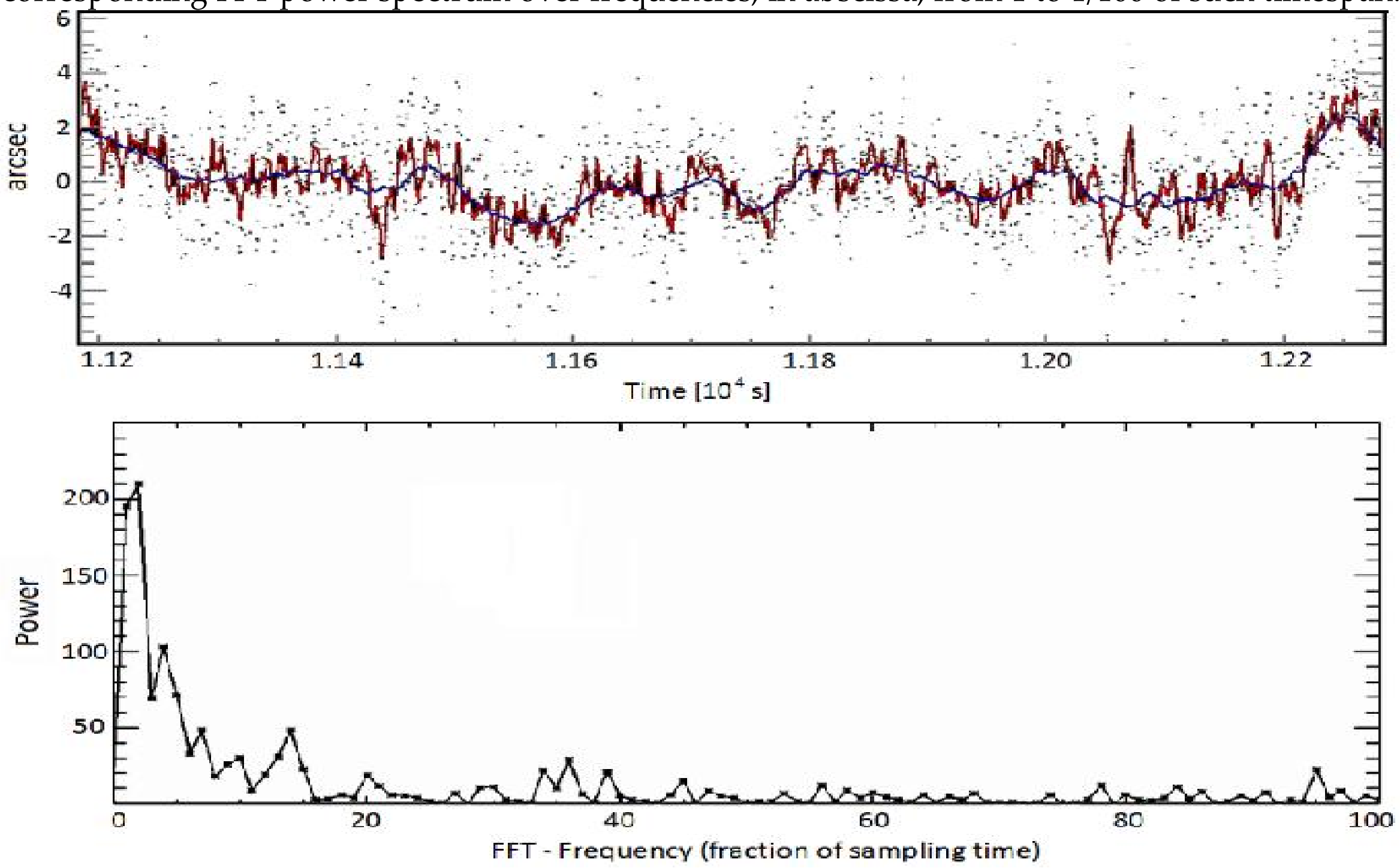

**Fig. 4.12 Variation of the center of the Sun over 1000 s at IRSOL Telescope.** The FFT shows clearly power at low frequencies, to be associated with slow image motion.

The Northern limb is considered as position angle P.A.=0 degrees. In this image, for the first time, the effect of "sub-Hertz" fluctuations is evidenced to the amplitude of one arcsecond.
This phenomenon explains the difference between two following measurement of the solar diameter made with drift-scan method, and this also explains the need of several measurements to be statistically averaged in order to give a reference value for the solar diameter.
But several measurements can occur under different meteorological conditions even in the same day, and the single measurements cannot be considered as statistically independent, as the Gaussian hypothesis requires.
That is the reason of the great scatter that the yearly averages published for Greenwich and Capitol Observatory both show.[190] Also modern CCD measurements show significant scatters within the same series of diameter measurements.[191]
The IRSOL telescope is larger than the turbulent cells, and rather independent of them, while all solar astrolabes have objective lenses of about 10 cm, comparable with the turbulent atmospheric cells.
From 2008 this method has been included in the framework of the **Project Clavius,** among Italian (Como and Rome Universities) and Swiss scientific Institutions, for the development of fast detector for physics and astronomy.
The drift-scan project at Locarno is being upgraded with fast imaging detectors, exploiting the idea of simultaneous measurement of the seeing and of the solar diameter. We started with a commercial SANYO CG9 with CMOS detector at 60 frames per second, reaching a resolution of 0.25″ for a single diameter measurement, and we used either a CCD Baumer camera as a detector and the ultra fast MIMOTERA CMOS detector.
Christopher **Clavius** (1538-1612) described the 1567 hybrid (and "mesosphere-dominated") eclipse from Rome. This international project celebrated the 400$^{th}$ anniversary of his death. Later the Institute d'Astrophysique de Paris in 2010, the Observatorio Nacional of Rio de Janeiro and the Huairou Solar Observing Station of National Astronomical Observatories of China in 2011 joined our efforts.

[189] The max and min position are 16.91 and 11.43 arcsec, and the corresponding standard deviation in case of uniform distribution would be 1.58 arcsec [(b-a)/√12].

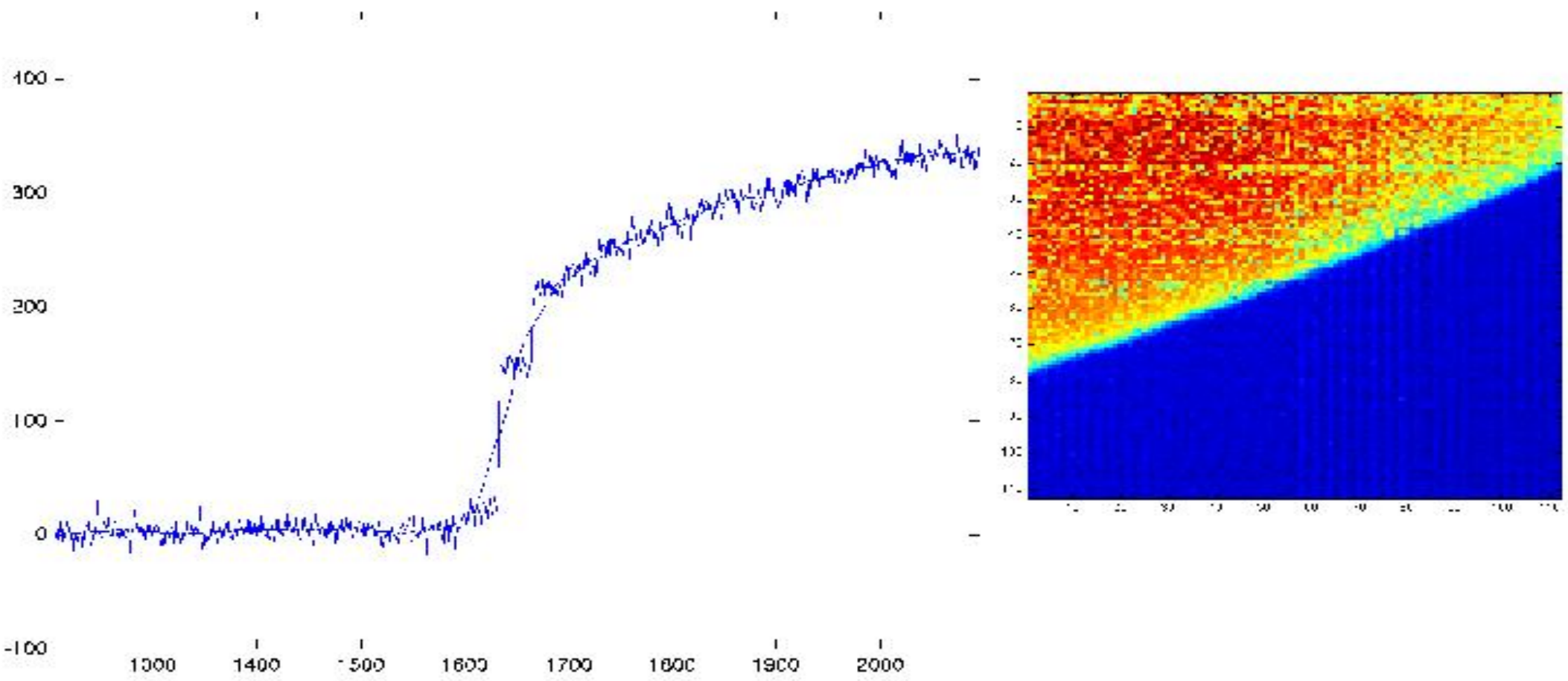

**Fig. 4.13 MIMOTERA's image and profile of the radial luminosity of the Sun.**

How the number of measurements improve the statistics? We can obtain more than 30 consecutive diameters (as in the 1990s) per day using the grid method. For this purpose we organized 4 observational sessions in coincidence with solar eclipses: August 1$^{st}$ 2008 partially visible from Locarno under clouds; July 22, 2009 not visible from Locarno; and January 4, 2011. Another observational session was conducted after July 10, 2010 eclipse, total over Easter Island and not visible from Locarno.
The conclusion of this study has been the following: the better determination possible with drift-scan method is to observe the Sun with two instruments in parallel: one wide field (about six solar radii) and another with narrow field of view (about 100/200 arcsec). With the wider field of view the image motion contribution can ben detected and isolated, in order to correct the diameter's measurement limb-to-limb.

[190]Gething P.J.D., *Greenwich observations of the horizontal and vertical diameters of the Sun*, Monthly Notices of the Royal Astronomical Society, **115**, p.558 (1955).
[191]Wittmann A. D., Alge E. and M. Bianda, *Detection of a significant change in the solar diameter, Solar Physics*, **145**, p. 205, 206 (1993).

## 4.4 Transits at the Carte du Ciel 33 cm refracting telescope

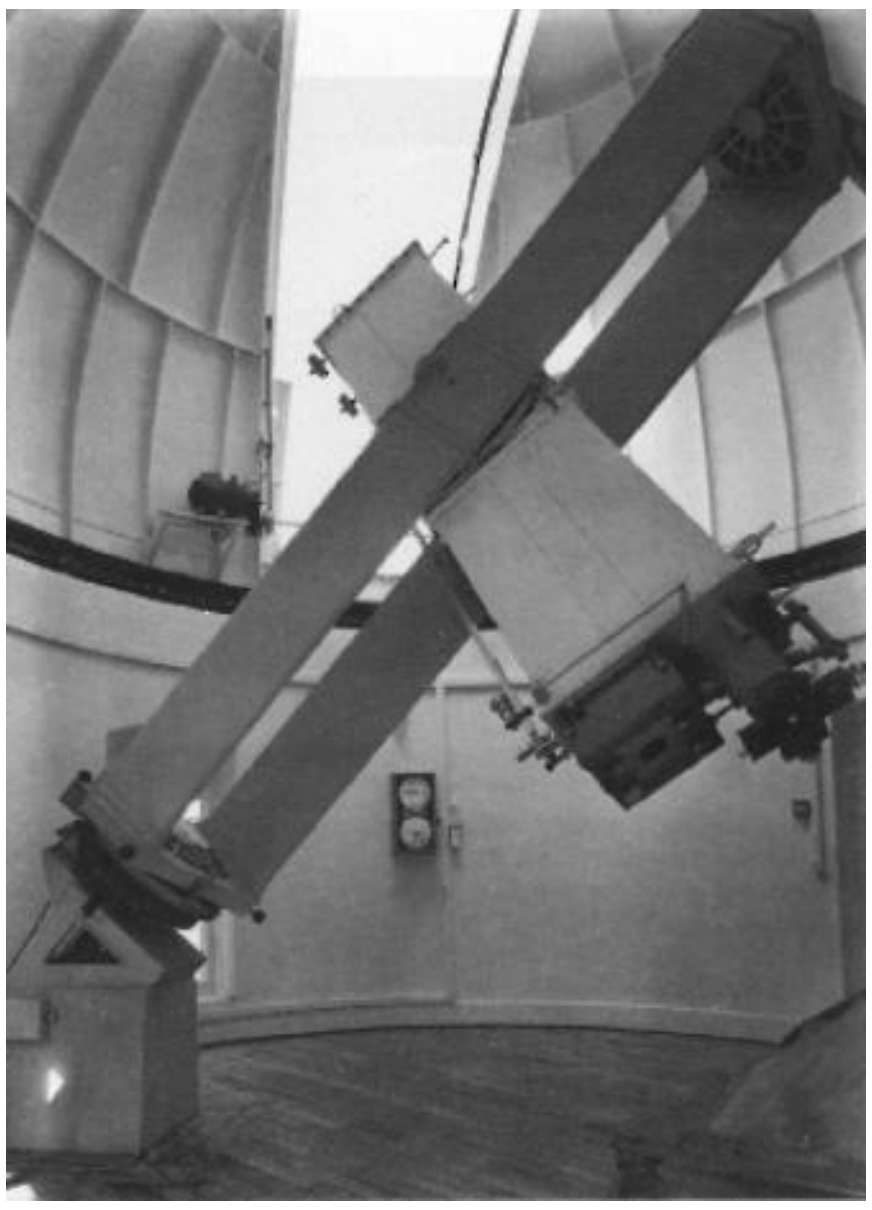

**Fig. 4.14 The Carte du Ciel: two refracting telescopes of 25 (visual) and 33 cm designed for astrophotography.**

This historical telescope (located at the Institute d'Astrophysique and Observatoire de Paris) used for astrophotography since 1885, has been equipped with a CCD camera Lumenera and screened with a panchromatic filter in astrosolar.

The Carte du Ciel 33 cm refracting telescope in Paris, used for our transit measurements, has a PSF particularly clean because of no obstructions in its optics, and a very small scattered light effect.

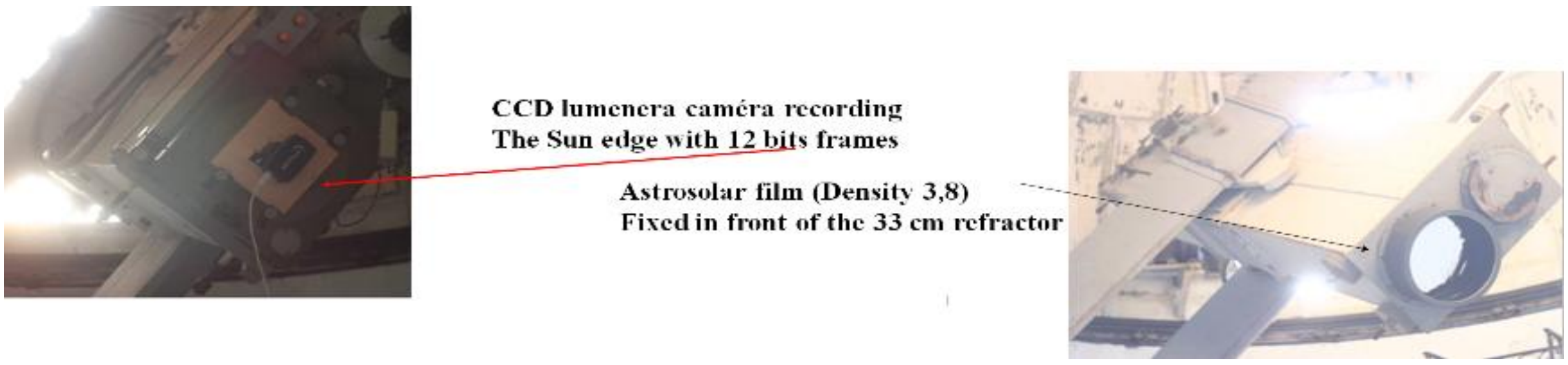

**Fig. 4.15 The Carte du Ciel: new experimental setup for solar observations.**

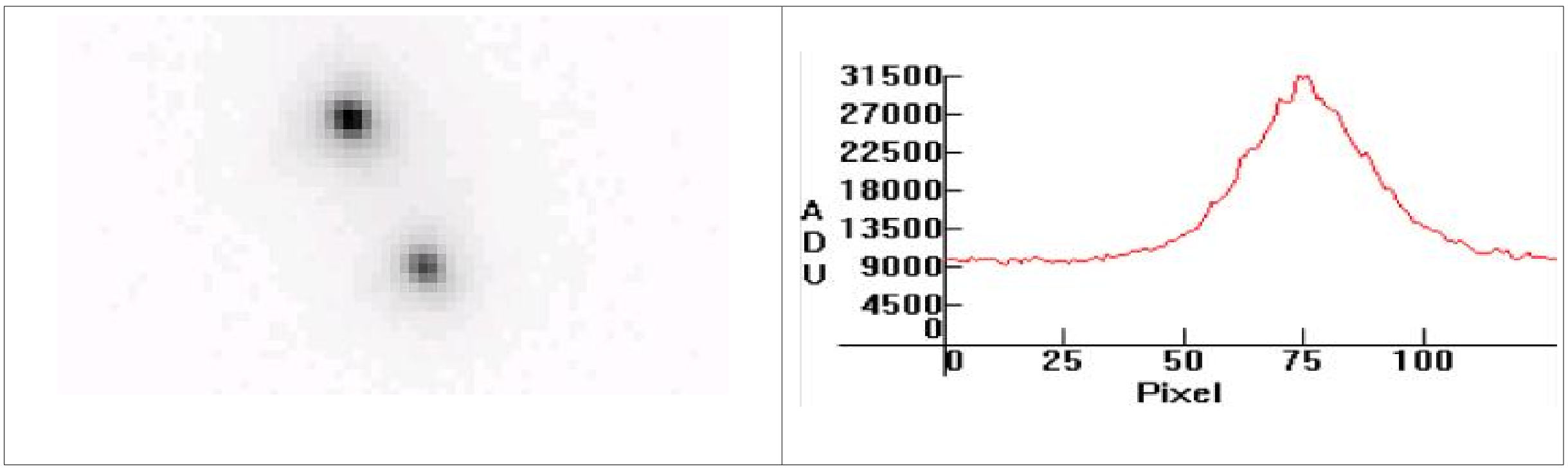

**Fig. 4.16 Left: Epsilon 2 Lyrae as seen with the 33 cm refractor of the Carte du Ciel**. The angular distance of the two components is 2.35 arcsec.

**The FWHM of the PSF <0.5 arcsec, i.e. the instrument is diffraction limited.**

**Right: Sirius spread function in daylight** (about 2.5 arcsec FWHM due to the seeing turbulence). A pixel corresponds to 0.1 arcsec.

The PSF test with epsilon 2 Lyrae has been realized with a blue filter of 460 nm, by the superposition of 66 selected images of 50 ms of exposure (22 october 2010). The same instrument, with an astrosolar filter, is used for the solar transits, under much more turbulence. The advantage of this instrument is the minimal irradiation. On the contrary, all other instruments used in the past, had an irradiation larger than 1.5 arcsec.
The Carte du Ciel is 3.6 meters F/21 telescope without central or off-axis obstructions and any secondary mirror, neither flat neither curved. It is well known that any secondary mirror introduces irradiation effects, and modifications of the PSF. Thanks to these considerations the Carte du Ciel is an ideal instrument for the transits to measure the solar diameter with the drift-scan method.
The exposure time in Sun light is much smaller than 50ms, but the turbulence is larger than night time and the final effect on PSF can be as large as seen for Sirius on October 1st 2010 (fig. 4.15 right).

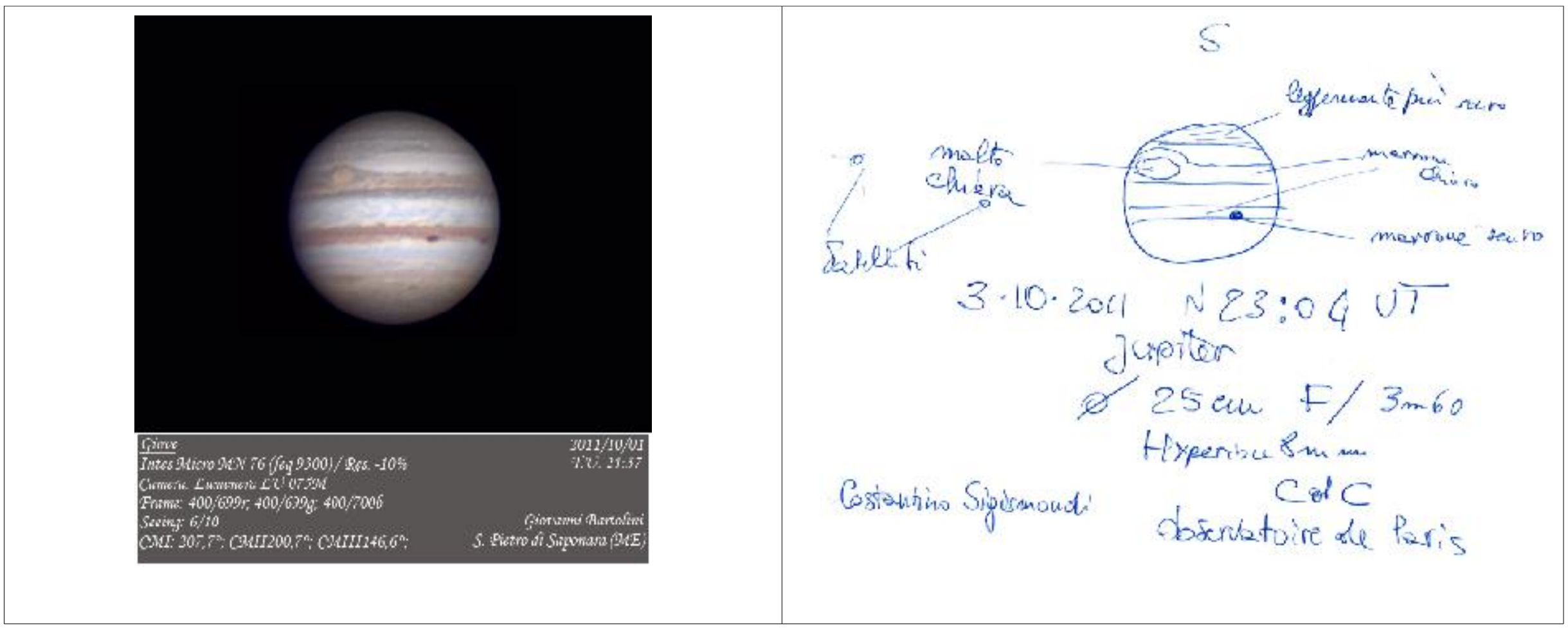

**Fig. 4.17 Jupiter Left: with CCD and best choice selection method (Oct 1, 2011); Right: drawing at the 25 cm refractor of Carte du Ciel on October 3, 2011. The PSF of this telescope yields stunning images.**

## 4.5 The operative definition of the irradiation effect

I have found the term *irradiation* in these studies of solar diameter in the year 2000, at Yale, when dealing for the first time with the Winoccult program of Dave Herald, made for simulating Baily's beads. After some time, in february 2001, when I was at NASA Goddard Space Flight Center to study the SDS, I suggested to eliminate this parameter.

The irradiation was a parameter corresponded to an increasement of the solar radius of 1.5 arcseconds and appeared without any further explication than “it happens in this way”. Moreover it was the same parameter for all instruments used to observe the eclipse, without distinction between naked eye and other instruments.

After reading many XIX century's papers on solar diameter subject that word and the “irradiation effect” came out several times, and the value adopted in Herald's software was an averaged value as in the cases published by Auwers in 1891.[192]

We decided to eliminate this parameter from that software.

Now it is clear that it depends on the instrument.

Irradiation (to not be confused with irradiance, which deals with the solar luminosity) is the effect of the Point Spread Function of the telescope, including all internal reflections, either on-axis and off-axis. Each optical element adds to an optical configuration some dispersed light, and consequently some irradiation effect.

The Carte du Ciel astrograph has an optical configuration particularly clean with the smallest Point Spread Function among all solar instruments used in my research.

While for the Point Spread Function it is rather easy to state a behaviour on-axis and its comparison with off-axis, in the irradiation effect we have the convoluted effect of all point-like sources which ideally compose the solar disk.

The contribution of stray light coming from off-axis directions is very large, since the solar radius is as large as about 16 arcminutes. This angular dimension is very large when compared with astronomy standards.

For example in Beppo SAX the X-Ray telescope qualification studies were extended up to these angles as limiting cases, with very large Point Spread Functions.[193]

The optical configuration of the IRSOL telescope has a field stop for avoiding to pour too much stray lights into the focal plane. This stop is a pinhole located below the elliptical Gregory secondary mirror.

---

[192] Auwers, A., *Der Sonnendurchmesser und der Venusdurchmesser nach den Beobachtungen an den Heliometern der deutschen Venus-Expeditionen*, Astronomische Nachrichten **128**, 361 (1891).

[193] Sigismondi C., P*erspectives on the observation of clusters of galaxies in X-ray band with SAX (X-ray Astronomy Satellite)*, Nuovo Cimento **B 112**, 501-515 (1997).

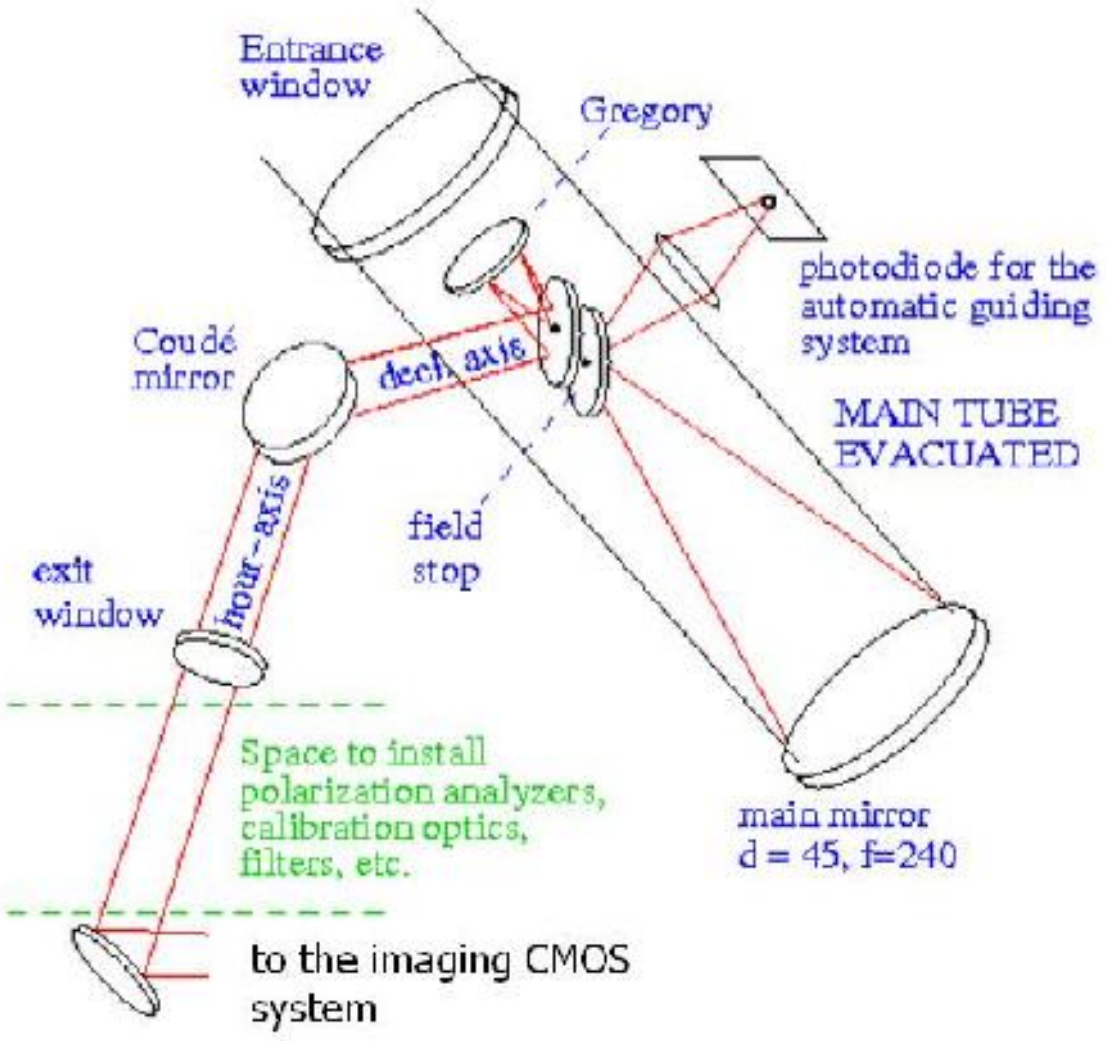

**Fig. 4.18 Scheme of 45 cm Gregory-Coudé evacuated solar telescope at IRSOL. The field stop allows to avoid 99% of diffuse light, without loosing in angular resolution. The field of view is 220″. The total focal length is 25 m.**[194]

The observation of the Sun made with smaller telescopes, as the ones used in eclipse field observations, is even more affected by this effect. Telescopes onboard of satellites, for their compact optical configuration and relatively long focal lengths, can have relevant irradiation effects.[195]

In conclusion the irradiation effect is a cumulative effect of on-axis and off-axis Point Spread Function and of the stray light, which is always abundant in the case of the Sun.

## 4.6 Evidence of 1″ slow image motion of the whole solar image

The diameters obtained in consecutive measurements are not consistent within all the experimental errors. The first results of the new measurements tests for both telescopes (2008-2010) are discussed: the effect of low frequency waves (0.01 Hz) in the atmosphere could explain the difference in the successive hourly circle transits' measurements. This finding can cast some light also on some ancient puzzling data (Rome - Campidoglio & Greenwich observations).

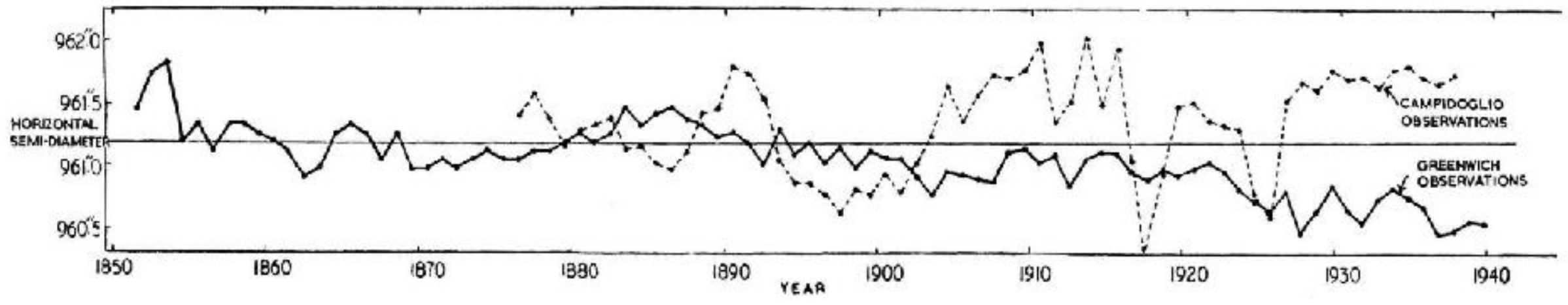

**Fig. 4.19 The series of observations of the solar diameter in Greenwich and in the Capitol's observatory in Rome.**

[194]Source www.irsol.ch cfr. also Sigismondi, C., Bianda M and J. Arnaud, *European Projects of Solar Diameter Monitoring,* in RELATIVISTIC ASTROPHYSICS: 5th Sino-Italian Workshop on Relativistic Astrophysics, AIP Conference Proceedings, p. 189-198 (2008).

[195]See e. g. the image in fig. 2.7, althought very sharp, from HINODE satellite, this image shows a ghost image in the disk of Venus.

There are various puzzling problems that this figure presents.
The meridian transits have been averaged over a whole year for each point. The large scatter between each value (up to 1.5 arcsec in the case of Capitol's observations in 1917-18, at the end of the First World War) suggests that the data are not normally distributed. If the N data for each year would be Gaussian, their average would be spread within $\sqrt{N}$, and so the large variations detected would be considered as real. So what is the origin of these large scatters between the yearly averages?
To avoid "personal equations" the roman observatory employed four observers at each transit, which was projected on the ground on a grid of evenly spaced lines. Each one evaluated the time of the contacts with the eye-ear method. It happened that in 1917-18 only one observer operated, while for the Greenwich observations we have a single observer along the years, and ageing processes in his vision could contribute to systematic effects.
The differences among the yearly averages remain still unexplained, under a statistical and an experimental point of view. I wish to remember that these astronomers were very skilled observers and fine physicists, devoted to obtain the most accurate measurement possible.
The discovery of the 0.01 Hz fluctuations of the atmosphere starts to unveil this mystery.
Also the consideration that the Airy's meridian circle in Greenwich was a 6 inches instrument, i.e. a 15.2 cm with its Point Spread Function effect, and the roman instrument was a 11 cm telescope can contribute to explain the larger scatter in the roman data.

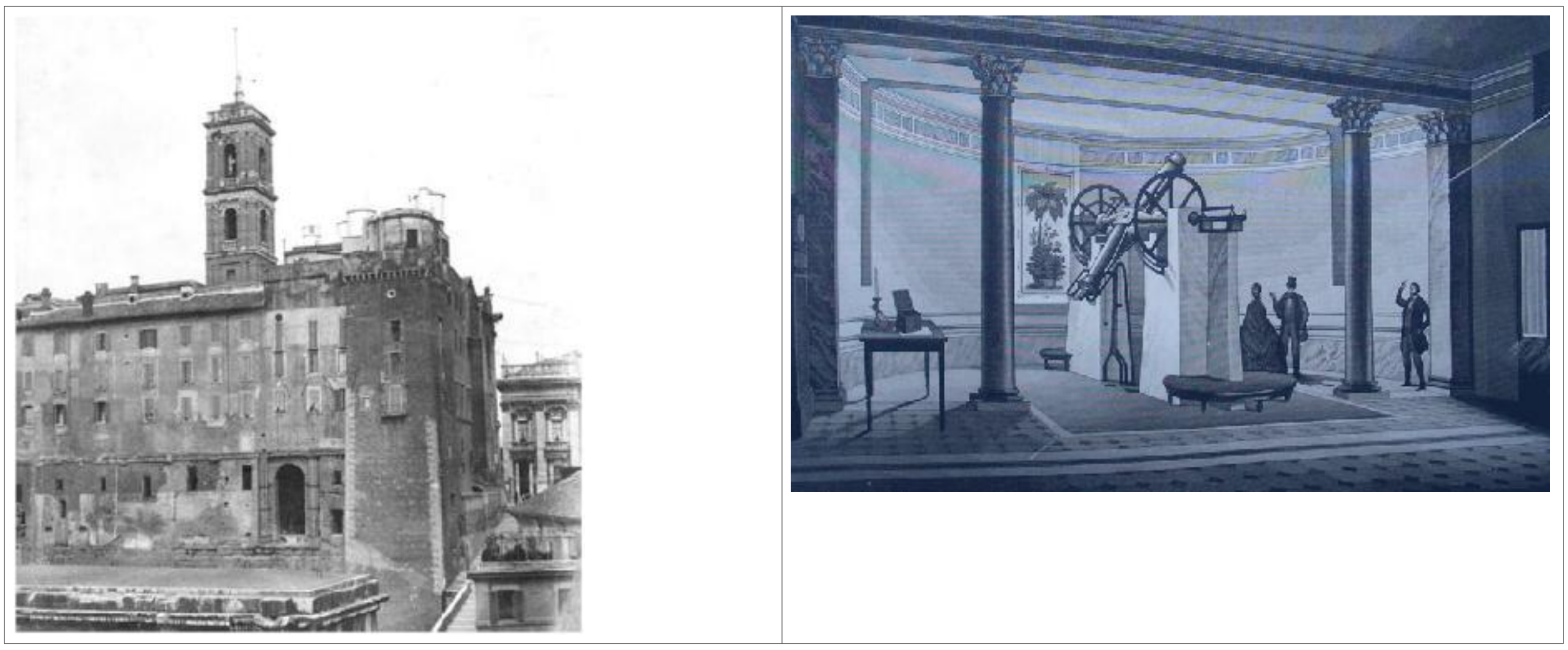

**Fig. 4.20** Left: ancient photo of **the observatory of Campidoglio in Rome**, as before 1937 when it was destroyed. It has been founded by pope Leo XII in 1829.
Right: the telescope used for the meridian transits in the observatory of Campidoglio in Rome, gift of pope Pius IX in 1853.

Presently the instrument used in the Capitol Observatory (no longer existing since 1937) is conserved at the entrance hall of Monteporzio Catone Observatory, with an incomplete label. The measurements taken in Locarno (fig. 4.11) have been repeated in Huariou Solar Station (China, National Observatories).[196] The observation of the full solar disk is performed with the Solar Magnetism and Activity Telescope (SMAT)[197] that is a telescope with a tele-centric optical system of 10 cm aperture and 77.086 cm effective focal length realized to investigate the global magnetic configuration and the relationship with solar activities. The birefringent filter for the measurement of vector magnetic field is centered at 5324.19 A (Fe) and its bandpass is 0.1 A. The detector is a CCD camera, Kodak KAI-1020. it is used for the measurement of full disk. The image size of the telescope is 7.4mm×7.4mm, and the size of CCD is 992×1004 pixels. The frame rate of the CCD camera is 30 frame/s and its maximum transmission rate is 60 Mbyte/s. The exposure time is 2 ms.

[196] A. Raponi , X. Wang, G. De Rosi, M. Bianda, R. Ramelli, *The power spectrum of the seeing during solar observations*, 4 meeting Franco-Chinois, Nice 15-18 November 2011.
[197] Zhang, H. et al., *Solar Magnetism and the Activity Telescope at HSOS,* Chinese Journal of Astronomy and Astrophysics **7**, pp. 281-288 (2007).

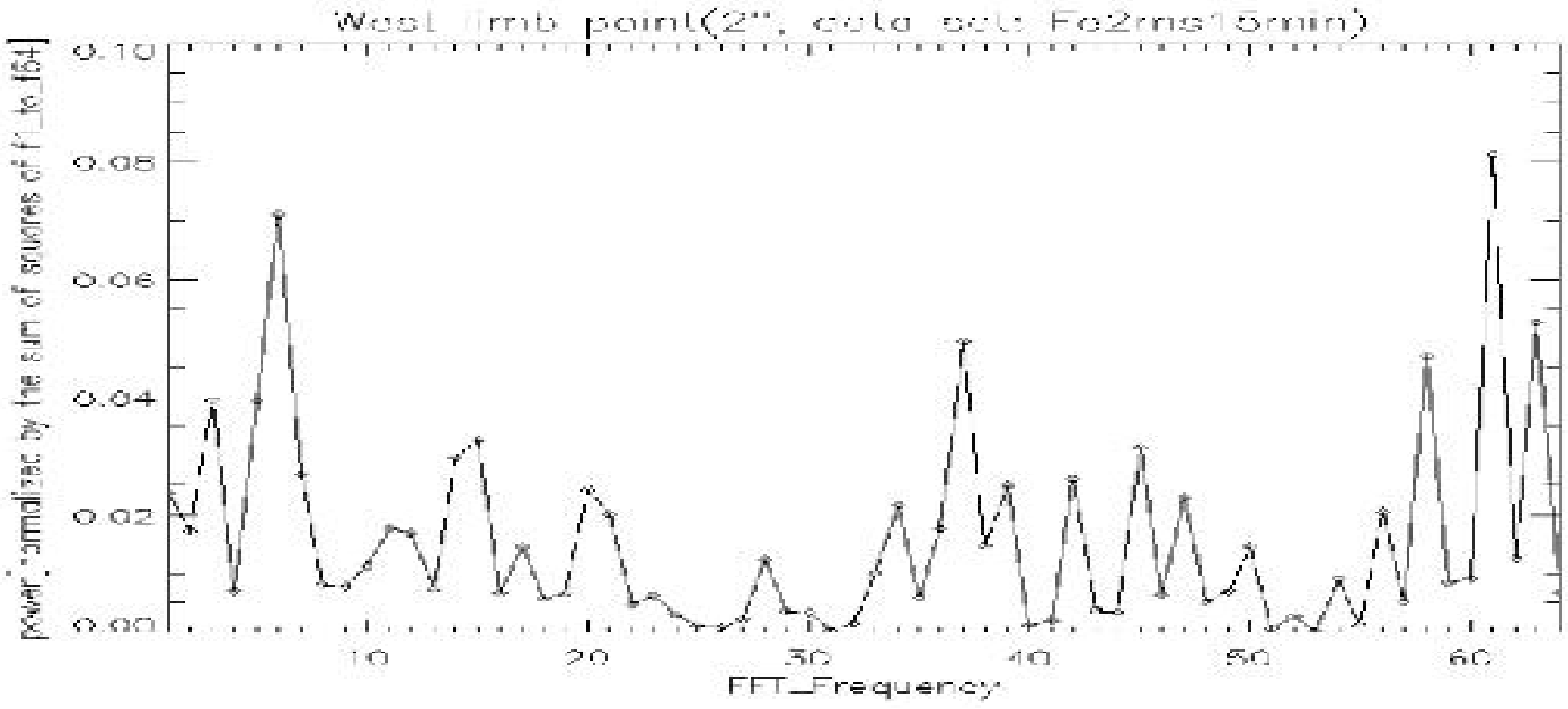

**Fig. 4.21 The power spectrum of the seeing at Huairou Solar Station (China).** The timespan is 15 minutes, 60 corresponds to 1/60 of this time, i.e. 15 s; 30 to 30 s; 15 to 1 minute and 5 to 3 minutes. There is power at all these timescales.

## 4.7 Space calibration of transits measurements

Combined observations space-ground: eclipses and drift-scan. In the occasions of two total eclipses of 1st August 2008 and 22nd July 2009 we have made drift-scan observations at Locarno observatory, and a third session was planned for 11 July 2010 eclipse. A single day of observations made with DORAYSOL on 23 September 2006 can be considered as overlapped with the annular eclipse of 22 September, observed in French Guyana.

A similar strategy is planned for SDS and Picard. A new flight of SDS (after the one still unpublished occurred in october 2009), shall be done during the Picard mission to compare the space measurements with the balloon-borne ones.

These observations, after the discovery of sub-Hertz image motion, have to be done with two telescopes: one for the details with 100 arcsecond field of view, and the other in parallel with 4 solar radii field of view to see the image motion.

## 4.8 Before and after this work

The situation of this filed of research before and after this work is described, putting into evidence the new contributions.

The role of seeing fluctuations under 1/10 and 1/100 of Hertz is crucial in drift-scan measurements of the solar diameter, either meridian transits or almucantarat transits. This study firstly evidenced this effect in a clear way. The fluctuation's scale is not defined here, since we did not apply the analysis to a single point only, but also on solar limb arches of several arcseconds (from 200 for IRSOL to the whole disk for Huairou Solar Station). Consequently the scale of the seeing (composed by blurring + image deformation + image motion) depends also on the algorithm used to define the solar figure.

An average made with N points distributed over all 360° of position angles (Huairou Solar Station), is different by the one made over 12° (IRSOL), and by the one, discrete, made on the preceding and following limb at Santa Maria degli Angeli Lucernaria by visual inspection (about 20° of the solar limb involved).

All these measurements have in common the detection of significant energy at the low frequencies regions around 0.01 Hz of the power spectrum.

These effects have to be taken into consideration for a fruitful analysis of the solar diameter measured with these methods. A full Sun imager would help to monitor the image motion occurred during a single transit observed with the drift-scan method.

# 5 Didactic Outreach

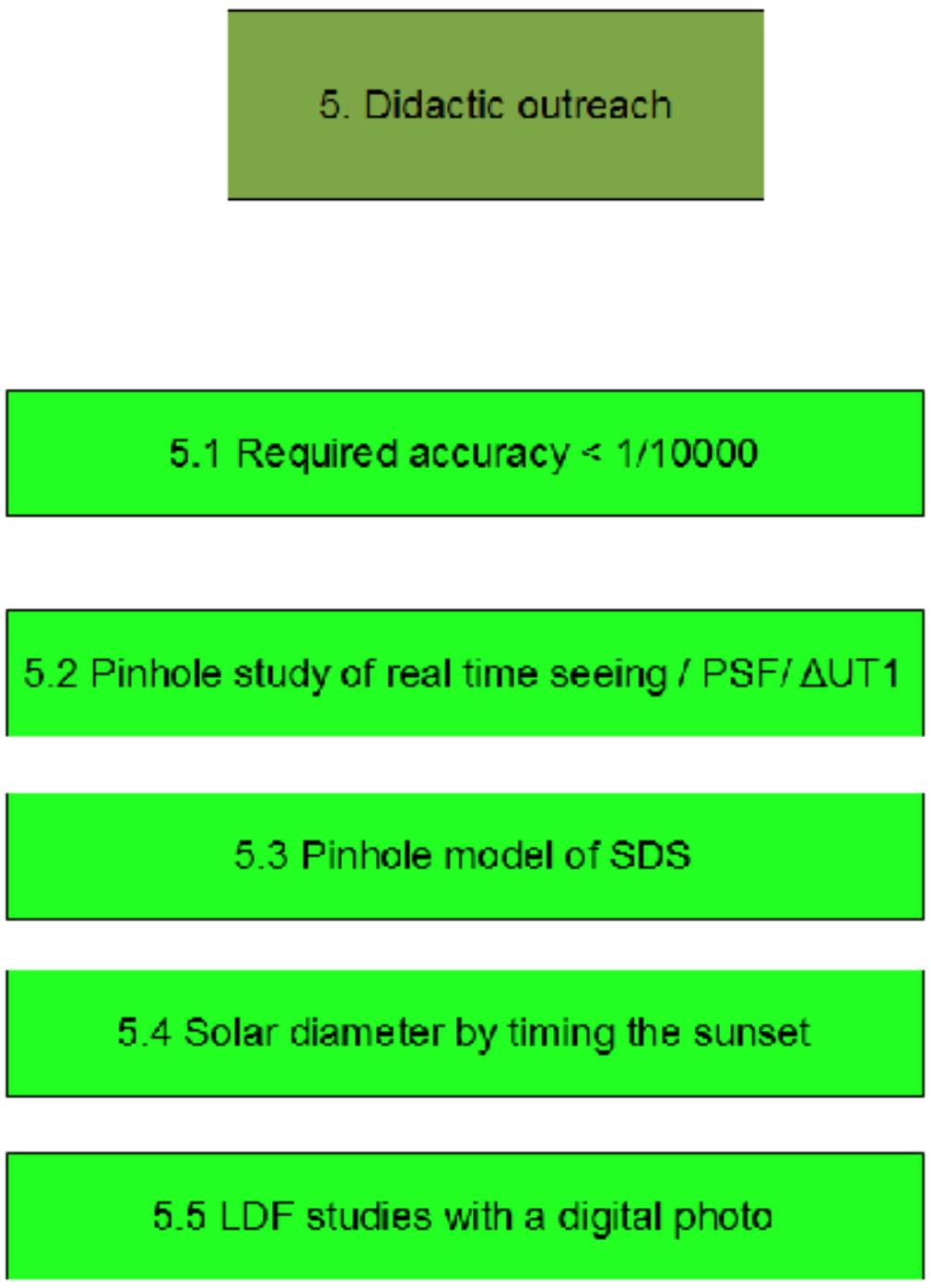

## 5.1 Required accuracy < 1/10000

The statement on human energy output versus solar input is “1 year of humankind energy production corresponds to 1 hour of Sun”. This is the starting discussion of a series of units on solar astrometry, in order to motivate this study in a framework of climate studies.

Knowing the evolution of our star, and therefore of its diameter, up to 1 part over 10.000 is necessary if we want to understand really the problems of global warming and climate changes.

The first thing on which to reflect is the intrinsic difficulty to obtain such a precision even in measuring a rod of 1 meter. The accuracy required is 0.1 mm, already out of the possibility of the unaided eye.

## 5.2 Pinhole studies of real time seeing, PSF and ΔUT1

A pinhole is a "lens-less telescope", completely free from spherical aberration and perfectly suitable for astrometric studies of the Sun. It has been exploited since 1475[198] by Paolo Toscanelli for a giant pinhole telescope built in Florence's Cathedral as pinhole camera. Measuring the solar declination with the pinhole telescope in Rome, Santa Maria degli Angeli (the Clementine gnomon of 1702) I have demonstrated its extraordinary accuracy, down to one arcsecond, for all the range of solar declinations, and the possibility to be used for monitoring the ΔUT1 evolution.

The observation of the Sun with a pinhole in order to check the first and the second Kepler's law has been realized for high school students. The settling of the experiment in order to minimize the errors of measure has permitted to verify the Rayleigh formula for obtaining sharpest images given the diameter of the pinhole.[199]

Pope Clement XI (1700-1721) ordered Francesco Bianchini (1662-1729) to build a Meridian Line. Bianchini was the Secretary of the Commission for the Calendar. He chose the Basilica of Santa Maria because of the stability over centuries of the ancient walls where the pinhole is located is a requirement for making high precision astrometry, such as the measurement of the inclination of the Earth axis over its orbit plan. In the 18th century it was possible to open the window holding the southern pinhole, and, even in daylight, stellar transits were recorded and precisely timed with pendulum mechanical clocks. The accuracy of such clocks was better than 1 s per day, and the observations of stellar transits allowed their synchronization with sidereal time. This "hybrid feature" of the Clementine Gnomon to measure solar and stellar transits allowed Bianchini to accomplish in 1703 the whole measurement of the duration of the tropical year, which was usually made by comparing observations very widely spread in time. The small deviation of the Line from true North of ~ 4'28.8 arcsec Eastwards[200] has been measured comparing the delays of transits at both solstices with respect to the ephemerides.[201]
The Clementine Gnomon is basically a solar meridian telescope dedicated to solar astrometry operating as a giant pinhole dark camera, being the basilica of Santa Maria degli Angeli the dark room. The solar images produced on the floor of the Basilica are free from distortions, excepted atmospheric refraction, because the pinhole is lensless.

Similar historical instruments are in Florence (at Duomo, made first by Toscanelli and later by Leonardo Ximenes SJ), in Bologna (in San Petronio, made by Giovan Domenico Cassini), in Milan (at Duomo, made by Antonio De Cesaris) and in Palermo (Cathedral, made by Giuseppe Piazzi). I have recently referenced the azimut of the Clementine Gnomon with respect to the celestial North pole, and it is 4' 28.8±0.6 arcsec, a comparison with similar coeval instruments is presented. Also the local deviations from a perfect line are known with an accuracy better than 0.5 mm. With these calibration data we used the Gnomon to measure

---

[198] Ulugh Begh in Samarkand (1437) measured the solar declination exploiting the same principle. His idea was probably arrived in Costantinoples – Istanbul around the fall of Eastern Roman Empire (1453), but the connection between Ulugh Begh and Toscanelli is not demostrated. It is more realiable to consider that the technique of the pinhole projection should appear sooner or later here and there because of its simplicity. A similar opinion is expresse by the Jesuit astronomer Father Angelo Secchi on the first discoverer of the solar spots. A. Secchi, *Il Sole*, Firenze (1884).

[199] Sigismondi, C., *Introduction to pinhole astronomy*, Proc. of "Einstein 120" conference, Bishkek, Kyrgyzsthan, 12 September 1999, V. Gurovich ed., Kyrgyz State University, Bishkek (2000) also on arXiv:1107.0820v1 (2011). Sigismondi, C. and L. Cesario, *Solar Astrometry with Pinholes*, in *Effective Teaching and Learning of Astronomy*, 25th meeting of the IAU, Special Session 4, 24-25 July, 2003 in Sydney, Australia, meeting abstract (2003).

[200] see figure 0.2 in Chapter 0.

[201] Sigismondi, C., *Astronomy in the Church: the Clementine Sundial in Santa Maria degli Angeli*, Rome, Il Cigno GG Editor, Roma - Italy (2009), also on arXiv:1106.2976v1 (2011).

Sigismondi, C., *Lo Gnomone Clementino: Astronomia Meridiana in Chiesa dal '700 ad oggi*, Astronomia UAI **3** 56 (2011) also on arXiv:1106.2498 (2011).

the delay of the solar meridian transit with respect to the time calculated by the ephemerides (ΔUT1). The growth of this astronomical parameter is compensated by the insertion of a leap second ad the end of the year in order to keep the Universal Time close to astronomical phenomena within less than a whole second. On December 31, 2008 at 23:59:59 there was one of those leap seconds leading to 23:59:60 before the new year's midnight 00:00:00, being ΔUT1~0.7 s at that date. Another one has been inserted on June 30, 2012. ΔUT1 has been measured with an accuracy of ±0.3 s. [202]

Using an evenly spaced grid at the focal plane in a drift-scan method gives the opportunity to have several diameters in a single transit[203] and it can be used for measuring the seeing in real time.[204]
A pinhole camera has the advantage of an undistorted field of view. Its imaging capability is limited by random (diffraction and atmospheric seeing) and systematic (penumbra) effects. The Pinhole Solar Monitor, PSM, measures the solar angular diameter by timing meridian transits. Meridian transits have been videorecorded with UTC synchronization at the pinhole gnomon of Santa Maria degli Angeli in Rome. The tarature of this Clementine Gnomon is outlined with its accuracy as PSM. On the Moon an array of such PSM equipped with 1000 lines for parallel transits could monitor 0.1 arcsec variations of solar diameter.
This apparatus was called Pinhole Solar Monitor.[205]
All students already have videocameras in their phones, and they are ready to start solar astronomy experiments.

## 5.3 Pinhole model of the Solar Disk Sextant

A flat mirror projects sunlight onto a framed pinhole. The pinhole produces the solar disk's image of diameter *Di* on a screen parallel to the frame posed at focal distance *f* better in darkened a room. For an ideal point-like pinhole the angular solar diameter is *tan(D)=Di/f* and can be measured:

7) like Kepler comparing the disk with pre-drawn disks of known diameter;
8) measuring *Di* on the image (possible errors: identifying the true diameter among chords; problems due to the motion of the image; limb darkening);
9) timing the passage of the disk perpendicularly to a profile posed before the screen (error sources: not perpendicular path; uncertainty on the limb's contact times).
10) Using two equal pinholes built on the same frame at distance d between centers measuring the focal *fc* where the disks are in contact; *tan(D)=d/fc* (main error source: *fc* uncertainty).

All methods have to deal with systematic errors due to diffraction and to the finite opening of the pinhole.
They are experiments with easy to find and low cost material good to make indoor demonstrations offering to students different levels of complexity in setup strategies and data analysis.

---

[202] Sigismondi, C., *Misura del ritardo accumulato dalla rotazione terrestre, DUT1, alla meridiana clementina della Basilica di Santa Maria degli Angeli in Roma*,in *"Mensura Caeli" Territorio, città, architetture, strumenti*, a cura di M. Incerti, p. 240-248, UnifePress, Ferrara, (2010). available also on: arXiv:1109.3558v1
[203] Sigismondi, C., *Pinhole Solar Monitor tests in the Basilica of Santa Maria degli Angeli in Rome, Solar Activity and its Magnetic Origin*, Proceedings of the 233rd Symposium of the International Astronomical Union held in Cairo, Egypt, March 31 - April 4, 2006, Edited by Volker Bothmer; Ahmed Abdel Hady. Cambridge: Cambridge University Press, pp.521-522 (2006).
[204] Sigismondi, C., *Daytime Seeing and Solar Limb Positions,* IAGA-II Proceedings, Cairo University Press Luc Damé and Ahmed Hady (eds.) p. 207-212 (2010), also on arXiv:1106.2539
Sigismondi, C., *Misure quantitative del seeing atmosferico*, Proceedings of the 42nd UAI congress, Padova - Italy, 24-27 September 2009, to appear in Astronomia UAI, also on arXiv:1106.2520
[205] Sigismondi, C. and C. Contento, *Pinhole Solar Monitor to Detect 0.01"RADIUS Variations*, in *Solar and Solar-Like Oscillations: Insights and Challenges for the Sun and Stars*, 25th meeting of the IAU, Joint Discussion 12, 18 July 2003, Sydney, Australia (2003).

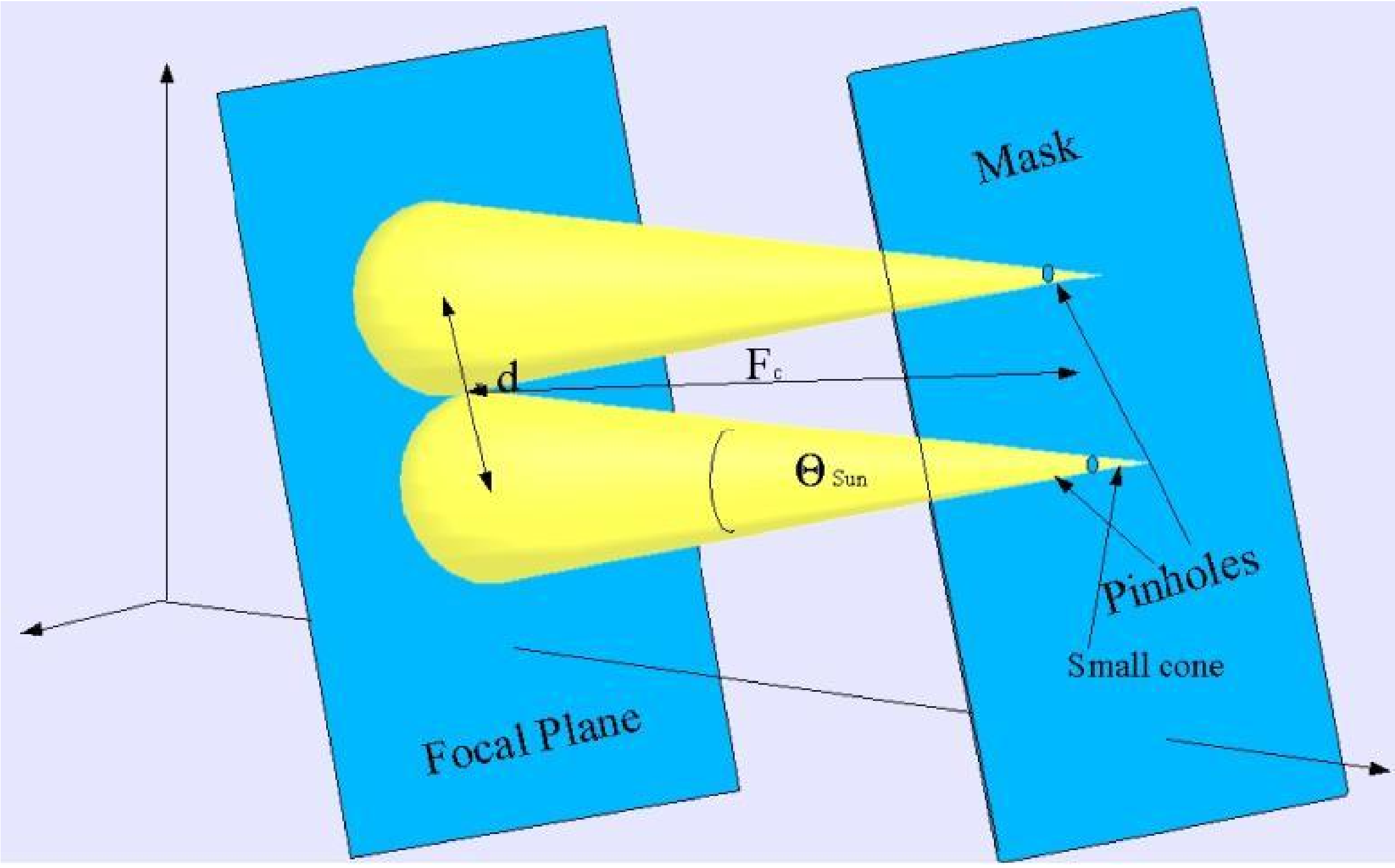

**Fig. 5.1 The two-pinholes heliometer, a simple device to measure the annual angular variations, ±1.67% i.e. ±32 arcsec, of the solar diameter, due to the eccentricity of Earth's orbit.**

A two-pinholes heliometer[206] can be designed to simulate the situation of having two images at the focal plane. Two scientific museums (Coimbra Museum of Science in Portugal and Robert Hooke Institute in Nice, France) have contacted me for implementing this instrument after the publication on the American Journal of Physics, annexed to this thesis.

At the turn of the sixteenth century Tycho Brahe and Johannes Kepler made single pinhole measurements of the solar diameter. Their accuracy was limited by the diffraction (unknown to them) and by the motion of the image on the screen. We discuss how two pinholes built on the same mask can be used to bypass all the problems inherent in the single pinhole approach. The distance at which the two images of the Sun are in contact is the only measurement needed, and the experimental accuracy is much better than measuring the diameter of a single moving image. We obtained 0.5% accuracy, sufficient to follow the angular variations of the solar diameter due to the motion of the Earth in its orbit.

## 5.4 Sunsets over the sea horizon and solar diameter

The measurement of the solar diameter is possible also from timing the sunset over the sea horizon.[207] The arcsecond level of accuracy is reached with video.[208]

[206] Sigismondi, C., Measuring the angular solar diameter using two pinholes, American Journal of Physics, **70**,1157-1159 (2002).
[207] Sigismondi, C., *Misura del diametro solare ad almucantarat zero*, Proceedings of the 42nd UAI congress, Padova - Italy, 24-27 September 2009; Astronomia UAI **6** (2012), also on arXiv:1106.2514 (2011).
[208] Sigismondi, C., *Sunsets and solar diameter measurement*, Proc. of 2nd Galileo-Xu Guangqi Meeting, Ventimiglia - Villa Hanbury, Italy, 11-16 July 2010, also on arXiv:1106.2201 (2011).

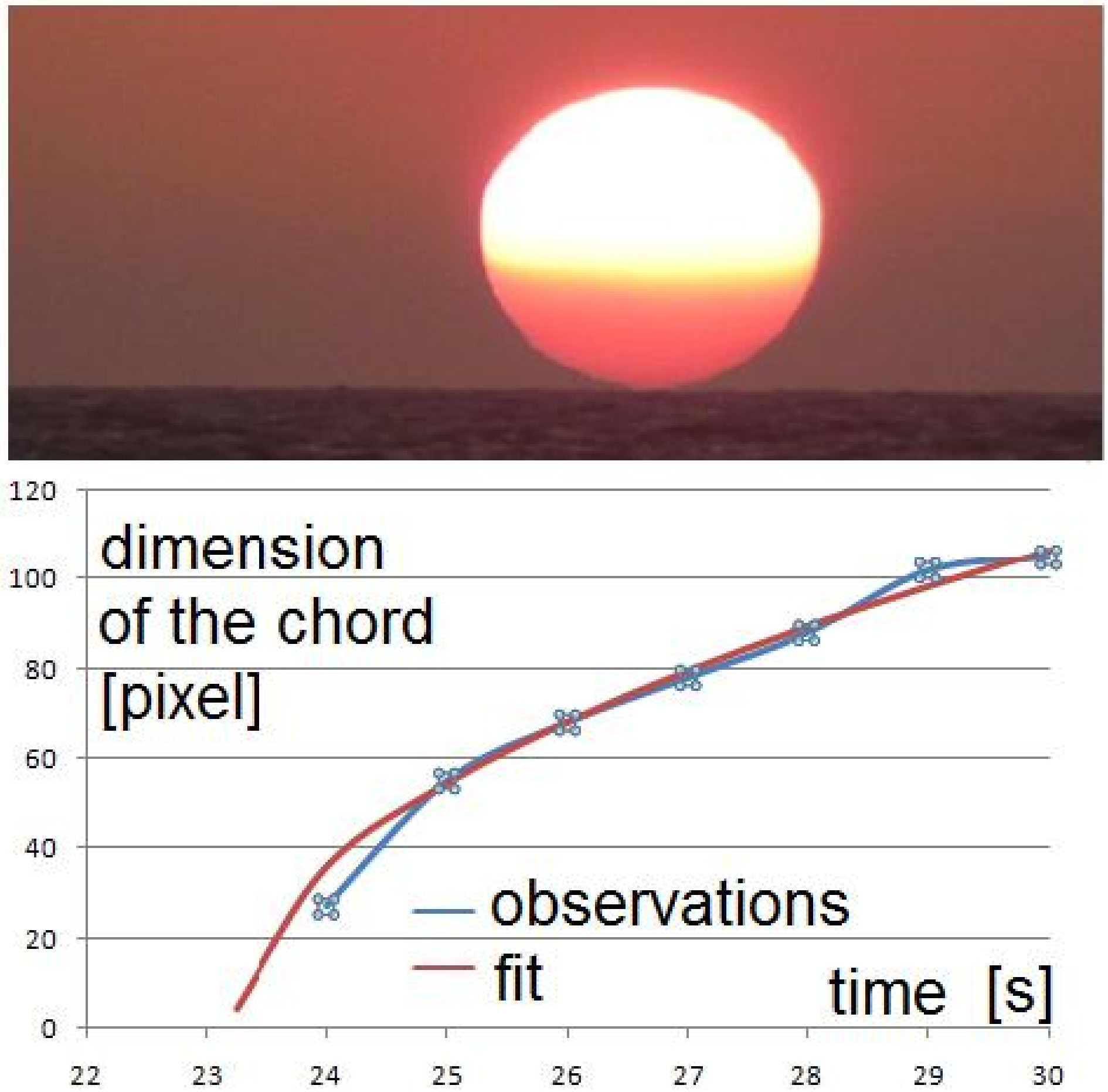

**Fig. 5.2 The method of chords applied to the intersection between the solar disk and the sea horizon. An accuracy better than 1 arcsec in the diameter's measurement has been achieved.**

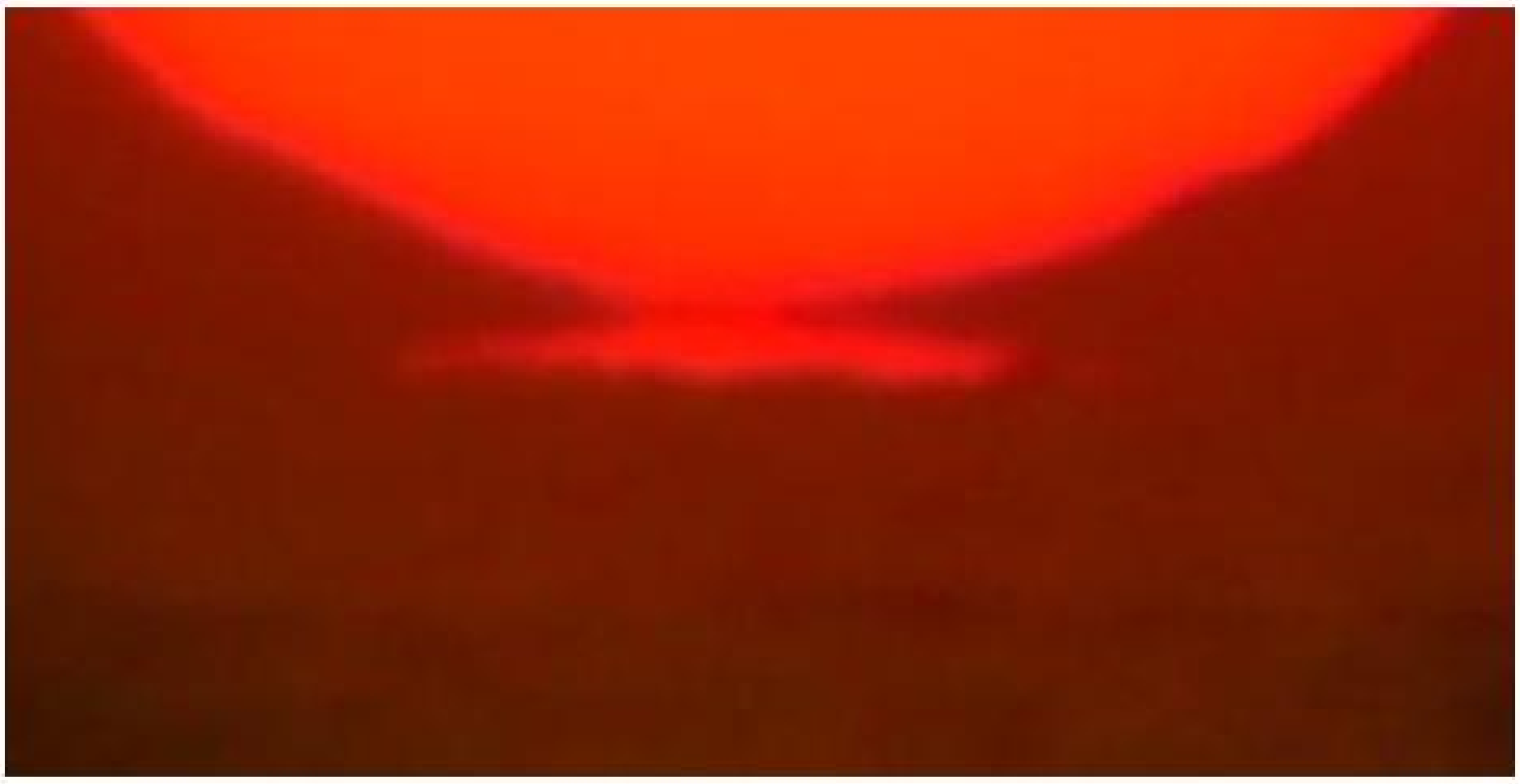

**Fig. 5.3 The contact between the solar disk with its reflected image occurs with connecting lights. This is a "black drop" phenomenon, which could affect the measurement of the solar diameter if it is not bypassed with the chords' fit.**

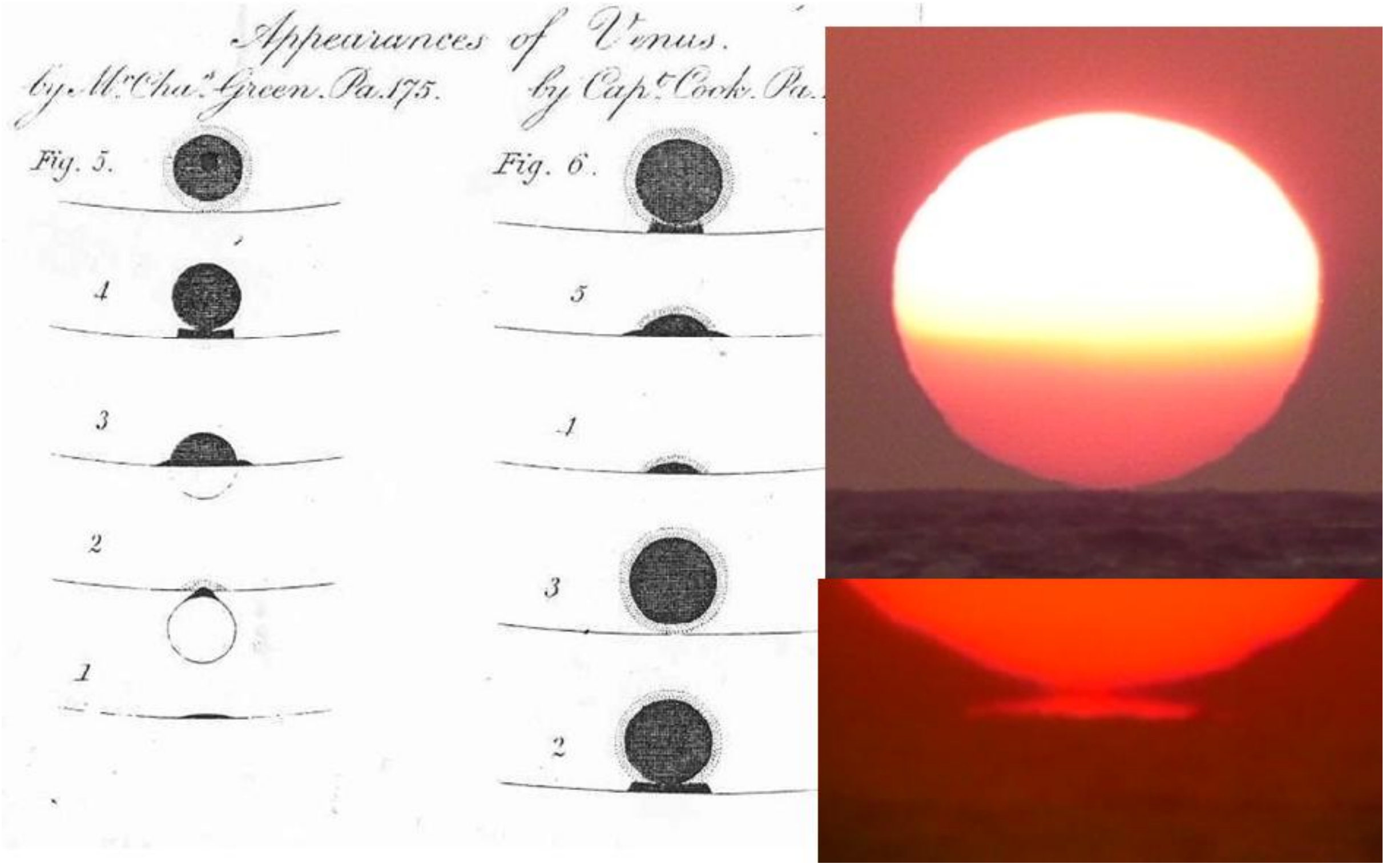

**Fig. 5.4 Black drop phenomena: the transit of Venus of 1761 and a sunset.**

## 5.5 LDF studies from a simple projected image of the Sun

Using an ordinary pair of binoculars or a normal telescope, and even a pinhole, it is possible to study the Limb Darkening Function of the Sun by a single digital photo of the image of the Sun projected on a white paper. It is to remember that every commercial videocamera is like a non-linear detector, because it is programmed in order to simulate the eye response.

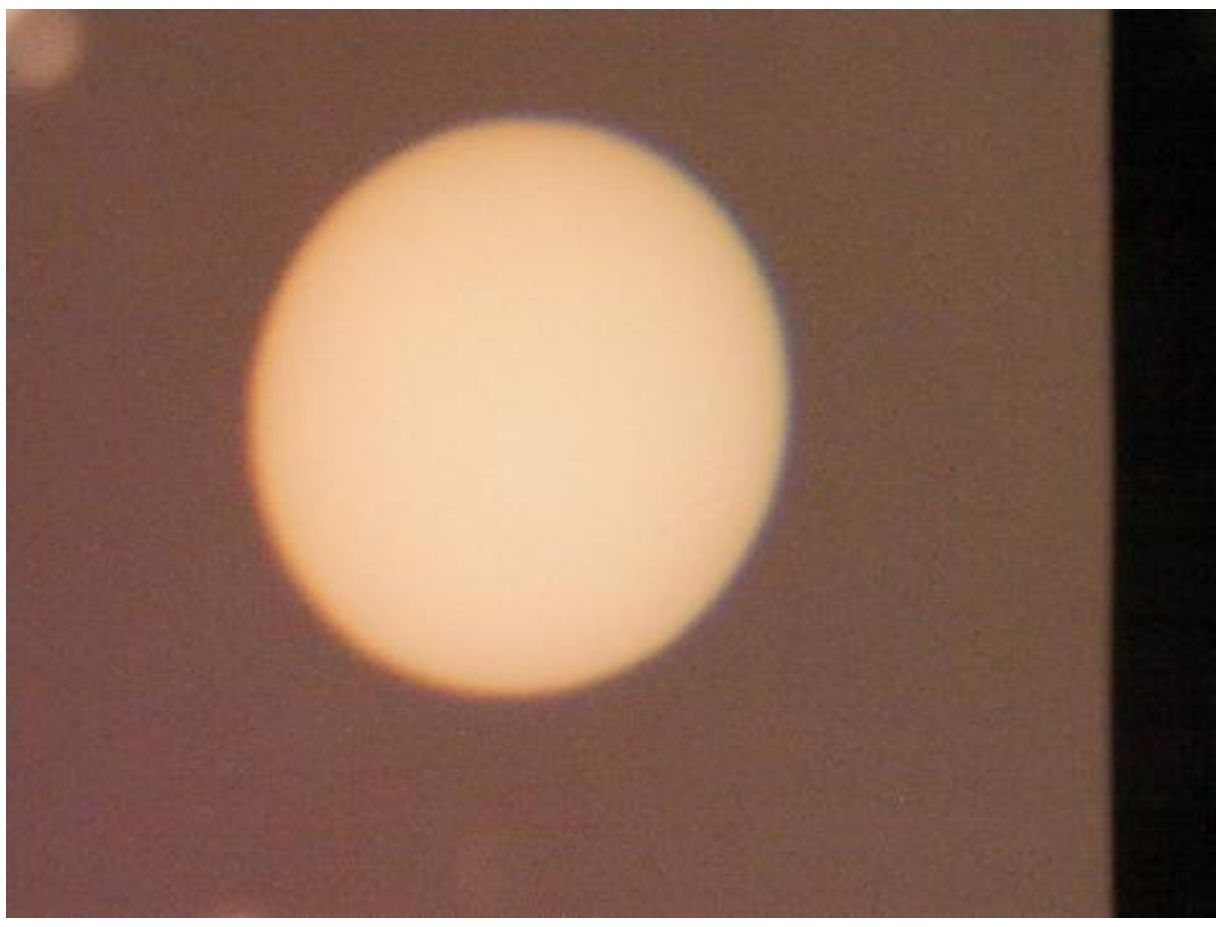

**Fig. 5.5 An image of the Sun projected by a binocular 7x50 onto a white screen (photo made at Yale University on November 3rd, 2010, during a demonstration).**

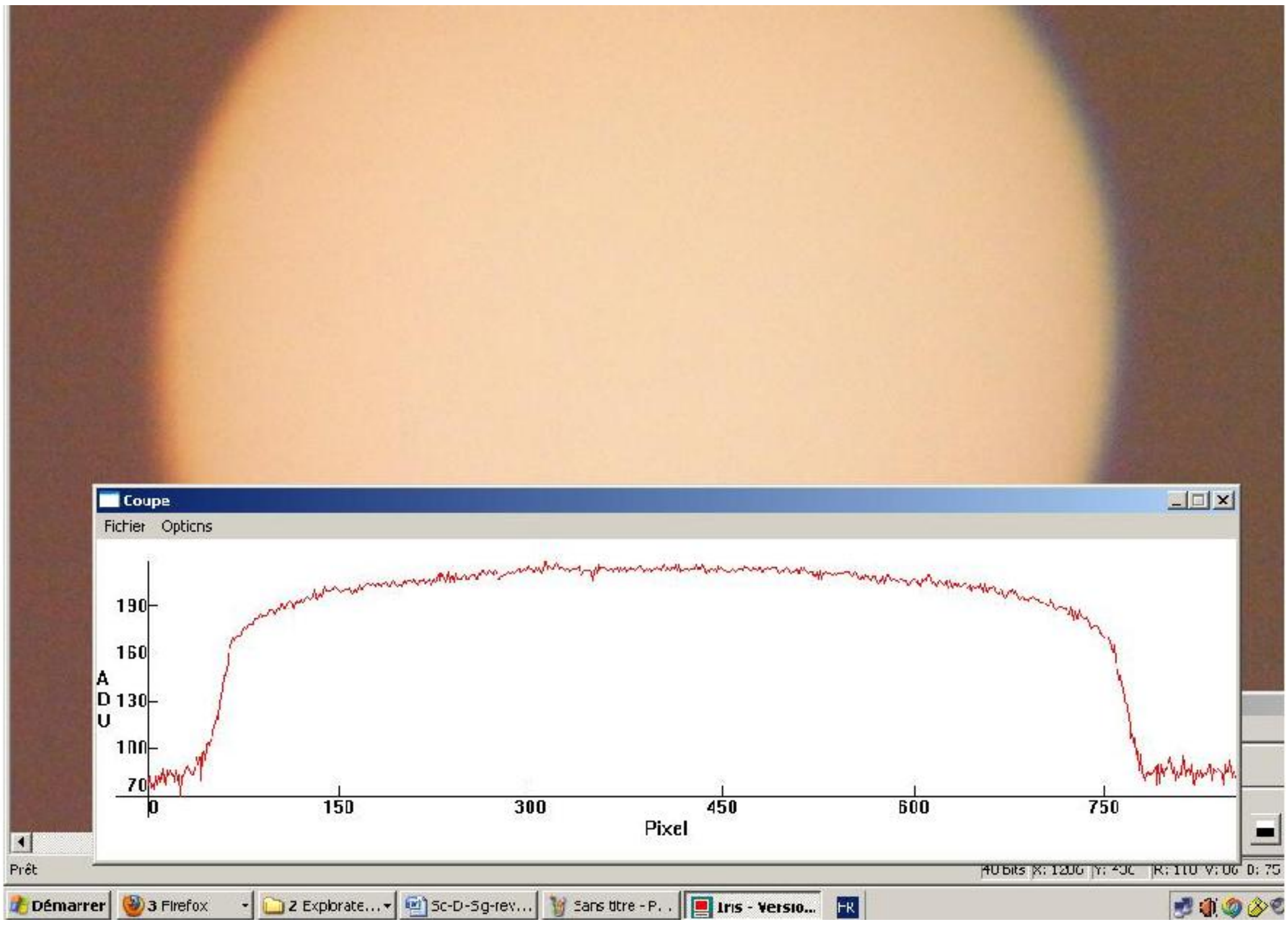

**Fig. 5.6 the Limb Darkening Function obtained on the same digital photo of fig. 5.4 in jpg, with IRIS software[209] analysis.**

[209] Buil, Christian, http://www.astrosurf.com/buil/us/iris/iris.htm (2010). Iris is a copyrighted freeware.

## Chapter 6: Conclusions and Perspectives

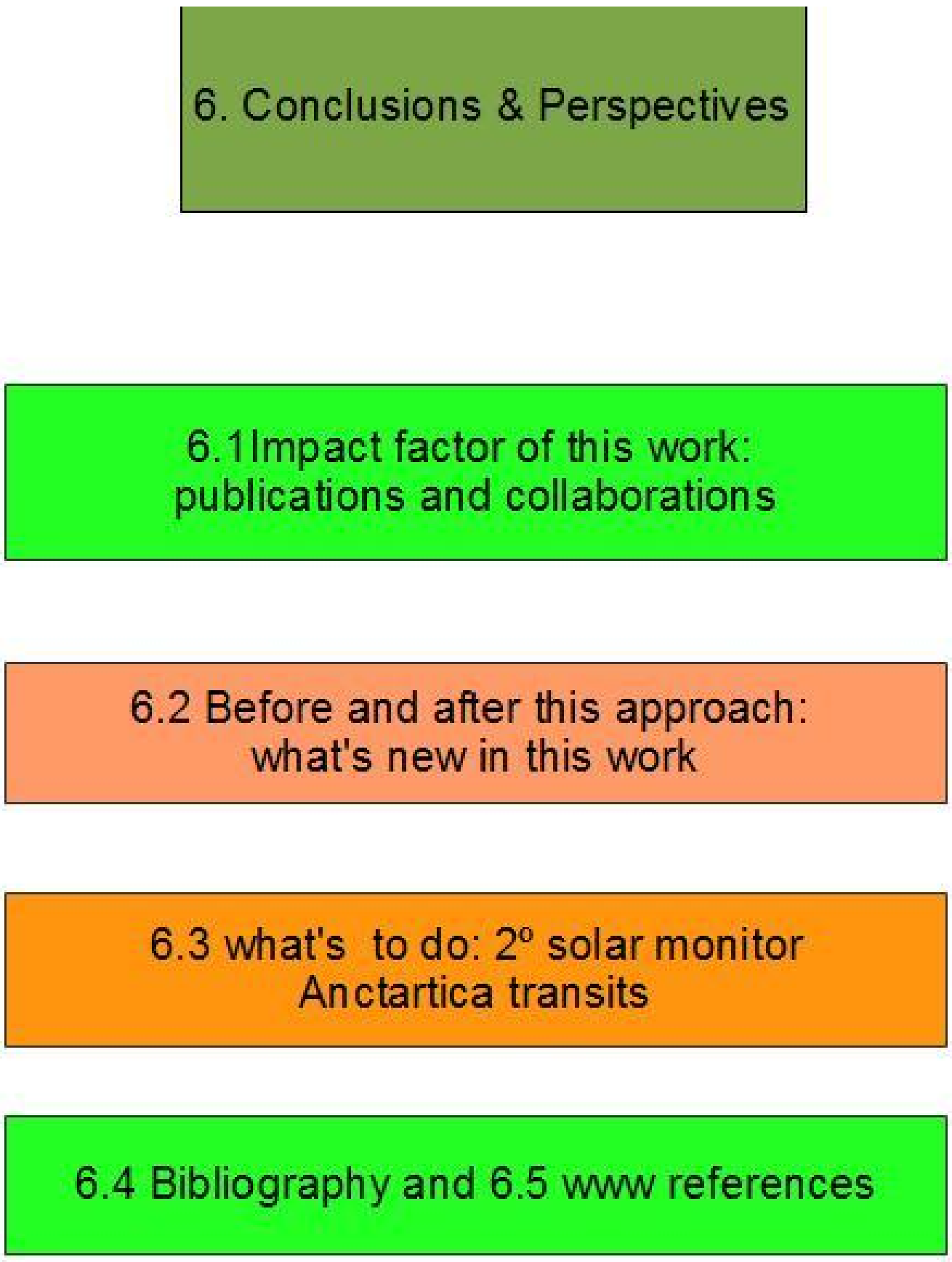

This book is published when the first data from Picard satellite are still unavailable.

My contribution to the measurement of solar diameter with high accuracy and from ground has been

1. to have shown the need of a more precise definition of solar limb during eclipses, taking into account the phenomena of tiny emission lines in the solar mesosphere.

2. to have individuated the low frequency motions of the atmosphere as one of the causes of the inconsistency between seeing errorbars of the daily transit's measurements of diameter and the corresponding errorbar of their averages.

This is a starting point for many relevant scientific issues on the physics of the Sun. The task to bridge together eclipses observations and direct solar astrometry methods has been started, and the procedures of automatization of transits measurements are is still in progress in Locarno where we realized preliminary tests at the same time of the eclipses of 2008, 2009, 2010 and 2011.

3. Qualified observers belonging to IOTA, International Occultation Timing Association, have contributed to

the publication of the data of Baily's beads of recent eclipses (2005-2008 and 2010). The problem of filters adopted in eclipse observations, firstly risen by me, found a partial solution with a standardization of the IOTA filter. The first atlas of Baily's beads has been published.

4. Historical eclipses have been re-analized with the latest lunar satellite data, the Japanese Kaguya. The measurements of solar diameter using different telescopes observing Baily's beads give different results: a new definition of solar limb is crucial, to better take into account the effects in the solar mesosphere.

## 6.1 Impact factor of this work

The field of the measurement of solar diameter using eclipses had an outburst of activity in the early years 1980s with a series of paper dealing with ancient observations also of Mercury's transits.
In 1994 the paper of Dunham, Sofia and Fiala posed a standard in this domain of research.
Only Dunham, through the direction of IOTA, International Occultation Timing Association, continued the series of measurements, no more funded by the US government, and presented in a COSPAR meeting of 2005 a new approach to the data. The polar beads were considered more stable with respect to the equatorial ones, because these region were better covered by the grazing stellar occultations, and because libration effects are reduced.
No new publications on this subject appeared on major journals until our two papers appeared in Solar Physics in 2009.
From that time on, we continued to rise the interest on this subject, which is now considered as a solid alternative to the satellite, and a valid reference for the solar diameter.
Our last achievement on the Limb Darkening Function recovered from Baily's beads light curve analysis is already matching in perfect timing, the organization of CNES (Centre National des Études Spatials, France) mission to Cairns, Australia, to observe the total eclipse of November 14, 2012 with a network of photometers.
In term of quotations in other works our number is still low, but this is due mainly to the fact that the field of eclipses and solar diameter has a restricted number of experts, and that to have new data it is necessary to wait for new unclouded total eclipses, each two years in average.[210]

The discovery of the slow fluctuations of the atmosphere which make inconsistent[211] two following measurements of the solar diameter made with hourly circle transits in drift-scan mode is another great achievement in this field.
This fact suggests to put in parallel with the main telescope another wider field telescope, in order to evaluate the motion of the whole figure of the Sun during the transit.
In this case the impact is directed on several ancient publications, where are presented large statistical errors on the determination of solar diameter. This effect explains, at least at the first order, these large fluctuations.
The authors of such works, published from 60 to 90 years ago, cannot react, but I am sure they would appreciate that discovery, that has been done in the observatory of Locarno, which participated to the project on solar diameter measurements since more than 35 years.

Finally on the field of didactic outreach, the high school's teachers are not prolific in scientific publications, due to their load of daily work with students.

---

[210] I have observed 4 eclipses, 2 of them I was clouded out.

[211] i.e. their values are more separated than the seeing would allow.

I have found some quotations of my works in various educational websites,[212] and also on refereed journals.[213]
The experiences I have suggested are not suitable for a further original publication, unless one says "yes it works".
The number of collaborations activated during these years is the real index of the impact factor of this work. Stable contacts in France at IAP, Institute d'Astrophysique de Paris, in Switzerland at Locarno IRSOL facility, in Brasil at the Observatorio Nacional in Rio de Janeiro are started independently one from another from congresses and directly from the publications. In these institutions I have spent several periods of research, with many publications still ongoing.

## 6.2 Before and after this approach: what's new in this work

The original achievements of this work are here drafted. They are the new results in this field, and they bring new understandings and points of view.

All the methods of measuring the solar diameter have been reviewed, compared and classified according to a logical order. The following table, presented at the Friedmann Seminar held in Rio de Janeiro on May 30-June 3, 2011, is summarizing also using different colors the concepts developed in this book.

| **Ground-based** | | | **Above the atmosphere** | |
|---|---|---|---|---|
| **Drift-scan** | **Heliometer** | **Astrolabe R3S2 d~9cm** | **Balloon borne** | **Satellite** |
| CLAVIUS project: IRSOL d= 45 cm (CH) | Koenigsberg Fraunhofer, Bessel 1824 d=15cm | **Rio de Janeiro (1999 - now )** | SDS (1992-2009) Diameter and Oblateness d = 20 cm | SOHO MDI (1999-2010) Diameter and oblateness |
| CLAVIUS project: Carte du Ciel d=33 cm (IAP Paris) | Goettingen (1895) | **São Paulo (1975-1990)** | **Eclipses & Planetary Transits** | HINODE diameter RHESSI (2008) oblateness |
| Greenwich d=15 cm (1850-1955) | **Rio de Janeiro (2009-now)** | Calern (FR) (1975 - 2008) | Baily's beads (1973 – 2012) | SDO Diameter, magnetic field |
| Rome–Capitol d=11cm (1877-1937) | Antalya (TK) Santiago (Chile) 1975-1995 | San Fernando (ES) | Mercury transits (1832 -2006) Venus transits (2004-2012) | Picard (2010-2013) Diameter, oblateness, irradiance |

Table 6.1 Different techniques of solar diameter measurements. Heliometers, balloon borne and satellite made instantaneous measurements, but are subject to optics errors. The others suffer from atmospheric variations during the measure.

[212] As an example I present http://faraday.physics.uiowa.edu/optics/6J10.80.htm
[213] Susman, K. and M. Čepič, Physics Education, **45**, 469 (2010).

Some problems in the four centennial activity of solar diameter measurements have been recognized and the solution has been approached.
The following scheme helps us to understand our main achievements presented in this book, with respect to some of the open problems still present in solar astrometry.

| ***Phenomenon*** | ***Interpretation*** |
| --- | --- |
| Long historical series on **meridian transits** (70-100 years Rome/Capitol and Greenwich) show big annual fluctuations ΔR/R>0.1% but COSPAR data show ΔL/L< 0.1% then ΔR/R<0.05% (according to Stefan-Boltzmann law ΔL/L≈2ΔR/R when ΔT/T=0). | Eye ageing process (**systematic errors in data acquisition**). Different cloudy days seasons from year to year (**data analysis problem**). Irradiance **PSF/diffraction effects** on small telescopes (Rome/Capitol d=11 cm, Greenwich d=15cm). |
| Different values in consecutive **drift-scan measurements** at IRSOL (d=45cm) and IAP Carte du Ciel (d=33cm). | **Atmospheric fluctuations at 0.01 Hz** (firstly measured by us in 2010 in CLAVIUS project), 10000 times slower than usual seeing frequencies. |
| During eclipses a **residual white light** is visible **after** the **photosphe**ric Baily's bead disappears. Beads of the same eclipse produce **different results** when observed **with different telescopes**. | Signal-to-noise ratio and the thin layer of **emitting lines** above the photosphere: the **solar mesosphere**. No **inflection points** detected **in limb darkening** from **Baily's beads light curves** (data analysis problem). |
| Different solar diameters are obtained with **Mercury transits** (Gambart d=7cm & Bessel d=15cm, 1832). | **Irradiance PSF and diffraction** enlarge the perceived diameter of the planet with the smaller telescope. |

Table 6.2 Phenomena and their interpretation. New solutions are in the second and third row: the detection of 0.01 Hz atmospheric variation, and the use of Limb Darkening Function also during eclipse observations.
The following scheme (already in fig. 3.19) has been proposed at the European Week of Astronomy and Space Science EWASS 12 held in Rome at Lateran University from 1 to 6 July 2012.[214]

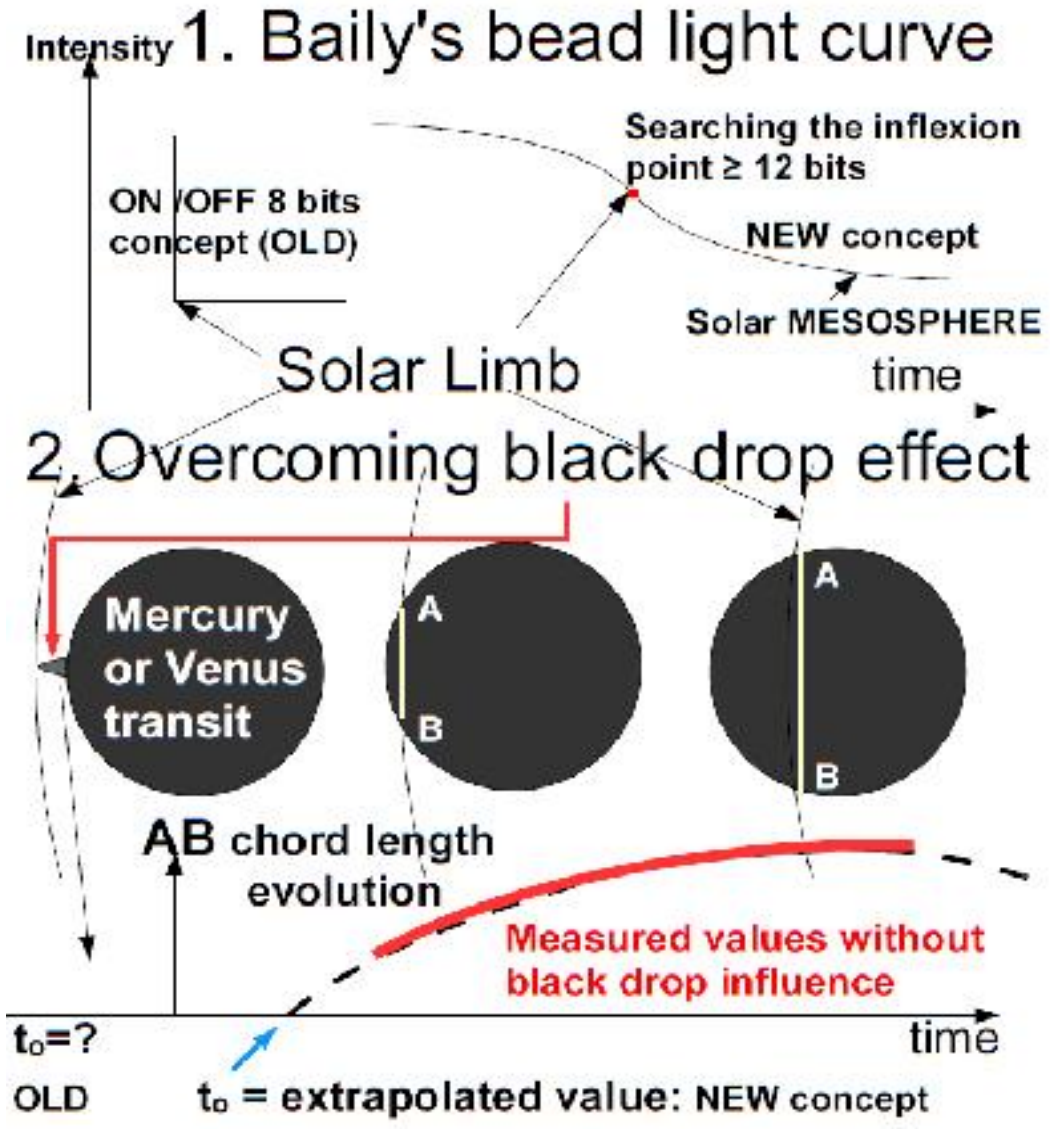

It explains the conceptual and practical improvements obtained in the approaches of solar eclipses (1) and planetary transits (2) for the measurement of the solar diameter.
The observers who travel for recording Baily's beads during total eclipses can upgrade their equipments in order to have a photometric resolution of 12 bits, and the possibility to record two or three color channels, in addition to the white light.
Also the time resolution should be increased to 60-100 images per second, or more, as such devices are already available in the market.
The observation in multiple color channels is not possible in case of using strong neutral density filters as Mylar filters, but with smaller densities, or in projection mode,[215] or even in direct light, this procedure can be possible.
The amount of information arising from a few beads observed with these new standard, will make worth the observational mission, capable of some millisecond of accuracy[216] in the solar diameter's determination. The scan of the Limb Darkening Function near the solar limb is made by the progress of the lunar limb at the angular speed determined by the eclipse: 0.3-0.5 arcsec/s, this is the main reason to exploit all this information increasing the photometric resolution.

[214] Sigismondi, C., *Solar diameter, eclipses and transits: the importance of ground-based observations*, submitted to MEMORIE DELLA SOCIETÀ ASTRONOMICA ITALIANA SUPPLEMENTI http://arxiv.org/abs/1211.4394 (2012).

This book ends with a didactic application of many of these principles developed in the most advanced solar laboratories around the World. The request of high accuracy in the quest of solar variability can be transmitted to the young generations through simple but thoughtful experiments here presented.

## 6.3 What's to do: 2° solar monitor, Antarctica, space and the Moon

The evidence of slow image motions in the atmospheric seeing, involving the whole figure of the Sun, suggests realizing a 2° solar monitor in order to correct the transits' data for this random-like motion.
This second full-disk monitor, during the drift-scan, could verify the motions of the whole disk.
The other solution, without that, should be to make a great statistics and analyze as many transits as possible in order to get an average value.
But this solution is not easy and not converging, since the data are not distributed as a Gaussian.
From Antarctica, at Dome C, there are very good and steady seeing conditions as good as 0.36 arcsec,[217] and it should be the ideal place of an experience of solar diameter's monitoring with small scale telescopes (10 cm of objective) and also with bigger ones.
A 10 cm class instrument would be easily transportable, even on top of a tower to avoid the ground level turbulence.
Moreover it can be robotized, as it happened for the IRAIT funded by the Italian Progetto Antartide.
IRAIT (International Robotic Antarctic Infrared Telescope) is a telescope with an 80 cm aperture, installed at Dome C. Equipped with AMICA (Antarctic Multiband Infrared CAmera), the main focal plane instrument, it observes in near (1.5 μm) and mid infrared regions (5.28 μm), benefiting from the exceptional site characteristics.[218] The small IRAIT has been tested by Runa Briguglio, who developed his project in astrophysics laboratory with prof. Alessandro Cacciani and myself at the Sapienza University of Rome in 2003,[219] during the Antarctic night in 2008.[220]

[215] The late Alan Fiala (+2010) of the US Naval Observatory told me that in projection nothing is lost of the information of the Baily beads, nevertheless it is necessary to realize this projection in a dark camera. This camera can be simply realized even on field observations since the background luminosity around the totality is significantly lower than the full daylight. I personally observed the total eclipse of 2006 in Egypt and the annular eclipse of 2006 in French Guyana with this method, and I strongly encourage to follow this method since it is possible to realize easily videos with modern camcorders capable of 60 to 300 frames per second (e.g. SANYO CG9 or HD1010 models). The same camcorders have to be implemented to acquire data directly from the telescope as well.
[216] Raponi, A., C. Sigismondi, K. Guhl, R. Nugent, A. Tegtmeier, *The Measurement of Solar Diameter and Limb Darkening Function with the Eclipse Observations*, Solar Physics **278**, 269 (2012).
[217] Gredel, R., *Site Characterization at Dome C – The ARENA Work*, 3rd ARENA Conference: *An Astronomical Observatory at CONCORDIA (Dome C, Antarctica)* L. Spinoglio and N. Epchtein (eds) EAS Publications Series **40**, 11-20 (2010).

The Pinhole Solar Monitor (PSM)[221] is a small space-device with two pinholes projecting two identical images on the focal plane. It has the advantage to be an astigmatic heliometer, without optics, compact, without need of accurate pointing and costless. On the payload of a microsatellite PSM exploits the side facing the Sun and monitors the secular variations of the angular solar diameter D.
Two pinholes of equal radius r are built on the same rigid platform at distance d between the centers; they project two images on the focal plane a flat screen parallel to the platform at distance f. When those images are in contact tan(D)=d/f.

The contact between the two solar limbs is simulated with a raytracer and studied numerically.

The focal length of contact, 1 meter, can be measured with a system of encoder-actuators within 10μm of precision. Although d and r are measurable within 1μm of accuracy they introduce errors in the calculation of D which are systematic due to thermal stability in the space.

With such a compact device an accuracy of 0.01 arcseconds is expected evaluating the variations of the mean angular solar radius D/2~959.63 arcseconds.
The same kind of apparatus could be used one the Moon, where the drift-scan method would profit of an angular velocity 1/27 smaller than on the Earth.[222]

---

218 di Varano, I., et al., EAS Publications Series, **33**, 279 (2008).
219 http://www.icra.it/solar/lab2002/fotoastro.html
220 http://astro.fisica.unipg.it/dome_c_news.htm site with photos of the small IRAIT
221 Sigismondi, C. and C. Contento, *Pinhole Solar Monitor to Detect 0.01"RADIUS Variations*, in *Solar and Solar-Like Oscillations: Insights and Challenges for the Sun and Stars*, 25th meeting of the IAU, Joint Discussion 12, 18 July 2003, Sydney, Australia (2003).
222 Sigismondi, C., *Pinhole Solar Monitor tests in the Basilica of Santa Maria degli Angeli in Rome*, in *Solar Activity and its Magnetic Origin*, Proceedings of the 233rd Symposium of the International Astronomical Union held in Cairo, Egypt, March 31 - April 4, 2006, Edited by Volker Bothmer; Ahmed Abdel Hady. Cambridge: Cambridge University Press, pp.521-522 (2006).

## 6.4 Bibliography

A list of other useful books and papers, to integrate the notes in text.

## 6.5 www references

A list of other useful web links.

ACRIM - Active Cavity Radiometer Irradiance Monitor - http://www.acrim.com/

Big Bear Solar Observatory – BBSO, New Jersey Institute of Technology.

IAGA web site http://www.iugg.org/IAGA/

Marshall Sapace Flight Center – NASA –
http://solarscience.msfc.nasa.gov/SunspotCycle.shtml

National Geophysical Data Center – NGDC, http://www.ngdc.noaa.gov/

Picard website http://smsc.cnes.fr/PICARD/Fr/

RHESSI website http://hesperia.gsfc.nasa.gov/hessi/index.html

Predictions of Solar Spot Number - http://www.solarcycle24.com/

Royal Greenwich Observatory – USAF/NOAA
http://solarscience.msfc.nasa.gov/greenwch.shtml

SIDC website http://sidc.oma.be/

SDO website http://sdo.gsfc.nasa.gov/

SOHO website http://sohowww.nascom.nasa.gov/home.html